\documentclass[aps,prd,preprint,tightenlines,preprintnumbers,showpacs,superscriptaddress,notitlepage,nofootinbib,eqsecnum,floatfix]{revtex4-1}

\usepackage{bm}
\usepackage{graphicx}
\usepackage[pdftex]{color}
\usepackage{amsmath, amssymb}
\usepackage{hyperref}
\hypersetup{
    colorlinks=true,     
    linkcolor=blue,      
    citecolor=blue,      
    filecolor=blue,      
    urlcolor=blue        
}
\usepackage{rotating}
\usepackage{mathtools}
\usepackage{multirow}

\newcommand{\FerMILC}{Fermilab/MILC}

\newcommand{\bi}{\begin{itemize}}
\newcommand{\ei}{\end{itemize}}

\newcommand{\ben}{\begin{enumerate}}
\newcommand{\een}{\end{enumerate}} 

\newcommand{\be}{\begin{equation}}
\newcommand{\ee}{\end{equation}}

\newcommand{\bea}{\begin{eqnarray}}
\newcommand{\eea}{\end{eqnarray}}

\newcommand{\spp}{\vphantom{\Big(}} 

\newcommand{\kp}{\ensuremath{\kappa}}

\newcommand{\BR}{\ensuremath{\mathcal{B}}}            
\newcommand{\oBR}{\ensuremath{\overline{\mathcal{B}}}}            
\newcommand{\Det}{\ensuremath{\mathop{\text{Det}}}}   
\newcommand{\Dslash}{\ensuremath{D\kern-0.6em/\kern0.15em}}
\newcommand{\tr}{\ensuremath{\mathop{\text{tr}}}}     
\newcommand{\Tr}{\ensuremath{\mathop{\text{Tr}}}}     
\newcommand{\order}{\text{O}}                         
\newcommand{\op}{\mathcal{O}}
\newcommand{\latop}{O}
\newcommand{\me}[1]{\ensuremath{\langle \op_{#1}^q  \rangle}} 
\newcommand{\methree}[1]{\ensuremath{\langle \op_{#1}^q  \rangle / M_{B_q}  }}  

\newcommand{\MSbar}{\ensuremath{\overline{\rm MS}}}

\newcommand{\cpt}{$\chi$PT}

\newcommand{\hmrscpt}{HMrS$\chi$PT}

\newcommand{\gev}{\ensuremath{\text{GeV}}}   

\newcommand{\LamQCD}{\ensuremath{\Lambda_\text{QCD}}}  
\newcommand{\LamHQ}{\ensuremath{\Lambda_\text{HQ}}}    

\newcommand{\ZVbb}{\ensuremath{Z_{V^4_{bb}}}}
\newcommand{\ZVll}{\ensuremath{Z_{V^4_{ll}}}}

\newcommand{\alV}{\ensuremath{\alpha_V}}        
\newcommand{\crit}{\ensuremath{\kappa_\text{crit}}}

\begin{document}


\title{\texorpdfstring{\boldmath$B^0_{(s)}$}{B}-mixing matrix elements from lattice QCD for the 
Standard Model and beyond}



\author{A.~Bazavov}
\affiliation{Department of Physics and Astronomy, University of Iowa, \\ Iowa City, Iowa, 52242, USA}

\author{C.~Bernard} 
\affiliation{Department of Physics, Washington University, St.~Louis, Missouri, 63130, USA}

\author{C.~M.~Bouchard}\email{cmbouchard@wm.edu}
\affiliation{Physics Department, The College of William and Mary, Williamsburg, Virginia, 23185, USA}

\author{C.~C.~Chang}
\altaffiliation{Present address: Lawrence Berkeley National Laboratory, Berkeley, California, 94720, USA}
\affiliation{Department of Physics, University of Illinois, Urbana, Illinois, 61801, USA}

\author{C.~DeTar} 
\affiliation{Department of Physics and Astronomy, University of Utah, \\ Salt Lake City, Utah, 84112, USA}

\author{Daping~Du}
\affiliation{Department of Physics, Syracuse University, Syracuse, New York, 13244, USA}

\author{A.~X.~El-Khadra}\email{axk@illinois.edu}
\affiliation{Department of Physics, University of Illinois, Urbana, Illinois, 61801, USA}

\author{E.~D.~Freeland}\email{eliz@fnal.gov}
\affiliation{Liberal Arts Department, School of the Art Institute of Chicago, \\ Chicago, Illinois, 60603, USA}

\author{E.~G\'amiz}
\affiliation{CAFPE and Departamento de F\'{\i}sica Te\'orica y del Cosmos, Universidad de Granada,
18071, Granada, Spain}

\author{Steven~Gottlieb}
\affiliation{Department of Physics, Indiana University, Bloomington, Indiana, 47405, USA}

\author{U.~M.~Heller}
\affiliation{American Physical Society, Ridge, New York, 11961, USA}

\author{A.~S.~Kronfeld}
\affiliation{Fermi National Accelerator Laboratory, Batavia, Illinois, 60510, USA}
\affiliation{Institute for Advanced Study, Technische Universit\"at M\"unchen, 85748 Garching, Germany}

\author{J.~Laiho}
\affiliation{Department of Physics, Syracuse University, Syracuse, New York, 13244, USA}

\author{P.~B.~Mackenzie}
\affiliation{Fermi National Accelerator Laboratory, Batavia, Illinois, 60510, USA}

\author{E.~T.~Neil}
\affiliation{Department of Physics, University of Colorado, Boulder, Colorado 80309, USA}
\affiliation{RIKEN-BNL Research Center, Brookhaven National Laboratory, \\ Upton, New York 11973, USA}

\author{J.~Simone}
\affiliation{Fermi National Accelerator Laboratory, Batavia, Illinois, 60510, USA}

\author{R.~Sugar}
\affiliation{Department of Physics, University of California, Santa Barbara, California, 93016, USA}

\author{D.~Toussaint}
\affiliation{Department of Physics, University of Arizona, Tucson, Arizona, 85721, USA}

\author{R.~S.~\surname{Van de Water}}
\affiliation{Fermi National Accelerator Laboratory, Batavia, Illinois, 60510, USA}

\author{Ran Zhou}
\affiliation{Fermi National Accelerator Laboratory, Batavia, Illinois, 60510, USA}

\collaboration{Fermilab Lattice and MILC Collaborations}
\noaffiliation

\date{\today} 

\preprint{FERMILAB-PUB-16-030-T }

\begin{abstract}
We calculate---for the first time in three-flavor lattice QCD---the hadronic matrix elements of all five
local operators that contribute to neutral $B^0$- and $B_s$-meson mixing in and beyond the Standard Model.
We present a complete error budget for each matrix element and also provide the full set of correlations
among the matrix elements.
We also present the corresponding bag parameters and their correlations, as well as specific combinations of
the mixing matrix elements that enter the expression for the neutral $B$-meson width difference.
We obtain the most precise determination to date of the SU(3)-breaking ratio $\xi = 1.206(18)(6)$, where the
second error stems from the omission of charm sea quarks, while the first encompasses all other
uncertainties.
The threefold reduction in total uncertainty, relative to the 2013 Flavor Lattice Averaging Group results, tightens the constraint from $B$ mixing on the
Cabibbo-Kobayashi-Maskawa (CKM) unitarity triangle.
Our calculation employs gauge-field ensembles generated by the MILC Collaboration with four lattice spacings
and pion masses close to the physical value.
We use the asqtad-improved staggered action for the light valence quarks, and the Fermilab method for the
bottom quark.
We use heavy-light meson chiral perturbation theory modified to include lattice-spacing effects to
extrapolate the five matrix elements to the physical point.
We combine our results with experimental measurements of the neutral $B$-meson oscillation frequencies to
determine the CKM matrix elements
$|V_{td}| = 8.00(34)(8) \times 10^{-3}$,
$|V_{ts}| = 39.0(1.2)(0.4) \times 10^{-3}$, and
$|V_{td}/V_{ts}| = 0.2052(31)(10)$,
which differ from CKM-unitarity expectations by about 2$\sigma$. 
These results and others from flavor-changing-neutral currents point towards an emerging tension between weak processes that are mediated at the loop and tree levels.

\end{abstract}

\pacs{
    12.15.Mm,  
    12.15.Hh,   
    14.40.Nd,   
    12.38.Gc   
    12.15.Ff}    
\maketitle

\section{Introduction}
    \label{sec:intro}  
    
The search for new physics lies at the heart of high-energy physics research.
Following the discovery of the Higgs boson at the Large Hadron Collider (LHC) in 2012, experiments at the
LHC continue to search for new heavy particles which may be directly produced in high-energy collisions.
Evidence for new physics, however, may also be found in indirect searches at low
energies~\cite{Hewett:2012ns}, where it would appear as a discrepancy between Standard-Model expectations
and experimental measurements.
Indirect searches can probe, and in some cases are already probing, new-physics scales that are orders of
magnitude higher than those accessible through direct searches~\cite{Isidori:2010kg,Buras:2013ooa,%
Buras:2014zga}.
Because we do not know \emph{a priori} the properties (masses, couplings, quantum numbers) of the postulated
new particles, indirect searches are being pursued across all areas of particle physics to provide as broad
a search window as possible~\cite{deGouvea:2013onf,Dawson:2013bba,Butler:2013kdw,Albrecht:2013wet}.
The central challenge with this approach is the high precision required from both theory and experiment to
definitively interpret any deviations seen as evidence for new physics.

Neutral $B_{q}$-meson ($q=s,d$) mixing is a particularly interesting process for indirect new-physics
searches in the quark-flavor sector, since it is both loop and GIM suppressed in the Standard Model.
The physical observables are the mass differences $\Delta M_q$ and decay-width differences
$\Delta\Gamma_q$ between the heavy and light neutral $B_q$-meson mass eigenstates, and the
flavor-specific CP asymmetries $a^q_{\rm fs}$.
The $B_q$-meson mass differences have been measured at the sub-percent level~\cite{Amhis:2014hma}.
The measured width differences and CP asymmetries have much larger uncertainties~\cite{Amhis:2014hma}, but are expected to
improve in the next several years~\cite{Bediaga:2012py,Aushev:2010bq}.

Theoretical predictions of $B_q$-mixing observables in both the Standard Model and beyond depend upon the
hadronic matrix elements of local four-fermion operators in the effective weak Hamiltonian:
\begin{equation}
    \me{i} (\mu) = \langle \bar B^0_q |\op_i^q | B^0_q\rangle (\mu) , 
    \label{eq:MEdef}
\end{equation}
where $\mu$ is the renormalization scale.
These operators arise after integrating out physics at energy scales above~$\mu$.
Their matrix elements can be calculated in lattice QCD with standard methods.
In the Standard Model, only one matrix element, $\langle\op_1\rangle$, contributes to the mass
difference~$\Delta M_q$. 
Beyond the Standard Model (BSM), however, $\Delta M_q$
can receive contributions from five distinct operators. These same  five operators also 
contribute to the Standard-Model width difference $\Delta \Gamma_q$.

In this paper, we calculate, for the first time in three-flavor lattice QCD, the matrix elements for all
five local operators in the $B_d$ and $B_s$ systems.
Only a few lattice-QCD results for $B_q$-mixing matrix elements exist
to-date~\cite{Dalgic:2006gp,Gamiz:2009ku,Albertus:2010nm,Bazavov:2012zs,Carrasco:2013zta,Aoki:2014nga}, with
uncertainties of about 5--15\% that are much larger than the corresponding experimental errors.
Most efforts have focused only on the Standard-Model $B_q$-mixing matrix
elements~\cite{Dalgic:2006gp,Gamiz:2009ku,Albertus:2010nm,Aoki:2014nga,Bazavov:2012zs}.
In Refs.~\cite{Albertus:2010nm,Aoki:2014nga}, the RBC and UKQCD collaborations treat the $b$ quark in the
static limit~\cite{Eichten:1989kb,Hasenfratz:2001hp,DellaMorte:2005yc}, which results in
$\order(\Lambda/m_b)$ errors.
This effect is included in the error budget of Refs.~\cite{Albertus:2010nm,Aoki:2014nga} via a
power-counting estimate and contributes significantly to the total error.
The HPQCD collaboration~\cite{Dalgic:2006gp,Gamiz:2009ku,Dowdall:2014qka} uses nonrelativistic QCD (NRQCD)
for the $b$-quark action~\cite{Lepage:1992tx,Gamiz:2008sk}.
Their earlier calculations include three dynamical sea quarks~\cite{Dalgic:2006gp,Gamiz:2009ku}.
Recently, however, they presented preliminary results~\cite{Dowdall:2014qka} (with perturbatively improved
NRQCD~\cite{Hammant:2013sca}) from the first calculation of the Standard-Model $B_q$-mixing matrix elements
with four flavors of sea quarks (up, down, strange, and charm), where the average $u,d$-quark mass is at its
physical value~\cite{Bazavov:2010ru,Bazavov:2012xda}.
In Ref.~\cite{Bazavov:2012zs}, the Fermilab Lattice and MILC collaborations (\FerMILC) presented a
calculation of the ratio of $B_s$-to-$B_d$ mixing matrix elements [$\xi$ defined in Eq.~(\ref{eq:xidef})]
using relativistic $b$ quarks with the Fermilab interpretation on a small subset of the three-flavor
ensembles generated by the MILC collaboration~\cite{Bernard:2001av,Aubin:2004wf,Bazavov:2009bb}.
The ETM collaboration~\cite{Carrasco:2013zta} published the first results for the complete set of
$B_q$-mixing matrix elements, based, however, on gauge-field configurations with only two flavors of sea
quarks.
Preliminary results from this project have been reported earlier~\cite{Bouchard:2011xj}; those results are
superseded by this work.

Given theoretical calculations of the hadronic matrix elements, neutral $B_q$-meson mixing can be used both
to determine Standard-Model parameters and to search for new physics.
In the Standard Model, the mass differences are proportional to the product of CKM matrix elements
$|V_{tq}^*V_{tb}|^{2}$.
Experimental measurements of $\Delta M_q$ can therefore be used to determine these CKM combinations,
assuming no new-physics contributions.
The ratio of CKM matrix elements $|V_{td}/V_{ts}|$ can be obtained especially precisely from $\Delta
M_d/\Delta M_s$, because several correlated uncertainties in the $B_d$- and $B_s$-mixing hadronic matrix
elements largely cancel.
For Standard-Model tests, $B_q$-mixing provides prominent constraints on the apex of the CKM unitarity
triangle.
For new-physics searches, it constrains BSM parameter spaces and in some cases enables discrimination
between models.
(For recent reviews, see, for example, Refs.~\cite{Hansmann-Menzemer:2015ija,Buras:2015nta,Buras:2014zga,%
Buras:2013ooa,Nierste:2013kca,Lenz:2014jya,Bobeth:2014bya,Borissov:2013yha} and for specific examples, see Refs.~%
\cite{Blanke:2015wba,Buras:2014yna,Buras:2013dea,Buras:2013raa,Lenz:2010gu,Lenz:2012az,Brod:2014bfa,%
Bobeth:2014rda,Shimizu:2013jia,Kim:2013ivd,Nelson:2013ula,Fortes:2013dba,Dekens:2014ina,Queiroz:2014pra,Arbey:2014msa,Enomoto:2015wbn}.) %
Further, several small tensions are seen between experiment and theory for $\Delta M_s/\Delta M_d$,
$\epsilon_K$, and the CP asymmetry $S_{\psi K_s}$~\cite{Lunghi:2008aa,Buras:2008nn}.
All of the above comparisons are presently limited by the theoretical uncertainties on the hadronic matrix
elements.
To sharpen them and potentially reveal new-physics effects, better lattice calculations of the hadronic
matrix elements that can leverage the impressive experimental precision are needed.

To that end, we improve upon the previous lattice $B_q$-mixing matrix element calculations in several ways.
We now compute the complete set of dimension-six $\Delta B=2$ four-fermion operators with three sea-quark
flavors.
We use the same valence- and sea-quark actions as in our earlier calculation of the ratio
$\xi$~\cite{Bazavov:2012zs} but employ a much larger subset of the MILC configurations, including ensembles
at four lattice spacings, covering a range of $a \approx 0.045$--$0.12$~fm.
We quadruple the statistics on the ensembles used in Ref.~\cite{Bazavov:2012zs}, while adding ensembles at
smaller lattice spacings and lighter sea-quark masses.
Although the average $u,d$-quark masses on the ensembles used in this work are all larger than in Nature,
they extend down to $m_l \approx 0.05 m_s$, which corresponds to a pion mass $M_\pi\approx175$~MeV that is
close to the physical value.
Finally, we have changed the chiral-continuum extrapolation to account for certain effects that arise from
the use of staggered light valence quarks~\cite{Bernard:2013dfa}, thereby eliminating a significant source
of systematic uncertainty in Ref.~\cite{Bazavov:2012zs}.
These effects were discovered after a preliminary report~\cite{Bouchard:2011xj} on the current work
appeared, so they were not included in the chiral extrapolations and error budget at that time.
We now include this source of uncertainty, as well as all others.
We present our matrix element results together with their correlations to facilitate their use in other
phenomenological studies beyond this work.
We also form several phenomenologically interesting combinations, including the SU(3)-breaking ratio $\xi$
and the corresponding bag parameters.

This paper is organized as follows.
Section~\ref{sec:background} presents the theoretical background and definitions of the hadronic matrix
elements.
Next, Sec.~\ref{sec:simulation} provides details of the numerical simulations, including the gauge-field
ensembles, the valence-quark actions, and the definitions of the two- and three-point lattice correlation
functions.
Section~\ref{sec:corr_analysis} describes the fit functions and analysis procedures used to
extract the desired matrix elements from the correlation functions.
Section~\ref{sec:PT} summarizes the perturbative matching of the lattice matrix elements to the continuum,
while Sec.~\ref{sec:kappa} describes how we correct \emph{a posteriori} for small mistunings of the
$b$-quark masses.
In Sec.~\ref{sec:ChPT}, we extrapolate the matrix elements to the physical light-quark masses and to the
continuum limit.
Section~\ref{sec:syserr} presents a detailed account of our systematic error analysis.
Our final results are discussed in Sec.~\ref{sec:results}, where we also explore the implications of our
results for Standard-Model phenomenology.
Section~\ref{sec:conclusions} provides a summary and some outlook.
The Appendices contain various details.
In Appendix~\ref{app:Results}, we tabulate our complete matrix-element and bag-parameter results for all
operators, along with the correlations between them.
In Appendix~\ref{app:pval}, we describe in detail our methods for measuring the goodness of fits performed
with Bayesian statistics.
Appendix~\ref{app:log} provides equations for translating the one-loop chiral logarithm functions between
the notation used in this work and in the original papers~\cite{Detmold:2006gh,Bernard:2013dfa}.

\section{Theoretical background}
    \label{sec:background}
\begin{figure}
\vspace{-2.2in}
\includegraphics[width=0.7\textwidth, angle=0]{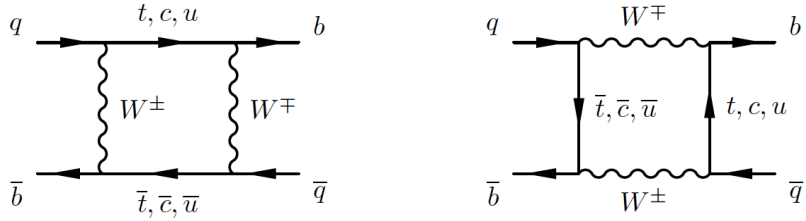}
\vspace{-2.4in}
\caption[Three point correlation function setup.]{Leading-order Feynman diagrams contributing to $B_q^0$
mixing in the Standard Model.}
\label{fig:box}
\end{figure}

    In the Standard Model, the leading-order electroweak interactions responsible for $B_q$-meson mixing occur
via the box diagrams depicted in Fig.~\ref{fig:box}.
The observables $\Delta M_q$ and $\Delta \Gamma_q$ are related to the off-diagonal elements of the time
evolution matrix, $M_{12}^q$ and $\Gamma_{12}^q$, as
\begin{equation}
    \Delta M_q \simeq 2 | M_{12}^q|, 
    \qquad
    \Delta \Gamma_q \simeq 2 |\Gamma_{12}^q| \cos{\phi_q} ,
\end{equation}
up to corrections of order $m_b^2/m_W^2 \sim 10^{-3}$, while the observable $a^q_{\rm fs}$ is given by
\begin{equation}
    a^q_\text{fs} = \frac{|\Gamma_{12}^q|}{|M_{12}^q|} \sin{\phi_q},
\end{equation}
where the CP violating phase $\phi_q = \arg \left[- M_{12}^q/\Gamma_{12}^q \right]$.
The mass and width differences provide complementary tests of the Standard Model.
The mass differences are calculated from the dispersive part of the box diagram and are therefore sensitive
to potential contributions from virtual heavy particles.
On the other hand, the decay-width differences are obtained from the absorptive part, which predominantly
receives contributions from light internal particles.
Even so, new-physics contributions can affect the width differences.

The energy scale accessible in the loop of the box diagram in Fig.~\ref{fig:box} is of order $M_{B_q}$, and
is far below the characteristic scale of the electroweak interactions, the $W$-boson mass $m_W$.
Using the operator-product expansion (OPE) to treat this disparity of scales leads to a local
effective four-quark operator description of $B_q$ mixing.
For extensions of the Standard Model that involve interactions mediated by new heavy particles at the
TeV~scale or above, the local effective four-quark operator remains a convenient description.
In this description, extending generically beyond the Standard Model, the effective Hamiltonian is
\begin{equation}
    \mathcal{H}_{\rm eff} = \sum_{i=1}^5 C_i \op_i^q + \sum_{i=1}^3 \tilde{C}_i \tilde{\op}_i^q,
    \label{eq:heff}
\end{equation}
where the Wilson coefficients $C_i$ contain information specific to the short-distance physics associated
with the flavor-changing interactions, and $\op_i^q$ are the effective local four-quark operators.
A basis of effective local four-quark operators is derived from the set of all Lorentz-invariant,
color-singlet current-current interactions among heavy-light quark bilinears, reduced via
discrete symmetries of QCD and Fierz rearrangement~\cite{Hagelin:1992tc,Gabbiani:1996hi,Bagger:1997gg}, to 
\begin{subequations} \label{eq:Oi}
\begin{align} 
    \op_1^q &=\bar{b}^\alpha\gamma_\mu L q^\alpha \, \bar{b}^\beta\gamma_\mu L q^\beta , \\
    \op_2^q &=\bar{b}^\alpha L q^\alpha \, \bar{b}^\beta L q^\beta , \\
    \op_3^q &=\bar{b}^\alpha L q^\beta \, \bar{b}^\beta L q^\alpha , \\
    \op_4^q &=\bar{b}^\alpha L q^\alpha \, \bar{b}^\beta R q^\beta ,  \\
    \op_5^q &=\bar{b}^\alpha L q^\beta \, \bar{b}^\beta R q^\alpha , \\
    \tilde{\op}_1^q&=\bar{b}^\alpha\gamma_\mu R q^\alpha \ \bar{b}^\beta\gamma_\mu R q^\beta ,  \\
    \tilde{\op}_2^q&=\bar{b}^\alpha R q^\alpha \ \bar{b}^\beta R q^\beta , \\
    \tilde{\op}_3^q&=\bar{b}^\alpha R q^\beta \ \bar{b}^\beta R q^\alpha ,
\end{align}
\end{subequations}
where $b$ and $q$ are continuum quark fields of given flavor, $R$ and $L$ are right- and
left-handed projection operators $(1\pm\gamma_5)/2$, Greek indices denote color, and Dirac indices are
implicit.
The operators $\tilde{\op}_i^q$ are the parity transforms of $\op_i^q$, $i=1,2,3$.
Because parity is a symmetry of QCD, the pseudoscalar-to-pseudoscalar matrix elements satisfy
$\langle\op_i^q\rangle=\langle\tilde{\op}_i^q\rangle$.
Below, we exploit this identity to increase statistics.

The $B_q^0$-mixing matrix elements have often been recast in terms of bag parameters $B^{(i)}_{B_q}$,
defined by~\cite{Gabbiani:1996hi}
\begin{align}
    \me{1}(\mu) &= c_1 f^2_{B_q} M^2_{B_q} B^{(1)}_{B_q}(\mu) ,
    \label{eq:Bq_1} \\
    \me{i}(\mu) &= c_i \left( \frac{M_{B_q}}{m_b(\mu)+m_q(\mu)} \right)^2
        f^2_{B_q} M^2_{B_q} B^{(i)}_{B_q}(\mu) ,  \quad i = 2,3,
    \label{eq:Bq_23} \\
    \me{i}(\mu) &= c_i \left[ \left( \frac{M_{B_q}}{m_b(\mu)+m_q(\mu)}\right)^2 + d_i \right]
        f^2_{B_q} M^2_{B_q} B^{(i)}_{B_q}(\mu) , \quad i = 4,5,
    \label{eq:Bq_45}     
\end{align}
where $c_i = \{2/3, -5/12, 1/12, 1/2, 1/6\}$, $d_4=1/6$, and $d_5=3/2$.
Before reliable calculations of the nonperturbative physics of the mixing matrix elements became available,
the bag parameters were introduced, motivated by the so-called vacuum saturation approximation
(VSA)~\cite{Lee:1972px} where $B_{B_q}^{(i)}=1$.
Other conventions for the $B$-parameters of the mixed-chirality operators are also used in the
literature~\cite{Lenz:2006hd}, for example in other recent lattice-QCD
calculations~\cite{Carrasco:2013zta,Dowdall:2014qka}.
With the definitions in Eq.~(\ref{eq:Bq_45}), however, $B^{(4)}_{B_q}$ and $B^{(5)}_{B_q}$ are indeed unity
in the VSA.

The Standard-Model $B_q$-meson oscillation frequency $\Delta M_q$ is often expressed in terms of the
renormalization-group-invariant version of the bag parameter $\hat B^{(1)}_{B_q}$, as in Eq.~(\ref{eq:DM})
below.
Following the notation of Ref.~\cite{Aoki:2013ldr}, we determine $\hat B^{(1)}_{B_q}$ from
$B^{(1)}_{B_q}(\mu)$ (evaluated in the $\MSbar$-NDR scheme), to two-loop order, by
\begin{equation}
    \hat B^{(1)}_{B_q} = \alpha_s (\mu)^{-\gamma_0/(2\beta_0)} 
        \left[1 + \frac{\alpha_s (\mu)}{4\pi} \left( \frac{\beta_1\gamma_0 - \beta_0\gamma_1}{2\beta_0^2}
        \right) \right] \ B^{(1)}_{B_q}(\mu),
    \label{eq:RGI}
\end{equation}
where the coupling $\alpha_s(\mu)$ is defined in the $\MSbar$ scheme and $\beta_0$ and $\beta_1$ are the
(scheme-independent) one- and two-loop beta-function coefficients.
The one- and two-loop anomalous dimensions of $\op_1$ are $\gamma_0 = 4$ and $\gamma_1=-7+\frac{4}{9}N_f$,
respectively; $\gamma_0$ is scheme-independent, while
$\gamma_1$ is given in the $\MSbar$-NDR scheme~\cite{Buras:1990fn}.

In the Standard Model, only the matrix element $\langle \op_1^q \rangle$ contributes to the mass difference:
\begin{equation}
    \Delta M_q = \frac{G_F^2 m_W^2 M_{B_q}}{6\pi^2}\, S_0 (x_t )\, \eta_{2B} \, |V_{tq}^* V_{tb}|^2\,
        f^2_{B_q} \hat B^{(1)}_{B_q} . 
    \label{eq:DM}
\end{equation}
Here, the Inami-Lim function $S_0(x_t)$~\cite{Inami:1980fz} describes the electroweak corrections and
depends on the mass of the top quark in the loop of Fig.~\ref{fig:box} through $x_t=m_t^2/m_W^2$, while
$\eta_{2B}$ is the perturbative-QCD correction factor known at next-to-leading order \cite{Buras:1990fn}.
For the Standard-Model decay-width difference $\Delta \Gamma_q$~\cite{Beneke:1996gn,Lenz:2006hd}, as well as
in general theories beyond the Standard Model, the mixing matrix elements (or equivalently bag parameters)
of operators $\langle \op_i^q \rangle$ ($i=2$--5) are also needed.
Together, the five matrix elements $\langle \op_i^q \rangle$ ($i=1$--5) are sufficient to parameterize the
hadronic contributions to $\Delta M_q$ in all possible BSM scenarios, and therefore enable model-specific
predictions related to mixing.
More precise mixing matrix elements, of course, provide stronger new-physics constraints.

In the Standard Model, both $\Delta M_q$ and $\Delta\Gamma_q$ receive contributions from higher-dimensional
operators beyond those in Eq.~(\ref{eq:Oi}) that are not considered in this work.
Corrections to the OPE used to derive Eq.~(\ref{eq:DM}) are negligible, because they are suppressed by 
$m_b^2/m_W^2$.
For $\Delta \Gamma_q$, however, a second OPE, the so-called heavy-quark expansion \cite{Lenz:2006hd}, is
needed to obtain a Standard-Model prediction in terms of local operators, yielding a joint power series in
$\Lambda/m_b$ and $\alpha_s$.
At leading order in the heavy-quark expansion, the Standard-Model expression for $\Delta \Gamma_q$ depends
only on $\langle \op_1^q\rangle$ and either $\langle \op_2^q\rangle$ or $\langle \op_3^q\rangle$.
At $\order{(1/m_b)}$, however, $\Delta \Gamma_q$ also receives contributions from the matrix elements
$\langle \op_{4,5}^q\rangle$.
Further, at this order, matrix elements of dimension-seven operators not calculated in this work enter
$\Delta \Gamma_q$; their contributions are numerically larger than those from the local matrix elements $\langle
\op_{4,5}^q \rangle$, and their uncertainties, after the reduction of errors on $\langle \op_{1,2,3}^q\rangle$ in this work, 
are the dominant source of error in the Standard-Model width differences~\cite{Artuso:2015swg}.

Certain combinations of the hadronic matrix elements $\langle\op_i^q\rangle$ are especially useful for
phenomenology.
As discussed in the previous section, the theoretical uncertainties on $\langle \op_1^q \rangle$ are
currently much larger than the experimental errors on $\Delta M_q$, and therefore limit the precision with
which one can obtain the CKM combinations $|V_{tq}^* V_{tb}|$.
Many of the theoretical errors cancel, however, in the ratio $\xi$, defined as
\begin{equation}
    \xi^2 = \frac{ f^2_{B_s}\hat B^{(1)}_{B_s} }{ f^2_{B_d}\hat B^{(1)}_{B_d}},
    \label{eq:xidef}
\end{equation}
thereby enabling a determination of the CKM-element ratio $|V_{td}/V_{ts}|$ from the
corresponding ratio of mass differences:
\begin{equation}
    \left| \frac{V_{td}}{V_{ts}} \right|^2 =
        \xi^2\, \frac{\Delta M_d}{\Delta M_s}\, \frac{M_{B_s}}{M_{B_d}} 
\end{equation}
that better leverages the experimental precision.

For Standard-Model calculations of the decay-width differences, 
it is useful to define the
$1/m_b$-suppressed combination $\langle R_0^q\rangle$~\cite{Beneke:1996gn}
\begin{equation} \label{eq:R0def}
    \langle R_0^q\rangle (\mu) = 2 \alpha_2(\mu) \me{1}(\mu) + 4\me{2}(\mu) + 4 \alpha_1(\mu) \me{3}(\mu) . 
\end{equation}
The coefficient functions $\alpha_{1,2} (\mu)$ are known at next-to-leading order (NLO) in
QCD~\cite{Beneke:1998sy} and are given in Eqs.~(\ref{eq:alpha1})--(\ref{eq:alpha2}) of Sec.~\ref{sec:results}.
Because the leading contributions in the heavy-quark expansion cancel by construction in the combination in
Eq.~(\ref{eq:R0def}), the calculation of $\langle R_0^q \rangle$ suffers from a larger uncertainty than for
the individual matrix elements $\me{1,2,3}$.

Finally, the decay-width differences are often parameterized in terms of the ratios of matrix elements
$\langle \op_i^q \rangle/\langle \op_1^q \rangle$ ($i$=2--5) because the theoretical uncertainties are
reduced.
These same ratios can also contribute to the mass differences in theories beyond the Standard Model.
Hence they are useful for Standard-Model and BSM calculations of the ratio $\Delta \Gamma_q/\Delta M_q$, as
well as for predictions of $B_q$-mixing observables in new-physics scenarios relative to their
Standard-Model values.

\section{Lattice simulation}
    \label{sec:simulation}
    Here we summarize the details of the numerical simulations.
First, in Sec.~\ref{sec:SimParams}, we describe the ensembles of gauge-field configurations and the light-
and valence-quark actions employed in the analysis.
Next, in Sec.~\ref{sec:corr_building}, we define the lattice two-point and three-point correlation functions
used to obtain the desired $B_q$-meson mixing matrix elements.

\subsection{Gauge Configurations and Valence Actions}
\label{sec:SimParams}

Our calculation employs gauge-field configurations generated by the MILC Collaboration with three dynamical
sea quarks~\cite{Bernard:2001av,Aubin:2004wf,Bazavov:2009bb}.
These ensembles use the Symanzik-improved gauge action
\cite{Weisz:1982zw,Weisz:1983bn,Luscher:1984xn,Luscher:1985zq} for the gluons and the asqtad-improved
staggered action \cite{Blum:1996uf,Orginos:1998ue,Lagae:1998pe,Lepage:1998vj,Orginos:1999cr,Bernard:1999xx}
for the quarks.
Generic discretization errors from the light-quark and gluon actions are of $\order(\alpha_sa^2\LamQCD^2)$.
In the numerical simulations, the fourth-root (square-root) of the strange-quark (light-quark) determinant
is taken to reduce the number of staggered-fermion species from four to one (two)~\cite{Hamber:1983kx}.
Although this procedure violates unitarity and locality at nonzero lattice
spacing~\cite{Prelovsek:2005rf,Bernard:2006zw,Bernard:2006ee,Bernard:2007qf,Aubin:2008wk}, a large body of
numerical and theoretical evidence indicates that the desired continuum-QCD theory is obtained when
the lattice spacing is taken to zero~\cite{Adams:2004mf,Shamir:2004zc,Durr:2005ax,Shamir:2006nj,%
Sharpe:2006re,Bernard:2006zw,Bernard:2007eh,Kronfeld:2007ek,Golterman:2008gt,Donald:2011if}.

\begin{figure}
	\includegraphics[width=0.3475\textwidth, angle=270]{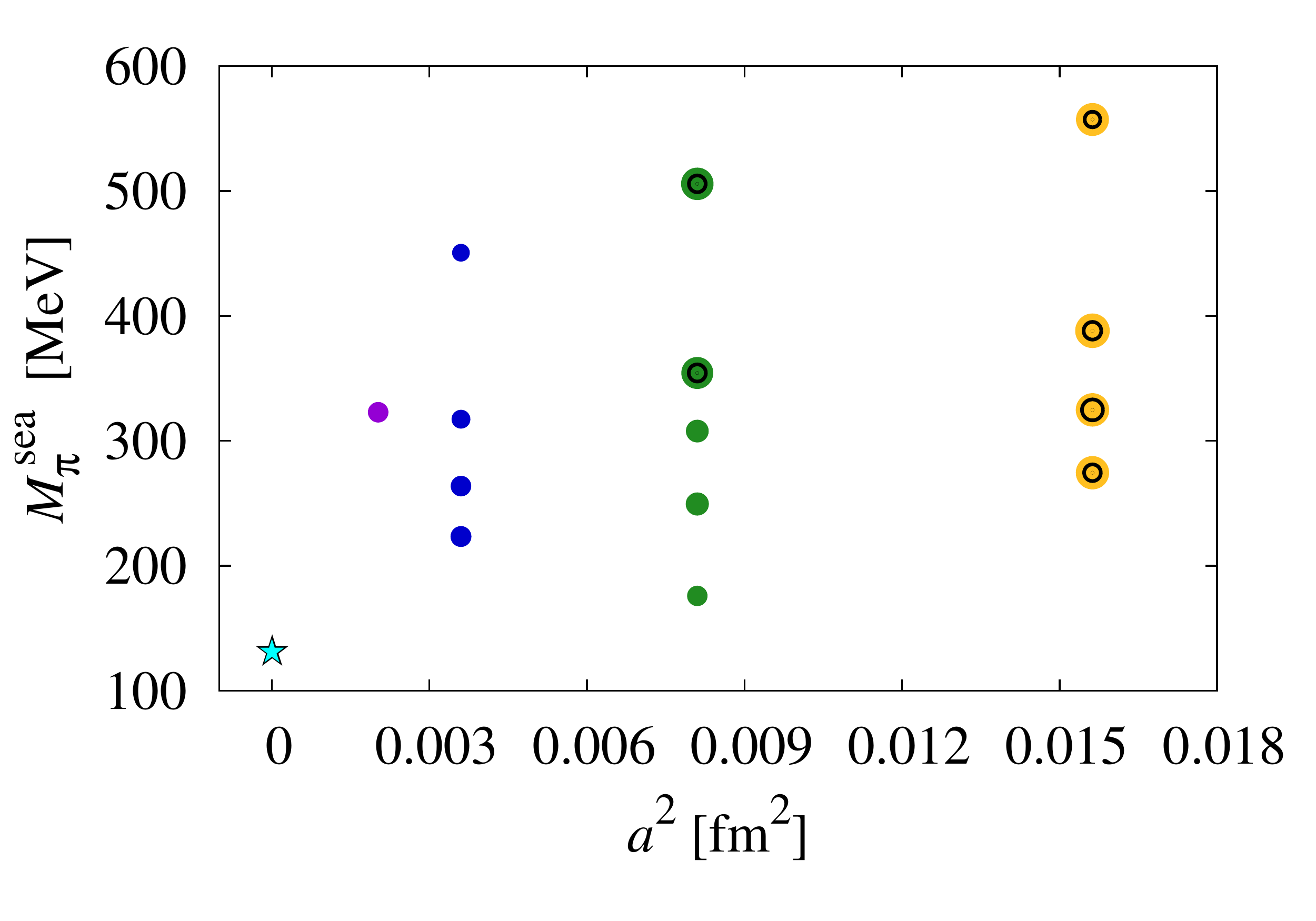}
	\hfill
	\includegraphics[width=0.3475\textwidth, angle=270]{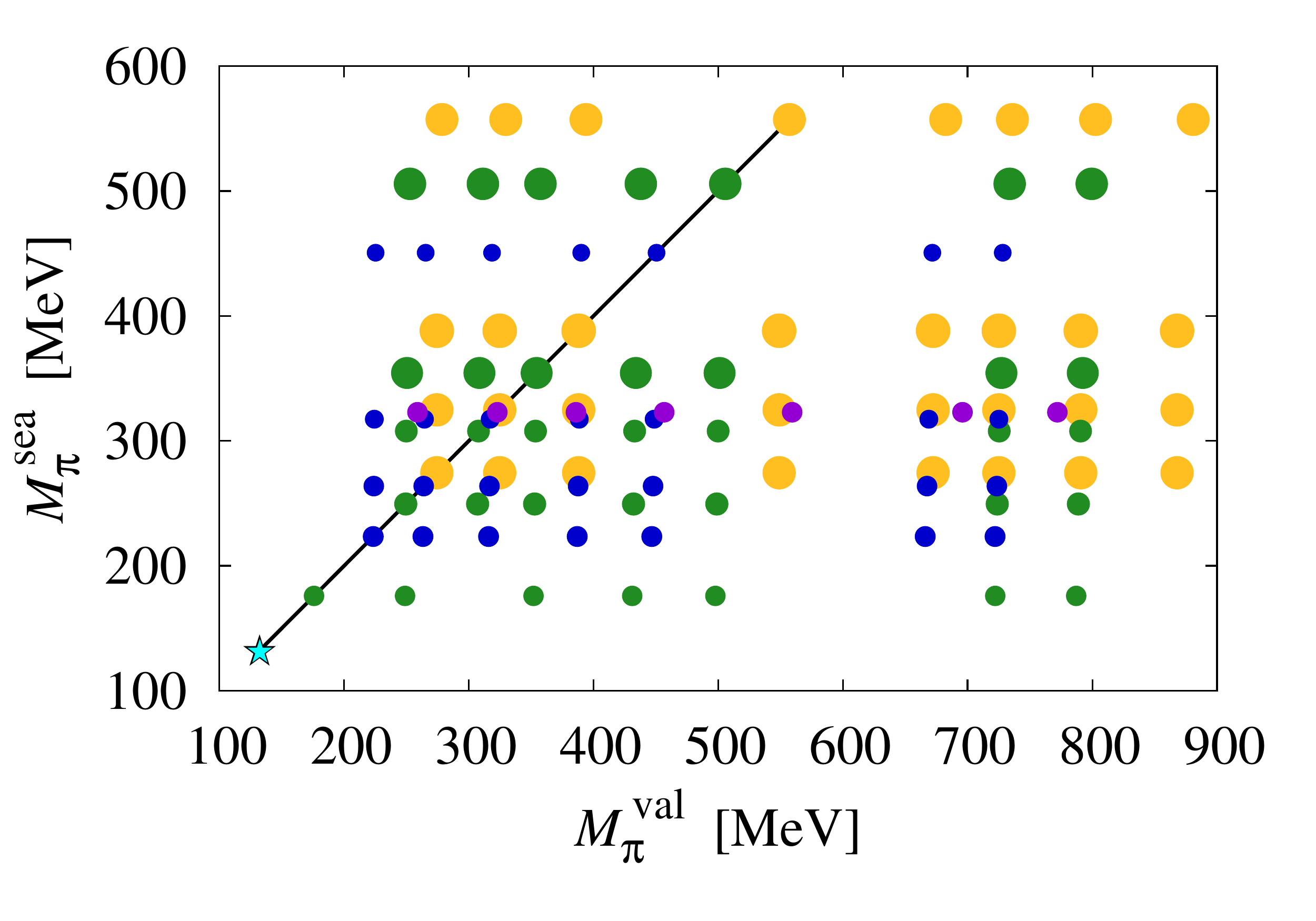}
    \caption{
    \vspace*{-0.25em}
    (\emph{left}) Distribution of lattice spacings and light sea-quark masses used in this 
    analysis (solid colored disks) and in our previous work~\cite{Bazavov:2012zs} (open black disks).
    (\emph{right}) Distribution of light sea- and valence-quark masses.
    The diagonal black line corresponds to the unitary point, $m'_l = m_q$.
    In each plot, the disk area is proportional to the statistical sample size
    $N_\text{conf}\times N_\text{src}$, and the cyan star corresponds to the physical point.}
	\label{fig:LatEns}
\end{figure}

We analyze fourteen ensembles of gauge-field configurations with a range of pion masses and lattice spacings.
Table~\ref{tab:LatEns} provides details of the numerical simulation parameters.
On each ensemble, the two light sea-quark masses are set equal.
The smallest simulated pion mass is 177~MeV, so only a short extrapolation to the value in
Nature is required.
For all ensembles, the strange sea-quark mass is tuned close to its physical value.
We analyze four lattice spacings ranging from $a \approx 0.12$--$0.045$~fm to guide the continuum
extrapolation.
The spatial lattice volumes are sufficiently large ($M_\pi L \gtrsim 3.8$ for all ensembles, where $L$ is
the lattice spatial extent) that finite-volume errors are expected to be at the sub-percent level for
heavy-light-meson matrix elements; we nevertheless include these effects in our systematic error analysis.
All ensembles have more than 500 configurations, and several contain as many as 2000.
Figure~\ref{fig:LatEns} visually summarizes the range of pion masses, lattice spacings, and the statistical
sample sizes.

\begin{table*}[tb]
\centering
\caption{Parameters of the QCD gauge-field ensembles used in this work~\cite{%
asqtad:en09a,*asqtad:en09b,*asqtad:en06a,*asqtad:en06b,*asqtad:en05a,*asqtad:en05b,*asqtad:en04a,
*asqtad:en17a,*asqtad:en15a,*asqtad:en15b,*asqtad:en14a,*asqtad:en13a,*asqtad:en13b,*asqtad:en12a,
*asqtad:en23a,*asqtad:en23b,*asqtad:en20a,*asqtad:en20b,*asqtad:en19a,*asqtad:en18a,*asqtad:en18b,
*asqtad:en24a}. 
From left-to-right we show the approximate lattice spacing $a$ in fm, the simulated light-to-strange
sea-quark mass ratio $am'_{l}/am'_{s}$, the ratio $r_1/a$ with uncertainties from the smoothing fit, the
lattice volume $N_s^3 \times N_t$, the taste-Goldstone pion mass $M_{\pi}$ and RMS mass $M_{\pi}^\text{RMS}$
in MeV, the dimensionless factor $M_{\pi}L$, and the number of configurations $N_\text{conf}$.
The primes on $m'_{l}$ and $m'_{s}$ distinguish the simulation values from the physical ones.}
\label{tab:LatEns}
\begin{tabular}{l@{\quad}c@{\quad}c@{\quad}c@{\quad}c@{\quad}c@{\quad}c@{\quad}c}
\hline\hline  
& & & & & & & \vspace{-1.0em}\\
$\approx a$~(fm)   &  $a{m}'_{l}/am'_{s}$  & $r_1/a$ &  $N_s^3 \times N_t$  & $M_{\pi}$~(MeV) &
$M_{\pi}^\text{RMS}$~(MeV) & $M_{\pi} L$ & $N_\text{conf}$ \\ 
& & & & & & &\vspace{-1.0em}\\ \hline
0.12  & 0.02/0.05     & 2.8211(28) & $20^3 \times 64$ &  555  &  670   & 6.2 & 2052 \\ 
0.12  & 0.01/0.05     & 2.7386(33) & $20^3 \times 64$ &  389  &  538   & 4.5 & 2259 \\ 
0.12  & 0.007/0.05    & 2.7386(33) & $20^3 \times 64$ &  327  &  495   & 3.8 & 2110 \\ 
0.12  & 0.005/0.05    & 2.7386(33) & $24^3 \times 64$ &  277  &  464   & 3.8 & 2099 \\ \hline
0.09  & 0.0124/0.031  & 3.8577(32) & $28^3 \times 96$ &  494  &  549   & 5.8 & 1996 \\ 
0.09  & 0.0062/0.031  & 3.7887(34) & $28^3 \times 96$ &  354  &  415   & 4.1 & 1931 \\ 
0.09  & 0.00465/0.031 & 3.7716(34) & $32^3 \times 96$ &  306  &  375   & 4.1 &  984 \\ 
0.09  & 0.0031/0.031  & 3.7546(34) & $40^3 \times 96$ &  250  &  330   & 4.2 & 1015 \\ 
0.09  & 0.00155/0.031 & 3.7376(34) & $64^3 \times 96$ &  177  &  280   & 4.8 &  791 \\ \hline
0.06  & 0.0072/0.018  & 5.399(17)  & $48^3 \times144$ &  450  &  467   & 6.3 &  593 \\
0.06  & 0.0036/0.018  & 5.353(17)  & $48^3 \times144$ &  316  &  341   & 4.5 &  673 \\
0.06  & 0.0025/0.018  & 5.330(16)  & $56^3 \times144$ &  264  &  293   & 4.4 &  801 \\
0.06  & 0.0018/0.018  & 5.307(16)  & $64^3 \times144$ &  224  &  257   & 4.3 &  827 \\ \hline
0.045 & 0.0028/0.014  & 7.208(54)  & $64^3 \times192$ &  324  &  332   & 4.6 &  801 \\
\hline\hline
\end{tabular}
\end{table*}

We also use the asqtad-improved staggered action for the light valence quarks.
On each ensemble, we simulate with several values of the valence light-quark mass.
These partially-quenched data are useful for constraining the fit coefficients in the chiral-continuum
extrapolation in Sec.~\ref{sec:ChPT}.
For the $b$ quarks, we use the isotropic clover action~\cite{Sheikholeslami:1985ij} with the Fermilab
interpretation~\cite{ElKhadra:1996mp}.
We fix the clover coefficient to the tadpole-improved tree-level value $c_\text{SW} = 1/u_0^3$, where
$u_0^4$ is from the average plaquette.
We tune the bare $b$-quark mass, or, equivalently, the hopping parameter $\kappa_b$, such that the
$B_s$-meson mass agrees with the experimental value, following the general approach described in
Ref.~\cite{Bernard:2010fr}.
For this work, we take the more recent and precise determinations of $\kappa_b$ from
Ref.~\cite{Bailey:2014tva}.
Table~\ref{tab:ValParams} provides details of the valence light- and $b$-quark simulation parameters.

\begin{table*}[tb]
\centering
\caption{Parameters of the valence-quark propagators used in this work.
From left-to-right, starting at the third column, we show the valence light-quark masses $am_q$, the
simulation clover coefficient $c_\text{SW}$ and hopping parameter $\kappa'_b$ in the $b$-quark action, the
rotation coefficient $d'_{1b}$ in the heavy-light current, and the number of time sources per configuration
$N_\text{src}$.
The same valence light-quark masses are used on all ensembles with the same approximate lattice spacing.
The primes on $m'_{l}$, $m'_{s}$, $\kappa'_b$, and $d'_{1b}$ distinguish the simulation values from the
physical ones.}
\label{tab:ValParams}
\begin{tabular}{l@{\quad}c@{\quad}c@{\quad}c@{\quad}c@{\quad}c@{\quad}c}
\hline \hline
& & & & & & \vspace{-1.0em} \\  
$\approx a$~(fm)   &  $a{m}'_{l}/am'_{s}$  &  $am_{q}$ & $c_\text{SW}$ & $\kappa'_b$ & $d_{1b}'$ &
    $N_\text{src}$ \\
& & & & & & \vspace{-1.0em} \\   \hline
0.12 & 0.02/0.05 & & 1.525 & 0.0918 & 0.09439 & 4 \\ 
0.12 & 0.01/0.05 & \{0.005, 0.007, 0.01, 0.02, & 1.531 & 0.0901 & 0.09334 & 4 \\ 
0.12 & 0.007/0.05 & 0.03, 0.0349, 0.0415, 0.05\} & 1.530 & 0.0901  & 0.09332 & 4 \\ 
0.12 & 0.005/0.05 & & 1.530  & 0.0901 &  0.09332 & 4 \\ \hline
0.09 & 0.0124/0.031 & &  1.473  & 0.0982  & 0.09681 & 4 \\ 
0.09 & 0.0062/0.031 & \{0.0031, 0.0047, 0.0062, & 1.476  & 0.0979  & 0.09677  & 4 \\ 
0.09 & 0.00465/0.031 & 0.0093, 0.0124, 0.0261, 0.031\} & 1.477  & 0.0977  & 0.09671 &  4\\ 
0.09 & 0.0031/0.031 & & 1.478  & 0.0976  & 0.09669 &  4 \\ 
\multirow{2}{*}{0.09} & \multirow{2}{*}{0.00155/0.031} & \{0.00155, 0.0031, 0.0062,  &
\multirow{2}{*}{1.4784}  & \multirow{2}{*}{0.0976} &  \multirow{2}{*}{0.09669} &  \multirow{2}{*}{4} \\ 
        &                       &    0.0093, 0.0124, 0.0261, 0.031\}                  &              &            &                &    \\ \hline
0.06 & 0.0072/0.018 & & 1.4276  & 0.1048  & 0.09636  & 4 \\
0.06 & 0.0036/0.018 &  \{0.0018, 0.0025, 0.0036, &  1.4287  & 0.1052  & 0.09631 & 8 \\
0.06 & 0.0025/0.018 &  0.0054, 0.0072, 0.016, 0.0188\} & 1.4293  & 0.1052  & 0.09633 & 4  \\
0.06 & 0.0018/0.018 & & 1.4298  & 0.1052  & 0.09635  & 4 \\ \hline
\multirow{2}{*}{0.045} & \multirow{2}{*}{0.0028/0.014}  & \{0.0018, 0.0028, 0.004, &
\multirow{2}{*}{1.3943}  & \multirow{2}{*}{0.1143}  & \multirow{2}{*}{0.08864} & \multirow{2}{*}{4} \\
	 &			    & 0.0056, 0.0084, 0.013, 0.16\} &     &             &               & \\
\hline\hline
\end{tabular}
\end{table*}

The lattice temporal extents are sufficiently large that we can increase the statistics of our simulations
by computing valence-quark propagators starting from multiple source locations on each configuration.
In practice, we use four equally-spaced time sources on the majority of ensembles, with the
exception of the $a \approx 0.06$~fm, $m'_{l}/m'_{s} =0.2$ ensemble, where we use eight.
We then average the results from all time sources on a single configuration.
To reduce autocorrelations between measurements computed on configurations close in Monte-Carlo simulation
time, we translate each gauge-field configuration by a random spatial shift $\bm{x}$ before calculating
valence-quark propagators.
After applying this random shift, we do not observe any significant remaining autocorrelations, as confirmed
by measurements of the autocorrelation time and simple binning studies shown in Sec.~\ref{sec:corr_analysis}.

We convert lattice quantities to physical units using the scale $r_1$, which is defined by the condition
$r_{1}^{2}F(r_{1})=1.0$, where $F(r)$ is the force between two static
quarks~\cite{Bernard:2000gd,Sommer:1993ce}.
The relative scale $r_1/a$ can be obtained precisely on each ensemble from the heavy-quark
potential~\cite{Bernard:2000gd,Bazavov:2009bb}.
To reduce the sensitivity of the lattice-spacing estimates to statistical fluctuations, we use $r_1/a$
values obtained from a fit of data on multiple ensembles to a smooth function of the coupling $\beta$
following Ref.~\cite{Bazavov:2009bb}.
The explicit function employed is based on expectations from perturbation theory~\cite{Allton:1996kr}, and
is given in Eqs.~(115)--(116) of Ref.~\cite{Bazavov:2009bb}, where additional details on the smoothing
procedure can be found.
Here we use the updated mass-independent, smoothed $r_1/a$ determinations listed in Table~\ref{tab:LatEns}.
(The $r_1/a$ values are mass-independent because they have been extrapolated at fixed $\beta$ to physical quark masses from the simulated quark masses of each ensemble.)
Compared with Ref.~\cite{Bazavov:2009bb}, the more recent analysis includes larger statistical
samples for some ensembles, and omits ensembles with strange-sea quark masses much larger than the physical
value.
The fit uncertainties are below 1\%, and are correlated between ensembles.

We multiply all lattice masses and matrix elements by the appropriate power of $r_1/a$ to make them 
dimensionless
before proceeding to the chiral-continuum extrapolation and further error analysis.  
At the end we obtain results in physical units using~\cite{Bazavov:2011aa}
\begin{equation}
    r_1 = 0.3117(22)~\text{fm},
    \label{eq:r1}
\end{equation}
fixed via the PDG~\cite{Agashe:2014kda} value of $f_\pi$.
The quoted uncertainty takes into account the difference in $r_1$ values obtained by the
MILC~\cite{Bazavov:2009fk} and HPQCD~\cite{Davies:2009tsa} collaborations.
Because the $r_1/a$ fit errors in Table~\ref{tab:LatEns} are smaller than the statistical uncertainties on
the $B_q$-mixing matrix elements, we do not include them in our central analysis.
We estimate the systematic errors due to both the relative scale $r_1/a$, including correlations between
ensembles due to the smoothing procedure, and absolute scale $r_1$ in Sec.~\ref{sec:syserr}, and add them to
the chiral-continuum-fit error \emph{a posteriori}.

\subsection{Lattice Operators and Correlation Functions}
\label{sec:corr_building}

We construct the two- and three-point correlation functions needed to obtain the matrix elements for neutral
$B_d$- and $B_s$-meson mixing using the same methods as in our earlier calculation of the SU(3)-breaking
ratio $\xi$~\cite{Bazavov:2012zs}.
In particular, we calculate them in a computationally efficient
manner using the ``open-meson propagators'' described in Ref.~\cite{Bazavov:2012zs}, which are general
objects with free spin and color indices.
The three-point correlation functions needed for all five operators can be obtained from the combination of
two open-meson propagators contracted with the appropriate Dirac structures.
The needed two-point correlation functions can be obtained from a single open-meson propagator.

Starting with the light $q$-flavored staggered field $\chi_q(x)$, we construct the naive field 
$\Upsilon_q(x)$ as in Refs.~\cite{Bernard:2013dfa,Wingate:2002fh}:
\begin{equation}
    \Upsilon_q(x) = \Omega(x) {\underline{\chi}}_q(x),
\end{equation}
where $\Omega (x) = \gamma_1^{x_1} \gamma_2^{x_2} \gamma_3^{x_3} \gamma_4^{x_4}$, and the underlined field
$\underline{\chi}_q$ denotes four copies of the staggered field, with the (suppressed) ``copy'' index
contracted with the right Dirac index of $\Omega$.
To reduce heavy-quark discretization effects, we rotate the $b$-quark field $\psi_b(x)$
via~\cite{ElKhadra:1996mp}
\begin{equation}
    \Psi_b(x) = [1+a d_{1b} \bm{\gamma}\cdot \bm{D}]\psi_b(x),
    \label{eq:d1}
\end{equation}
where $\bm{D}$ is a nearest-neighbor covariant distance operator.
The coefficient $d_{1b}$ is set to its value in tree-level tadpole-improved perturbation theory;
Table~\ref{tab:ValParams} gives the numerical values $d'_{1b}$ used in our simulations.

In analogy with Eqs.~(\ref{eq:Oi}), the lattice versions of the local $\Delta B=2$ four-quark operators are
constructed from the rotated $b$-quark field $\Psi_b$ and the naive $q$-flavored fields $\Upsilon_q$ defined
above:
\begin{subequations}\label{eq:latOi}
\begin{align}
    \latop_1^q(x) &= \bar{\Psi}_b^\alpha(x)\gamma^\mu L\Upsilon_q^\alpha(x) \,
        \bar{\Psi}_b^\beta(x)\gamma_\mu L\Upsilon_q^\beta(x), \\
    \latop_2^q(x) &= \bar{\Psi}_b^\alpha(x)L\Upsilon_q^\alpha(x)\,
        \bar{\Psi}_b^\beta(x)L\Upsilon_q^\beta(x), \\
    \latop_3^q(x) &= \bar{\Psi}_b^\alpha(x)L\Upsilon_q^\beta(x)\,
        \bar{\Psi}_b^\beta(x)L\Upsilon_q^\alpha(x), \\
    \latop_4^q(x) &= \bar{\Psi}_b^\alpha(x)L\Upsilon_q^\alpha(x)\,
        \bar{\Psi}_b^\beta(x)R\Upsilon_q^\beta(x), \\
    \latop_5^q(x) &= \bar{\Psi}_b^\alpha(x)L\Upsilon_q^\beta(x)\,
        \bar{\Psi}_b^\beta(x)R\Upsilon_q^\alpha(x),
\end{align}
\end{subequations}
and similarly for $\tilde{\latop}_i^q$ ($i=1,2,3$).
Again, $\alpha$ and $\beta$ are color indices.
With this choice, the leading discretization errors from the four-fermion operator are of
order $\alpha_s a\LamQCD$ and $(a\LamQCD)^2$~\cite{Bazavov:2012zs,GamizPT,Bernard:2013dfa}.

The continuum limit of the lattice four-quark operators in Eqs.~(\ref{eq:latOi}) is complicated by the spin
and taste components of the light quarks being ``staggered'' over a hypercube~\cite{Kawamoto:1981hw,%
Sharatchandra:1981si,Gliozzi:1982ib,KlubergStern:1983dg}, whereas all fields in Eqs.~(\ref{eq:latOi})
reside at the same site.
In the Symanzik effective field theory for staggered fermions~\cite{Lee:1999zxa}, the operators take the
form~\cite{Bernard:2013dfa}
\begin{equation}
    \bar{\Psi}_b\Gamma_i\Upsilon_q\,\bar{\Psi}_b\Gamma'_i\Upsilon_q \doteq \frac{1}{4}\sum_\Xi
        \bar{b}\Gamma_i\Gamma_\Xi q_{ck}\,\bar{b}\Gamma'_i\Gamma_\Xi q_{dl}\,\Gamma_\Xi^{ck}\Gamma_\Xi^{dl}
        + \text{opposite parity},
    \label{eq:wrongspin}
\end{equation}
where $\Xi$ runs over all 16 Dirac matrices, $c$ and $d$ are taste indices, and $k$ and $l$ are copy indices.
The fields on the left-hand side of Eq.~(\ref{eq:wrongspin}) are those in the lattice simulation, while
those on the right-hand side are defined in the continuum~\cite{Symanzik:1983dc}; the symbol $\doteq$ can be
read ``has the same matrix elements as.'' %
Compared with the continuum operators $\op_i^q$, the operators on the right-hand side of
Eq.~(\ref{eq:wrongspin}) have extra species indices (taste and copy) in addition to flavor.

On the right-hand side of Eq.~(\ref{eq:wrongspin}), the opposite-parity contribution is familiar from
heavy-light bilinears~\cite{Wingate:2002fh} and can be removed during the correlator fits, as discussed in
Sec.~\ref{sec:corr_analysis}.
In the sum given explicitly, only the terms with $\Gamma_\Xi=\openone$ and $\gamma^5$ return the Dirac
structure of the left-hand side.
Following Ref.~\cite{Bernard:2013dfa}, we refer to the others as ``wrong-spin operators.'' %
Because the $\op_i^q$ and $\tilde{\op}_i^q$ form a complete set, the same list of eight operators appears
(with extra species indices) after carrying out the sum in Eq.~(\ref{eq:wrongspin}), and the wrong-spin
terms do not contain any new Dirac structures.

The $B_q$-meson interpolating operators are similarly constructed from the fields $\Upsilon_q$ 
and~$\Psi_b$:
\begin{equation}
    {B^\dagger_q}(\bm{x},t) = \sum_{\bm{x}'}
        {\bar{\Upsilon}_q(\bm{x},t)S(\bm{x},\bm{x}')\gamma_5\Psi_b(\bm{x}',t),}
    \label{eq:Binterp}
\end{equation}
which creates (annihilates) a $B_q$ ($\bar B_q$) meson.
The spatial smearing function $S(\bm{x},\bm{x}')$ to the $b$-quark propagator is given by 
the ground-state 1S wavefunction of the Richardson potential~\cite{Richardson:1978bt,Menscher:2005kj}.
It provides good overlap of the interpolating operator with the $B_q$-meson ground-state
and suppresses unwanted contamination from excited states.
Further details on the smearing are given in Ref.~\cite{Bazavov:2011aa}.
In analogy with Eq.~(\ref{eq:wrongspin}), the interpolating operator becomes
\begin{equation}
    {\bar{\Upsilon}_q\gamma_5\Psi_b \doteq \frac{1}{2} \bar{q}_{ai}\gamma_5 b\,\delta^{ai}}
        + \text{opposite parity}
    \label{eq:OKspin}
\end{equation}
in the Symanzik effective theory after disentangling spin and taste.

The matrix elements can be extracted from three-point correlation functions with zero spatial momentum:
\begin{equation}
    C_{\latop_i^q}(t_x,t_y,t_0) = \sum_{\bm{x},\bm{y}} \langle B^\dagger_q(\bm{y},t_y+t_0)
        \latop_i^q (\bm{0},t_0) B^\dagger_q(\bm{x},t_x+t_0)\rangle .
    \label{eqn:corrQ}
\end{equation}
Despite the wrong-spin operators, the correlation function in Eq.~(\ref{eqn:corrQ}) is dominated by
contributions with the intended Dirac structure~\cite{Bernard:2013dfa}.
In the Symanzik-effective-theory notation, the three-point correlation function contains four terms of the form
\begin{equation}
    \langle\tr\left[\gamma^5 G(y,0)\Gamma_i\Gamma_\Xi q_c(0)\bar{q}_a(x)\right]
        \tr\left[\gamma^5 G(y,0)\Gamma_i\Gamma_\Xi q_d(0)\bar{q}_b(y)\right]\rangle
        \Gamma_\Xi^{ca}\Gamma_\Xi^{db},
\end{equation}
where $a,b,c,d$ are taste indices, $G$ is the heavy-quark propagator, and the trace is over color and spin.
When $\Gamma_\Xi=\openone$, the desired operator is recovered.
The other terms arise only when a hard taste-changing gluon with some momentum components near~$\pi/a$ 
is exchanged from one staggered-quark line to the other.
In the Symanzik effective field theory, this effect is described by a four-quark interaction (among light 
quarks).
These contributions are, thus, suppressed by a power of~$a^2$.
Like any taste-violating effect, the $\Gamma_\Xi\neq\openone$ terms lead to nonanalytic behavior in the
chiral limit that can be described in staggered \cpt~\cite{Bernard:2013dfa}.
As discussed in Sec.~\ref{sec:ChPT}, we can therefore account for them as part of the combined 
chiral-continuum extrapolation.

\begin{figure}
	\includegraphics[width=0.7\textwidth]{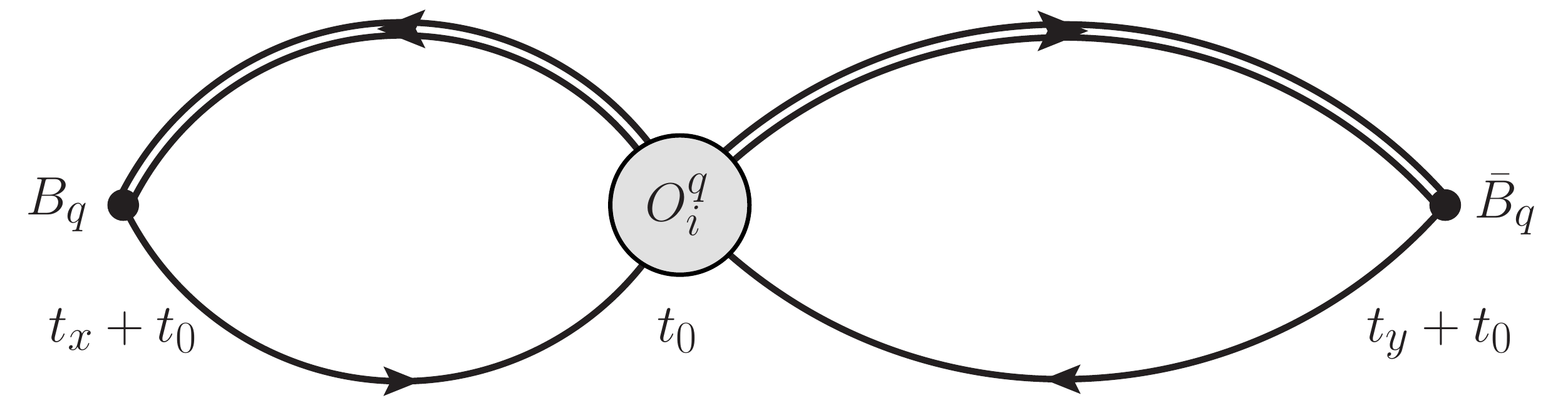}
	\caption{Lattice three-point correlation function $C_{\latop_i^q}(t_x,t_y,t_0)$.
    The double and single lines denote the bottom- and light-quark propagators, respectively.
    The $B_q$ meson created at $t_x + t_0 < t_0$ oscillates into a $\bar{B}_q$ meson via the
    $\Delta B=2$ four-fermion operator at time $t_0$.
    This $\bar{B}_q$ meson is subsequently annihilated at $t_y + t_0 > t_0$.}
    \label{fig:BBbar3pt}
\end{figure}

Figure~\ref{fig:BBbar3pt} shows the structure of $C_{\latop_i^q}(t_x,t_y,t_0)$.
The local four-fermion operator $\latop_i^q$ is placed at a fixed location $t_0$, where $t_0$ runs over the
time sources, while the $B_q$ mesons are placed at all possible spacetime points $x$ and $y$.
In practice, we construct the three-point correlators from two open-meson propagators, corresponding to the
${B}_q$ and $\bar{B}_q$ mesons at $t_x$ and $t_y$, respectively, combining the free spin and color
indices at $t_0$ as dictated by the spin-color structure of each operator~$\latop_i^q$.
Because we average data from multiple time sources at the outset of the
analysis, we henceforth drop the label $t_0$ in the three-point correlator
$C_{\latop_i^q}(t_x,t_y,t_0)$ and in the analogous two-point correlator (Eq.~(\ref{eqn:corr2pt}), below).
The three-point correlator $C_{\latop_i^q}(t_x,t_y)$ is now just a function of $t_x$ and $t_y$.

To isolate the mixing matrix elements from $C_{\latop_i^q}(t_x,t_y)$, we need to remove the
overlap of the $B_q$-meson operator with the ground state.
We obtain this normalization factor from the pseudoscalar two-point correlation function with zero spatial
momentum:
\begin{equation} \label{eqn:corr2pt}
    C_{B_q} (t) = \sum_{\bm{x}} \langle B_q(\bm{x},t) B^\dagger_q(\bm{0},0)\rangle,
\end{equation}
which is constructed by tying together the open end of an open-meson propagator per Eq.~(\ref{eq:Binterp}).

Finally, to reduce statistical uncertainties, we average over sets of physically equivalent, but not
numerically identical, data.
The periodic temporal boundary conditions of the lattice ensure that, in the limit of infinite statistics,
the two-point correlation functions are symmetric about the origin.
We therefore average the two-point correlator values $C_{B_q}(t)$ and $C_{B_q}(T-t)$ as well as the
three-point correlator values $C_{\latop_i^q}(t_x,t_y)$ and $C_{\latop_i^q}(T-t_y,T-t_x)$.
Further, parity conservation of QCD ensures that, in the limit of infinite statistics, the matrix elements
$\langle\tilde{\latop}_{1,2,3}\rangle$ are the same as $\langle\latop_{1,2,3}\rangle$.
We therefore calculate both $C_{\latop_i^q}(t_x,t_y)$ and $C_{\tilde{\latop}_i^q}(t_x,t_y)$ and average them
to improve the statistics.
Additional details on these issues can be found in Ref.~\cite{Bouchard:2011yia}.

\section{Correlator analysis}
    \label{sec:corr_analysis}
    Here we discuss our determinations of the $B$-mixing matrix elements from the two- and three-point
correlation functions defined in Eqs.~(\ref{eqn:corrQ}) and~(\ref{eqn:corr2pt}) of the previous subsection.
In this section, dimensionful quantities are given in units of the lattice spacing, where the explicit
factors of $a$ are suppressed to simplify expressions.

\subsection{Method} \label{sec:CorrFitMethod}
We estimate statistical uncertainties in the correlator analysis, and subsequently propagate them to the
chiral-continuum extrapolation, via bootstrap resampling.
In practice, on each ensemble we generate 600 bootstrap resamples; increasing the number does not change the
estimated statistical errors in the two- and three-point correlation functions.
We use the bootstrap distributions obtained from the two- and three-point fits to propagate the
statistical errors cleanly from the correlator fits to the chiral-continuum extrapolation discussed in
Sec.~\ref{sec:ChPT}.

\begin{figure}
	\vspace{-0.3in}
	\centering
	\hspace{-0.4in}
	\includegraphics[width=0.37\textwidth, angle=270]{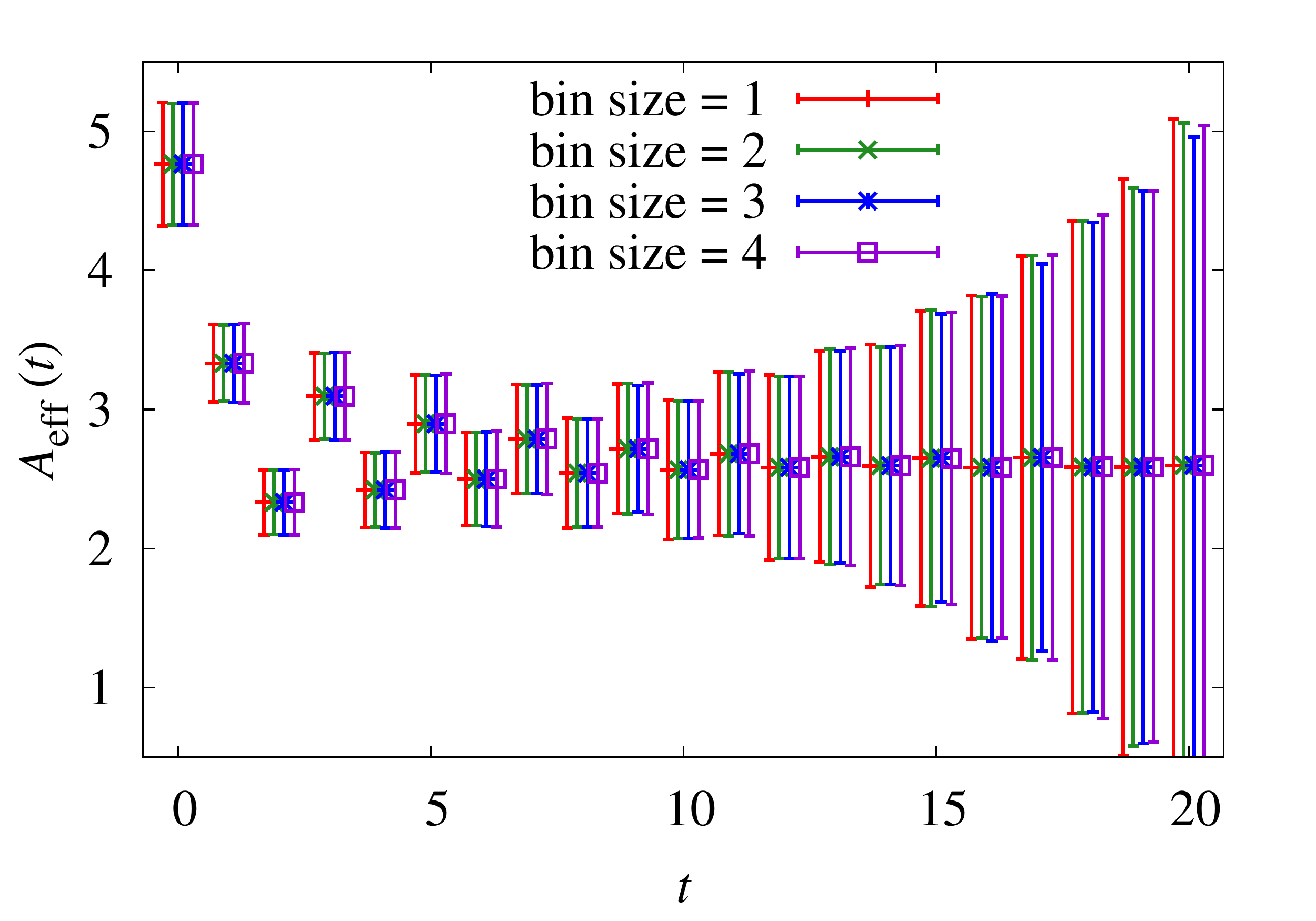}
	\hspace{-0.2in}
	\includegraphics[width=0.37\textwidth, angle=270]{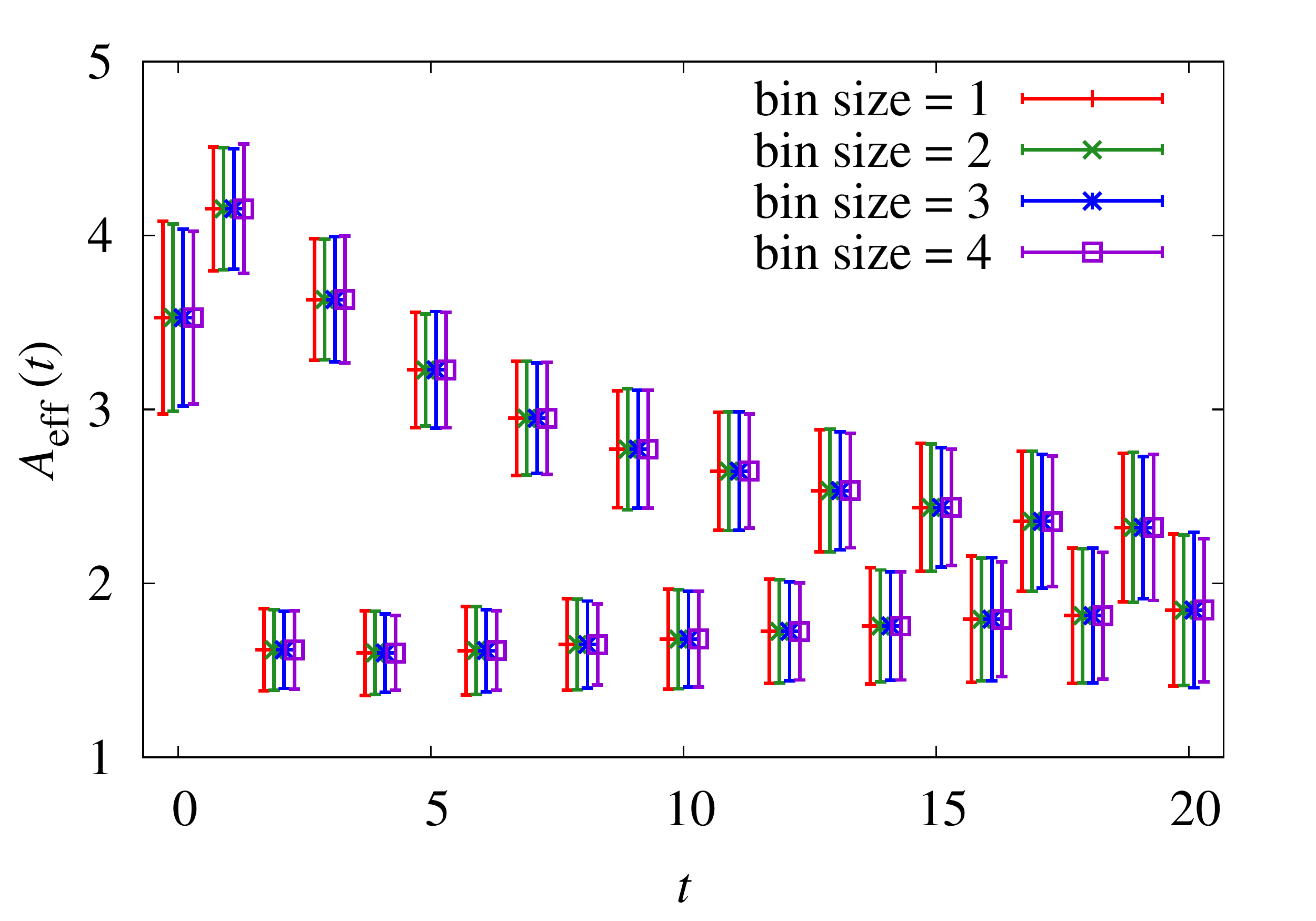}
	\hspace{-0.4in}
	\caption{Blocking studies with the effective amplitude [defined in Eq.~(\ref{eq:Aeff2pt})] of two-point 
    functions on the (\emph{left}) $a\approx 0.12$ fm, $am'_l/am'_s = 0.007/0.05$ ensemble with 
    $am_q=0.0349$ and (\emph{right}) the $a\approx 0.045$ fm ensemble with $am_q = 0.0018$.  
    Different colors and symbols denote different bin sizes.}
    \label{fig:bin}
\end{figure}

Before any fitting, we study the two-point correlation functions to assess the degree of autocorrelation
between successive configurations in the Monte-Carlo evolution.
We do this in several ways, each of which indicates that autocorrelations are negligible.
First, we block the data with varying bin sizes.
On all ensembles, we observe a negligible change in the errors of the correlator data when blocking is
applied.
This is illustrated in Fig.~\ref{fig:bin} which plots the effective amplitude [defined in
Eq.~(\ref{eq:Aeff2pt})] of two-point functions on one of our coarsest ensembles and on our finest ensemble
for different bin sizes.
In addition, we calculate the autocorrelation coefficient for configuration separation $\eta$, averaged over
timeslices,
\begin{equation}
    \mathcal{A}(\eta) = \frac{N_\text{conf}-1}{N_t(N_\text{conf}-\eta-1)} \sum_{t=0}^{N_t-1}
        \frac{ \sum_{i=0}^{N_\text{conf}-\eta-1}[C_{i+\eta}(t) - \overline{C(t)}] [C_i(t) - \overline{C(t)}]}
        { \sum_{i=0}^{N_\text{conf}-1} [C_i(t) - \overline{C(t)}]^2 }
    \label{eq:corrcoeff}
\end{equation}
where $\overline{C(t)} \equiv N_{\text{conf}}^{-1} \sum_{i=0}^{N_{\text{conf}}-1} C_i(t)$. 
The exponential autocorrelation time $\tau_\text{exp}$ is then obtained by comparing the calculated values
of $\mathcal{A}(\eta)$ to the expected behavior, $|\mathcal{A}(\eta)| = e^{-\eta / \tau_\text{exp}}$.
We analyze the autocorrelations directly on the two-point data and separately on the principal component of
the data covariance matrix.
Both analyses yield $\tau_\text{exp} \approx 0.15$, again indicating that no binning is required based on
the criterion that a reasonable bin size is twice the autocorrelation time~\cite{Montvay:1994cy}.
We therefore do not bin the correlator data in this work.
Reference~\cite{Bouchard:2011yia} provides further details on the autocorrelation 
studies.

We extract the hadronic matrix elements from simultaneous fits to two- and three-point correlation function
data, constraining the fit parameters with Gaussian priors~\cite{Lepage:2001ym} whose central values are drawn randomly from the Gaussian distribution during bootstrap resampling.
For all operators, ensembles, and valence-quark masses listed in Table~\ref{tab:ValParams}, we fix the
number of states included in the two- and three-point fit functions to be equal.
We also use comparable fit intervals in physical units on all ensembles.
We observe that the correlator fit results are stable against reasonable variations in the numbers of states
and timeslices of data included in the fit.
Details on the optimization of the correlation-function fits and determination of the $B$-mixing matrix
elements are given in the following subsections.

We implement the constrained fits in our analysis by minimizing the augmented $\chi^2_\text{aug}$ defined in
Eq.~(\ref{eq:augchi2}), which includes contributions from the priors.
We employ the least-squares-fitting software package {\tt lsqfit}~\cite{g_peter_lepage_2012_10236}, which
supports Bayesian priors and provides tools for correlated error propagation.
We evaluate the relative quality of our correlator fits using a statistic denoted $Q$, which is defined in
Eq.~(\ref{eq:Qval}).
Its definition is similar to the standard $p$~value~\cite{Agashe:2014kda} but is based on the minimum
of $\chi^2_\text{aug}$ and a counting of the degrees of freedom suited to Bayesian
analyses.
By construction, $Q$ lies in the interval $[0,1]$.
Larger $Q$~values indicate greater compatibility between the data and fit function given the prior
constraints.
Unlike a $p$~value, however, $Q$ is not expected to be uniformly distributed even when the hypothesis is
correct.
To test the influence of the priors on the best fit, we also examine a $p$~value based on the value of the
standard $\chi^2$ function evaluated at the parameter values that minimize $\chi^2_\text{aug}$.
Details and explicit formulas for $Q$ and $p$ are provided in Appendix~\ref{app:pval}.

\subsection{Two-point fits} \label{sec:2pts}

We fit the $B_q$ -meson two-point correlator data to the functional form
\be
    C_{B_q}(t) = \sum_{n=0}^{N^\text{2pt}-1} |Z_n|^2 (-1)^{n(t+1)}\left(e^{-E_nt} + e^{-E_n(N_t-t)} \right),
    \label{eq:2ptfitfn}
\ee
where $Z_n$ is the wavefunction normalization and $N_t$ is the temporal extent of the lattice.
This fit function includes the effects of normal- ($n$ even) and opposite-parity ($n$ odd) states present in
heavy-staggered meson correlation functions~\cite{Wingate:2002fh}, and accounts for the contribution from
backwards-propagating states associated with the use of periodic boundary conditions.

\begin{figure}
	\vspace{-0.3in}
	\hspace{-0.2in}
	\includegraphics[width=0.52\textwidth]{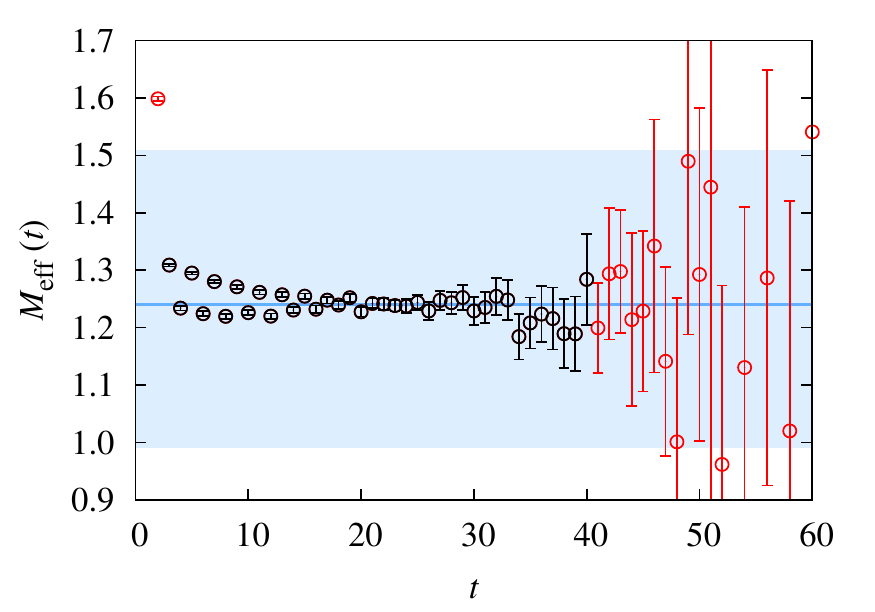}
	\hspace{-0.22in}
	\includegraphics[width=0.52\textwidth]{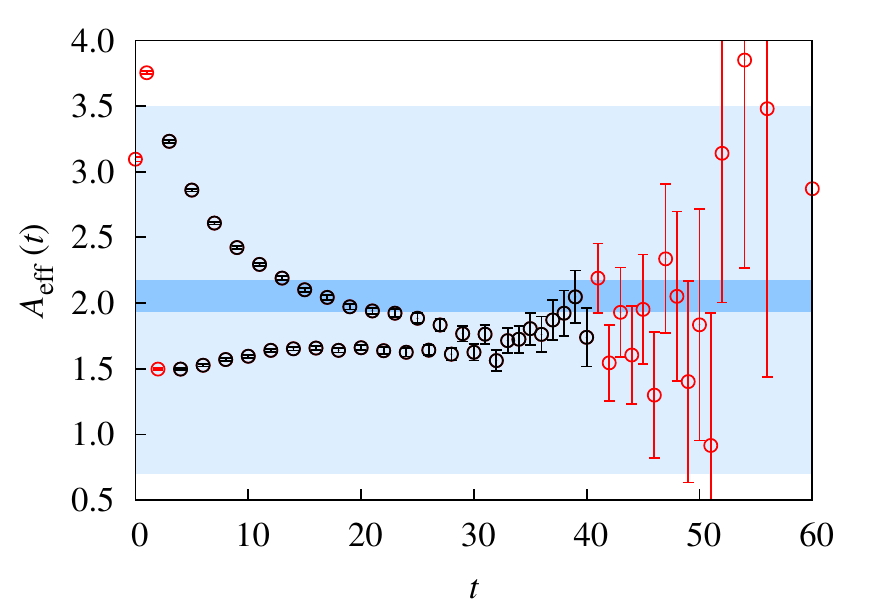}
	\hspace{-0.2in}
	\caption{$B_q$-meson effective mass (\emph{left}) and amplitude (\emph{right}) for the 
    $a\approx0.06$~fm, $am'_l/am'_s = 0.0018/0.018$ ensemble with $am_q=0.0018$.
	The wider light-blue bands show the ground-state prior choices for $E_0$ and $Z_0$.  
	The narrower dark-blue bands show the fit results using these priors to the data from $3 \leq t\leq40$, 
    which are denoted by black symbols.}
	\label{fig:Eff2pt}
\end{figure}

We loosely constrain the ground-state mass and amplitude using priors guided by the effective mass and
amplitude defined as
\begin{align}
    \label{eq:Meff2pt}
    M_\text{eff}(t) &= \cosh^{-1}\Big( \frac{C_{B_q}(t + 1) + C_{B_q}(t - 1)}{2C_{B_q}(t)} \Big), \\
    A_\text{eff}(t) &= C_{B_q}(t) e^{tM_\text{eff}} ,
    \label{eq:Aeff2pt}
\end{align}
where $M_\text{eff}$ in the exponent of Eq.~(\ref{eq:Aeff2pt}) is chosen by eye based on the plot of
$M_\text{eff}(t)$ in Eq~(\ref{eq:Meff2pt}), such that $A_\text{eff}(t)$ displays a plateau at large times.
At each lattice spacing, we examine the effective mass and amplitude for all light sea- and valence-quark
masses and choose common priors for $E_0$ and $Z_0$ with widths large enough to cover the observed
$M_\text{eff}$ and~$A_\text{eff}$.
Figure~\ref{fig:Eff2pt} shows samples of $M_{\text{eff}}(t)$ and $A_{\text{eff}}(t)$ with the ground-state
priors employed for all fits of two-point data at that lattice spacing.
The prior widths are more than $100\times$ and $10\times$ larger than the uncertainties on the fitted
ground-state energy and amplitude, respectively.
We choose the priors for the lowest-lying oscillating state and the excited-state energy splittings
guided by experimental measurements~\cite{Agashe:2014kda} and quark-model
predictions~\cite{PhysRevD.64.114004,Ebert2010}.
On all ensembles, we use priors that correspond to $E_1-E_0\approx0.3(7)$~GeV, $E_2-E_0\approx0.6(3)$~GeV,
and $E_{n+2}-E_n\approx0.5(3)$~GeV ($n>0$).
For the excited states ($n \geq 2$), we take the fit parameters to be the logarithms of the splittings
between energy levels, $\Delta_{n+2, n} \equiv\log{\left(E_{n+2}-E_{n}\right)}$.
This automatically imposes the ordering $E_{n+2} > E_{n}$.

\begin{figure}
	\vspace{-0.0in}
	\hspace{-0.15in}
	\includegraphics[width=0.52\textwidth]{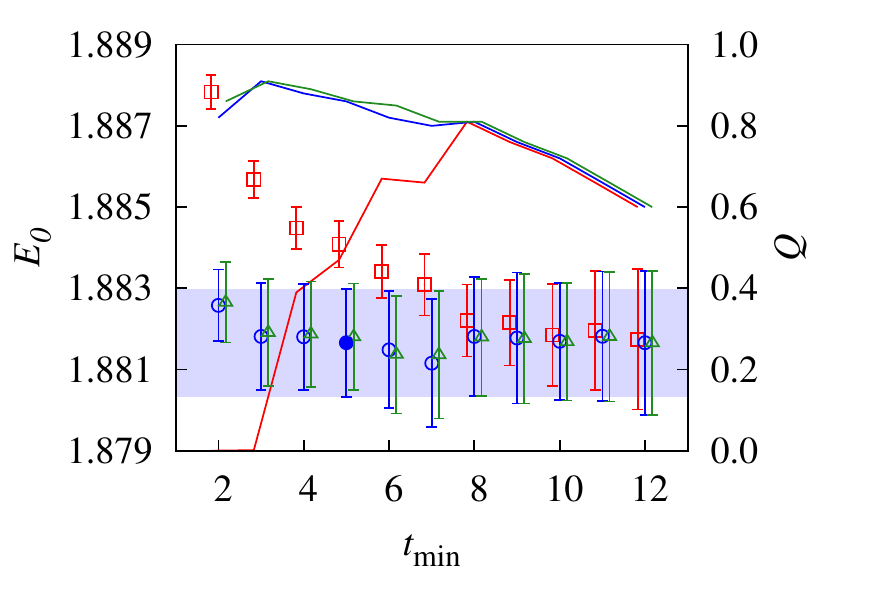}
	\hspace{-0.25in}
	\includegraphics[width=0.52\textwidth]{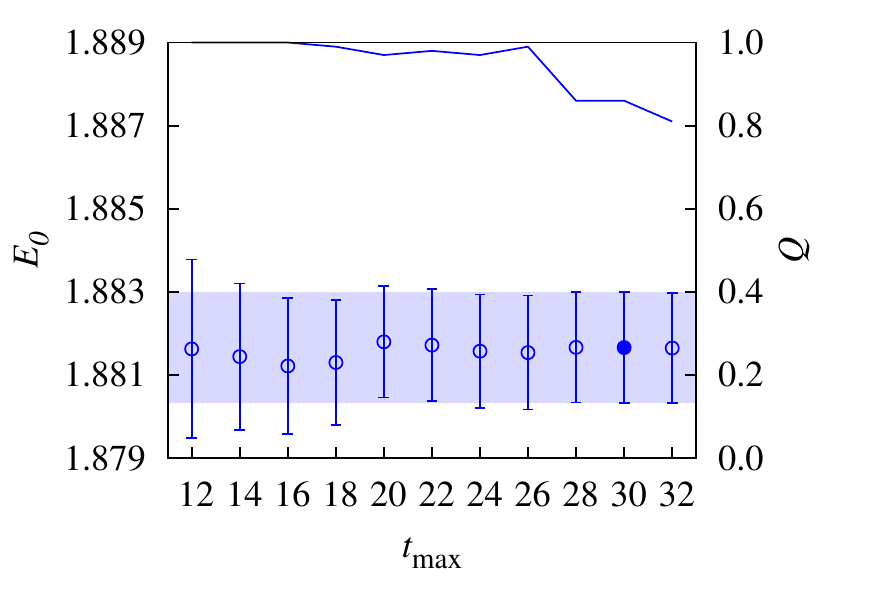}
	\hspace{-0.15in}
	\vspace{-0.15in}
	\caption{
	$B_q$-meson ground-state energy $E_0$ obtained from different two-point fits.  
	(\emph{left}) $E_0$ vs $t_\text{min}$ with fixed $t_\text{max} = 30$ obtained using $N^\text{2pt}=2$
    (red squares), 4 (blue circles), and 6 (green triangles).
	(\emph{right}) $E_0$ vs $t_\text{max}$ for fixed $t_\text{min} = 5$ and $N^\text{2pt}=4$.
	In both plots, the solid lines show the $Q$~values---defined in Eq.~(\ref{eq:Qval})---of the fits for 
    which the symbols have the same color.
	The optimal values of $t_\text{min}=5$ and $t_\text{max}=30$ for $N^\text{2pt}=4$ are denoted by filled 
    blue circles, and the corresponding fit result is depicted by the light blue bands.
	Data shown are from the $a\approx 0.12$ fm, $am'_l/am'_s = 0.007/0.050$ ensemble with $am_q=0.0349$.}
	\label{fig:Fit2pt}
\end{figure}

We fit the two-point correlation functions using fit intervals $t_\text{min} \leq t \leq t_\text{max}$,
chosen for each ensemble based on the emergence of plateaus and the onset of noise in the effective mass and
amplitude plots.
We first vary $t_\text{min}$ with fixed $t_\text{max}$, generally starting at $t_\text{min}=2$ and
increasing it until excited-state contributions have significantly decreased.
For example, for the correlation function shown in Fig.~\ref{fig:Eff2pt}, we consider $2\leq
t_\text{min}\leq 20$.
The ground-state masses obtained from fits including earlier times have greater precision, but also greater
contamination from excited states.
We therefore include several pairs of excited and oscillating states in our fits to enable us to include
early time slices.
Initial studies showed that including six states was sufficient to accommodate excited-state contributions
at early times, so we consider fits with $N^\text{2pt}\le6$.
We also vary $t_\text{max}$ with fixed $t_\text{min}$ and find that fit results are insensitive to the
addition of late-time-data for which statistical errors are larger.
For the data shown in Fig.~\ref{fig:Eff2pt} we consider $30\leq t_\text{max}\leq 60$.
We perform fits for all combinations of $N^\text{2pt}$, $t_\text{min}$, and $t_\text{max}$ within the ranges
described here.
Figure~\ref{fig:Fit2pt} demonstrates the stability of the fitted ground-state energy with respect to varying
the number of states and timeslices included in the fit.
The two-point correlator data and fit results presented in Figs.~\ref{fig:Eff2pt} and~\ref{fig:Fit2pt} are
representative of other valence-quark masses and sea-quark ensembles analyzed in this work.

Finally, after examining the stability plots for all light valence-quark masses and ensembles at a given
lattice spacing, we select an optimal fit interval $t_\text{min} \leq t \leq t_\text{max}$ for each choice
of $N^\text{2pt} = 2$, 4, and 6.
We employ ranges for all two-point correlator fits that correspond to approximately the same physical
distance on all lattice spacings.
For each ensemble, we choose the same $t_\text{min}$ for all valence masses.
The different ensembles have different numbers of configurations and, in some cases, different spatial
volumes.
As a result, the statistical precision of the data varies with ensemble.
We therefore allow for a small variation of $t_\text{min}$ with ensemble at a given lattice spacing.
In practice, $t_\text{min}$ differs between ensembles by no more than one timeslice.
For $N^\text{2pt} = 4$, which is the number of states used in our central fit for the
matrix elements, the fit intervals range from approximately $0.3~\text{fm}
\lesssim  t \lesssim 2.5~\text{fm}$ at our coarsest lattice spacing to $0.2
~\text{fm} \lesssim t  \lesssim 1.7~\text{fm}$ at our finest lattice spacing.

\subsection{Three-point fits} \label{sec:3pts}

\begin{figure}
	\vspace{-0.4in}
	\hspace{-0.8in}
	\includegraphics[width=0.62\textwidth]{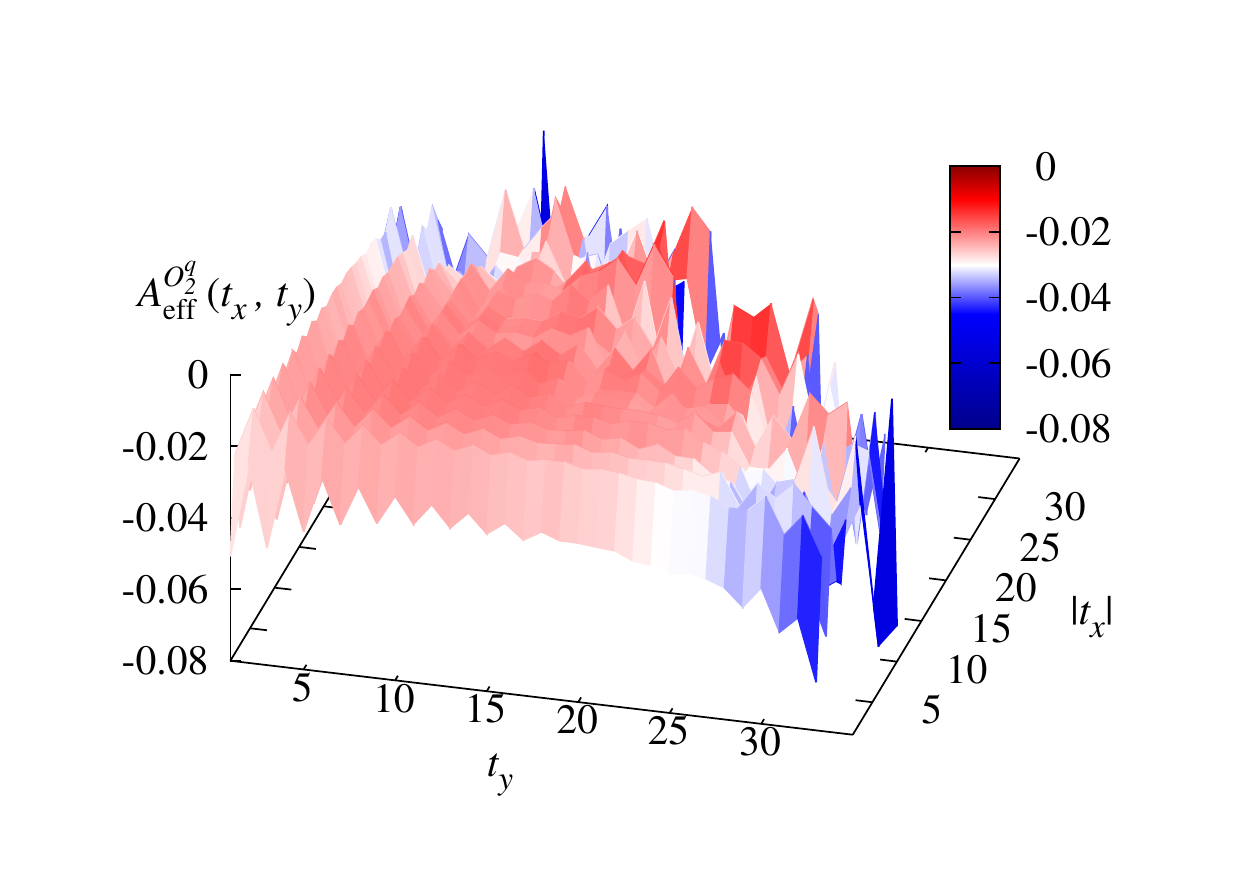} \hspace{-0.9in}
	\includegraphics[width=0.62\textwidth]{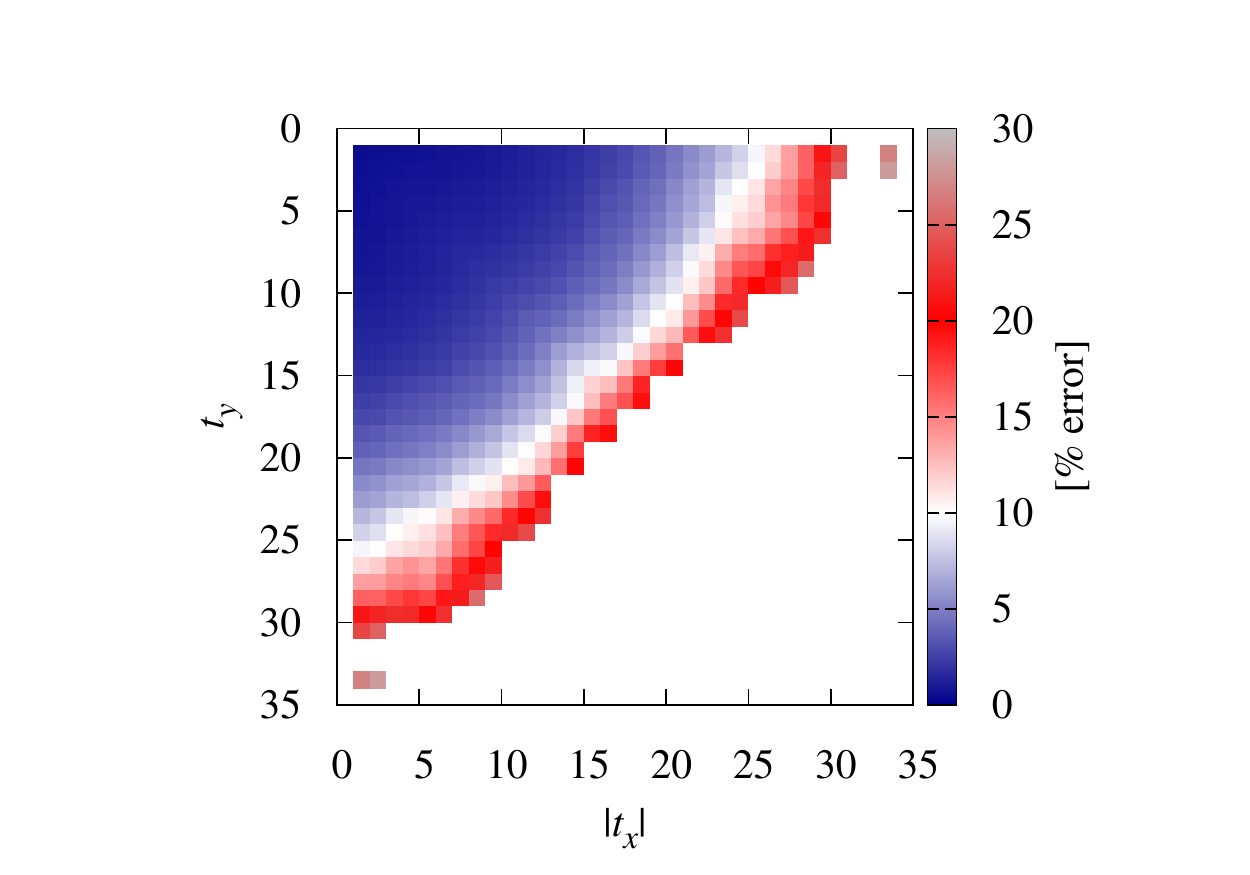}
    \hspace{-0.8in}
    \caption{$B_q$-mixing three-point effective amplitude $A_\text{eff}^{\latop_{2}^q}(t_x, t_y)$ central value (\emph{left}) and relative error (\emph{right}) on the $a\approx0.09$~fm, $am'_l/am'_s = 0.00465/0.031$ ensemble with $am_q=0.0062$.  Our preferred fit uses $t^\text{3pt}_\text{min}=4$ and $t^\text{3pt}_\text{max}=15$.
	\label{fig:Eff3pt}}
\end{figure}

In analogy with Eq.~(\ref{eq:2ptfitfn}), we fit the three-point
correlator data to the functional form
\begin{equation}
	C_{\latop_i^q}(t_x,t_y) = \sum_{n,m=0}^{N^\text{3pt}-1} \mathcal{Z}_{nm}^{\latop_i^q}
        \frac{Z_n Z_m^\dagger}{2\sqrt{E_nE_m}}(-1)^{n(t_x+1)+m(t_y+1)}\ e^{-E_n |t_x| -E_mt_y},
    \label{eq:3ptfitfn}
\end{equation}
where $t_x$ and $t_y$ are defined in Fig.~\ref{fig:BBbar3pt}.
Because we combine the data from all time sources~$t_0$, from now on we set $t_0=0$ for simplicity.
The desired hadronic matrix element $\langle\latop_i^q\rangle$, using notation introduced in
Eq.~(\ref{eq:MEdef}) for the physical matrix element, is proportional to the ground-state amplitude as
\begin{equation}
	\langle\latop_i^q\rangle / M_{B_q} = \mathcal{Z}_{00}^{\latop_i^q},
\end{equation}
where the factor of $M_{B_q}$ results from our use of nonrelativistic normalization of states in
Eq.~(\ref{eq:3ptfitfn}).
The effect of periodic boundary conditions is negligible in the three-point data and is omitted from
Eq.~(\ref{eq:3ptfitfn}).
As with the two-point correlation functions, we obtain loose priors for the ground-state amplitude by
examining the effective amplitude, defined here to be
\begin{equation}
    A_\text{eff}^{\latop_i^q}(t_x, t_y) = C_{\latop_i^q}(t_x, t_y) e^{(|t_x| + t_y)M_\text{eff}} 
    \label{eq:Aeff3pt}
\end{equation}
for all operators, ensembles and valence-quark masses on a given lattice spacing.
Figure~\ref{fig:Eff3pt}, left, shows an example three-point effective amplitude; plots for other operators, valence-quark masses and sea-quark ensembles look similar.
Because the temporal oscillations from opposite-parity states are clearly visible in the three-point data, their contributions are easily identified in the correlator fits and can be removed to obtain the desired ground-state energy and amplitude.
The three-point effective amplitudes display a plateau at moderately large $|t_x|$ and $t_y$ that has a
magnitude of order~$10^{-2}$.
We choose a prior for the ground-state three-point amplitude of $\mathcal{Z}_{00}^{\latop_i^q}=0\pm1$,
consistent with the large widths used for the ground state in the two-point data.

The three-point correlators are functions of both the $B_q$ and $\bar{B_q}$ timeslices, $t_x$ and $t_y$, and
we simultaneously fit the dependence on both times.
Charge-conjugation symmetry ensures that the $B_q$ and $\bar B_q$ mesons have identical energy eigenstates.
The construction of the three-point correlation functions using open-meson propagators as described in
Sec.~\ref{sec:corr_building}, in combination with the fact that the $B_q$ and $\bar B_q$ correlators decay
at the same rate, leads to an exact symmetry of the correlator data under the interchange
$|t_x|\leftrightarrow t_y$.
We choose the fit regions to reflect this symmetry, taking $|t_x|_\text{min}=t_{y, \rm min} \equiv
t^\text{3pt}_\text{min}$ and $|t_x|_\text{max}=t_{y, \rm max} \equiv t^\text{3pt}_\text{max}$ and only
including data for $t_y \geq |t_x|$ in the fit.
These requirements define a class of fit contours in the $|t_x|$-$t_y$ plane.
We have verified that the fitted ground-state amplitudes are independent of the exact contour choice,
however, as long as enough data are included.

We use slightly different fit regions for different operators.
For the matrix elements $\langle\latop_{3,4,5}\rangle$, we use a triangular region $\mathcal{S}$ in
the $|t_x|$-$t_y$ plane defined by
\begin{equation}
    \mathcal{S} = 
    \{(t_x,t_y)~|~t_\text{min}^\text{3pt}\leq|t_x|\leq t_\text{max}^\text{3pt},\;
    |t_x| \leq t_y \leq t_\text{max}^\text{3pt}\}.
    \label{eq:region345}
\end{equation}
For the matrix elements $\langle\latop_{1,2,3}\rangle$, we use a fan-shaped region
$\mathcal{S}\cap\mathcal{S}'$ with
\begin{equation}
    \mathcal{S}' =
    \{(t_x,t_y)~|~\sqrt{t_x^2 + t_y^2} \leq t_\text{max}^\text{3pt},\;
        |t_x|+t_y \geq 2 t_\text{min}^\text{3pt} + 3\}
     \label{eq:reqion123}
\end{equation}
cutting out points at large and small times.
The matrix element $\langle\latop_3\rangle$, which is extracted using both fit regions, provides a
consistency check.

We perform simultaneous fits of the two- and three-point correlation functions using the same number of
states for both, $N^\text{2pt}=N^\text{3pt}$, and with common fit parameters for the energies and
wave-function normalizations.
For the two-point data, we use the optimal timeslices determined in Sec.~\ref{sec:2pts}.
For the three-point data, we consider several fit ranges,
$t_\text{min}^\text{3pt}\leq|t_x|,\,t_y\leq t_\text{max}^\text{3pt}$, which are chosen based on inspection 
of the effective three-point amplitude and its relative error.
For example, for the correlator data shown in Fig.~\ref{fig:Eff3pt}, we examine the time ranges
$2\leq t^\text{3pt}_\text{min} \leq 14$ and $15 \leq t^\text{3pt}_\text{max} \leq 30$.
To select the preferred fit range and number of states, we first plot the fit results for the ground-state
three-point amplitude $\mathcal{Z}_{00}^{\latop^q_i}$ vs $t_\text{min}^\text{3pt}$ for $N^\text{3pt}=2$, 4,
and~6.
We find that $N^\text{3pt}=N^\text{2pt}=4$ leads to stable fits for all operators, light-quark masses, and
ensembles and, therefore, choose this number of states as the basis for the rest of the analysis.

For each ensemble and operator, we choose $t^\text{3pt}_\text{min}$ so that it is roughly constant for all
valence masses.
As with $t_\text{min}$ in the two-point correlator analysis, we vary $t_\text{min}^\text{3pt}$ slightly
among ensembles with the same lattice spacing.
We also allow a small, monotonic increase in $t_\text{min}^\text{3pt}$ with increasing valence mass to
accommodate an improving signal-to-noise ratio.
Further, because the three-point data for different operators $\latop_i$ have different signal-to-noise
ratios, we employ a slightly different $t_\text{min}^\text{3pt}$ for each operator.
For a given lattice spacing and operator, the variation in $t_\text{min}^\text{3pt}$ between ensembles
and valence mass is less than $\sim0.2$~fm.
The variation in the lowest time slice---when considering all operators, ensembles, and valence masses on a
single lattice spacing---is slightly greater, but still less than $\sim0.4$~fm.
The minimum time slices included for the three-point correlator fits are similar on all lattice spacings,
ranging from approximately~0.3--0.8~fm.

\begin{figure}
	\centering
	\vspace{-0.2in}
	\hspace{-0.3in}
	\includegraphics[width=0.53\textwidth]{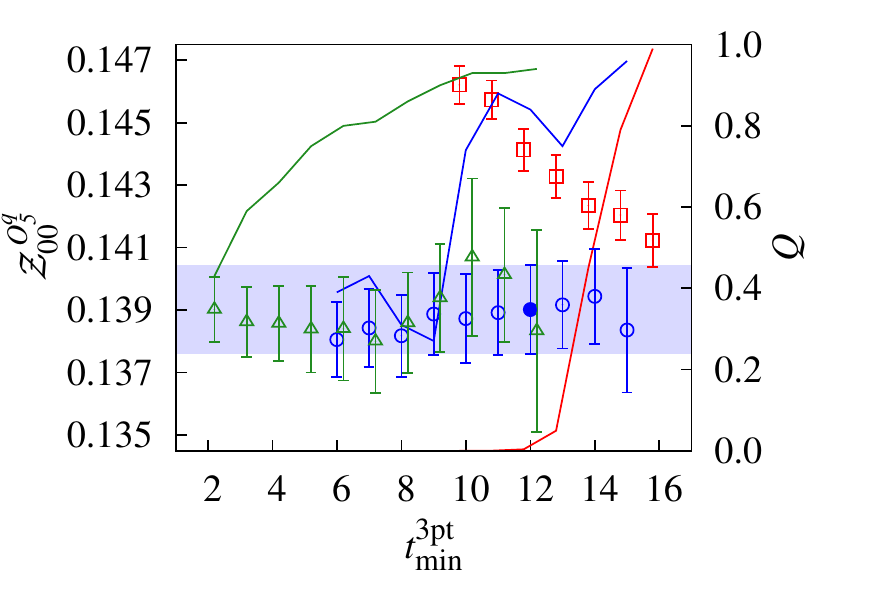}
	\hspace{-0.2in}
	\includegraphics[width=0.53\textwidth]{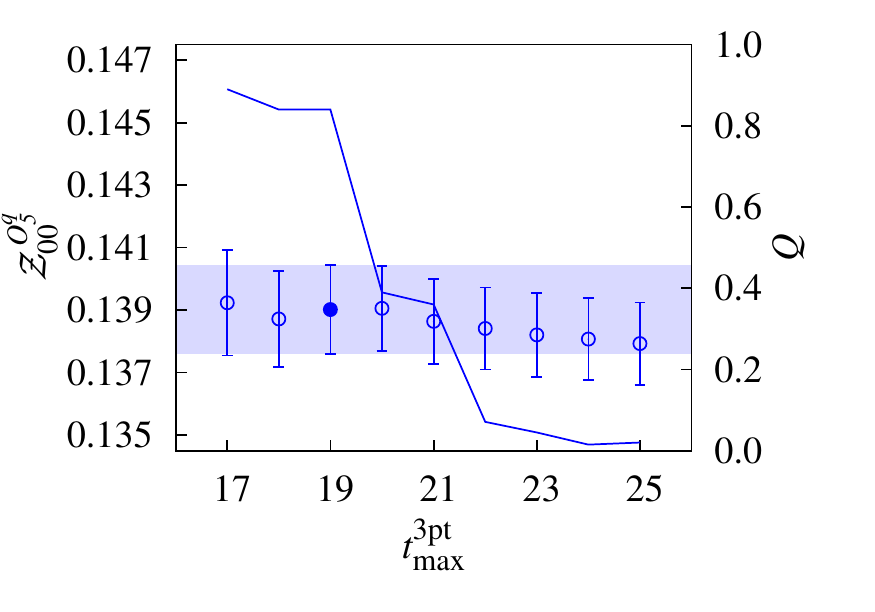}
    \hspace{-0.5in}
    \vspace{-0.2in}
    \caption{$B_q$-mixing ground-state amplitude obtained from different combined two- and three-point
    fits.
    (\emph{left}) $\mathcal{Z}_{00}^{\latop^q_5}$ vs $t_\text{min}^\text{3pt}$ with fixed 
    $t_\text{max}^\text{3pt}=19$ obtained using $N^\text{3pt}=2$ (red squares), 4 (blue circles), and 6 
    (green triangles).
    (\emph{right}) $\mathcal{Z}_{00}^{\latop^q_5}$ vs $t_\text{max}^{\text{3pt}}$ for fixed 
    $t_\text{min}^\text{3pt}=12$ and $N^\text{3pt}=4$.
    In both plots, the solid lines show the $Q$~values---defined in Eq.~(\ref{eq:Qval})---of the fits for 
    which the symbols have the same color.
    For fits with $N^\text{3pt}=4$, the optimal values of $t_\text{min}^{\text{3pt}}=12$ and 
    $t_\text{max}^\text{3pt}=19$ are shown as filled blue circles and the corresponding fit result is 
    depicted by the light blue bands.
    Data shown are from the $a\approx 0.045$ fm ensemble with $am_q = 0.016$.
    \label{fig:Fit3pt}}
\end{figure}

We then vary $t_\text{max}^\text{3pt}$ for the selected $N^\text{3pt}$ and $t_\text{min}^\text{3pt}$ to
verify that the fitted ground-state three-point amplitude is insensitive to reasonable variations of
$t_\text{max}^\text{3pt}$, as shown in Fig.~\ref{fig:Fit3pt}.
We choose a value of $t_\text{max}^\text{3pt}$  that maximizes the amount of data in the fit without degrading the $Q$ value.
The result is an approximately equal number of data points in each fit. 
Consequently, $t_\text{max}^\text{3pt}$ ranges from $t_\text{max}^\text{3pt} \sim 2.5~\text{fm}$ at our
coarsest lattice spacing to $t_\text{max}^\text{3pt} \sim 0.9~\text{fm}$ at our finest lattice spacing.
Figures~\ref{fig:Eff3pt} and~\ref{fig:Fit3pt}, and the choices inferred from inspecting them, represent well
what we find for other ensembles and valence masses.

In addition to verifying that fit results are stable under reasonable
changes to the number of states and time ranges in the
fit, we perform several additional checks.
We verify that ground-state energies obtained from the combined two- and three-point correlator fits are
consistent with those from the two-point-only fits.
We also ensure the fitted parameters are not in tension with the priors.

\begin{figure}
	\includegraphics[width=0.49\textwidth]{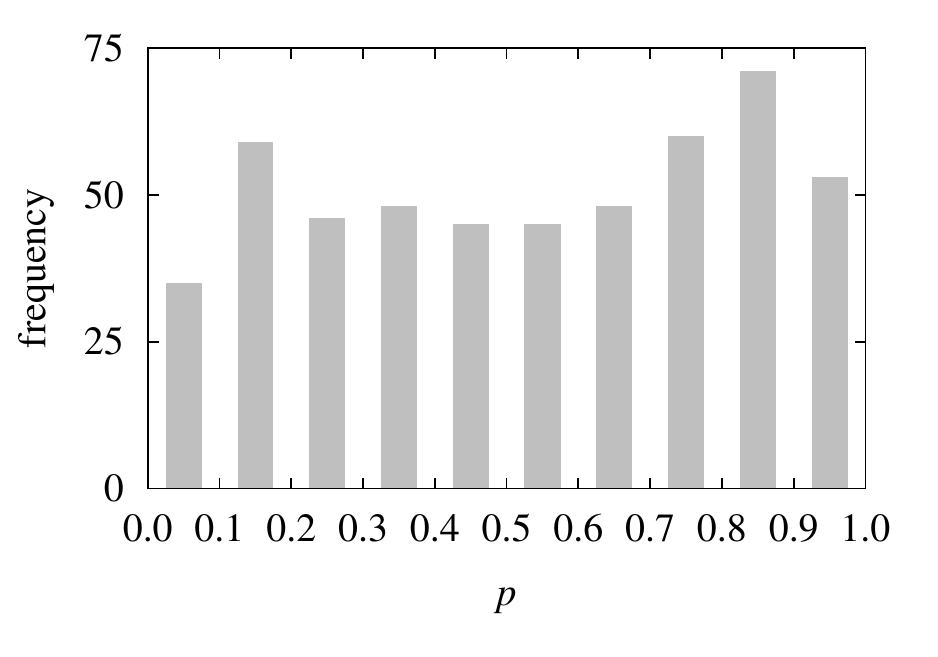}
	\includegraphics[width=0.49\textwidth]{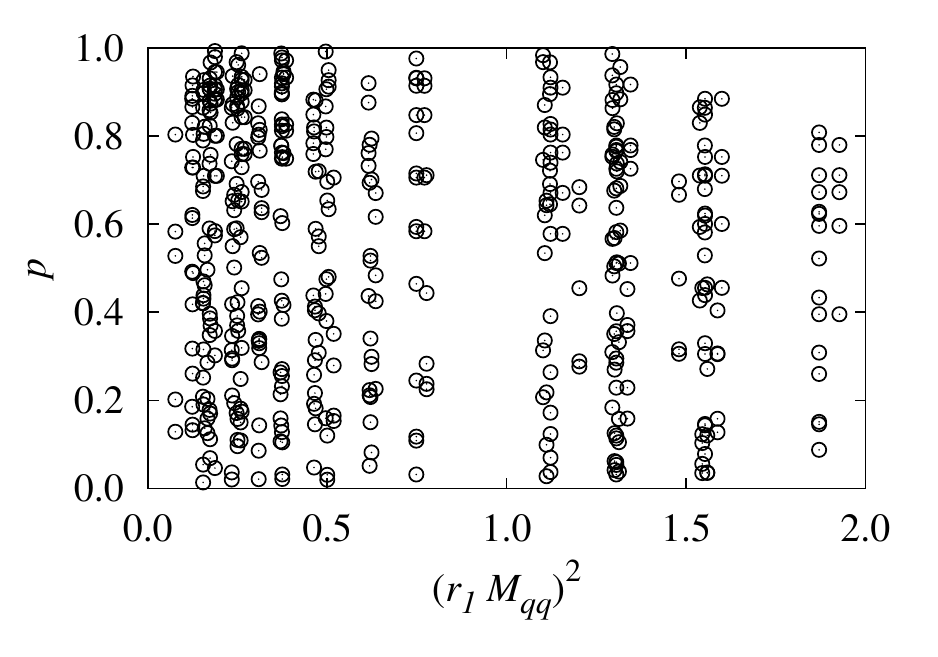} 
	\vspace{-0.2in}
	\caption{(\emph{left}) Histogram and (\emph{right}) scatter plot vs valence mass of $p$~values---%
    defined in Eq.~(\ref{eq:unaugpval})---for all 510 combined two- and three-point fits used to 
    obtain our final results for the matrix elements of operators $\latop_{i}^q$ ($i$=1--5).}
	\label{fig:pvalues}
\end{figure} 
In total, we carry out 510 separate fits of the two- and three-point correlators for the five operators,
fourteen ensembles, and seven or eight valence-light-quark masses per ensemble listed in
Tables~\ref{tab:LatEns} and~\ref{tab:ValParams}.
As shown in Fig.~\ref{fig:pvalues}, the distribution of $p$~values---defined in Appendix~\ref{app:pval}---is
approximately uniform, confirming that the chosen fit functions indeed describe our data.
Further, it shows that the priors on the fit parameters and the slight variation in fit regions between
operators, valence masses, and ensembles, have not introduced significant bias into our matrix-element
results.

\section{Operator Renormalization and Matching}
    \label{sec:PT}
    In this section we discuss the renormalization and matching needed to convert the bare lattice operators
$\latop_i(a)$ to renormalized operators $\bar{\latop}_i(\mu)$ evaluated at a common scale~$\mu$.
When applied to the matrix elements $\langle\latop_i\rangle$ obtained in the previous section, this
procedure yields the corresponding renormalized matrix elements $\langle\bar{O}_i\rangle(\mu)$ for every
light-valence quark mass and lattice ensemble included in our analysis.
The continuum matrix elements $\langle\bar{\op}_i\rangle(\mu)$ (evaluated at the physical light quark
masses) are then determined in Sec.~\ref{sec:ChPT} from a chiral-continuum extrapolation of the
$\langle\bar{O}_i\rangle(\mu)$.

The four-fermion lattice operators defined in Eqs.~(\ref{eq:latOi}) mix under renormalization.
Like their continuum counterparts defined in Eqs.~(\ref{eq:Oi}), they have nonzero anomalous dimensions.
It is convenient to carry out the renormalization and matching to the desired continuum scheme in one step.

In dimensional regularization, the Dirac algebra does not close, and one has to specify a physical operator
basis, distinct from the remaining ``evanescent'' operators~\cite{Dugan:1990df}.
The continuum one-loop renormalization coefficients have been calculated in the \MSbar-NDR scheme for two
different choices of evanescent operators~\cite{Gamiz:2008sk,Monahan:2014xra}.
We denote by BBGLN the scheme defined in Refs.~\cite{Beneke:1998sy,Beneke:2002rj} and by BMU the scheme
defined in Ref.~\cite{Buras:2000if}.
These two schemes are the ones most widely used in the literature.
These two choices of evanescent operators differ only in the renormalization of operators $\op_2$
and~$\op_3$.
The lattice operators can be renormalized and matched to one of the schemes mentioned above by a suitable
choice of dimension-7 operators $P_p$ and coefficient matrices~\cite{Evans:2009du,GamizPT}:
\begin{align}
    \bar{\latop}_i(\mu) &= Z_{ij}(a\mu) \latop_j(a) + ab_{ip}(a\mu) P_p(a), \nonumber \\
        &\doteq \bar{\op}_i(\mu) + \order(a^2),
    \label{eq:matching-all-orders}
\end{align}
where the bar denotes the chosen continuum scheme (e.g., \MSbar-NDR-BBGLN), and $\mu$ is the renormalization
scale.
On the second line $\doteq$ can be read ``has the same matrix elements as,'' as in Eq.~(\ref{eq:wrongspin}).
Using the rotated field $\Psi_b$ in Eqs.~(\ref{eq:latOi}) suffices to ensure that the coefficients $b_{ip}$
start at order~$\alpha_s$.
We therefore neglect them in the matching process, leading to an uncertainty of order $\alpha_sa\LamQCD$
commensurate with that stemming from the choice of $c_\text{SW}$ in the action.
We estimate the effects of these corrections, together with other heavy-quark discretization errors, as
explained in Section~\ref{sec:chiralfitfunc}.
When calculating the matching matrix~$Z_{ij}$, the quark/antiquark states select the same external tastes 
as do the $B$-meson interpolating operators in the three-point correlation function, Eq.~(\ref{eqn:corrQ}),
and the loop integration automatically includes the hard taste-changing gluons.
Further details of this matching procedure will be presented elsewhere~\cite{GamizPT}.

In this paper, we carry out the renormalization and matching with perturbation 
theory, expanding 
\begin{equation}
    Z_{ij}(a\mu) = 2\mathcal{C}\left[\delta_{ij} + \sum_{l=1}\alpha_s^l Z_{ij}^{[l]}(a\mu) \right],
    \label{eq:matching}
\end{equation}
where the factor of $2$ and $\mathcal{C}=2\kappa'_b(1+a m_0)$ are normalization factors related to
conventions for the staggered and clover fermion fields, respectively.
Here the mass $m_0$ in $\mathcal{C}$ is given by $am_0=1/(2\kappa_b')-1/(2\crit)$, where the critical
hopping parameter $\crit$ is the value at which the rest mass vanishes. 
Only the one-loop coefficient in Eq.~(\ref{eq:matching}) is available at present~\cite{GamizPT}. 

From experience, we expect the coefficients $Z_{ij}^{[l]}$ to be large, because of contributions from
external-leg tadpole diagrams~\cite{Lepage:1992xa}.
We use two modifications to Eq.~(\ref{eq:matching}) to improve convergence.
One method is to absorb the large perturbative corrections into a redefinition of the basic parameters of
the lattice action by dividing the gauge links in the action by a typical measure of the tadpole
contributions via $u_0$~\cite{Lepage:1992xa}.
The tadpole-improved renormalization is defined by:
\begin{align}
    Z_{ij}     &= 2u_0 \tilde{\mathcal{C}} \zeta_{ij},  \label{eq:matchingPT} \\
    \zeta_{ij} &= \delta_{ij} +  \sum_{l=1}\alpha_s^l \zeta_{ij}^{[l]}(a\mu) ,
\end{align}
where we factor out $u_0$ for the asqtad quarks and $\tilde{\mathcal{C}}=2\kappa'_bu_0(1+\tilde{m}_0)$,
$\tilde{m}_0=m_0/u_0$, for the heavy quarks.
We call this the ``tadpole-improved perturbative matching.'' %
The values of $\crit$ and $u_0$ are calculated nonperturbatively on each ensemble and collected in
Table~\ref{tab:ZV}.
The one-loop tadpole-improved coefficients above are given by
\begin{equation} 
    \zeta_{ij}^{[1]} = Z_{ij}^{[1]} - 
        \delta_{ij} u_0^{[1]}\left(\frac{9}{4}+ \frac{1}{1+\tilde{m}_0}\right), \label{eq:tadZs}    
\end{equation}
where the one-loop value of $u_0$ is defined by $u_0=1+\alpha_s u_0^{[1]}+\order(\alpha^2_s).$
For the fourth root of the plaquette, $u_{0\text{P}}^{[1]}=-0.76708(2)$, while for the link in 
Landau gauge, $u_{0\text{L}}^{[1]}=-0.750224(3)$; here we examine tadpole improvement with both.

The other method is to factor out the renormalization associated with the vector
currents~\cite{ElKhadra:2001rv,Harada:2001fi}, defining $\rho_{ij}$ by
\begin{align}
    Z_{ij}    &= \ZVbb\ZVll\rho_{ij},
    \label{eq:matchingmNPR} \\
    \rho_{ij} &= \delta_{ij} + \sum_{l=1} \alpha_s^l \rho_{ij}^{[l]}(a\mu),
\end{align}
and computing the matching factors \ZVbb\ and \ZVll\ nonperturbatively.
Then
\begin{equation}
    \rho_{ij}^{[1]} = Z_{ij}^{[1]}-\delta_{ij}\left(\ZVbb^{[1]}+\ZVll^{[1]}\right),
    \label{eq:ziimNPR}
\end{equation}
where, perturbatively,
\begin{align} 
    \ZVbb &= \tilde{\mathcal{C}}\left[1+\alpha_s\ZVbb^{[1]}+ \order(\alpha_s^2)\right],
    \label{eq:ZVbb1loop}\\
    \ZVll &= 2 u_0 \left[1+\alpha_s\ZVll^{[1]}+ \order(\alpha_s^2)\right].
    \label{eq:ZVll1loop}
\end{align}
Below we call the method based on Eq.~(\ref{eq:matchingmNPR}) ``mostly nonperturbative matching'' (mNPR).

Table~\ref{tab:rhos} gives the results~\cite{GamizPT} for $\rho_{ij}^{[1]}$ and
$\zeta_{ii}^{[1]}-\rho_{ii}^{[1]}$ [which is the same for all $i$, as is clear from
Eqs.~(\ref{eq:tadZs}) and (\ref{eq:ziimNPR})] at the lattice spacings and bottom-quark
masses employed in this work.
\begin{sidewaystable}
    \centering
    \caption{One-loop renormalization coefficients in Eq.~(\ref{eq:matchingmNPR}) at the renormalization 
        scale $\mu$ corresponding to the tree-level pole mass.
        The entries correspond to the BBGLN~\cite{Beneke:1998sy,Beneke:2002rj} choice of evanescent 
        operators.
        The BMU~\cite{Buras:2000if} evanescent scheme can be obtained from Eqs.~(\ref{eq:BBGLN2BMU}).
        The right-most column gives the difference between the tadpole-improved perturbative method 
        using $u_{0\text{P}}$ and the mostly nonperturbative approach; the 
        difference is the same for all diagonal elements (and zero off the diagonal).
        The errors from the VEGAS numerical integration are a few in the last digit shown and, thus, 
        negligible compared with the uncertainty from truncating the perturbative series expansion.
        The three sets of entries with $a\approx 0.12$~fm and $am'_l/am'_s = 0.01/0.05$ correspond, 
        from top to bottom, to $\kappa'_b=0.0901, 0.0860, 0.0820$, which are the values used in the 
        bottom-quark mass-correction analysis discussed in Sec.~\ref{sec:kappa}.}
    \label{tab:rhos}
    \setlength{\tabcolsep}{5pt}
        \begin{tabular}{cr@{/}l rrrrr rrrrr c}
        \hline\hline
        $\approx a$~(fm) & $am'_l$&$am'_s$ & $\rho^{[1]}_{11}$~~ & $\rho^{[1]}_{12}$~~ & $\rho^{[1]}_{22}$~~ & $\rho^{[1]}_{21}$~~ & $\rho^{[1]}_{33}$~~ & $\rho^{[1]}_{31}$~~ & $\rho^{[1]}_{44}$~~ & $\rho^{[1]}_{45}$~~ & $\rho^{[1]}_{55}$~~ & $\rho^{[1]}_{54}$~~ & $\zeta^{[1]}_{ii}-\rho^{[1]}_{ii}$\\
        \hline
        0.12  & 0.02&0.05 & $-0.2684$ & $-0.3115$ & $-0.0300$ &  $0.0211$ &  $0.3641$ & $-0.0280$ & $-0.1421$ & $-0.2974$ & $-0.0353$ & $-0.2290$ &  $0.3167$ \\
              & 0.01&0.05 & $-0.2733$ & $-0.3224$ &  $0.0072$ &  $0.0213$ &  $0.3492$ & $-0.0215$ & $-0.0818$ & $-0.2952$ & $-0.0370$ & $-0.2025$ &  $0.3214$ \\
              & 0.01&0.05 & $-0.2835$ & $-0.3703$ &  $0.1601$ &  $0.0199$ &  $0.2943$ &  $0.0032$ &  $0.1701$ & $-0.2829$ & $-0.0333$ & $-0.0916$ &  $0.3398$ \\
              & 0.01&0.05 & $-0.2806$ & $-0.3470$ &  $0.0868$ &  $0.0210$ &  $0.3194$ & $-0.0082$ &  $0.0490$ & $-0.2896$ & $-0.0368$ & $-0.1453$ &  $0.3313$ \\
              & 0.007&0.05 & $-0.2736$ & $-0.3225$ &  $0.0076$ &  $0.0213$ &  $0.3490$ & $-0.0214$ & $-0.0813$ & $-0.2952$ & $-0.0369$ & $-0.2024$ &  $0.3215$ \\
              & 0.005&0.05 & $-0.2735$ & $-0.3225$ &  $0.0076$ &  $0.0212$ &  $0.3490$ & $-0.0214$ & $-0.0812$ & $-0.2952$ & $-0.0367$ & $-0.2024$ &  $0.3215$ \\
        \hline
        0.09  & 0.0124&0.031 & $-0.2323$ & $-0.2659$ & $-0.1989$ &  $0.0186$ &  $0.4394$ & $-0.0612$ & $-0.4149$ & $-0.3036$ & $-0.0195$ & $-0.3471$ &  $0.2944$ \\
              & 0.0062&0.031 & $-0.2349$ & $-0.2681$ & $-0.1903$ &  $0.0188$ &  $0.4353$ & $-0.0594$ & $-0.4020$ & $-0.3034$ & $-0.0207$ & $-0.3413$ &  $0.2956$ \\
              & 0.00465&0.031 & $-0.2363$ & $-0.2694$ & $-0.1851$ &  $0.0189$ &  $0.4330$ & $-0.0583$ & $-0.3937$ & $-0.3033$ & $-0.0212$ & $-0.3375$ &  $0.2963$ \\
              & 0.0031&0.031 & $-0.2371$ & $-0.2702$ & $-0.1820$ &  $0.0190$ &  $0.4316$ & $-0.0577$ & $-0.3892$ & $-0.3033$ & $-0.0214$ & $-0.3357$ &  $0.2967$ \\
              & 0.00155&0.031 & $-0.2371$ & $-0.2703$ & $-0.1816$ &  $0.0190$ &  $0.4315$ & $-0.0576$ & $-0.3890$ & $-0.3033$ & $-0.0217$ & $-0.3355$ &  $0.2968$ \\
        \hline
        0.06  & 0.0072&0.018 & $-0.1666$ & $-0.2160$ & $-0.4038$ &  $0.0121$ &  $0.5438$ & $-0.1064$ & $-0.7425$ & $-0.3051$ &  $0.0169$ & $-0.4864$ &  $0.2645$ \\
              & 0.0036&0.018 & $-0.1626$ & $-0.2135$ & $-0.4149$ &  $0.0116$ &  $0.5495$ & $-0.1089$ & $-0.7604$ & $-0.3050$ &  $0.0192$ & $-0.4935$ &  $0.2628$ \\
              & 0.0025&0.018 & $-0.1629$ & $-0.2136$ & $-0.4140$ &  $0.0117$ &  $0.5491$ & $-0.1088$ & $-0.7588$ & $-0.3050$ &  $0.0191$ & $-0.4931$ &  $0.2629$ \\
              & 0.0018&0.018 & $-0.1629$ & $-0.2136$ & $-0.4137$ &  $0.0117$ &  $0.5490$ & $-0.1087$ & $-0.7588$ & $-0.3051$ &  $0.0189$ & $-0.4930$ &  $0.2630$ \\
        \hline
        0.045 & 0.0028&0.014 & $-0.0115$ & $-0.1454$ & $-0.7594$ & $-0.0061$ &  $0.7490$ & $-0.1947$ & $-1.3044$ & $-0.2974$ &  $0.1084$ & $-0.7197$ &  $0.2066$ \\
        \hline\hline
    \end{tabular}

\end{sidewaystable}
The values for $\zeta_{ii}^{[1]}-\rho_{ii}^{[1]}$ in Table~\ref{tab:rhos} are obtained using $u_0$ from
the plaquette; the analogous factors using the Landau-link $u_0$ can be deduced from Eqs.~(\ref{eq:tadZs})
and (\ref{eq:ziimNPR}) using the $Z$ factors and tadpole-improvement factors in Table~\ref{tab:ZV}, and the
values of $u_{0\text{P}}^{[1]}$ and $u_{0\text{L}}^{[1]}$ given above.
The entries in Table~\ref{tab:rhos} correspond to the BBGLN~\cite{Beneke:1998sy,Beneke:2002rj} choice of
evanescent operators.
To match to the BMU~\cite{Buras:2000if} evanescent scheme, only a few entries
change~\cite{Becirevic:2001xt,Monahan:2014xra}:
\begin{subequations}\label{eq:BBGLN2BMU}
\begin{align}
    \rho^{[1]}_{22} &\to \rho^{[1]}_{22} - \frac{1}{\pi}  , \\  
    \rho^{[1]}_{21} &\to \rho^{[1]}_{21} - \frac{1}{24\pi}, \\  
    \rho^{[1]}_{33} &\to \rho^{[1]}_{33} + \frac{1}{3\pi} , \\  
    \rho^{[1]}_{31} &\to \rho^{[1]}_{31} - \frac{1}{24\pi},     
\end{align}
\end{subequations}
with the same changes for the corresponding $\zeta^{[1]}_{ii}$ (because no lattice diagrams enter in the
difference between BBGLN and BMU).

When matching the lattice regulator to \MSbar, the scale $\mu$ enters only via $\mu a$.
We choose $\mu a$ equal to the tree-level pole mass of the bottom quark, in lattice units, which is computed
from the (dimensionless) quark-mass parameters of the lattice action.
We need not, and do not, specify $\mu$ in physical units.
In Sec.~\ref{sec:results}, however, we need to specify a scale to obtain several useful results and find
$\mu=\overline m_b$, namely the \MSbar\ mass, to be convenient.
We do not distinguish between $\overline m_b$ and the tree-level lattice pole mass, because the difference
in the matching factor is of order~$\alpha_s^2$.

For the strong coupling, we set $\alpha_s=\alpha_V(q^*)$, where $\alpha_V$ is the renormalized coupling in
the $V$~scheme~\cite{Lepage:1992xa} and $q^*$ is a typical gluon loop momentum in the
process~\cite{Lepage:1992xa,Hornbostel:2002af}.
We use $q^*=2/a$, the same choice made for heavy-light currents with the same actions and
ensembles~\cite{Bailey:2014tva,Lattice:2015tia,Bailey:2015dka}.
Other reasonable choices of $q^*$ would lead to differences of order $\alpha_s^2$, which are incorporated
into the functional form for the chiral-continuum extrapolation, described in Sec.~\ref{sec:ChPT}.
The values of $\alpha_V(q^*)$ are obtained as in Ref.~\cite{Mason:2005zx} and are listed in
Table~\ref{tab:ZV}.
\begin{table}[tp]
\centering
\caption{Strong coupling in the $V$~scheme at the scale $\mu=2/a$, and renormalization factors for
heavy-heavy and light-light vector currents used in the matching relation Eq.~(\ref{eq:matchingmNPR}).
The errors shown on \ZVbb\ and \ZVll\ are statistical only.
Note that these renormalization factors are not expected to be close to 1 because of
the normalization conventions in Eqs.~(\ref{eq:ZVbb1loop}) and (\ref{eq:ZVll1loop}).  Also shown are the nonperturbatively determined critical hopping parameter and tadpole-improvement factors from the fourth root of the plaquette and (for some ensembles) the link in Landau gauge. }
\label{tab:ZV}
\setlength{\tabcolsep}{5pt}
\begin{tabular}{cclllcccc}
\hline\hline
$\approx a$~(fm) & $am'_l/am'_s$&~~$\kappa'_b$&~~\crit&~~$u_{0\text{P}}$ & $u_{0\text{L}}$ &$\alV(2/a)$& \ZVbb    & \ZVll    \\
\hline
0.12             &    0.02/0.05  & 0.0918 & 0.14073  & 0.8688 & 0.837 & 0.3047 & 0.4928(5) & 1.734(3) \\
                 &    0.01/0.05  & 0.0901 & 0.14091  & 0.8677 & 0.835  & 0.3108 & 0.5031(5) & 1.729(3) \\
                 &    0.01/0.05  & 0.0860 & 0.14091  & 0.8677 & 0.835  & 0.3108 & 0.5266(6) & 1.729(3) \\
                 &    0.01/0.05  & 0.0820 & 0.14091  & 0.8677 & 0.835  & 0.3108 & 0.5494(6) & 1.729(3) \\ 
                 &   0.007/0.05  & 0.0901 & 0.14095  & 0.8678  & 0.836 & 0.3102 & 0.5030(5) & 1.730(3) \\
                 &   0.005/0.05  & 0.0901 & 0.14096  & 0.8678 & 0.836 & 0.3102 & 0.5030(5) & 1.729(3) \\
\hline
0.09             &  0.0124/0.031 & 0.0982 & 0.139052 & 0.8788  & -- & 0.2582 & 0.4511(5) & 1.768(4) \\ 
                 &  0.0062/0.031 & 0.0979 & 0.139119 & 0.8782 & 0.854  & 0.2607 & 0.4531(5) & 1.766(4) \\ 
                 & 0.00465/0.031 & 0.0977 & 0.139134 & 0.8781 & --  & 0.2611 & 0.4543(5) & 1.766(4) \\
                 &  0.0031/0.031 & 0.0976 & 0.139173 & 0.8779  & -- & 0.2619 & 0.4550(5) & 1.765(4) \\
                 & 0.00155/0.031 & 0.0976 & 0.139190 & 0.877805 & -- & 0.2623 & 0.4550(5) & 1.765(4) \\
\hline
0.06             &  0.0072/0.018 & 0.1048 & 0.137582 & 0.8881 & --  & 0.2238 & 0.4088(5) & 1.798(5) \\  
                 &  0.0036/0.018 & 0.1052 & 0.137632 & 0.88788 & -- & 0.2245 & 0.4065(5) & 1.797(5) \\
                 &  0.0025/0.018 & 0.1052 & 0.137667 & 0.88776 & 0.869 & 0.2249 & 0.4066(5) & 1.797(5) \\
                 &  0.0018/0.018 & 0.1052 & 0.137678 & 0.88764 & 0.869 & 0.2253 & 0.4066(5) & 1.796(5) \\
\hline
0.045            &  0.0028/0.014 & 0.1143 & 0.136640 & 0.89511 & 0.8797 & 0.2013 & 0.3502(4) & 1.818(8) \\
\hline\hline
\end{tabular}
\end{table}

We define the flavor-conserving quantities \ZVbb\ and \ZVll\ nonperturbatively through the standard
charge-normalization conditions:
\begin{align}
    \langle B_s|B_s\rangle &= \ZVbb a^3\sum_{\bm{x}} \langle B_s|\bar{\Psi}\gamma^4\Psi(x)|B_s\rangle,
    \label{eq:ZVbb} \\
    \langle D_l|D_l\rangle &= \ZVll a^3\sum_{\bm{x}}
        \langle D_l|\bar{\Upsilon}_l\gamma^4\Upsilon_l{(x)}|D_l\rangle.
    \label{eq:ZVll}
\end{align}
The factors \ZVbb\ and \ZVll\ are computed from two- and three-point functions, as discussed in
Ref.~\cite{Bazavov:2011aa}, on the ensembles used in this work.
The results are listed in Table~\ref{tab:ZV}.
We use the same light-light renormalization factor \ZVll\ for all light valence-quark masses (on the same
ensemble) because its dependence on $m_q$ is very mild.
At tree level, the product of these two factors is equal to 
$2u_0\tilde{\mathcal{C}}$ 
in Eq.~(\ref{eq:matchingPT}), since it is simply the combined normalization of the lattice fields.

The mostly nonperturbative method trades some pieces that contribute to the one-loop renormalization with
corresponding nonperturbative pieces.
In particular, the heavy- and light-quark wavefunction renormalizations cancel at one loop in the difference
in Eq.~(\ref{eq:ziimNPR}), and they are instead included nonperturbatively in the matching relation
Eq.~(\ref{eq:matchingmNPR}) via the factors \ZVbb\ and \ZVll.
For operators $\latop_1$, $\latop_3$, and $\latop_5$, the one-loop coefficients are generally smaller, but
for $\latop_2$ and $\latop_4$ that is not the case.
That said, the mNPR method is our preferred choice.
In particular, on the finer lattices it works the best for $\langle\op_1\rangle$, the matrix element with
the strongest phenomenological motivation to achieve high precision.

\section{Bottom-quark mass correction}
    \label{sec:kappa}
    
As is customary with Wilson fermions, the Fermilab action uses the hopping parameter $\kappa$ to parametrize
the bare quark mass.
We rely on the $\kappa$-tuning analysis discussed in detail in Ref.~\cite{Bailey:2014tva}, which yielded the
physical values for $\kappa_b$ listed in Table~\ref{tab:m2diff}.
\begin{figure}[tb]
	\vspace{-0.2in}
	\includegraphics[width=0.5\textwidth, angle=270]{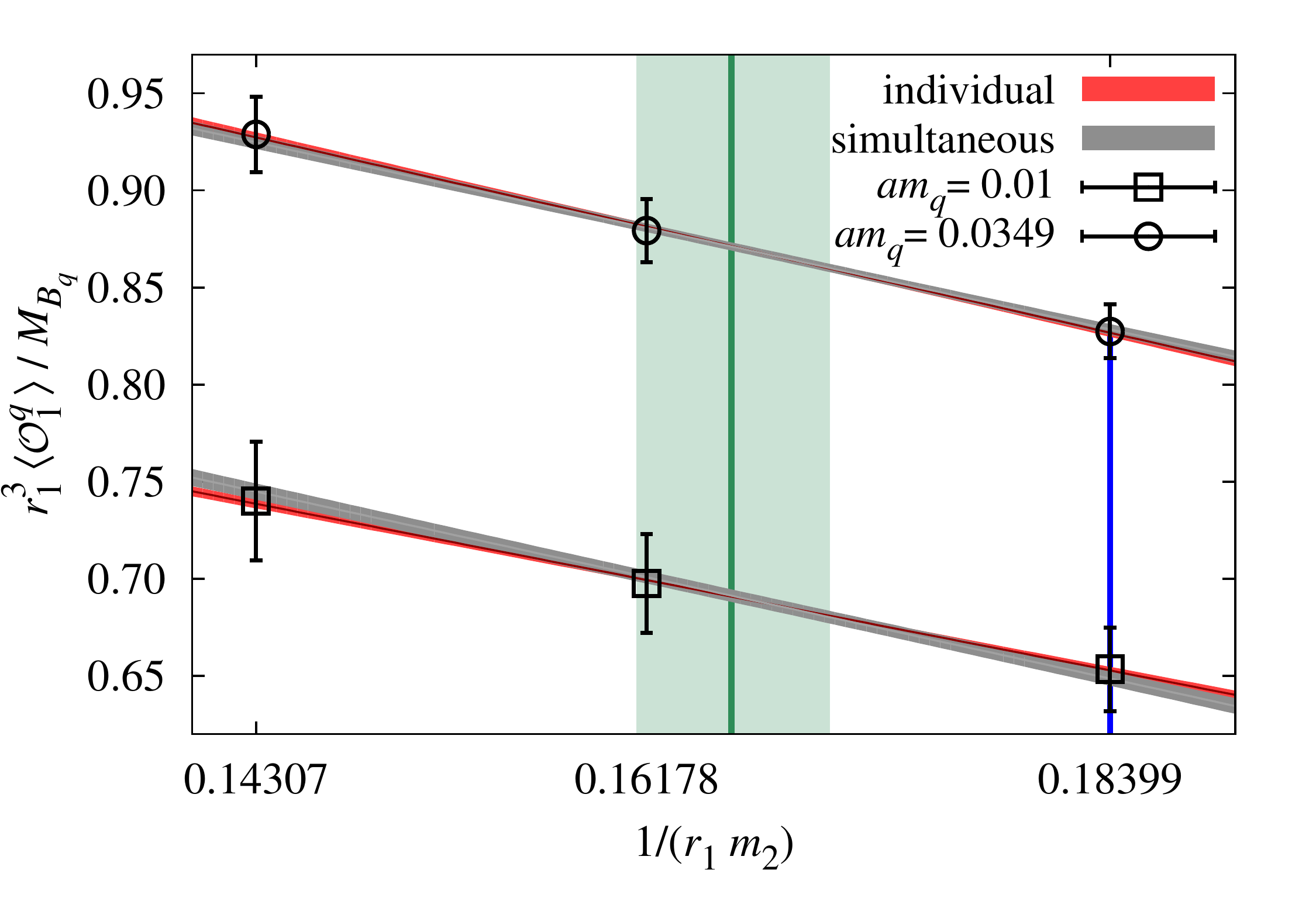}
    \caption{The variation of $r_1^3 \langle {\op}^q_1 \rangle / M_{B_q}$ with $1/(r_1 m_2)$.
    The blue vertical line indicates the simulated $\kappa'_b$.
    The solid green vertical line indicates the value of $1/(r_1 m_2)$ corresponding to the physical
    $\kappa_b$ obtained in Ref.~\cite{Bailey:2014tva}, while the filled band shows the error from
    statistics plus fitting systematics and the lattice-scale uncertainty.
    Results from both separate fits of the two valence-quark masses $am_q = 0.01, 0.0349$ (red lines) and a
    simultaneous fit with a common slope (gray bands) are shown.}
	\label{fig:kappa}
\end{figure}
In brief, low statistics runs were initially used to select the simulation values $\kappa'_b$ that were
subsequently used in the full-statistics runs.
After production running, the physical values $\kappa_b$ were obtained by requiring the simulation
$B_s$-meson kinetic mass to match the PDG value.

\begin{table}
\centering
\setlength{\tabcolsep}{4pt}
\caption{Tuned $\kappa_b$ values obtained in Ref.~\cite{Bailey:2014tva}, and differences 
$\Delta(1/(r_1m_2))$ between the simulated and physical inverse $b$-quark kinetic masses on each ensemble. 
For $\kappa_b$, the first error is from statistics and fitting, and the second is due to the uncertainty in 
$r_1$. For $\Delta(1/(r_1m_2))$ the error is the quadrature sum of the uncertainties in $\kappa_b$ and in 
$r_1$.}
\label{tab:m2diff}
\begin{tabular}{l c c r}\hline\hline
 & & \vspace{-1.0em}\\
$\approx a$~(fm)   &  $a{m}'_l/am'_s$  &  ~~$\kappa_b$ & $\Delta(1/(r_1m_2))\,$   \\ 
 & & \vspace{-1.0em}\\ \hline
0.12 & 0.02/0.05 & 0.0879(9)(3) & $-$0.0221(50) \\ 
0.12 & 0.01/0.05 & 0.0868(9)(3) & $-$0.0181(49)\\ 
0.12 & 0.007/0.05 & 0.0868(9)(3) & $-$0.0181(49)\\
0.12 & 0.005/0.05 & 0.0868(9)(3) & $-$0.0181(49) \\ \hline
0.09 & 0.0124/0.031 & 0.0972(7)(3) & $-$0.0061(45)\\
0.09 & 0.0062/0.031 & 0.0967(7)(3) & $-$0.0073(45)\\
0.09 & 0.00465/0.031 & 0.0966(7)(3) & $-$0.0067(45)\\
0.09 & 0.0031/0.031 & 0.0965(7)(3) & $-$0.0067(45)\\
0.09 & 0.00155/0.031 & 0.0964(7)(3) & $-$0.0073(45)\\ \hline 
0.06 & 0.0072/0.018 & 0.1054(5)(2) & 0.0041(38)\\
0.06 & 0.0036/0.018 & 0.1052(5)(2) & 0.0000(37)\\
0.06 & 0.0025/0.018 & 0.1051(5)(2) & $-$0.0007(37)\\
0.06 & 0.0018/0.018 & 0.1050(5)(2) & $-$0.0014(37)\\ \hline 
0.045 & 0.0028/0.014  & 0.1116(3)(2) & $-$0.0251(30)\\ \hline\hline
\end{tabular}
\end{table}

Here, we describe how we correct the matrix elements \emph{a posteriori} to account for the fact that they
were not computed at the physical $\kappa_b$.
We simulate all five operators at three values of $\kappa'_b$ on the $a\approx0.12$~fm,
$am'_l/am'_s=0.010/0.05$ ensemble.
The three $\kappa'_b$ values include the simulation value and two others chosen to straddle the physical
value.
We obtain the bare lattice matrix elements $\langle\latop_i^q\rangle/M_{B_q}$ as described in
Sec.~\ref{sec:3pts} for each of the three $\kappa'_b$ values, and at two values of the valence light-quark
mass, $am_q = 0.01$ and $0.0349$, which correspond roughly to the $B_d$ and $B_s$ mesons.
We then convert the matrix elements to $r_1$ units using the $r_1/a$ values listed in Table~\ref{tab:LatEns}
and match them to a continuum scheme via the mNPR method described in Sec.~\ref{sec:PT}.
In the rest of the paper, we simply denote the renormalized lattice operators by $\op_i$, because their
matrix elements differ from the continuum ones only by discretization and perturbative truncation errors
that the systematic error estimate takes into account.

For each operator, we fit the data for $r_1^3\langle\op_i^q\rangle/M_{B_q}$ using a function
linear in the inverse tree-level kinetic quark mass, $1/(r_1 m_2)$, to obtain the slope
\begin{equation}
	\mu_i^q \equiv
        \frac{ \Delta(r_1^3 \langle \op_i^q \rangle / M_{B_q}) } {\Delta(1/(r_1m_2))},
    \label{eq:slope}
\end{equation}
with an accompanying uncertainty from the fit.
This form is chosen because heavy-quark
physics suggests a mild dependence on $1/m_Q$, which is identified with the inverse kinetic
mass $1/m_2$ in the Fermilab interpretation.
We try both separate fits for each set of valence light-quark data, and simultaneous fits to all data.
Figure~\ref{fig:kappa} shows these fits for $r_1^3 \langle\op_1^q\rangle/M_{B_q}$.

For all operators, we find the slopes to be independent of $am_q$.
We therefore use $am_q$-independent slopes $\mu_i$ from the simultaneous fits to
determine the $\kappa_b$-correction for all valence-quark masses and ensembles.
The slopes $\mu_i$ for each of the five operators are given in Table~\ref{tab:slopes}.

On each ensemble, we calculate the differences in inverse kinetic masses, $\Delta (1/(r_1 m_2))$, from the
difference between the simulated and tuned values of $\kappa_b$, incorporating the uncertainties in
$\kappa_b$ and $r_1$, as listed in Table~\ref{tab:m2diff}.
Finally, we obtain the needed corrections to the matrix elements from the product
$\mu_i\times\Delta(1/(r_1m_2))$.
The corrections are small, with relative shifts on the matrix elements of a few percent or less on the
$a\approx 0.09$ and 0.06~fm ensembles, and about 10\% or less on the $a \approx 0.12$ and 0.045~fm
ensembles, where the differences between the tuned and simulation $\kappa$ values are largest.

The way we incorporate the $b$-quark mass corrections into the chiral-continuum fits is described in
Sec.~\ref{sec:ChPT}.
Briefly, we include the shifts in the matrix elements as fit parameters constrained by priors with central
values set to the calculated shift size and widths set to the computed errors.
This approach allows us to propagate the uncertainties in our determinations of the slopes and in the
physical values of $\kappa_b$ directly into the chiral-continuum fit error.

\begin{table}
\centering
\caption{Slopes $\mu_i$ defined in Eq.~(\ref{eq:slope}) for the $B_q$-mixing matrix elements
renormalized in the continuum \MSbar-NDR-BBGLN~\cite{Beneke:1998sy,Beneke:2002rj} and
\MSbar-NDR-BMU~\cite{Buras:2000if} schemes, where the last acronym refers to the choice of evanescent
operators.
For each operator and scheme, the slope is obtained from a simultaneous fit to data with three $\kappa'_b$
values and two light valence-quark masses on the $a\approx$ 0.12 fm, $am'_l/am'_s = 0.01/0.05$ ensemble. 
}
\label{tab:slopes}
\begin{tabular}{l@{\quad}c@{\quad}cc@{\quad}cc@{\quad}c@{\quad}c}\hline\hline
 & \vspace{-1.0em}\\
$i$ & 1 & \multicolumn{2}{c}{2} & \multicolumn{2}{c}{3} & 4 & 5 \\ 
	& & BMU & BBGLN & BMU & BBGLN \\
& \vspace{-1.0em} \\ \hline
$\mu_i$ & $-$2.35(11)  	& 1.252(70) 	& 1.288(70)	& 0.130(24)	& 0.115(24)	& $-$2.74(27)	& $-$1.86(18) \\ \hline\hline  
\end{tabular}
\end{table}

\section{Chiral-continuum extrapolation}
    \label{sec:ChPT}
    We extrapolate the $B_q$-mixing matrix elements to the physical light-quark masses and the continuum limit
using SU(3), partially-quenched, heavy-meson, staggered chiral perturbation theory
(\hmrscpt)~\cite{Aubin:2005aq}.
The \hmrscpt\ expressions for $B_q$-mixing were derived in Ref.~\cite{Bernard:2013dfa} by generalizing the
continuum calculation of Ref.~\cite{Detmold:2006gh} to include the taste-symmetry-breaking effects of the
staggered light quarks.
References~\cite{Detmold:2006gh,Bernard:2013dfa} work to one loop in \cpt\ and use an effective Lagrangian
for the $B_q$ meson derived at leading order in~$1/m_b$.
As noted below, however, some effects of order~$1/m_b$ are included in our chiral-continuum fit function,
both explicitly and implicitly.

Every ensemble has a different lattice spacing, even those with the same nominal value (\emph{i.e.}, the
``$\approx a$'' listed in many tables).
As explained in Sec.~\ref{sec:simulation}, we multiply the matrix elements by the appropriate power of
$r_1/a$ to bring them into dimensionless, but physical, units before performing the chiral-continuum
extrapolation.
The \cpt\ expressions require some external inputs, which we also bring into $r_1$ units with the value in
Eq.~(\ref{eq:r1}).

Section~\ref{sec:chiralfitfunc} provides the expressions for the chiral fit function employed, both
nonanalytic chiral logarithms and additional analytic terms, while Sec.~\ref{sec:chptfit} discusses
specifics of the fixed inputs and priors used in the chiral fit.
We will refer to this fit, which includes uncertainty contributions associated with the choice of
fit function and inputs, as our \emph{base} fit.
The results of the base fit are given in Sec.~\ref{sec:basefit}.
Below, in Sec.~\ref{sec:syserr}, we study the stability of our fit results under reasonable modifications to
the fit function and input data, showing that the errors quoted for the base fit include the uncertainties
due to truncating the chiral and heavy-quark expansions and encompass the range suggested by the other cross
checks.

\subsection{Chiral fit function}        
\label{sec:chiralfitfunc}

Schematically, the fit function for each matrix element takes the form
\begin{equation}
    F_i = F_i^\text{logs} + F_i^\text{analytic} + F_i^\text{HQ disc} +
        F_i^{\alpha_sa^2~\text{gen}} + F_i^\kappa + F_i^\text{renorm},     
    \label{eq:chiralfitfunc-schematic}
\end{equation}
in which the individual terms are functions of the heavy and light meson masses, the lattice spacing, and
the low-energy constants (LECs) of \hmrscpt.
The first term denotes the expression for the one-loop \hmrscpt\ chiral logarithms, which contains
nonanalytic dependence on the light-quark masses and lattice spacing.
The next term, $F_i^\text{analytic}$, represents analytic terms in the chiral expansion, namely, a
polynomial in the light-quark masses and lattice spacing; this term is needed to cancel the scale dependence
in $F_i^\text{logs}$.
The remaining terms lie outside \hmrscpt\ and parametrize other sources of systematic uncertainty.
We account for heavy-quark discretization errors via $F_i^\text{HQ disc}$ and generic light-quark and gluon
discretization errors of order $\alpha_sa^2$ via $F_i^{\alpha_sa^2\,\text{gen}}$.
The next term $F_i^\kappa$ accounts for the uncertainty in the adjustment of the matrix elements from their
values at the simulated heavy-quark mass $\kp'_b$ to the physical~$\kp_b$.
Finally, $F_i^\text{renorm}$ models neglected higher-order contributions to the operator renormalization,
including off-diagonal terms in~$Z_{ij}$.
Below, each of these terms is defined in detail.

\subsubsection{Chiral logarithms}
\label{sec:chptLogs}

The complete NLO \hmrscpt~expressions for the chiral logarithms are given in Ref.~\cite{Bernard:2013dfa}.
For matrix element $\langle\op_1^q\rangle$,
\begin{align}
    F_1^\text{logs} &= \beta_1 \left( 1 +
        \frac{\mathcal{W}_{q\bar{b}}+\mathcal{W}_{b\bar{q}}}{2} +
        \mathcal{T}_q^\text{(1,2,3)} + \tilde{T}_q^{(\rm{1a})} + 
        \mathcal{Q}_q^{(i)} + \tilde{Q}_q^{(\text{1a})}  \right)
    \nonumber \\ & \hspace*{0.8em}
        + (\beta_2+\beta_3) \tilde{T}^\text{(1b)}_q + (\beta'_2+\beta'_3) \tilde{Q}^\text{(1b)}_q.
    \label{eq:ME1chipt}
\end{align}
For $\langle\op_{2,3}^q\rangle$,
\begin{align}
    F_i^\text{logs} &= \beta_i \left( 1 +
        \frac{\mathcal{W}_{q\bar{b}}+\mathcal{W}_{b\bar{q}}}{2}
        + \mathcal{T}_q^\text{(1,2,3)} + \tilde{T}_q^{(\rm{23a})} \right)
        + \beta'_i \left( \mathcal{Q}_q^{(i)} + \tilde{Q}^{(\text{23a})}_q  \right)
    \nonumber \\
    & \hspace*{0.8em}
        + \beta_1 \tilde{T}^{(\rm{23b})}_q + \beta_j \tilde{T}^{(\rm{23c})}_q
        + \beta_1 \tilde{Q}^{(\rm{23b})}_q + \beta'_j \tilde{Q}^{(\rm{23c})}_q 
   \label{eq:ME23chipt}
\end{align}
where $i,j = 2,3$ and $j\neq i$.
Finally, for $\langle\op_{4,5}^q\rangle$, 
\begin{align}
    F_i^\text{logs} &= \beta_i \left( 1 +
        \frac{\mathcal{W}_{q\bar{b}}+\mathcal{W}_{b\bar{q}}}{2}
        + \mathcal{T}_q^{(\rm 4,5)} + \tilde{T}^{(\rm{45a})}_q  \right) 
        + \beta'_i \left( \mathcal{Q}_q^{(i)} + \tilde{Q}^{(\text{45a})}_q \right)
    \nonumber \\
    & \hspace*{0.8em} + \beta_j \tilde{T}^{(\rm{45b})}_q + \beta'_j \tilde{Q}^{(\rm{45b})}_q,
    \label{eq:ME45chipt}
\end{align}
where $i,j = 4,5$ and, again, $j\neq i$.
In Eqs.~(\ref{eq:ME1chipt})--(\ref{eq:ME45chipt}), the coefficients $\beta_i$ and $\beta'_i$ are the
leading-order LECs for the matrix elements $\langle\bar{B}|\op_i|B\rangle$ and
$\langle\bar{B}^*|\op_i|B^*\rangle$, respectively, and they are the same as in the \cpt\ description of
continuum QCD~\cite{Detmold:2006gh}.
The terms $\mathcal{W}_{q\bar{b}}=\mathcal{W}_{b\bar{q}}$, $\mathcal{T}_q$, and $\mathcal{Q}_q$ are
standard contributions~\cite{Detmold:2006gh} from wave-function renormalization, tadpole, and sunset
diagrams, respectively.
The terms $\tilde{T}_q$ and $\tilde{Q}_q$ stem from tadpole and sunset diagrams of the wrong-spin operators
discussed in Sec.~\ref{sec:simulation}.
Our notation for the wrong-spin terms in Eqs.~(\ref{eq:ME1chipt})--(\ref{eq:ME45chipt}) separates the LECs
from the loop-diagram functions differently from Ref.~\cite{Bernard:2013dfa}; Appendix~\ref{app:log}
contains a dictionary to translate.
Note that two additional LECs enter the one-loop expressions: the tadpole functions are proportional to
$1/f^2_\pi$, while the self-energy and sunset functions are proportional to $g_{B^*B\pi}^2/f^2_\pi$,
where $f_\pi$ is the pion decay constant and $g_{B^*B\pi}$ is the $B^*$-$B$-$\pi$ coupling.

The chiral logarithms depend on the ratio of the light pseudoscalar meson masses to the \cpt\
renormalization scale $\Lambda_\chi$.
At nonzero lattice spacing, taste-symmetry breaking splits the squared masses $M_{ab,\xi}^2$ for mesons of
different taste $\xi$:
\begin{equation}
    M_{ab,\xi}^2 = B_0 (m_a + m_b) + a^2 \Delta_\xi,
    \label{eq:piontypemass}
\end{equation}
where $m_a$ and $m_b$ are the masses of constituents $a$ and $b$, $B_0$ is the leading \cpt\ LEC, and
$a^2\Delta_\xi$ ($\xi = P, A, T, V, I$) denote taste splittings~\cite{Lee:1999zxa}.
Taste violations also give rise to quark-disconnected hairpin diagrams at one loop in \cpt, whose
contributions are suppressed by $\alpha_s^2a^2$~\cite{Aubin:2003mg}.
The wrong-spin functions $\tilde{T}_q$ and $\tilde{Q}_q$ are also of order $\alpha_s^2a^2$ after
cancellations among several terms, each of which is of order unity.

We evaluate the loop-diagram functions with nonzero hyperfine and flavor splittings, $\Delta^*=M_{B^*}-M_B$
and $\delta_{sq}=M_{B_s}-M_{B_q}$, in the propagators, as in Ref.~\cite{Bazavov:2011aa} for $B_q$-meson
decay constants.
We take the hyperfine splitting $\Delta^*$ to be independent of the light valence flavor~$q$.
We obtain the flavor splitting for arbitrary $q$ from the physical $\delta_{sd}$ as follows.
At lowest order in \hmrscpt, the flavor splitting is proportional to the quark-mass difference:
\begin{equation}
    \delta_{sq} \approx 2 \lambda_1 B_0 (m_s - m_q) \approx \lambda_1 (M_{ss}^2 - M_{qq}^2),
    \label{eq:deltasq}
\end{equation}
where $m$ and $M$ are quark and meson masses, respectively.
We then obtain the quantity $\lambda_1$ from setting $q$ to $d$ and taking $\delta_{sd}$ from the
PDG~\cite{Agashe:2014kda}.
The use of nonzero hyperfine and flavor splittings formally introduces contributions of order~$1/m_b$.

Our matrix-element data, of course, include contributions to all orders in the chiral expansion and in
$1/m_b$.
We therefore choose not to impose heavy-quark spin-symmetry relations among the $\beta_i$ and $\beta'_i$ in
order to allow the fit parameters to absorb the renormalization parts of higher-order corrections.
Because our data are all generated with a close-to-physical $b$-quark, every 
parameter implicitly absorbs some $1/m_b$ dependence.

\subsubsection{Analytic terms in the chiral expansion}
\label{sec:chptAnalytic} 

The analytic terms 
\begin{equation}
    F_i^\text{analytic} = F_i^\text{NLO} + F_i^\text{NNLO} + F_i^{\rm N^3LO} 
\end{equation}
are simple polynomials in the light-quark masses and lattice spacing.
In practice, we use the dimensionless variables
\begin{equation}
    x_q \equiv \frac{M_{qq}^2}{(8 \pi^2 f_\pi^2)},
    \label{eq:xq}
\end{equation}
and similarly for $x_l$ and $x_s$, and
\begin{equation}
    x_{\bar{\Delta}} \equiv \frac{a^2\bar{\Delta}}{(8 \pi^2 f_\pi^2)},
    \label{eq:xDelta}
\end{equation}
where
\begin{equation}
    \bar{\Delta} = \frac{1}{16} \left(
        \Delta_P  +  4\Delta_A  +  6\Delta_T  + 4 \Delta_V  +  \Delta_I \right)
    \label{eq:avetaste} 
\end{equation}
is the taste-averaged splitting, and each $\Delta_\xi$ is the taste splitting defined via
Eq.~(\ref{eq:piontypemass}).
The quantity $x_{\bar\Delta}$ is a convenient proxy for taste-breaking discretization effects.
The NLO, next-to-next-to-leading order (NNLO), and next-to-next-to-next-to-leading order (N$^3$LO) analytic
expressions contain terms linear, quadratic, and cubic, respectively, in $x_q$, $x_l$, $x_s$,
and~$x_{\bar\Delta}$:
\begin{align}
    F_i^\text{NLO}  &= \left[ c_0^i x_q + c_1^i (2 x_l + x_s)+ c_2^i x_{\bar\Delta} \right] \beta_i,
    \label{eq:FiNLO} \\[0.7em]
    F_i^\text{NNLO} &= \left[ d_0^i x_q x_{\bar\Delta} + 
        d_1^i (2x_l + x_s) x_{\bar \Delta} + d_2^i x_q (2x_l + x_s) + d_3^i x_q^2 \right.
    \nonumber \\
        & \hspace{0.4em} + \left. d_4^i (2x_l + x_s)^2 + d_5^i x_{\bar \Delta}^2 +
        d_6^i (2x_l^2+x_s^2) \right] \beta_i,
    \label{eq:FiNNLO} \\[0.7em]
    F_i^\text{N$^3$LO} &= \left[ e_0^i x_q^2 x_{\bar \Delta} + 
        e_1^i x_q (2 x_l + x_s) x_{\bar \Delta} + e_2^i x_q x_{\bar \Delta}^2  +
        e_3^i x_q^2  (2 x_l + x_s)  + e_4^i  x_q^3 \right. \nonumber\\
        & \hspace{0.4em} + e_5^i  x_q (2 x_l + x_s)^2 + e_6^i (2 x_l + x_s)^2 x_{\bar \Delta} +
        e_7^i (2 x_l + x_s) x_{\bar \Delta}^2  \nonumber \\
        & \hspace{0.4em} + e_8^i (2 x_l + x_s)^3 + e_9^i (2 x_l + x_s) (2 x_l^2 + x_s^2) +
        e_{10}^i x_{\bar \Delta}^3  \nonumber \\
        & \hspace{0.4em} + \left. e_{11}^i (2 x_l^2 + x_s^2) x_{\bar \Delta} +
        e_{12}^i (2 x_l^3 + x_s^3) + e_{13}^i  x_q  (2 x_l^2 + x_s^2) \right] \beta_i.
    \label{eq:FiNNNLO}
\end{align}
When these terms are expressed as polynomials in $x_q$, $x_l$, $x_s$, and $x_{\bar{\Delta}}$, the 
coefficients $c_n$, $d_n$, and $e_n$ are expected to be of order~1.
The N$^3$LO analytic terms $F_i^\text{N$^3$LO}$ are not included in our base fit, but are added to study the
effect of truncating the chiral expansion, as discussed in Sec.~\ref{sec:chiral-LECs-error}.

\subsubsection{Heavy- and light-quark discretization terms}
\label{sec:chptHQ}

We add the term $F_i^\text{HQ disc}$ to account for discretization errors in the $b$-quark action and the
four-quark operators.
The HQET description of lattice gauge theory~\cite{GamizPT,Kronfeld:2000ck,Harada:2001fi} 
gives for the leading terms
\begin{equation}
    F_i^\text{HQ disc} = F_i^{\alpha_s a~\text{HQ}} + F_i^{a^2~\text{HQ}} + F_i^{a^3~\text{HQ}},
\end{equation}
where each term has a different dominant lattice-spacing dependence:
\begin{align}
    F_i^{\alpha_s a~\text{HQ}} &= (r_1 \Lambda_\text{HQ})^3 (a \Lambda_\text{HQ})\, \left[
        z_B^i f_B(m_0 a) + z_3^i f_3(m_0 a)  \right], \\
    F_i^{a^2~\text{HQ}} &= (r_1 \Lambda_\text{HQ})^3 (a \Lambda_\text{HQ})^2 \left[
        z_E^i f_E(m_0 a) + z_X^i f_X(m_0 a) + z_Y^i f_Y(m_0 a) \right], \\
    F_i^{a^3~\text{HQ}} &= (r_1 \Lambda_\text{HQ})^3 (a \Lambda_\text{HQ})^3 \; z_2^i f_2(m_0a).
\end{align}
The overall factors $(r_1 \Lambda_\text{HQ})^3$ capture the typical size of the matrix elements
$\methree{i}$, the functions $f_a(m_0a)$ describe the difference between lattice and continuum
QCD~\cite{Oktay:2008ex}, and the fit parameters $z_a^i$ and the powers of $a\LamHQ$ model HQET matrix
elements.
The explicit forms of the ``mismatch'' functions $f_a(m_0a)$ are given, for example, in Appendix~A of
Ref.~\cite{Bazavov:2011aa}.

When studying the stability of our fit results in Sec.~\ref{sec:LQ-disc-error}, we also consider the term
\begin{equation}
    F_i^{\alpha_s a^2~\text{gen}} = h^{i}_0 \frac{\alpha_sa^2}{r_1^2} \beta_i
    \label{eq:gena2}
\end{equation}
to account for generic discretization errors from the asqtad-improved staggered
light-quark~\cite{Lepage:1998vj} and Symanzik-improved gluon~\cite{Luscher:1985zq} actions.
We do not, however, include $F_i^{\alpha_sa^2~\text{gen}}$ in our base fit.
This term is similar to the term $c_2^ix_{\bar\Delta}$ in Eq.~(\ref{eq:FiNLO}), which is proportional to
$\alpha_s^2a^2$ instead of~$\alpha_sa^2$.

\subsubsection{Renormalization and \texorpdfstring{$m_b$}{m\_b} correction}
\label{sec:chptRenorm}

Before carrying out the chiral-continuum extrapolation, we renormalize the matrix elements with the mostly
nonperturbative method described in Sec.~\ref{sec:PT}.
Because the matching matrix $\rho_{ij}$ in Eq.~(\ref{eq:matchingmNPR}) is available only at one-loop order
in perturbation theory, we incorporate renormalization effects of order~$\alpha_s^2$ and~$\alpha_s^3$ into
the chiral-continuum fit by adding the terms
\begin{align}
    F_i^{\alpha_s^2~\text{renorm}} &= \alpha_s^2 \rho^{[2]}_{ij} \beta_j,
    \label{eq:renorm-alpha2} \\
    F_i^{\alpha_s^3~\text{renorm}} &= \alpha_s^3 \rho^{[3]}_{ij} \beta_j,
    \label{eq:renorm-alpha3}
\end{align}
summing over $j$ to incorporate higher-order perturbative mixing.
The base fit allows $\rho^{[2]}_{ij}$ to float and sets $\rho^{[3]}_{ij}=0$, but we also examine a fit
with $\rho^{[3]}_{ij}$ floating.
When we carry out the matching with tadpole-improved perturbation theory---as a further estimate of
truncation effects---we only consider $F_i^{\alpha_s^2~\text{renorm}}$.

Section~\ref{sec:kappa} describes how we determine the shift in the matrix elements between the simulated
and physical $b$-quark mass.
We apply this correction by adding the term
\begin{equation}
    F_i^\kappa = \mu_i\, \Delta\left( \frac{1}{r_1 m_2} \right)
    \label{eq:base-tune}
\end{equation}
to our fit function.
While this term represents a correction to the data, both factors carry uncertainty.
We therefore find it convenient to introduce the slopes (for each $i$) and the shifts in $1/(r_1m_2)$ (for
each ensemble) as fit parameters with Gaussian priors.

\subsection{Chiral fit inputs}      
\label{sec:chptfit}

In the base analysis used to obtain our final results, we fit all five matrix elements simultaneously 
because they share common parameters.
Our base fit consists of the terms
\begin{equation}
    F_i^{\text{base}} = F_i^\text{logs} + F_i^\text{NLO} + F_i^\text{NNLO} +
        F_i^{\alpha_sa~\text{HQ}} + F_i^{a^2~\text{HQ}} + F_i^{a^3~\text{HQ}} + F_i^\kappa +
        F_i^{\alpha_s^2~\text{renorm}},
    \label{eq:chiralfitfunc-base}
\end{equation}
and we require shared fit parameters to be equal for all five matrix elements.
In the base fit, we employ the finite-volume expressions for the NLO logarithms, in which the one-loop
integrals become discrete sums.

The parameters common to the expressions for all operators are $f_\pi$, $g_{B^*B\pi}^2$, $\Lambda_\chi$,
$\Delta^*$, $\lambda_1$, $\delta_V'$, $\delta_A'$ in the chiral logarithms; the LECs $\beta_i$ for $i$=1--5
and $\beta_j'$ for $j$=2--5; and $\Lambda_\text{HQ}$.
The parameters that are distinct for each operator are $c_n$, $d_n$, $e_n$, in the \cpt\ analytic
terms; $z_E$, $z_2$, $z_B$, $z_3$, $z_X$, $z_Y$ in the heavy-quark discretization terms; and $h_0$ in the
generic light-quark and gluon discretization term.
The indices of the higher-order renormalization parameters, $\rho^{[2]}_{ij}$ and $\rho^{[3]}_{ij}$, make
their dependence on the operators explicit.
The $b$-quark mass tuning term $F_i^\kappa$ contains a slope $\mu_i$ for each operator and a difference
$\Delta(1/(r_1m_2))$ for each ensemble.

Every parameter is constrained with a Gaussian prior~\cite{Lepage:2001ym}.
In Secs.~\ref{sec:inputs-priors-loose} and~\ref{sec:inputs-priors}, we explain the prior central values and
widths used for each one.
In addition, several inputs are fixed in our fits; we present their numerical values in
Sec.~\ref{sec:inputs-fixed}.
When converting lattice quantities to $r_1$ units, we multiply by the appropriate power of $r_1/a$ from
Table~\ref{tab:LatEns}, but treat the associated uncertainty outside the chiral-continuum fit, as discussed 
in Sec.~\ref{sec:SystErrorB}.
On the other hand, when converting external, dimensionful quantities to $r_1$ units, we
include the uncertainty in $r_1$ in Eq.~(\ref{eq:r1}) in the prior widths.

\subsubsection{Loosely constrained fit parameters}
\label{sec:inputs-priors-loose}

The most important parameters in the chiral-continuum extrapolation are constrained only loosely, to 
provide stability.

From our previous calculation of the SU(3)-breaking ratio $\xi$ with a smaller data
sample~\cite{Bazavov:2012zs}, and from \cpt\ power-counting, we expect the LECs $r_1^3\beta_i$ to be of
order~1.
For the leading LECs $\beta_i$, we use priors $\beta_i = 1(1)$ for $i = 1,3,4,5$ and
$\beta_2 = -1(1)$.
For the $\beta'_i$, which appear first at NLO, we use $\beta'_i = 0(1)$ for $i = 2,3,4,5$ (there is no
$\beta'_1$).
Note that the prior widths on $\beta_i$ are two order of magnitude larger than the statistical errors on matrix
elements obtained from the correlator fits.

As discussed in Sec.~\ref{sec:chiralfitfunc}, the chiral analytic terms are written such that their
coefficients $c_n^i$, $d_n^i$, and $e_n^i$ are expected to be of order~1 or smaller.
For all of these coefficients, we take the prior central value to be zero.
For the NLO coefficients, we use a prior width of 10 because we expect them to be well-determined by the
data.
For the NNLO and N$^3$LO coefficients, we use a width of 1, guided by \cpt\ power-counting.

We take $h_0^i=0(1)$ in the generic light-quark and gluon discretization term $F_i^{\alpha_sa^2~\text{gen}}$.
The priors for the heavy-quark discretization terms are chosen as follows~\cite{Bazavov:2011aa}.
By HQET power counting, we expect each parameter $z_a^i$ to be of order unity.
Because we do not \emph{a priori} know the signs,  we take 0 for every prior central value.
To choose the prior widths, we start by noting that each mismatch function is associated with a
higher-dimension operator in HQET, although some operators share the same mismatch function.
The width of each prior is chosen so that the width-squared equals the number of terms sharing the same
mismatch function.
Thus, the priors are 0(2) for $z_2^i$, $z_B^i$, $z_Y^i$ and $0(\sqrt{8})$ for $z_E^i$, $z_X^i$,
for all $i=1$--5.
The four-fermion operators are subject to several dimension-seven corrections, denoted $P_j$ in
Eq.~(\ref{eq:matching-all-orders}).
When the two Dirac matrices are the same, as they are for $i=1,2,3$, there are five distinct $P_j$; when
they are not, as for $i=4,5$, there are ten~\cite{Evans:2009du,GamizPT}.
Therefore, we take $z_3^i=0(\sqrt{5})$, $i=1,2,3$, and $z_3^i=0(\sqrt{10})$, $i=4,5$.

For the unknown perturbative coefficients $\rho^{[2]}_{ij}$, we use priors 0(1).
When terms of order~$\alpha_s^3$ are included, we again take priors $\rho^{[3]}_{ij}=0(1)$.

\subsubsection{Constrained fit parameters}      
\label{sec:inputs-priors}

Other inputs to the chiral-continuum extrapolation are taken from experiment or from other lattice-QCD
calculations.
In each case, the prior central value and width are precisely those suggested by the external information.

In heavy-meson \cpt, the NLO chiral logarithms 
are multiplied by the pion decay constant~$f{_\pi}$
and the $B^*$-$B$-$\pi$ coupling $g_{B^*B\pi}$.
Because the value of $r_1$ used in this paper is set from the PDG value of $f_{{\pi^{\pm}}}$, we take
\begin{equation}
    r_1 f_{{\pi^{\pm}}} = 0.2060(15) \,,
    \label{eq:r1fpi} 
\end{equation}
from the \cpt\ analysis of light pseudoscalar mesons \cite{Bazavov:2009fk}.
For $g_{B^*B\pi}$, we use unquenched lattice-QCD
calculations~\cite{Detmold:2011bp,Detmold:2012ge,Bernardoni:2014kla,Flynn:2015xna}, taking
\begin{equation}
    g_{B^*B\pi} = 0.45(8),
    \label{eq:gBBpi}
\end{equation}
with an error covering the spread among the different results.

To choose a value for $\Delta^*$, we could consider the hyperfine splitting in the $B^0$ system, the
$B_s$ system, or our lattice-QCD data.
We showed earlier~\cite{Bernard:2010fr} that the asqtad ensembles reproduce the measured $B_s^*$-$B_s$
hyperfine splitting, within lattice-QCD uncertainties. %
The experimental average for the hyperfine splitting for the $B^0$ system is more accurate than that
for the $B_s$ system, although they are close to each other.
Therefore, we choose the $B^0$ system, for which $\Delta^*=45.78(35)$~MeV~\cite{Agashe:2014kda}, or
\begin{equation}
    r_1 \Delta^* =  0.07231(75),
\end{equation}
including the error on $r_1$.

The experimental value for the flavor splitting is
$\delta_{sd}=M_{B_s}-M_{B_d}=87.19(29)$~MeV~\cite{Agashe:2014kda}.
Plugging this number with $M_{\pi^0}=134.9766(6)~\rm{MeV}$~\cite{Agashe:2014kda} and 
$M_{\eta_s}=685.8(4.0)$~MeV~\cite{Davies:2009tsa} into Eq.~(\ref{eq:deltasq}) gives 
$\lambda_1=0.1929(35)~\text{GeV}^{-1}$, or
\begin{equation}
    \lambda_1/r_1 = 0.1221(18) 
\end{equation}
for the parameter that appears in the chiral fit function.

We take the hairpin parameters on the $a \approx 0.12$~fm lattices to be $r_1^2 a ^2 \delta'_A = -0.28(6)$
and $r_1^2 a ^2 \delta'_V = 0.00(7)$ from chiral fits to light pseudoscalar-meson masses and decay constants
on a subset of the MILC asqtad ensembles~\cite{Aubin:2004fs}.
Because the hairpin contributions arise from taste-symmetry breaking, they are expected to scale with the
lattice spacing in the same way as the taste splittings $\Delta_\xi$.
Thus, to obtain values for the hairpin parameters at other lattice spacings, we scale them by the ratio of
the weighted average of the taste splitting, Eq.~(\ref{eq:avetaste}), between the coarse and target lattice
spacings.

The priors for the slopes $\mu_i$ and mass differences $\Delta(1/(r_1m_2))$ in $F_i^\kappa$ are taken from
 the central values and errors obtained in the $\kappa$-tuning analysis given in
Tables~\ref{tab:m2diff} and~\ref{tab:slopes} of Sec.~\ref{sec:kappa}.

\subsubsection{Fixed inputs}  \label{sec:inputs-fixed}

In our fits, we fixed the values of the leading-order, light-meson \cpt\ LECs $\Delta_\xi$
($\xi=A,T,V,I$) and $B_0$ in Eq.~(\ref{eq:piontypemass}) because they can be obtained from the light
pseudoscalar-meson spectrum with uncertainties negligible for the present purpose.
The results from simple, linear fits in the valence-quark mass are given in Table~\ref{tbl:tastesplittings}.

\begin{table}
\caption{Taste splittings and the leading-order LEC $B_0$ used in this work~\cite{Bailey:2014tva}.
The second through fifth columns show the taste splittings for the taste axial-vector, tensor, vector, and scalar mesons, respectively.
The parameters $r_1^2a^2\Delta_\xi$ and $B_0$ enter the tree-level expression for the squared pseudoscalar
meson mass, Eq.~(\ref{eq:piontypemass}).}
\label{tbl:tastesplittings}
\begin{tabular}{c  c c c c  c}
\hline\hline
    $\approx a$ (fm) & $r_1^2a^2\Delta_A$ & $r_1^2a^2\Delta_T$ & $r_1^2a^2\Delta_V$ & $r_1^2a^2\Delta_I$ &  $r_1B_0$ \\ 
    \hline 
    0.12    &  0.2270               & 0.3661                & 0.4803                & 0.6008                & 6.832 \\       
    0.09    &  0.0747               & 0.1238                & 0.1593                & 0.2207                & 6.639  \\
    0.06    &  0.0263               & 0.0430                & 0.0574                & 0.0704                & 6.487  \\
    0.045   &  0.0104               & 0.0170                & 0.0227                & 0.0278                & 6.417  \\
\hline\hline    
\end{tabular}
\end{table}

We fix the two effective-field-theory cutoff scales, $\Lambda_\text{HQ}$ and $\Lambda_\chi$, in the fits.
We use $\Lambda_\text{HQ} = 800$~MeV, based on studying the lattice-spacing dependence of matrix 
elements shifted to common light-quark masses via the chiral-continuum fit, and have  
explicitly checked that our fit results are insensitive to reasonable changes to its value.
At a fixed order in \cpt, the chiral expression is independent of the scale $\Lambda_\chi$ that enters the
chiral logarithms.
We use $\Lambda_\chi=1$~GeV and have verified that the fit results change by less than one tenth of the fit
error for $\Lambda_\chi=0.5$ and 0.75~GeV.

Finally, we fix the relative scales on each ensemble, $r_1/a$, to the values listed in
Table~\ref{tab:LatEns} in our base fit.
In Sec.~\ref{sec:syserr}, we discuss how we estimate the error from $r_1/a$, which we add to the fit error
{\it a posteriori}.

\subsection{Chiral fit results}
\label{sec:basefit}

\begin{figure}
    \centering
    \vspace{-0.30in}
    \hspace{-0.295in}
    \includegraphics[width=0.281\textwidth,angle=270]{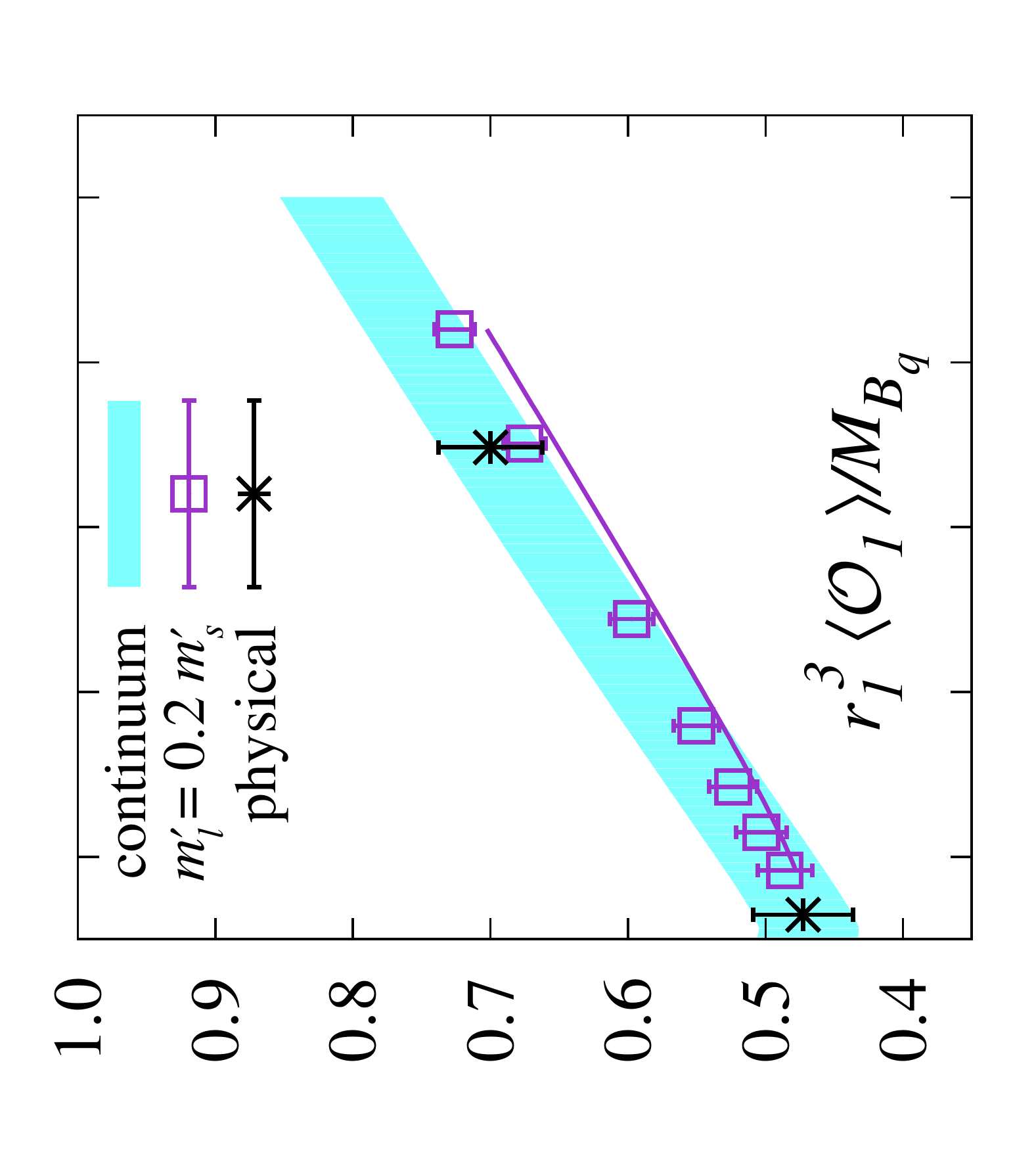}\hspace{-0.375in}
    \includegraphics[width=0.281\textwidth,angle=270]{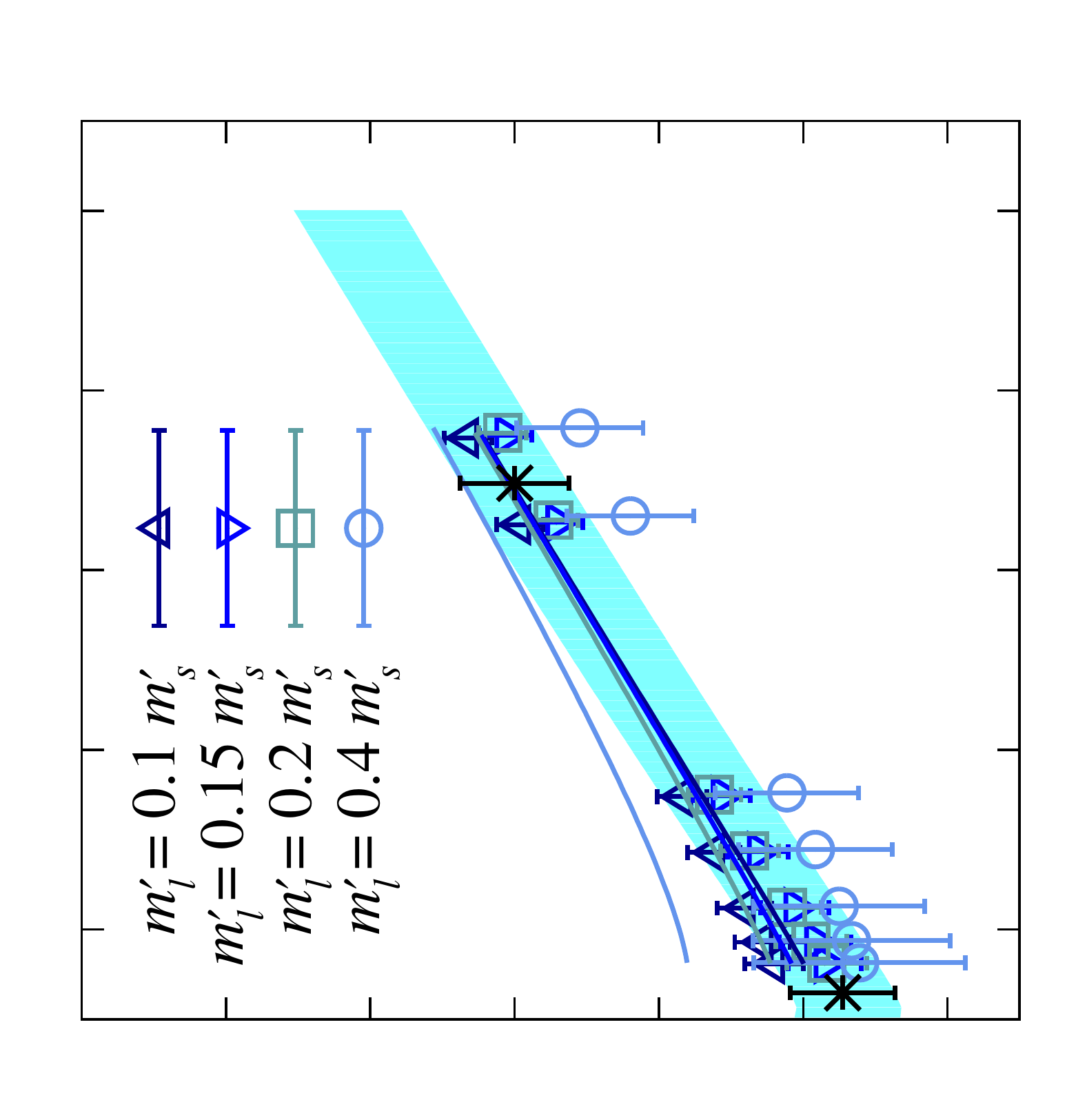}\hspace{-0.375in}
    \includegraphics[width=0.281\textwidth,angle=270]{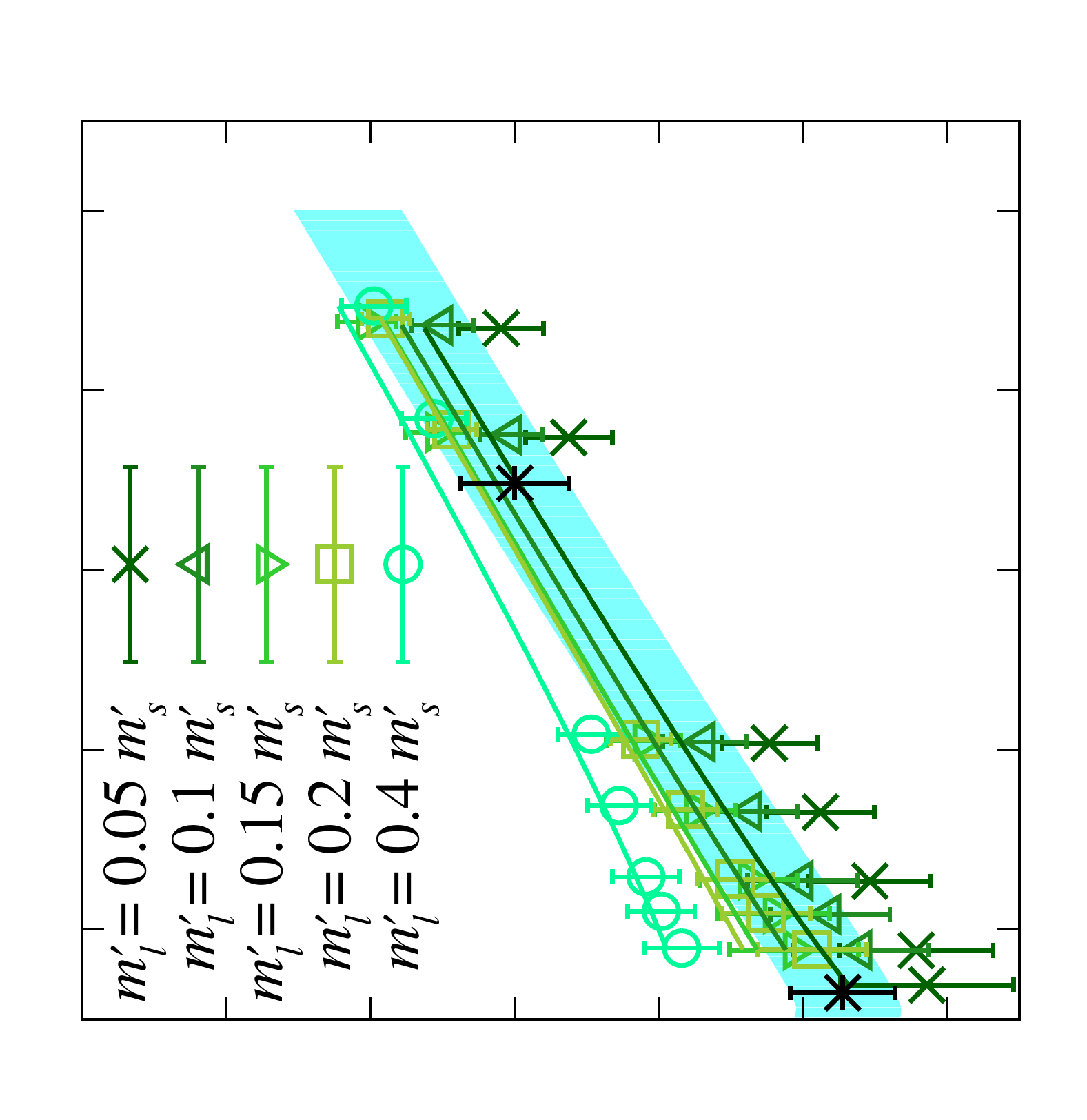}\hspace{-0.375in}
    \includegraphics[width=0.281\textwidth,angle=270]{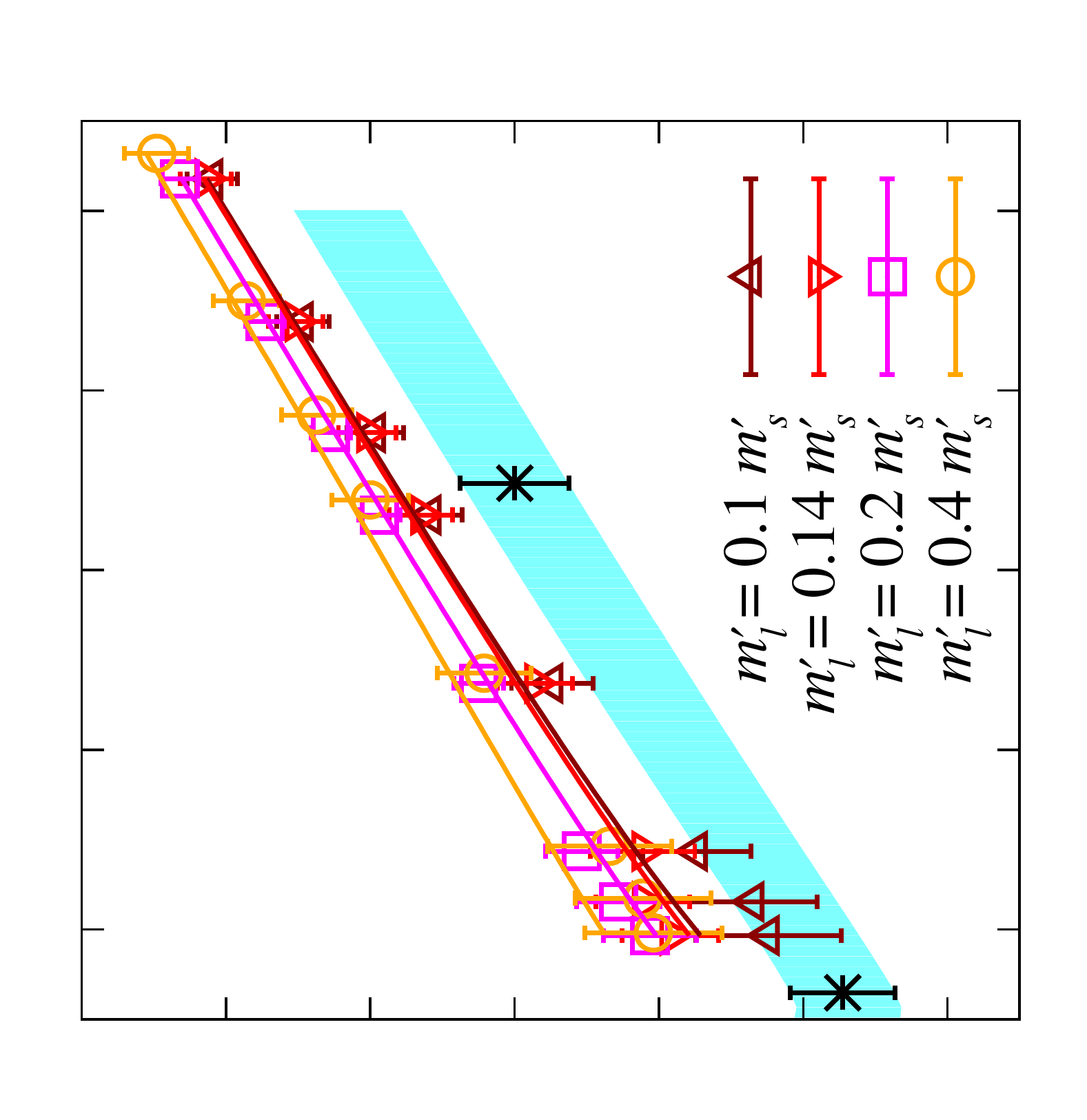}\hspace{-1.5in}\\
    \vspace{-0.22in}
    \hspace{-0.360in}
    \includegraphics[width=0.2816\textwidth,angle=270]{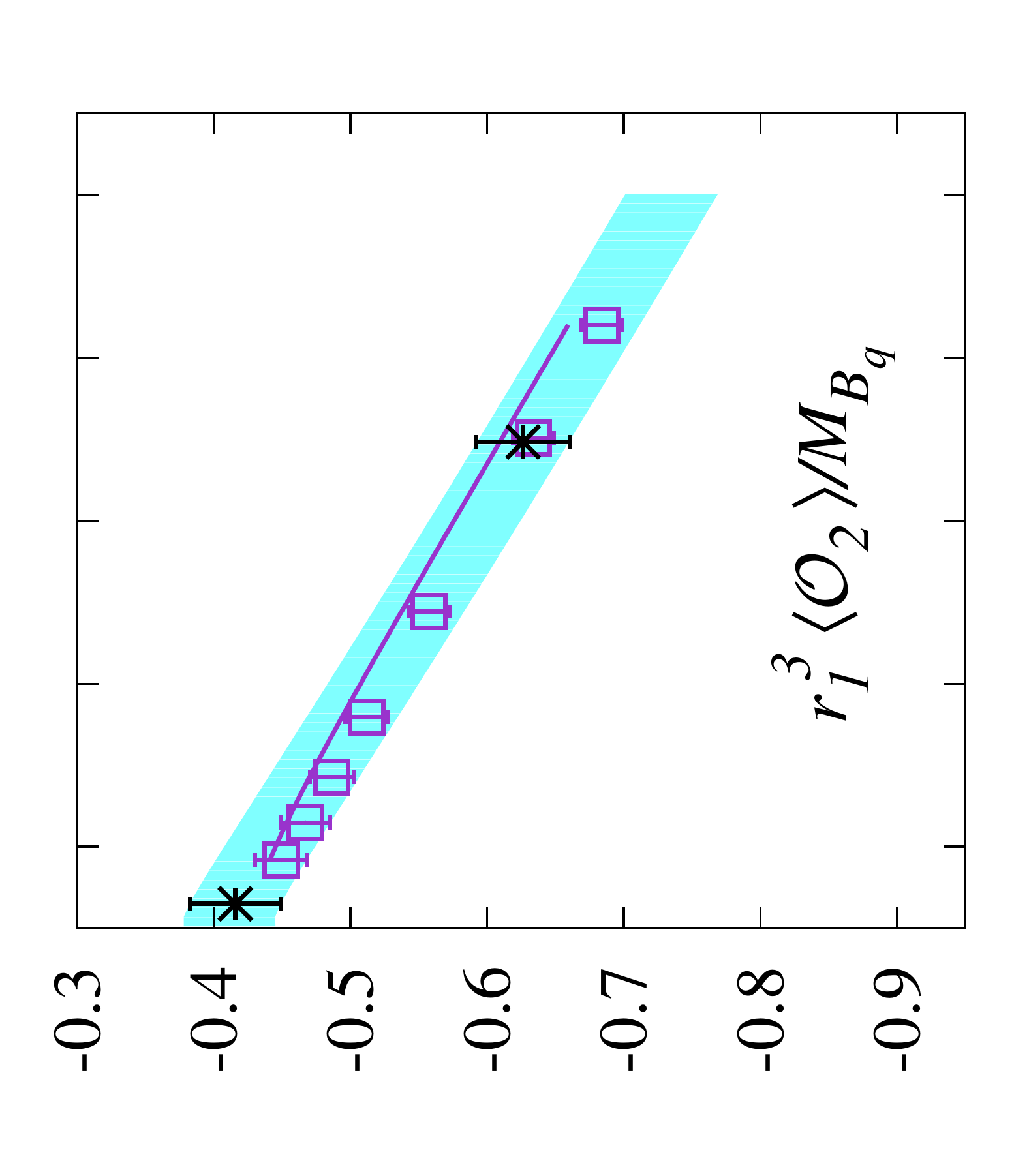}\hspace{-0.375in}
    \includegraphics[width=0.2816\textwidth,angle=270]{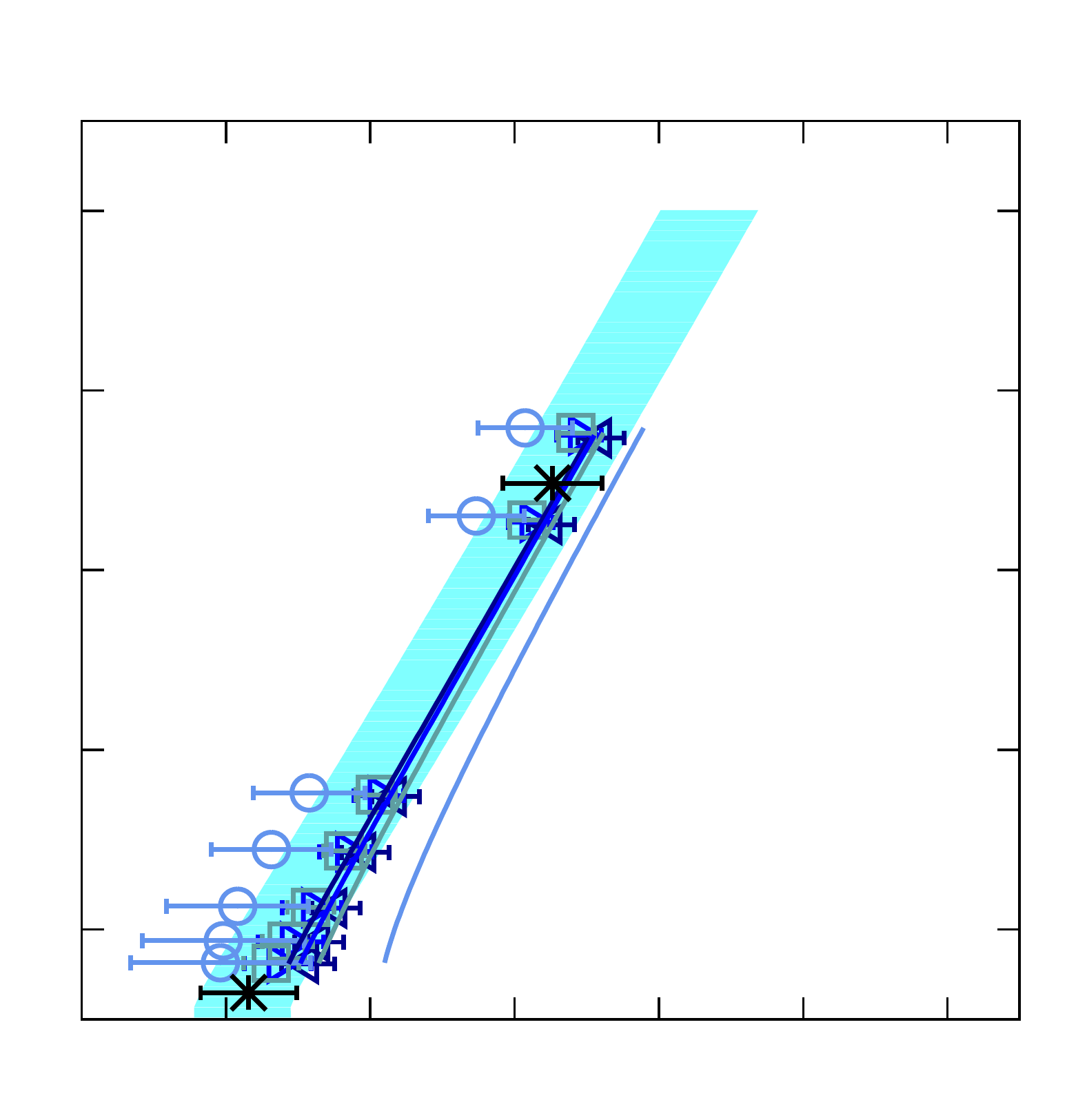}\hspace{-0.375in}
    \includegraphics[width=0.2816\textwidth,angle=270]{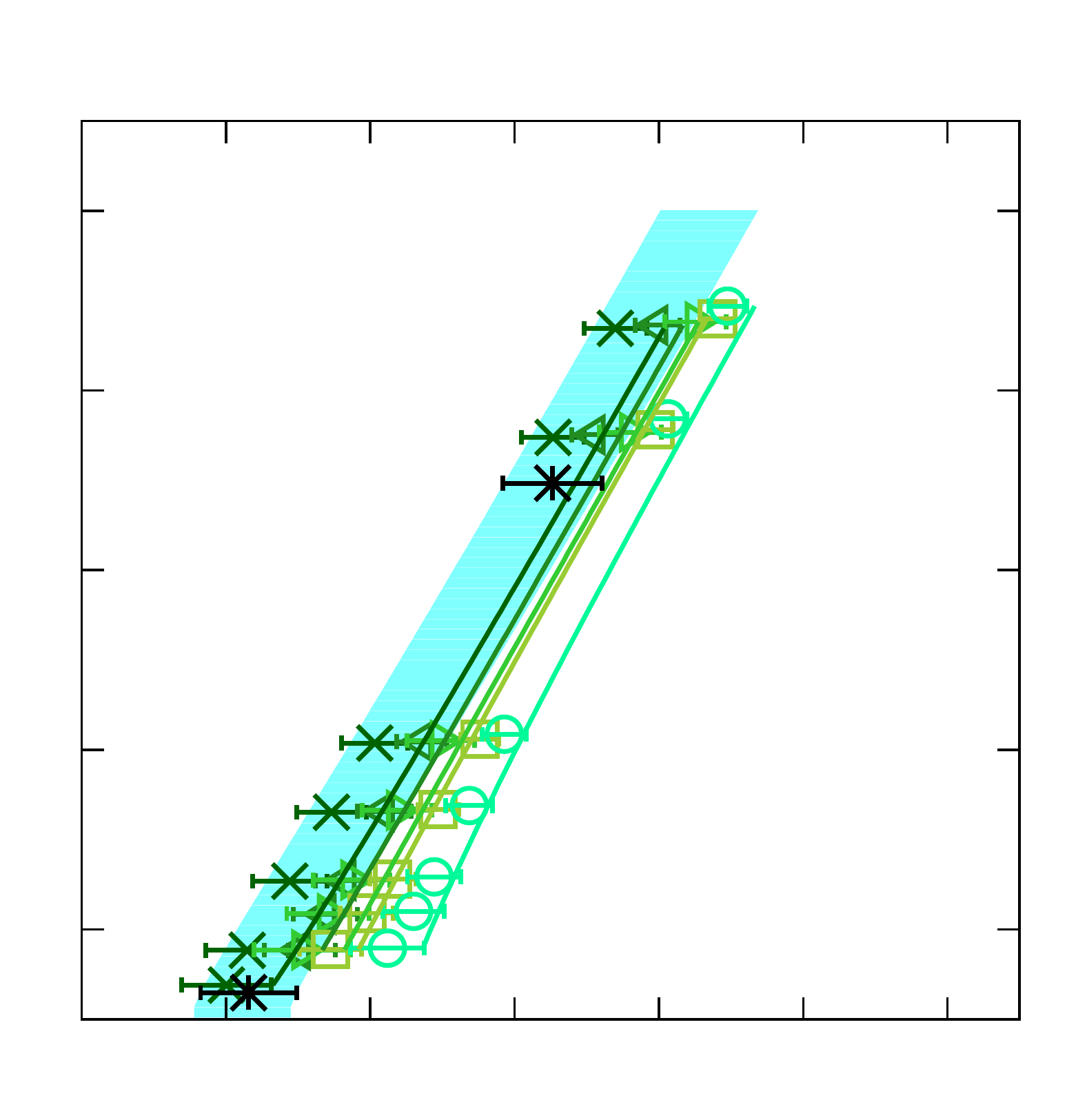}\hspace{-0.375in}
    \includegraphics[width=0.2816\textwidth,angle=270]{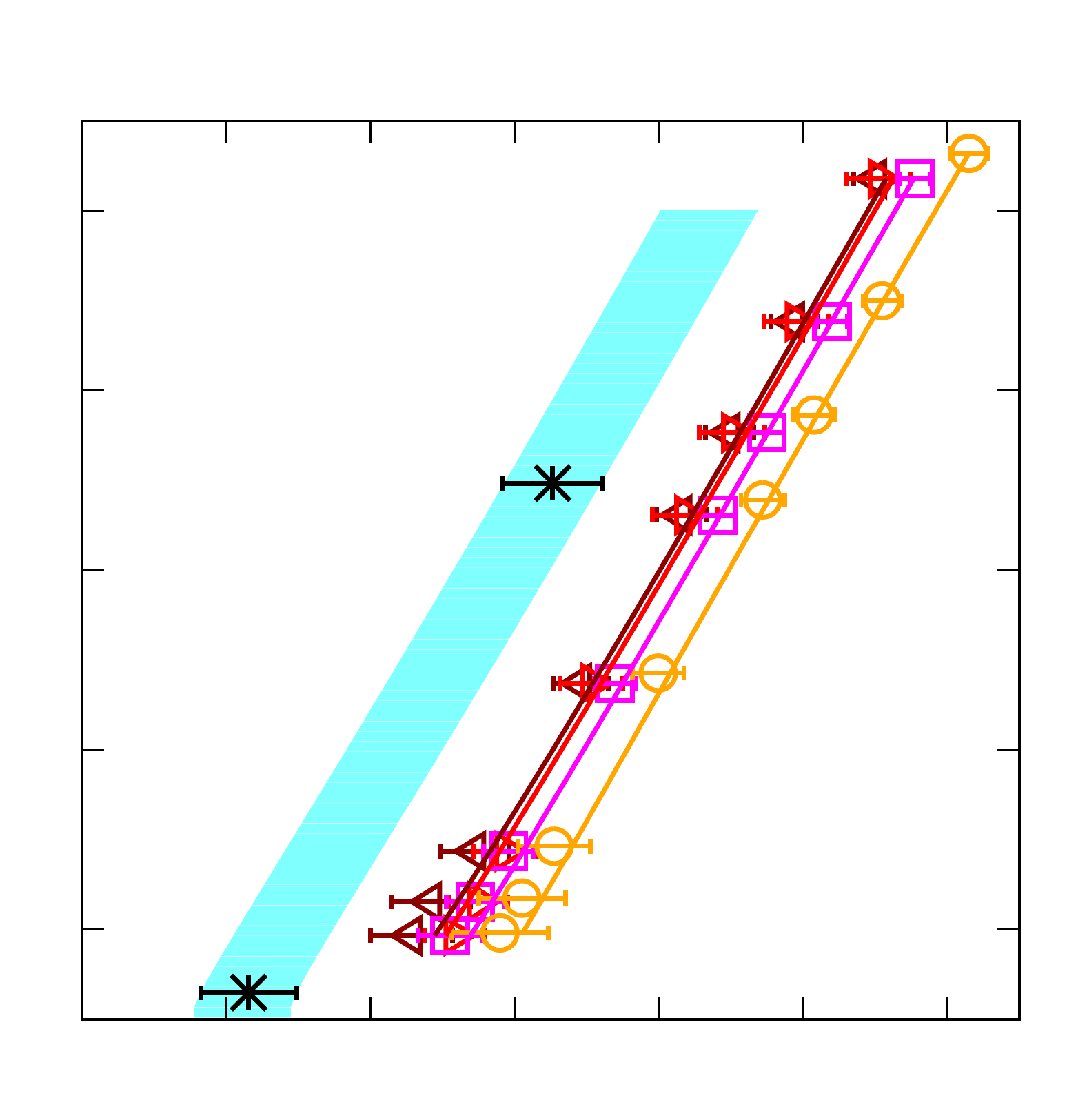}\hspace{-1.0in}\\
    \vspace{-0.22in}
    \hspace{-0.360in}
    \includegraphics[width=0.2816\textwidth,angle=270]{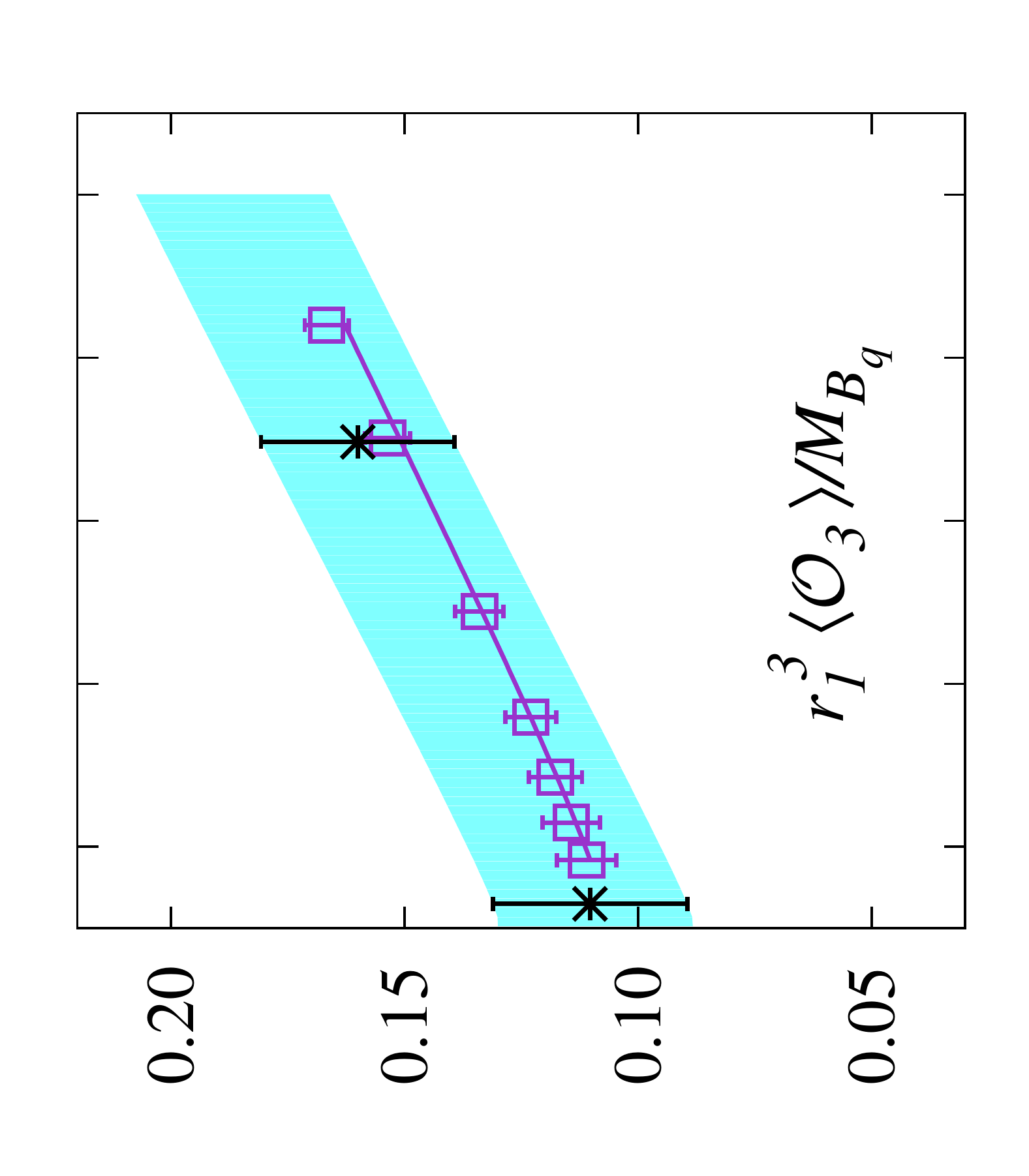}\hspace{-0.375in}
    \includegraphics[width=0.2816\textwidth,angle=270]{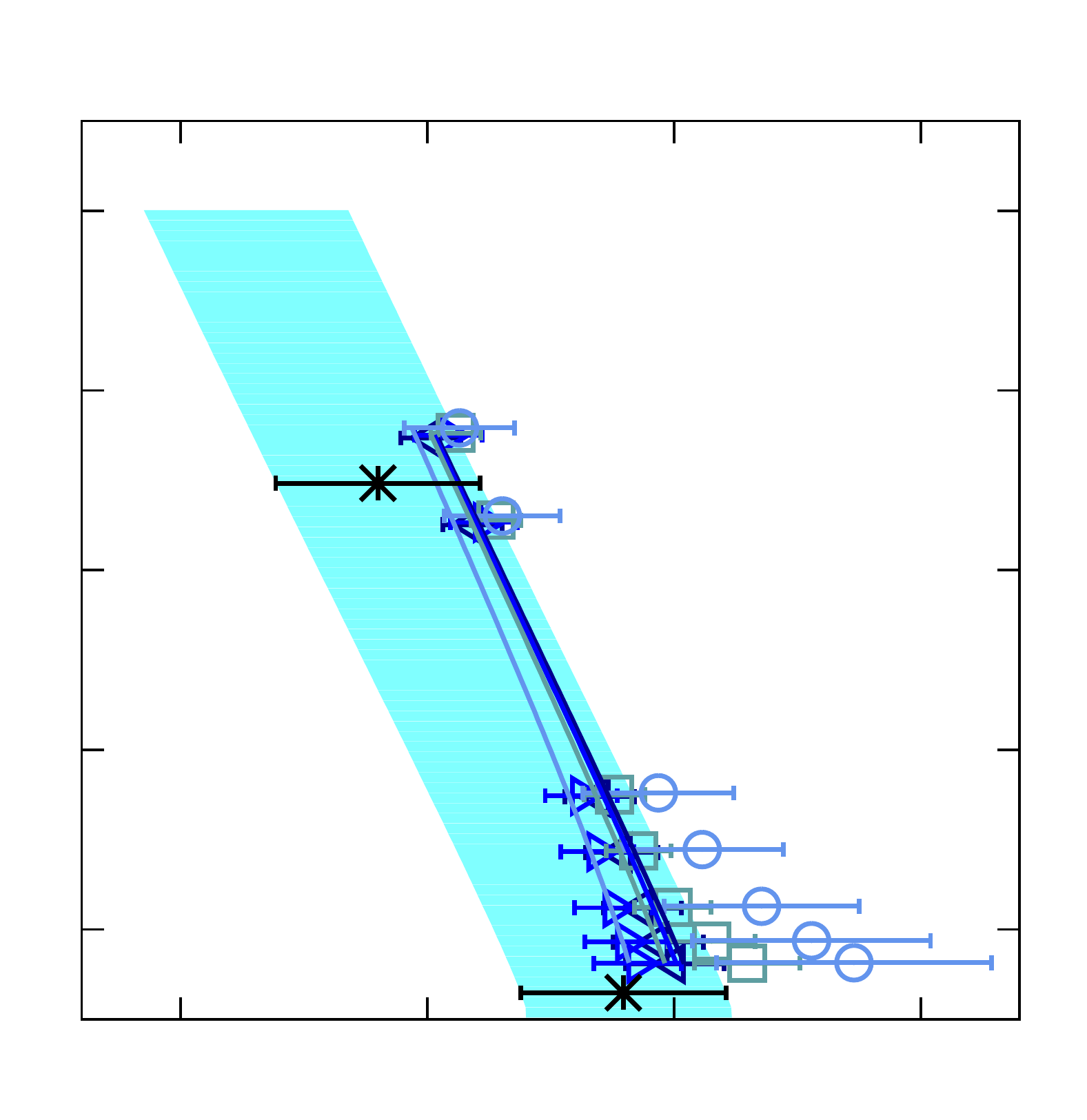}\hspace{-0.375in}
    \includegraphics[width=0.2816\textwidth,angle=270]{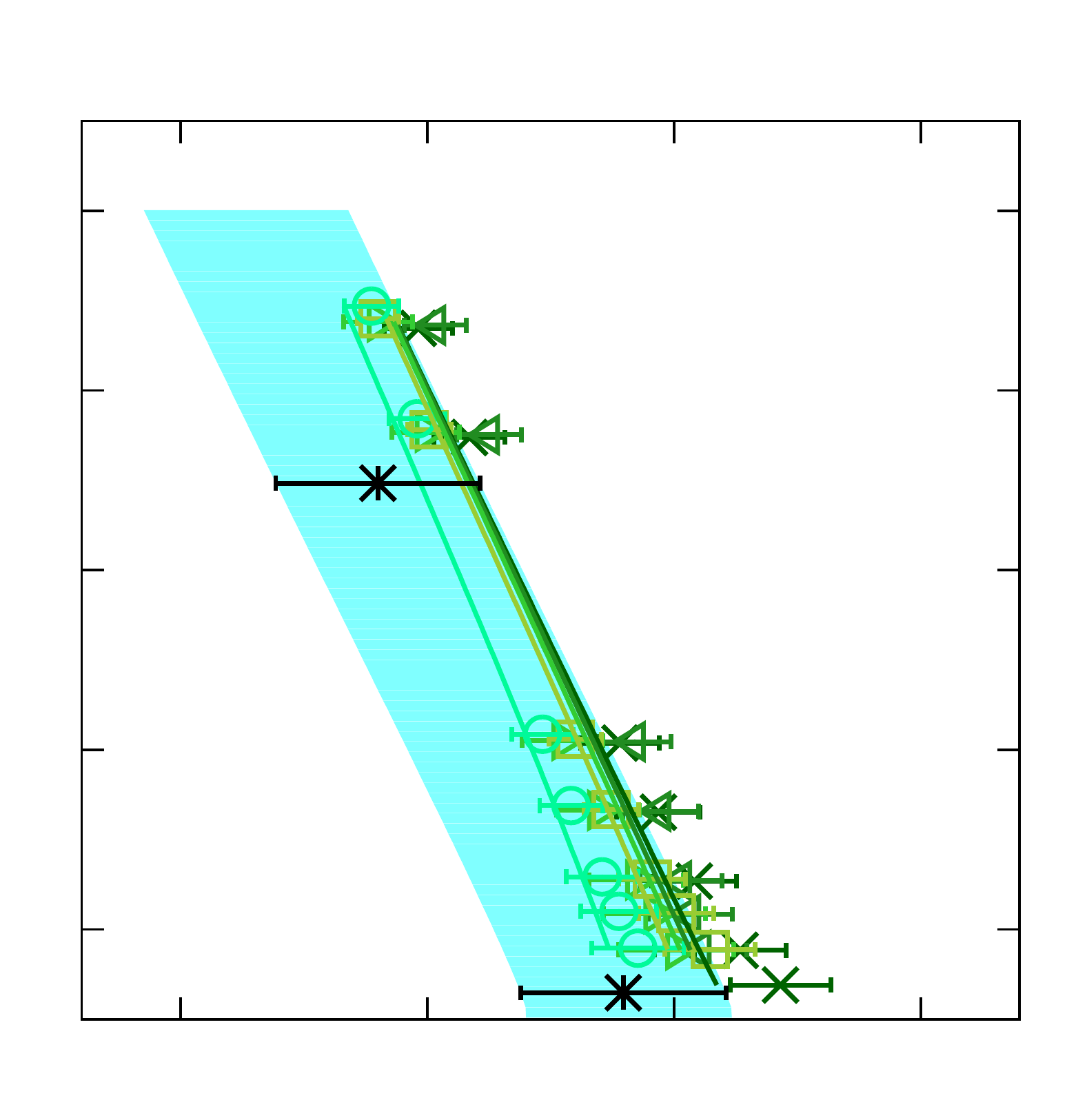}\hspace{-0.375in}
    \includegraphics[width=0.2816\textwidth,angle=270]{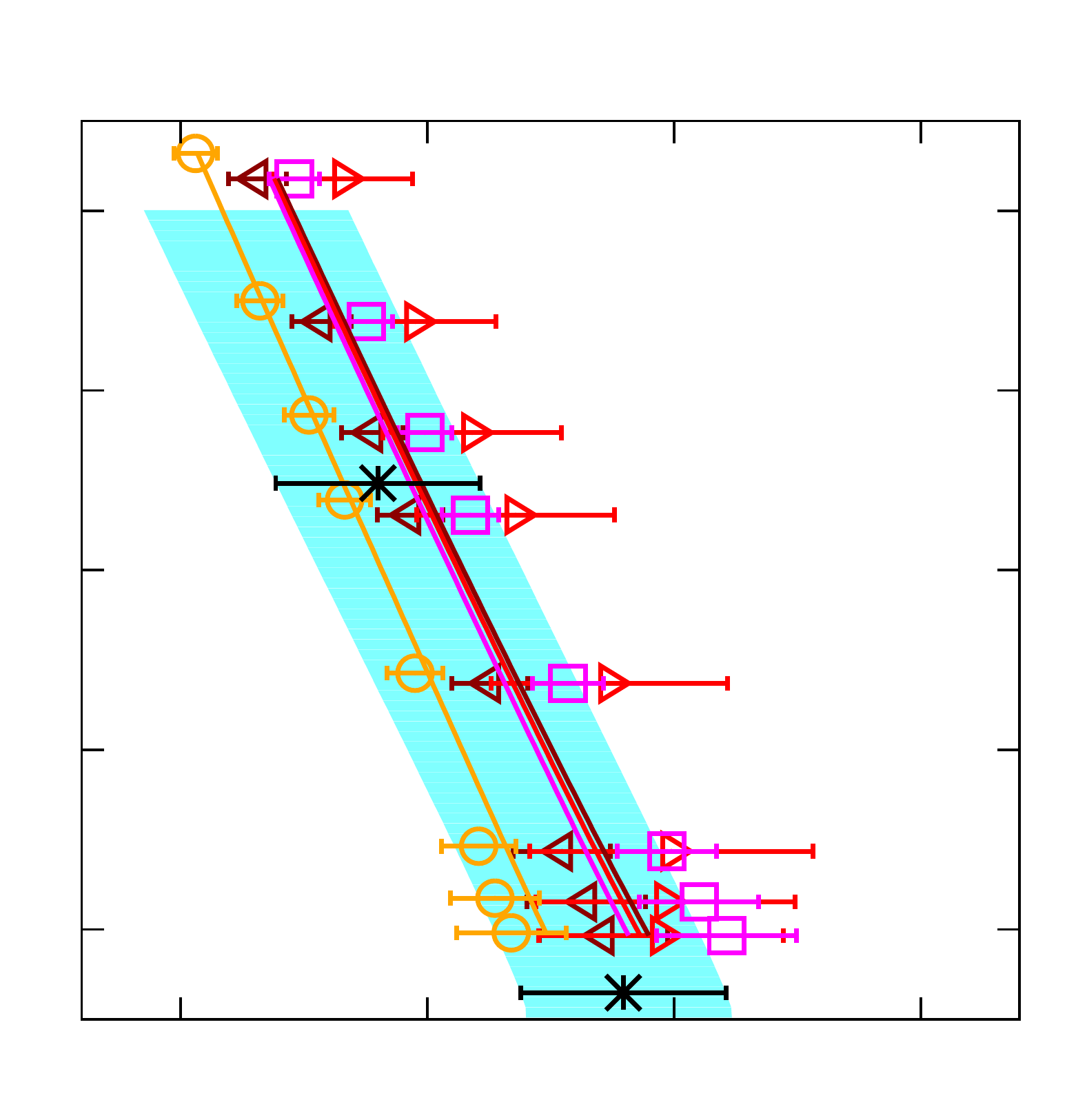}\hspace{-1.0in}\\
    \vspace{-0.22in}
    \hspace{-0.2880in}
    \includegraphics[width=0.2816\textwidth,angle=270]{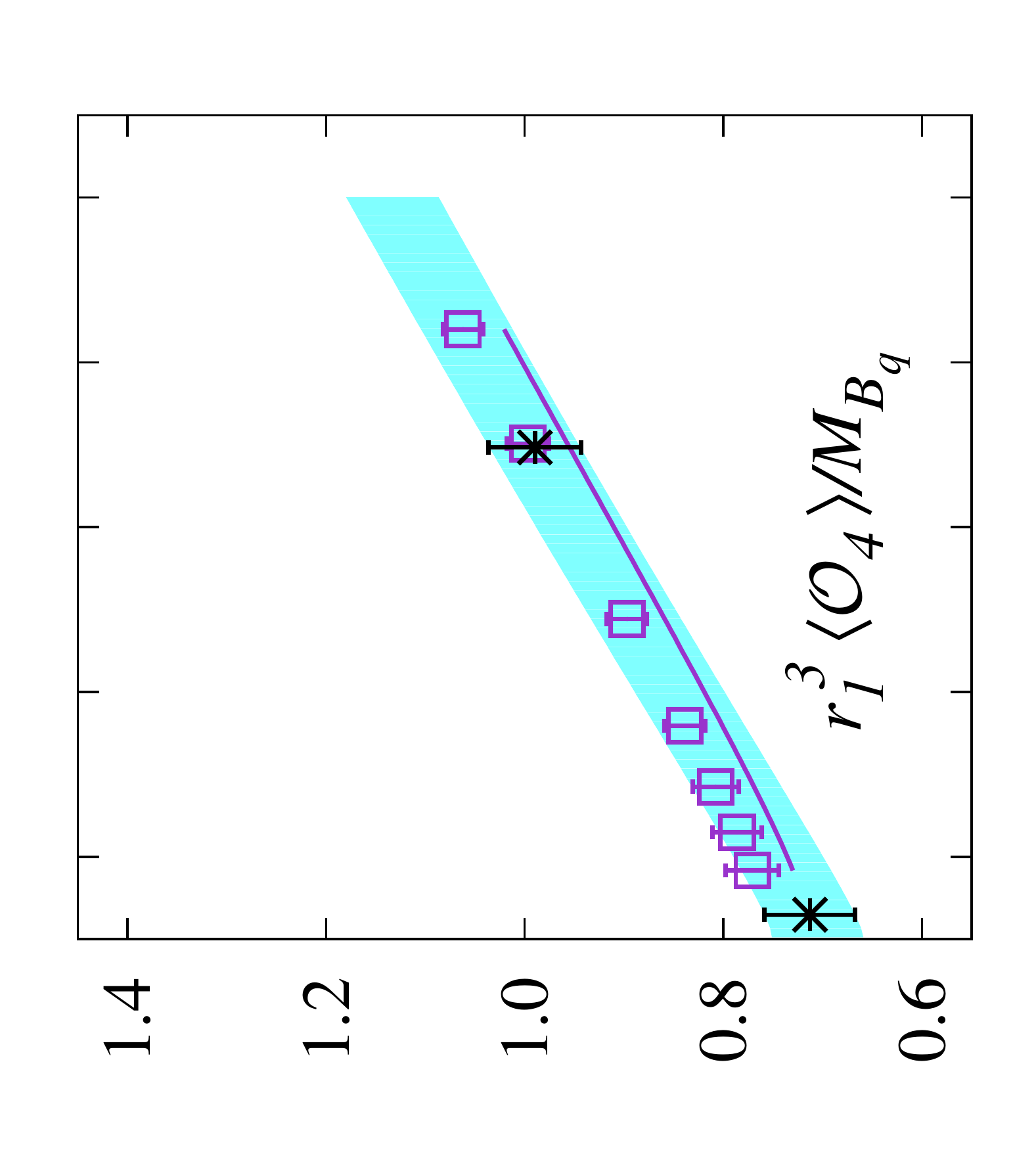}\hspace{-0.375in}
    \includegraphics[width=0.2816\textwidth,angle=270]{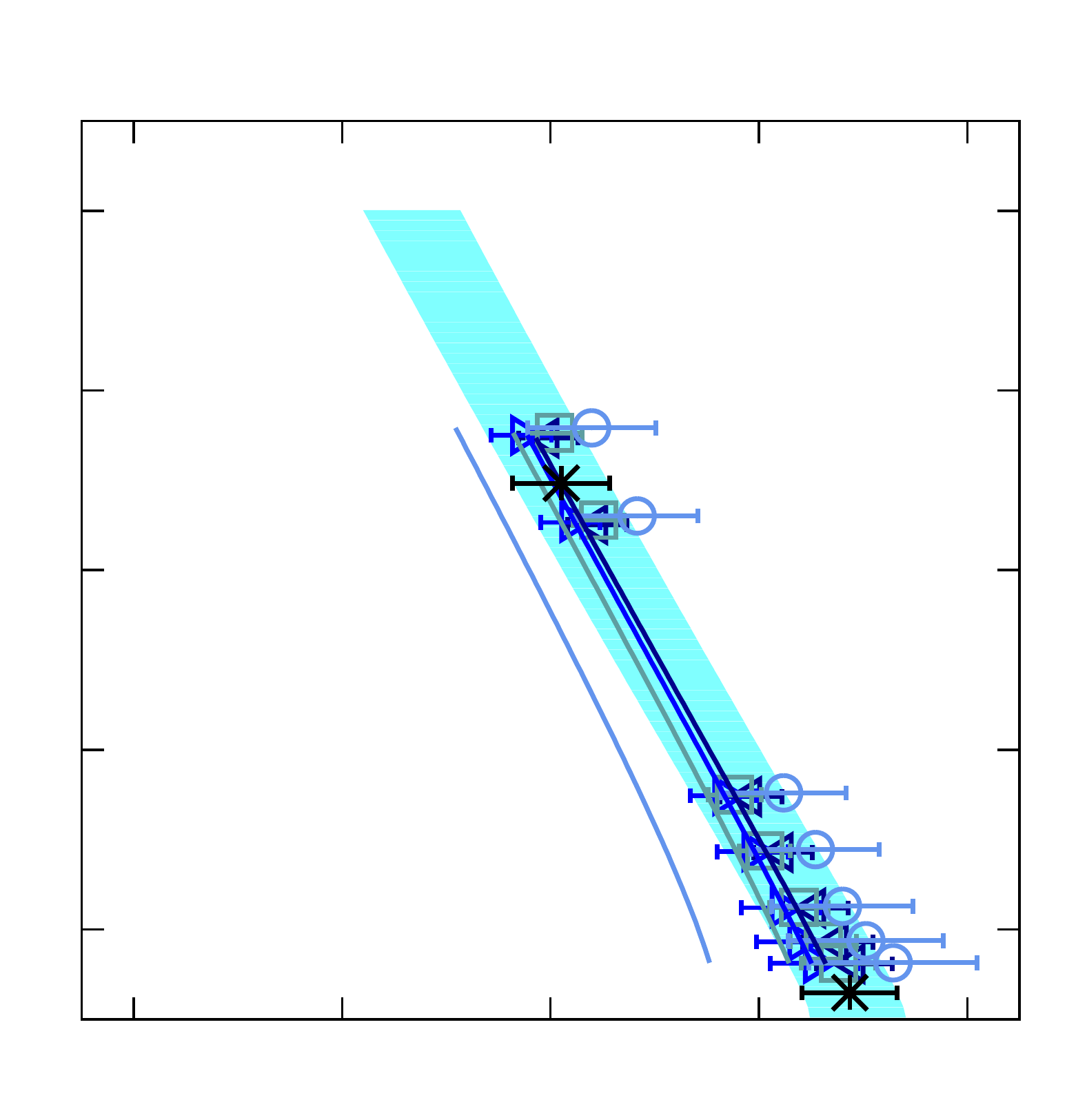}\hspace{-0.375in}
    \includegraphics[width=0.2816\textwidth,angle=270]{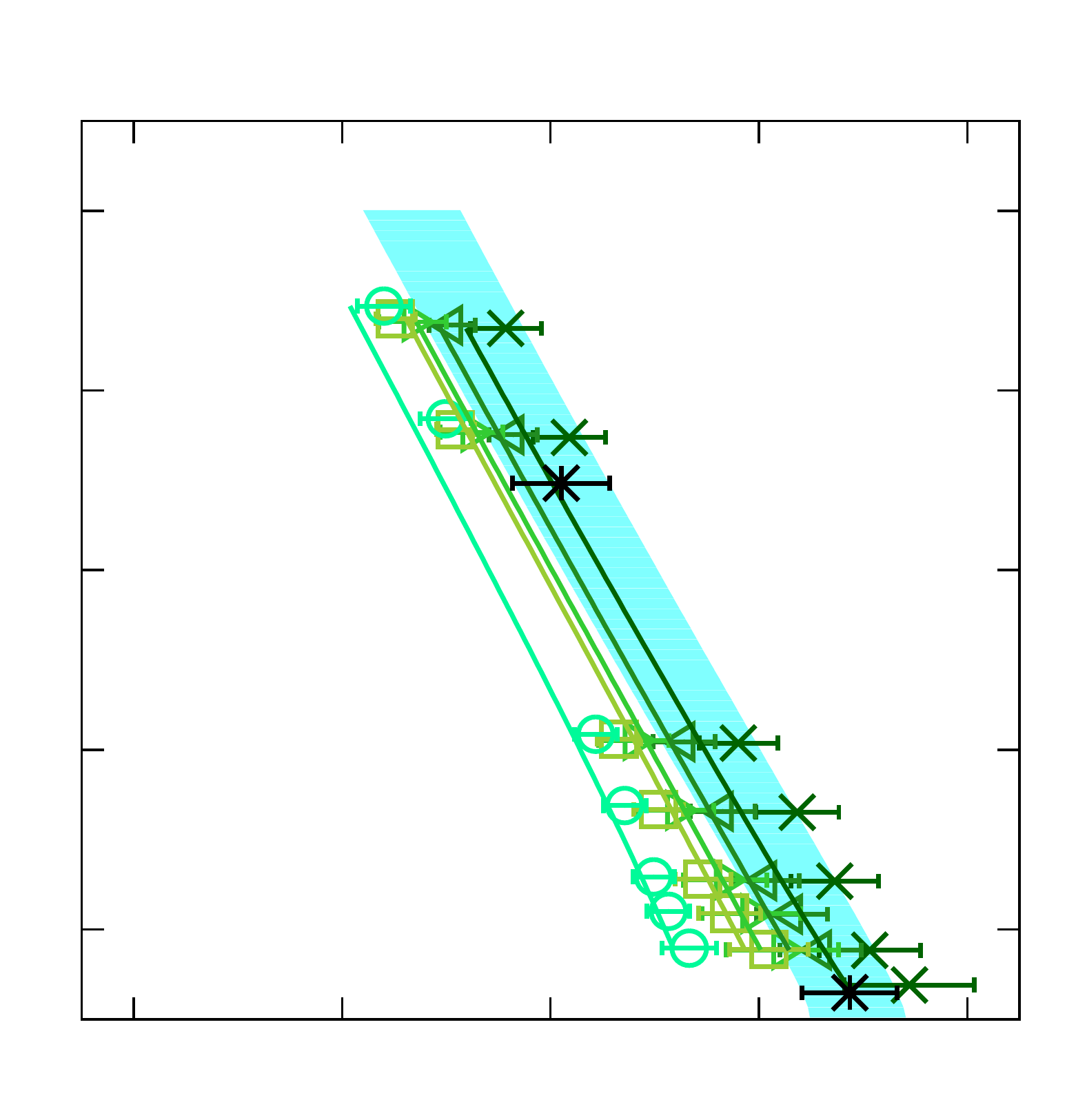}\hspace{-0.375in}
    \includegraphics[width=0.2816\textwidth,angle=270]{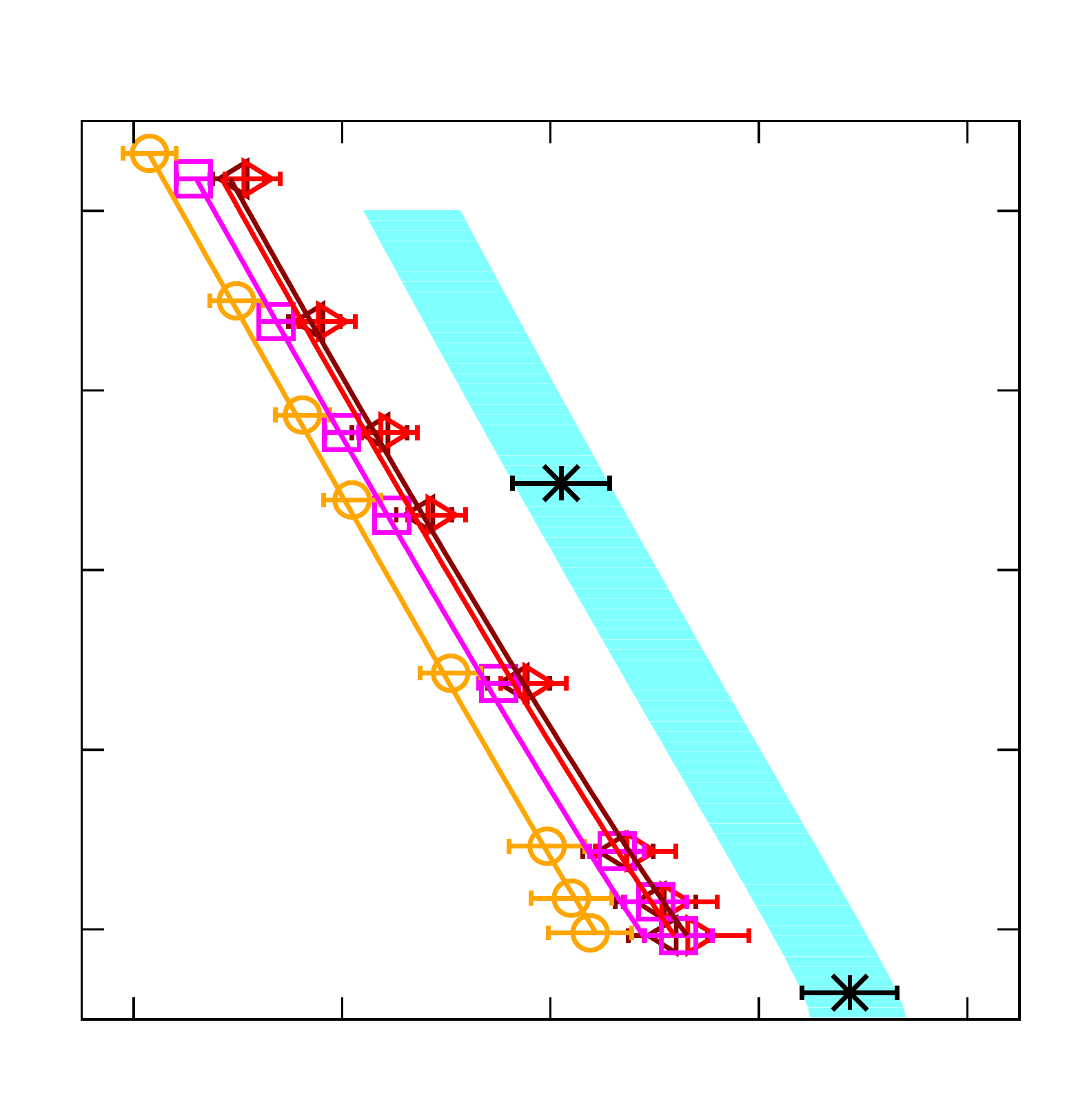}\hspace{-1.0in}\\
    \vspace{-0.22in}
    \hspace{-0.330in}
    \includegraphics[width=0.328\textwidth,angle=270]{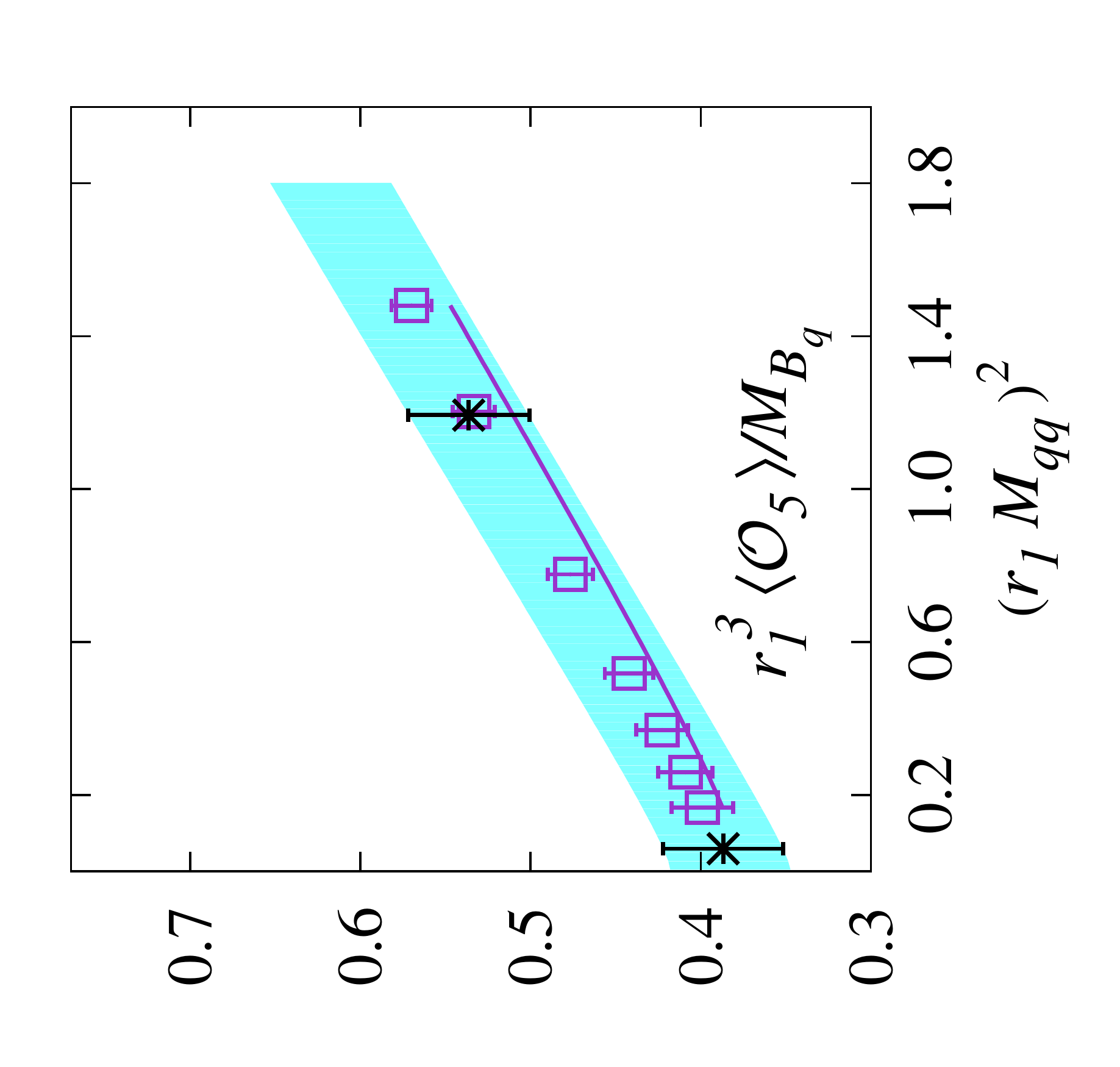}\hspace{-0.375in}
    \includegraphics[width=0.328\textwidth,angle=270]{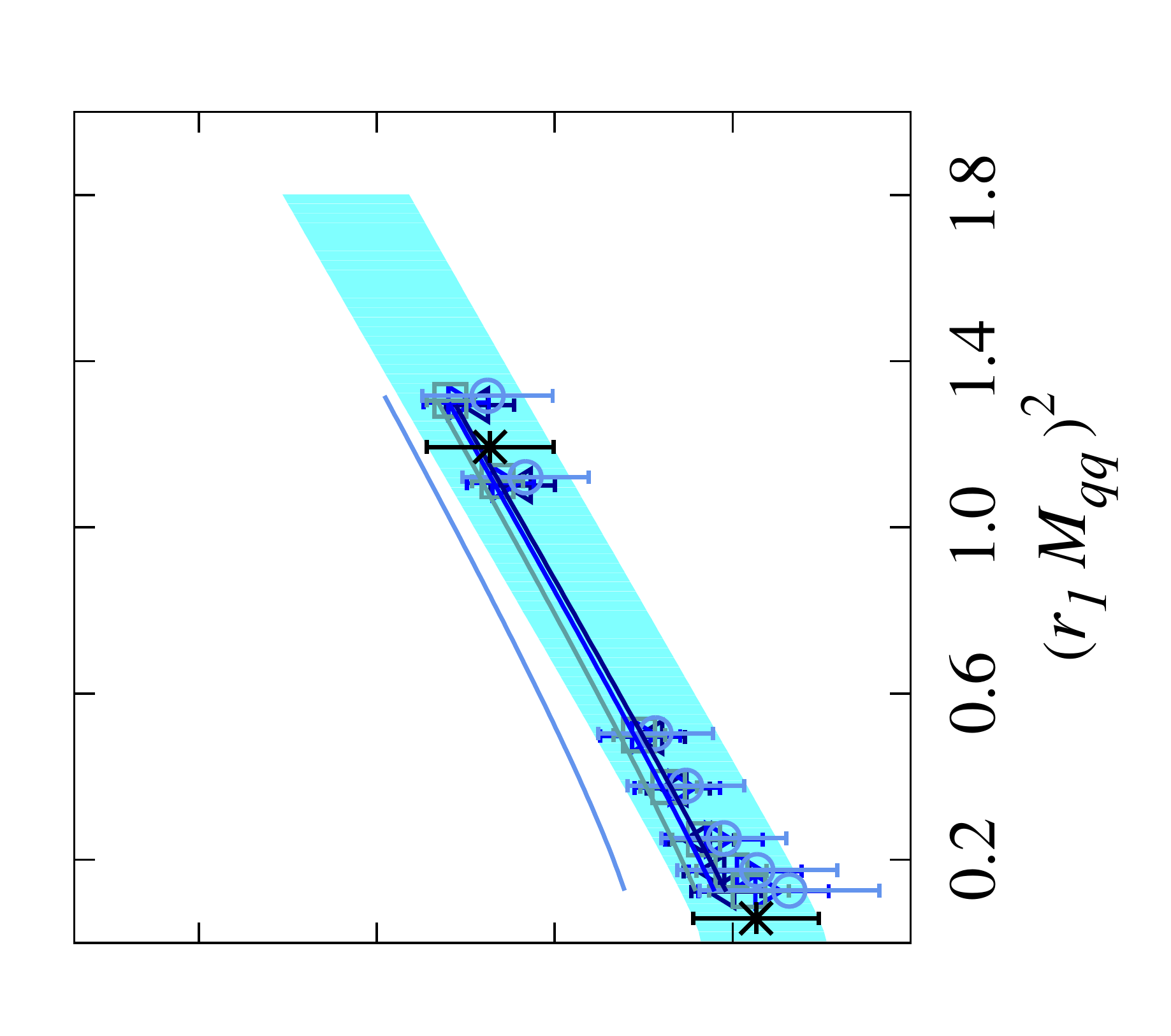}\hspace{-0.375in}
    \includegraphics[width=0.328\textwidth,angle=270]{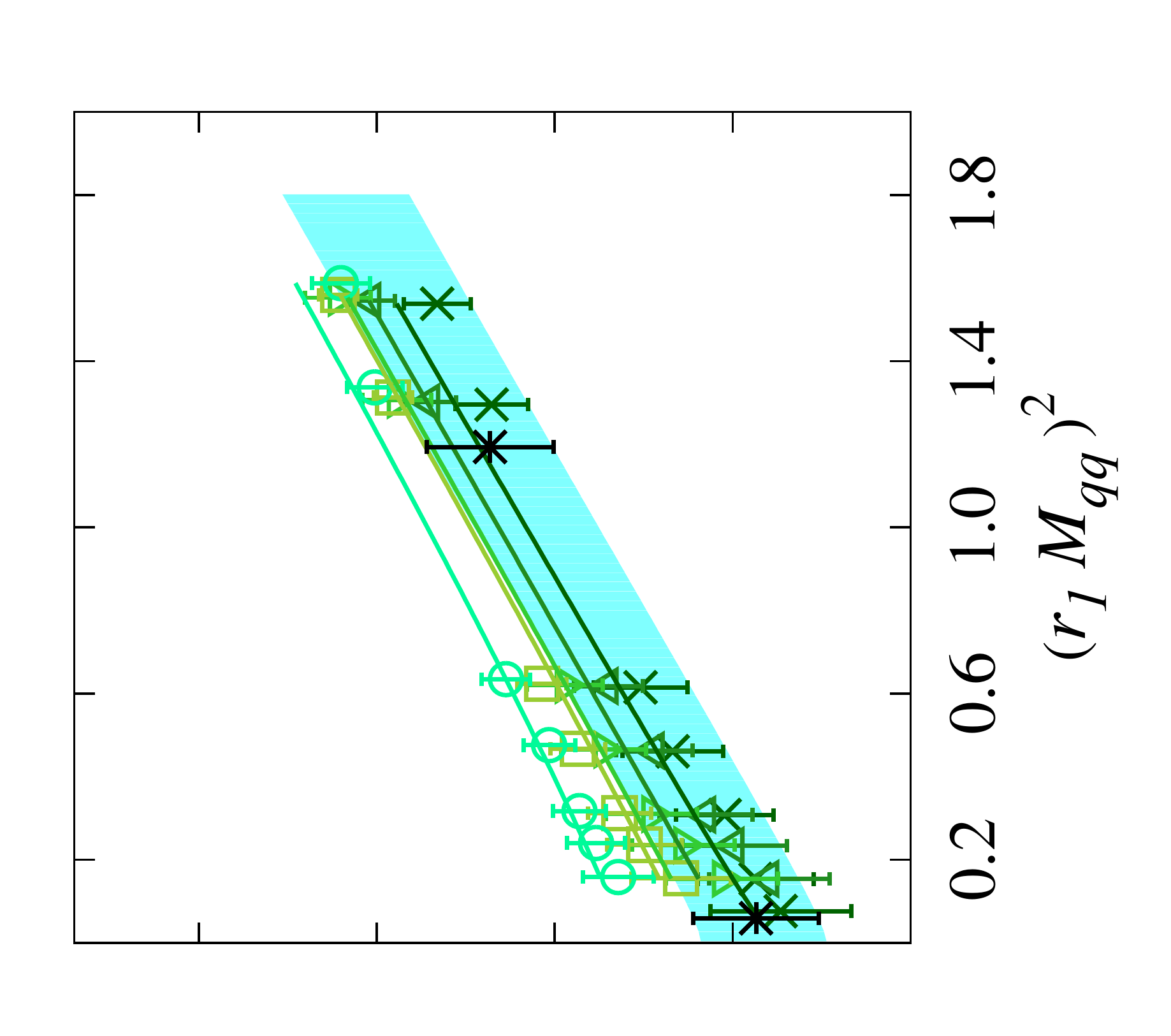}\hspace{-0.375in}
    \includegraphics[width=0.328\textwidth,angle=270]{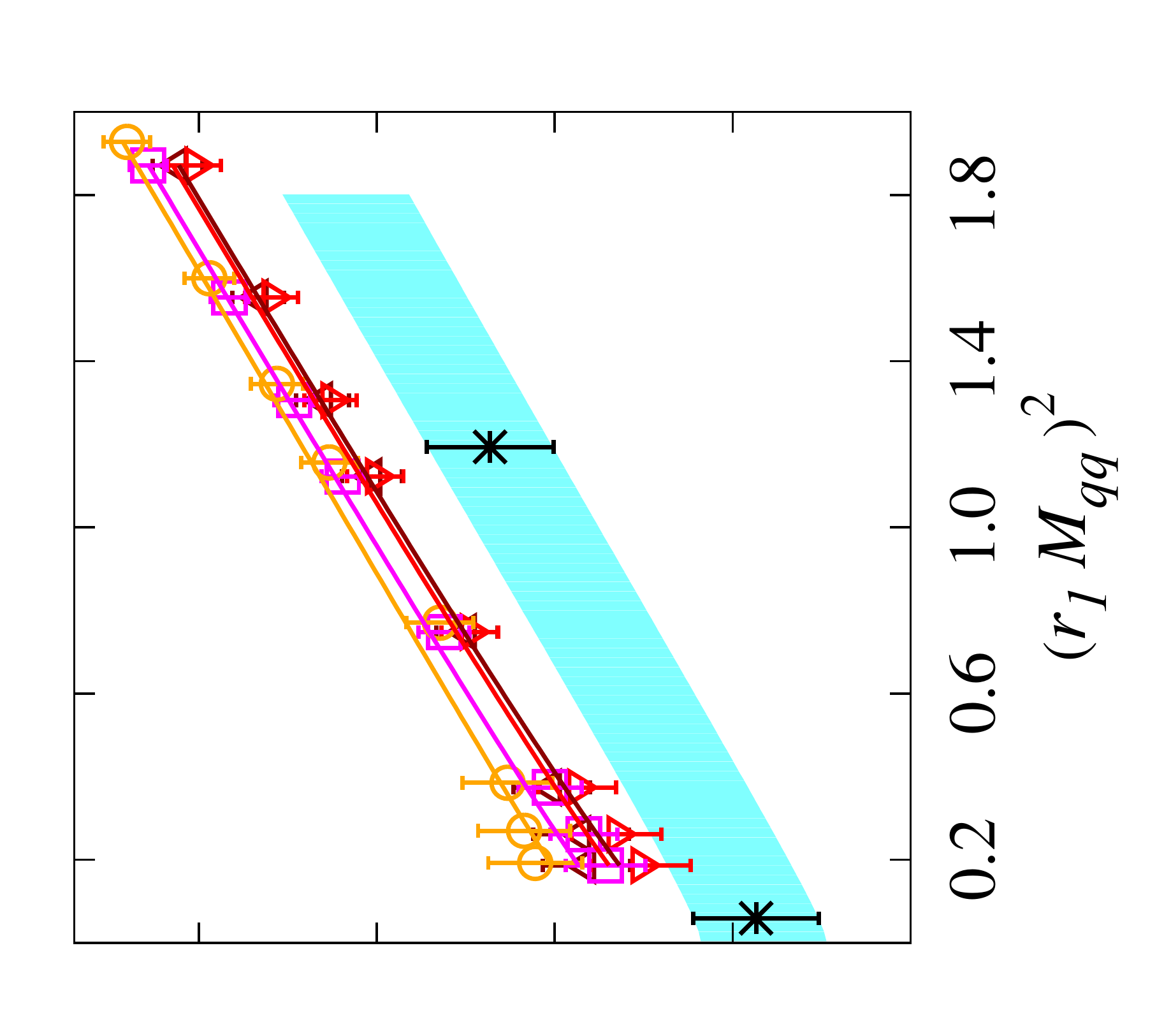}\hspace{-0.1in}
    \vspace{-0.0in}
    \caption{(color online) Chiral-continuum extrapolation of the $B_q$-mixing matrix elements from a 
        combined fit to all data.
        From top-to-bottom: results for operators $\langle\op_1^q\rangle$--$\langle\op_5^q\rangle$.
        From left-to-right: results on lattice spacings $a \approx $ 0.045--0.12~fm.  
        The correlated $\chi^2_\text{aug}/\text{dof}$ = 134.9/510.}
    \label{fig:ChPTfits}
\end{figure}

Using the fit function in Eq.~(\ref{eq:chiralfitfunc-base}) and the inputs summarized in the previous
section, we construct an augmented $\chi^2$ function,
\begin{equation}
    \chi^2_\text{aug} = \sum_{\alpha,\beta}
        \left[F^\text{base}-Z\langle O\rangle/M_B\right]_\alpha \left(\sigma^2\right)_{\alpha\beta}^{-1}
        \left[F^\text{base}-Z\langle O\rangle/M_B\right]_\beta +
        \sum_m \frac{(P_m - \tilde{P}_m)^2}{\tilde{\sigma}_m^2}
    \label{eq:thefit}
\end{equation}
where the multi-indices $\alpha$ and $\beta$ run over all data for all (renormalized) operators, ensembles,
and valence quark masses, and the $P_m$ run over all fit parameters discussed above.
The data covariance $\sigma^2_{\alpha\beta}$ is obtained from the bootstrap of the matrix elements, and the
priors $\tilde{P}_m(\tilde{\sigma}_m)$ have been explained above.
After finding the parameter set that minimizes $\chi^2_\text{aug}$, we extrapolate the renormalized matrix
elements to the physical point by setting the light-quark masses to those in Table~\ref{tbl:extrap-point}
and the lattice spacing to zero.
\begin{table}
    \caption{Physical, renormalized, light-quark masses and the meson mass LEC in $r_1$
        units~\cite{Bailey:2014tva}.
        Errors on the quark masses include statistics and the systematic uncertainties from the 
        chiral-continuum extrapolation and the uncertainty on the physical scale $r_1$; other sources of 
        uncertainty are negligible.
        The value of $m_u$ is used only for the estimate of isospin-breaking errors in
        Sec.~\ref{sec:isospin}.
        The continuum $r_1B_0$ quoted here is 11\% smaller than an incorrect value given in Table~VIII of 
        Ref.~\cite{Bailey:2014tva}.
        	(Reference~\cite{Bailey:2014tva} handled $r_1B_0$ in a way that avoided any significant impact on the final result for $B\to D^*\ell\nu$.)}        
         \label{tbl:extrap-point}
    \begin{tabular}{c@{\quad}  c@{\quad} c@{\quad} c@{\quad} c@{\quad} c}
    \hline\hline
        $r_1m_d\times10^3$ & $r_1m_s\times10^3$ & $r_1\hat{m}\times10^3$ & $r_1m_u\times10^3$ & $r_1B_0$ \\ 
        \hline 
        4.94(19) & 99.2(3.0) & 3.61(12) & 2.284(97) & 6.015 \\
    \hline\hline
    \end{tabular}
\end{table}
This procedure is repeated for alternative fits by modifying the fit function $F^\text{base}$ in
Eq.~(\ref{eq:thefit}).
Figure~\ref{fig:ChPTfits} shows our base-fit results for each matrix element vs the squared meson mass
$M_{qq}^2=2B_0m_q$, which is proportional to the light valence-quark mass.
We obtain a correlated $\chi^2_\text{aug}/\text{dof}$ = 134.9/510.

\section{Systematic Error Budget}
    \label{sec:syserr}
    

We now consider all sources of uncertainty in our results and estimate their contributions to the total
error.
As discussed in the previous section, we perform a combined extrapolation to the physical light-quark mass
and the continuum limit to extract the physical matrix elements from our lattice data.
The chiral-continuum fit function of Eq.~(\ref{eq:chiralfitfunc-base}), used for our base fit, includes
higher-order terms constrained with Gaussian priors.
In Sec.~\ref{sec:SystErrorA} we discuss systematic error contributions that are included in the
chiral-continuum fit error, and the extensive tests we perform to check for residual truncation effects.
Next, in Sec.~\ref{sec:SystErrorB}, we discuss the remaining contributions that are added to the
chiral-continuum fit error \emph{a posteriori}, with the exception of the error due to omitting charm sea
quarks, which is considered separately in Sec.~\ref{sec:charm}.
We also comment on the effect of correlations between the data, and compare the results of fits using
various subsets of data.
Last, in Sec.~\ref{sec:errsum}, we summarize the sources of uncertainty considered and provide 
comprehensive error budgets for the matrix elements and $\xi$.

In Secs.~\ref{sec:SystErrorA} and~\ref{sec:SystErrorB}, we study the stability of our fit results and their
relative goodness-of-fit, respectively, against reasonable variations of the fit function, input parameters
and prior widths, and data included.
We show stability plots only for the matrix elements $\me{1}$ and $\me{4}$ ($q=s,d$), and for the ratio
$\xi$; these plots are representative of the behavior observed for all matrix elements and ratios.
We evaluate the relative fit quality using the augmented $\chi^2_{\rm aug}$ defined in
Eq.~(\ref{eq:augchi2}), which includes contributions from the prior constraints on the fit parameters, and
counts the number of prior constraints as additional data points when computing the degrees of freedom.

All of the fit variations used to test stability and estimate remaining systematic uncertainties employ the
infinite-volume expressions for the chiral logarithms.
As discussed in Sec.~\ref{sec:FV-err}, we find that finite-volume errors are negligible, which allows us to
repeat the additional fits much more quickly without changing the conclusions about the errors.

\subsection{Errors  encompassed by the base chiral-continuum fit}\label{sec:SystErrorA}
 
As described in Sec.~\ref{sec:inputs-priors}, we use Gaussian priors to constrain the fit parameters and
most of the external inputs in our chiral-continuum fit.
This allows for the uncertainties in the input parameters to be automatically included in the total fit
error of the resulting matrix elements.
Further, it enables the inclusion of higher-order terms in the chiral and heavy-quark expansions such that
the fit error incorporates possible truncation errors.
We track the error contributions of each data point and prior via the dependence of the best-fit parameters
on each piece of information, including correlations among them, following the approach described in 
Appendix~A of Ref.~\cite{Bouchard:2014ypa}.
The statistical error is given by the quadrature sum of the errors from all data points.
This procedure allows us to separate the total fit error ($\sigma_{\rm fit}$) into approximate sub-errors.
The sub-errors give a useful picture of the most (and least) important sources of uncertainty in the matrix
elements and $\xi$, but are not used in our final error budgets.
Here we discuss the results of this estimated breakdown and also the fit variations that we perform to test
the robustness of our error estimate.
The stability of the central values and errors of our results under these fit variations, illustrated in
Figs.~\ref{fig:stabO1ANDxi}--\ref{fig:stabO4ANDchisq}, indicates that the corresponding uncertainties are
indeed encompassed by the base-fit error.

Table~\ref{tbl:ME_fit_error} shows for each matrix element the approximate breakdown of the total fit error
into the sub-errors.
The first column shows the statistical error, which gives the largest contribution to the total fit error
for all quantities listed.

\subsubsection{Parametric inputs}
\label{sec:chiral-log-error}

The ``inputs'' column of the error budget in Table~\ref{tbl:ME_fit_error} shows the uncertainty
contributions from all parametric inputs that are constrained by Gaussian priors.
It includes the parameters $f_\pi$, $g_{B^*B\pi}^2$, $\Delta^*$, $\lambda_1$, $\delta_V'$, $\delta_A'$ that
appear in the chiral logarithms, Eqs.~(\ref{eq:ME1chipt})--(\ref{eq:ME45chipt}).
It also includes the error on the physical $d$- and $s$-quark masses that are used in the chiral
extrapolation (interpolation) of the matrix elements.
The pion taste splittings $\Delta_\xi$, and the tree-level LEC $B_0$ are fixed; their errors are negligibly
small and not included.

Although the error on the parameter $f_\pi$ is already included in our fit, one can study the dependence of
our fit result on reasonable variations to the decay constant, which provides one measure of the uncertainty
due to truncating the chiral expansion.
For the base fit, we use the PDG value given in Eq.~(\ref{eq:r1fpi}).
To test the dependence of our result on reasonable variations of the decay constant, we perform a fit in
which the decay constant is set to the PDG value 
$f_{K^\pm}=(156.2\pm0.2\pm0.6\pm0.3)~\text{MeV}$~\cite{Agashe:2014kda}, or
\begin{equation}
    r_1 f_{{K^{\pm}}} = 0.2467(21)  .
    \label{eq:r1fK} 
\end{equation}
The result of this fit is labeled ``$f_K$~vs.~$f_\pi$'' in
Figs.~\ref{fig:stabO1ANDxi}--\ref{fig:stabO4ANDchisq}.
The largest observed change in central value is for the SU(3)-breaking ratio $\xi$, which is still less than
a half of $\sigma_{\rm fit}$.

\subsubsection{Bottom-quark mass uncertainty}

Section~\ref{sec:kappa} describes how we adjust the matrix elements calculated at the simulated $b$-quark
masses to the physical value.
The uncertainty in this adjustment is included via priors on the difference between the simulated and tuned
kinetic mass, $\Delta (1/(r_1 m_2))$, and on the slope, $\mu_i$.
The resulting error is given in the column labeled ``$\kappa$ tuning'' in
Table~\ref{tbl:ME_fit_error}.

\subsubsection{Renormalization and matching uncertainty} 
\label{sec:renorm-err}

The results of the base fit are obtained with matrix elements that are renormalized with the mNPR expression
in Eq.~(\ref{eq:matchingmNPR}) where the $\rho_{ij}$ are calculated at one-loop order in perturbation theory
as in Eq.~(\ref{eq:ziimNPR}).
We take mNPR as our preferred approach because it includes all-orders and nonperturbative contributions
from wave-function renormalization that are omitted with tadpole-improved perturbation theory.
The chiral-continuum fit function for the base fit includes the generic terms of order $\alpha_s^2$ in
Eq.~(\ref{eq:renorm-alpha2}), where the unknown higher-order coefficients $\rho_{ij}^{[2]}$ are constrained
with Gaussian priors.
The contribution to the error on each matrix element and $\xi$ from these terms is shown in the column
labeled ``matching'' in Table~\ref{tbl:ME_fit_error}.

We consider a number of fit variations to investigate the robustness of our error estimate and to test for residual effects of truncating the perturbative expansion in $\alpha_s$.  First, we study the change in the matrix elements that results from varying the scale $q^*$ at which the strong coupling is evaluated in the range $1/a$--$3/a$, finding differences that are commensurate with the estimated matching errors in Table~\ref{tbl:ME_fit_error}.  Second,
we remove the $\order(\alpha_s^2)$ terms in Eq.~(\ref{eq:renorm-alpha2}) from the base fit function.
Third, we include higher order corrections through $\order(\alpha_s^3)$, adding the terms in
Eq.~(\ref{eq:renorm-alpha3}) to the base fit function.
These two previous fits are labeled ``mNPR'' and ``$\text{mNPR} + \alpha_s^3$", respectively, in Figs.~\ref{fig:stabO1ANDxi}--\ref{fig:stabO4ANDchisq}. The changes in central values and error bars are negligibly small.  Last, we perform two fits using the same fit function as for the base fit, but renormalize the matrix elements using tadpole-improved, one-loop perturbation theory taking two different definitions of $u_0$ in Eq.~(\ref{eq:matchingPT}): the fourth root of the average plaquette and the average link in Landau gauge. These final fits are labeled ``$\text{PT}_\text{P}+\alpha_s^2$'' and ``$\text{PT}_\text{L}+\alpha_s^2$'', respectively, in Figs.~\ref{fig:stabO1ANDxi}--\ref{fig:stabO4ANDchisq}. Here we see more significant changes in the central values that are still within two $\sigma_{\rm fit}$.
Notably, however, the variations between the ``$\text{PT}_\text{P}+\alpha_s^2$'' and
``$\text{PT}_\text{L}+\alpha_s^2$'' results indicate a systematic uncertainty associated with the choice of
tadpole-improvement factor, bolstering our view that mNPR should be more reliable.
Moreover, the stability of the mNPR results when adding terms of order $\alpha_s^2$ and $\alpha_s^3$ suggests
that the errors on the renormalized matrix elements properly include the uncertainty from residual
perturbative truncation effects.

\subsubsection{Truncation of the chiral and heavy-meson expansions}
\label{sec:chiral-LECs-error}
We estimate the error due to the truncation of the chiral expansion from the contributions of the
leading-order \cpt\ coefficients \{$\beta_i$, $\beta'_i$\} and the LECs $\{c_n, d_n\}$ of all analytic terms
that do not depend on the lattice spacing.
This error is shown in the column labeled ``chiral'' in Table~\ref{tbl:ME_fit_error}.
We also investigate the size of residual truncation effects in both the chiral and the heavy-meson
expansions through fit variations with fewer or additional terms; these are described below.


As discussed in Section~\ref{sec:chptfit}, the chiral-continuum base fit includes all NLO and NNLO analytic
terms of the chiral expansion.
To test the robustness of our error estimate and check for the size of residual truncation effects, we
consider two separate fit variations, the first including only NLO analytic terms, and the second 
including all analytic terms through N$^3$LO.
In the first case, where we omit the NNLO analytic terms, we also exclude the heaviest data points with
$r_1m_q\gtrsim 0.65\, r_1m_s$, since they are above the expected range of validity of NLO \cpt.
The results of this fit are labeled ``NLO $(m_q < 0.65 m_s)$'' in
Figs.~\ref{fig:stabO1ANDxi}--\ref{fig:stabO4ANDchisq}.
We observe only small changes to the results, but a relatively large change in the $\chi^2_{\rm aug}$/dof
due to the substantial alteration of both data set and fit function.
The second case, where we include the complete set of N$^3$LO analytic terms, yields the fit results that
are labeled ``N$^3$LO'' in Figs.~\ref{fig:stabO1ANDxi}--\ref{fig:stabO4ANDchisq}.
We observe negligible changes in both the central value and error bars.
Comparing the result of the base fit with NNLO analytic terms to those from the fit variations with only NLO
analytic terms and with analytic terms through N$^3$LO, we conclude that our base fit correctly accounts for
the error due to the truncation of the chiral expansion, and that residual truncation effects are negligibly
small.

Finally, we study the impact on our fit of our choices for prior widths on the \cpt\ LECs, which are based
on power-counting expectations.
We perform three different fits in which we separately widen the priors by a factor of two on (1) the LO
($\beta_i$ and $\beta'_i$) parameters; (2) the NLO LECs ($c_{n}$); or (3) the NNLO LECs ($d_{n}$).
These fits are labeled ``LO $\times$ 2'', ``NLO $\times$ 2'', and ``NNLO $\times$ 2'', respectively in
Figs.~\ref{fig:stabO1ANDxi}--\ref{fig:stabO4ANDchisq}.
For all variations, we observe negligible changes with respect to the base fit.
This confirms that our prior widths on the LECs are not unduly affecting our base-fit results.

As discussed in Section~\ref{sec:chptLogs}, we include in our base fit hyperfine- and
flavor- splitting effects in the heavy-light mesons that appear in the \cpt\ loop integrals, which are the
leading corrections in the $1/M_B$ expansion.
To test for heavy-meson truncation effects, we set both splittings to zero.
The result of this fit is labeled ``no splitting'' in Figs.~\ref{fig:stabO1ANDxi}--\ref{fig:stabO4ANDchisq}.
For this fit variation we see small-to-negligible changes in the matrix element results and in~$\xi$.
Because the effects of the leading $1/M_B$ corrections included in the base fit
are already so small, we conclude that residual heavy-meson truncation effects are negligible.

\subsubsection{Light-quark discretization errors}
\label{sec:LQ-disc-error}

Using the same procedure as for the ``chiral" error, we estimate the uncertainty associated with light-quark
discretization effects from the contributions of the \cpt\ LECs $\{c_n,d_n\}$ of all analytic terms that
depend on the lattice spacing.
This error is shown in the column labeled ``LQ disc'' in Table~\ref{tbl:ME_fit_error}.

The base chiral-continuum fit function incorporates taste-symmetry breaking effects in the chiral
logarithms, plus the corresponding taste-breaking analytic terms needed to maintain independence on 
the chiral scale.
The base fit does not however, include a separate analytic term for generic discretization contributions
from the light-quark and gluon actions.
The results from a fit in which the generic discretization term $F_i^{\alpha_s a^2\; {\rm gen}}$ [defined in
Eq.~(\ref{eq:gena2})] is added to the base fit are labeled ``generic $\order(\alpha_s a^2)$'' in
Figs.~\ref{fig:stabO1ANDxi}--\ref{fig:stabO4ANDchisq}.
We observe negligibly small changes in the central values and errors for the fitted matrix elements and for
$\xi$, and no discernible change in the $\chi^2_{\rm aug}$/dof.
This indicates that the analytic terms already included in the base-fit function are sufficient to describe
the lattice-spacing dependence of the data, and that the base-fit error properly accounts for the
contribution from light-quark discretization effects.

\subsubsection{Heavy-quark discretization errors}
\label{sec:HQ-disc-err}

We include six terms of order $\alpha_sa$, $a^2$, and $a^3$ in our base chiral-continuum fit to account for
discretization effects from the heavy-quark action and four-quark operators.
As in the previous sections, we estimate the error due to heavy-quark discretization effects from the
contributions of the coefficients of these terms.
This error is shown in the column labeled ``HQ disc'' in Table~\ref{tbl:ME_fit_error}.

To test for residual heavy-quark discretization effects, we also perform alternate fits including fewer
heavy-quark discretization terms.
The results of these fits are shown in Figs.~\ref{fig:stabO1ANDxi}--\ref{fig:stabO4ANDchisq}.
Our first fit variation includes only the two $\order(\alpha_s a)$ terms.
The results, which are labeled ``HQ $\order(\alpha_s a)$ -- only'' show only small deviations from the base
fit.
For some matrix elements, the resulting errors are slightly smaller than for the base fit, indicating that
additional terms are needed to saturate the error.
Our second fit variation includes the two $\order(\alpha_s a)$ terms and the three at $\order(a^2)$.
Differences between the results of this fit, which are labeled ``HQ $\order(\alpha_s a, a^2)$ -- only'', and
the base fit are imperceptible.
Although including heavy-quark terms through $\order(\alpha_s a, a^2)$ already saturates the error from
heavy-quark discretization effects, we include a sixth term of $\order(a^3)$ in the base fit because it
appears formally at the same order in the heavy-quark expansion as the other five.

The heavy-quark discretization terms depend upon a cut-off scale $\LamHQ$, which is fixed in the
chiral-continuum fit.
Reasonable variations in this parameter are absorbed by changes in the fitted coefficients and therefore do
not affect our results.

\begin{table}[tp]
\caption{Breakdown of the chiral-continuum fit error.
The labels and estimation procedure are described in the text.
Entries are in percent.
\label{tbl:ME_fit_error}}
\begin{tabular}{c@{\quad}c@{\quad}c@{\quad}c@{\quad}c@{\quad}c@{\quad}c@{\quad}c@{\quad}r}\hline\hline
           & statistics & inputs &$\kappa$ tuning & matching & chiral & LQ disc & HQ disc & fit total  \\
\hline
$\langle\op_1^d\rangle$ & 4.2 & 0.4    & 2.1            & 3.2     & 2.3    & 0.6    & 4.6     &  7.7 \\
$\langle\op_2^d\rangle$ & 4.6 & 0.3    & 1.1            & 3.7     & 2.6    & 0.6     & 4.6    &  8.0 \\
$\langle\op_3^d\rangle$ & 8.7  & 0.2    & 2.1            & 12.6     & 4.8   & 1.2    & 9.9    & 19.0 \\
$\langle\op_4^d\rangle$ &3.7 & 0.4    & 1.7            & 2.2      & 1.9   & 0.5     & 3.9   &  6.4 \\
$\langle\op_5^d\rangle$ & 4.7 & 0.5   & 2.5    & 4.7  & 2.7    & 0.8     & 4.9   &  9.1 \\
$\langle\op_1^s\rangle$ & 2.9 & 0.4    & 1.5  & 2.1   & 1.6    & 0.4  & 3.2   &  5.4 \\
$\langle\op_2^s\rangle$ & 3.1& 0.3    & 0.8  & 2.5 & 1.6    & 0.4 & 3.1 & 5.5 \\
$\langle\op_3^s\rangle$ & 5.9 & 0.3    & 1.4 & 8.6  & 3.0 & 0.7  & 6.9     & 13.0 \\
$\langle\op_4^s\rangle$ & 2.7 & 0.4    & 1.2 & 1.6 & 1.3  & 0.3 & 2.9    &  4.8 \\
$\langle\op_5^s\rangle$ & 3.4 & 0.4    & 1.8 & 3.4 & 1.9 & 0.5 & 3.6     &  6.7 \\
$\xi$                               & 0.8 & 0.4 & 0.3  & 0.5 & 0.4    & 0.1    & 0.7 & 1.4 \\
\hline\hline
\end{tabular}
\end{table}

\subsection{Errors considered after the base chiral-continuum fit}
\label{sec:SystErrorB}

For some sources of uncertainty, we estimate the error contributions to the matrix elements and $\xi$ after
the chiral-continuum extrapolation. To do so, we perform additional fits and compare the results to that
of the base fit.
This includes the errors from the relative scale $r_1/a$ and finite-volume effects.
Further, although we include the indirect contributions of the error from the physical scale $r_1$ through
its effects on physical input parameters in the chiral-continuum fit, here we consider its impact on the
final conversion of the matrix elements from $r_1$ units to GeV.
The error contributions from these uncertainty sources are listed in Table~\ref{tbl:ME_tot_error} and our
methodology for their estimation is described below.
We add these errors to the fit error \emph{a posteriori} to obtain the ``Total'' error in
Table~\ref{tbl:ME_tot_error}.

\subsubsection{Finite-volume effects}
\label{sec:FV-err}

Our base fit employs the finite-volume expressions for the NLO chiral
logarithms~\cite{Bernard:2013dfa}, which are sums over
the discrete momenta allowed with periodic boundary conditions. 
We estimate the error due to omitted higher-order finite-volume corrections by performing a second fit using
the infinite-volume NLO chiral logarithms.
We take half the shift in the central value between the base fit and the infinite-volume fit as an estimate
of the finite-volume error.
We expect this difference between NLO fits to be a conservative estimate of the finite-volume error, since
the omitted finite-volume corrections are in fact of NNLO.
The finite-volume error is listed in the ``FV'' column of Table~\ref{tbl:ME_tot_error}, and is negligible
when added in quadrature to the fit error.

\subsubsection{\texorpdfstring{Relative scale $r_1/a$}{r1/a}}
\label{sec:r1a}

We use the relative lattice spacings $r_1/a$ from Table~\ref{tab:LatEns} to convert our data from lattice
units to $r_1$ units before adjusting for the $b$-quark mass and performing the chiral-continuum fit.
We must then account for the uncertainties in the $r_1/a$ values in our total error budgets for the matrix
elements and $\xi$.

We have considered including the values of $r_1/a$ as constrained parameters in the
chiral-continuum fit in order to incorporate their uncertainties and correlations into the fit error.
When we do so, we find that a fit that includes the errors on $r_1/a$ and their
correlations (via Gaussian priors) returns fitted $r_1/a$ values on some ensembles outside of their prior
constraints.
We therefore fix the values of $r_1/a$ in our base fit to prevent the matrix-element data, which have much
larger statistical errors than those for the heavy-quark potential, from significantly changing the relative
lattice spacings.
We take the increases in the errors on the matrix elements and $\xi$ from the fit in which we constrain the
$r_1/a$ values with Gaussian priors as the errors due to $r_1/a$ uncertainties.
To separate this error, we subtract in quadrature the base-fit error from the constrained-$r_1/a$-fit error.
The resulting error estimate is shown in the column labeled ``$r_1/a$'' in Table~\ref{tbl:ME_tot_error}.

\subsubsection{Absolute scale (\texorpdfstring{$r_1$}{r1}) uncertainty }

We take the values for input meson masses and the pion decay constant in our chiral-continuum fit from the
PDG.
We constrain each of these parameters with Gaussian priors, and include the error from the conversion from
GeV to $r_1$ units by adding the 0.7\% error on the physical value of $r_1$ [Eq.~(\ref{eq:r1})] to the PDG
error to obtain the prior width.
Similarly, the quark masses, which are also input parameters to the chiral-continuum fit function, are
determined from experimentally measured meson masses, and are therefore also affected by the uncertainty in
Eq.~(\ref{eq:r1}).
As discussed in Sec.~\ref{sec:chiral-log-error}, all of the indirect effects of the $r_1$ uncertainty on the
error on the matrix elements and $\xi$ from the physical input parameters are already included in the
chiral-continuum fit error.

After our chiral-continuum fit, however, we must still convert our final results for the matrix elements
from $r_1$ units to GeV.
This introduces an additional error due to the uncertainty in the physical value of $r_1$ which is listed in
the column ``$r_1$'' of Table~\ref{tbl:ME_tot_error}.
Note that this direct ``$r_1$" error does not enter dimensionless quantities such as the bag parameters and
the ratios of matrix elements.

\subsubsection{Isospin breaking and electromagnetism} \label{sec:isospin}

We now consider the systematic effects on the mixing matrix elements and $\xi$ due to $m_u \neq m_d$. With our partially quenched analysis we can vary the valence- and sea-quark masses independently.  We obtain the physical $B_d$- and $B_s$-meson mixing matrix elements after the chiral-continuum fit by fixing the valence-quark masses to $m_d$ and $m_s$ given in Table~\ref{tbl:extrap-point}, and fixing the light sea-quark mass to the average up-down quark mass $\hat{m} = (m_u + m_d)/2$.
This accounts for the dominant isospin-breaking effects from the valence sector.
Because the light sea-quark masses are degenerate in all our ensembles, however, the effects of isospin breaking in the sea are a source of systematic error in our calculation.

In order to estimate the error introduced by neglecting isospin-breaking effects in the sea we consider the dependence of the chiral-continuum fit function on the light sea-quark masses.
Because the expressions for the $B_q$-mixing matrix elements are symmetric under the interchange
$m_u^\text{sea} \leftrightarrow m_d^\text{sea}$, the leading contributions from isospin-breaking in the sea
sector are of $\order{((m_d^\text{sea} - m_u^\text{sea})^2)}$, and are similar in size to NNLO terms in the
chiral expansion.
In practice, isospin-breaking errors arise from two types of terms in the chiral-continuum fit function:
logarithms containing mixed valence-sea mesons, and analytic terms containing the sum of the squares of
sea-quark masses.
We estimate the numerical size of the errors in the matrix elements to be $(x_d - x_u)^2 \sim 0.01\%$, where
$x_q$ is the dimensionless $\chi$PT expansion parameter defined in Eq.~(\ref{eq:xq}), and we have taken the
coefficient to be unity based on power-counting expectations.
Isospin-breaking errors in the ratio $\xi$ are expected to be further suppressed by the SU(3)-breaking
factor $(m_s - m_d) / \LamQCD \sim 1/5$.  The error introduced by neglecting isospin breaking in the sea sector is negligible compared with the other sources of uncertainty discussed above, so it is not shown in Table~\ref{tbl:ME_tot_error}.

The MILC asqtad gauge-field ensembles employed in this work do not include electromagnetism.
This introduces errors of order $\alpha_\text{EM}$ from the omission of one-loop diagrams such as those with
a photon connecting the valence $b$ and $d(s)$ quarks, or a photon connecting one valence quark to a
sea-quark loop.%
\footnote{In $\Delta M_q$, further EM corrections stem from adding photons to the box diagrams.} %
We estimate the numerical size of the omitted electromagnetic contributions to the matrix elements to be
$\alpha_\text{EM}/\pi \sim 0.2\%$, where we include a loop-suppression factor of $\pi$ in the denominator.
This is consistent with the size of the electromagnetic contribution to the proton-neutron mass
difference, which has been found to be 2--3~MeV in QCD+QED
simulations~\cite{Borsanyi:2014jba,Horsley:2015eaa}.
Again, errors in the SU(3)-breaking ratio $\xi$ are yet smaller.  The error from omitting electromagnetism is shown in the ``EM'' column in Table~\ref{tbl:ME_tot_error}.

Other recent dynamical lattice-QCD calculations of the $B_d$-mixing matrix elements~\cite{Gamiz:2009ku,Carrasco:2013zta,Aoki:2014nga} do not account for isospin breaking, which is currently a subleading source of error compared to other uncertainties.  It is interesting, however, to estimate the size of the omitted valence isospin-breaking corrections to these results with our data.  We do so by taking the difference between the matrix elements evaluated with the correct valence-quark mass and those with the valence-quark mass fixed to the isospin average, {\it i.e.}:
\begin{equation}
	\delta_i^{\rm isospin} \equiv \frac{\langle \op_i^d \rangle - \langle \op_i^{ud} \rangle}{\langle \op_i^{ud} \rangle} \,, \label{eq:val_isospin}
\end{equation}
with the sea-quark mass set to the average light-quark mass in both.  Table~\ref{tab:val_isospin} gives the valence isospin-breaking corrections for all five matrix elements $i$=1--5 and also for the ratio $\xi$.  The corrections to the matrix elements are positive, and are about a half a percent.  They can in principle be used to adjust lattice results for $B_d$-mixing matrix elements from isospin-symmetric full-QCD simulations, although in practice this is not yet necessary since the isospin-breaking shifts are much smaller than the total errors currently quoted. Nevertheless, they indicate that the inclusion of valence isospin breaking will be important once the errors on the $B$-mixing matrix elements reach the few percent level.  

Further, because the matrix elements $\langle \op_i^d \rangle$ are proportional to the square of the $B_d$-meson decay constant, the valence isospin-breaking corrections in Table~\ref{tab:val_isospin} can provide a very rough guide as to the size of the analogous shifts that might be expected for $f_{B_d}$.  If one assumes that valence isospin breaking is negligible in the bag parameters and takes the entire correction to the matrix elements to be from $f_{B_d}^2$, the approximately 0.5\% shifts in $\langle \op_i^d \rangle$ would correspond to about a 0.25\% shift in $f_{B_d}$, or about 2.5~MeV.  Numerically, a correction of this size would be important given the uncertainties currently quoted on the most precise lattice $B$-meson decay constant calculations~\cite{Na:2012kp,Dowdall:2013tga}, which are performed in isospin-symmetric full QCD.  We will compute the valence isospin-breaking correction to $f_{B_d}$ (and $f_{D_+}$) directly in our forthcoming $B$- and $D$-meson decay-constant analysis~\cite{Kronfeld:2015xka}, which includes partially-quenched data.

\begin{table}
\caption{Valence isospin-breaking corrections defined in Eq.~(\ref{eq:val_isospin}) for the $B_d$-mixing matrix elements and $\xi$.}
\label{tab:val_isospin}
\begin{tabular}{l@{\quad}c@{\quad} c@{\quad}c@{\quad} c @{\quad}c@{\quad}  c}
\hline\hline
    $i$ & 1 & 2 & 3 & 4 & 5 & $\xi$ \\ 
    \hline 
    $\delta_i^{\rm isospin}$ (\%)    &  0.51(11) 	& 0.56(12) 	& 0.49(15)		& 0.423(93)	& 0.409(99) &  $-0.257(56)$ \\       
\hline\hline    
\end{tabular}
\end{table}

\subsubsection{Other consistency checks}    \label{sec:altdatasets}

\begin{figure}[tbp]
    \vspace{-0.2in}
    \includegraphics[width=0.38\textwidth,angle=0,origin=c]{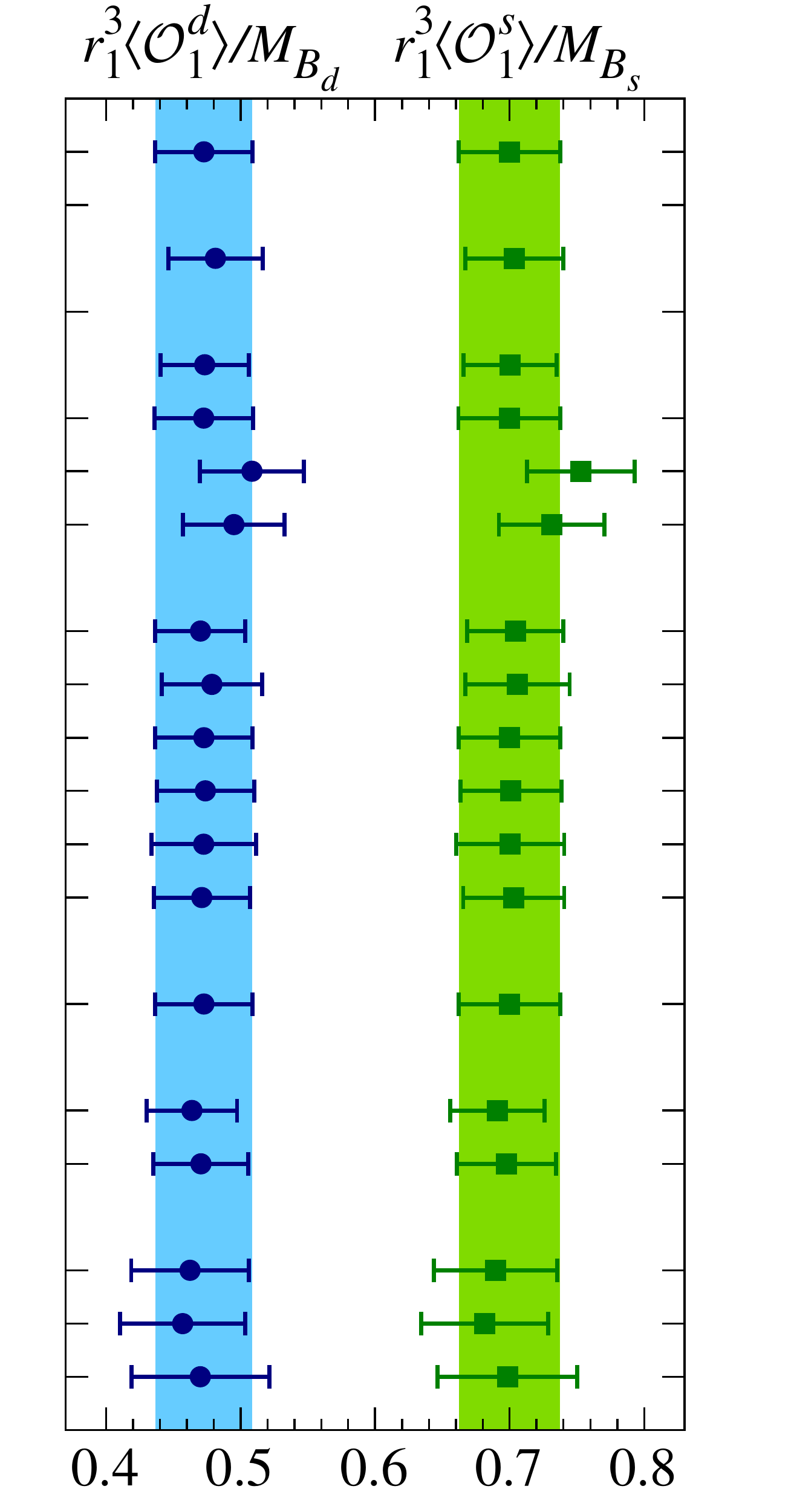}
    \hspace{-0.8in}
    \includegraphics[width=0.38\textwidth,angle=0,origin=c]{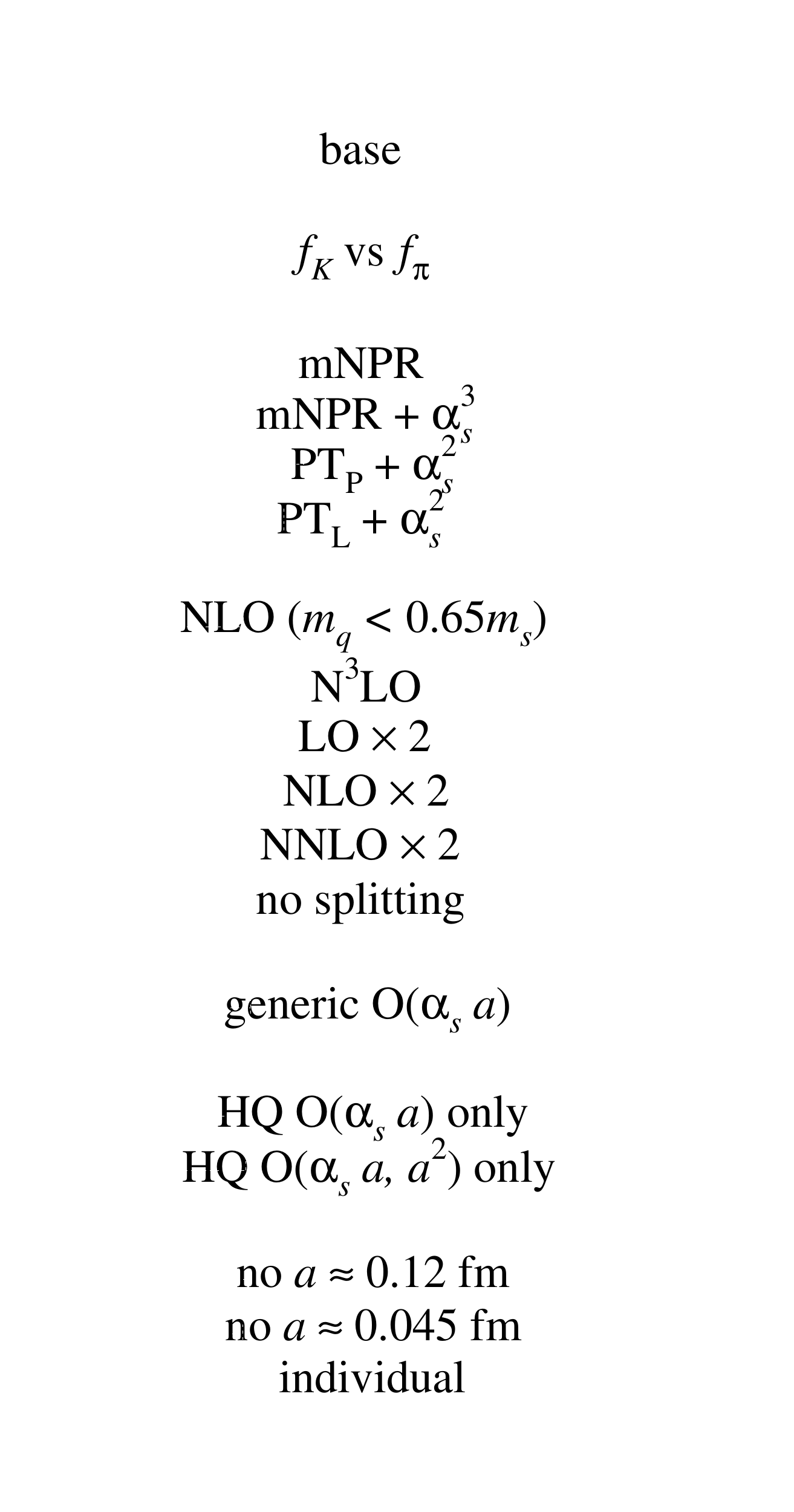}
    \hspace{-0.8in}
    \includegraphics[width=0.38\textwidth,angle=0,origin=c]{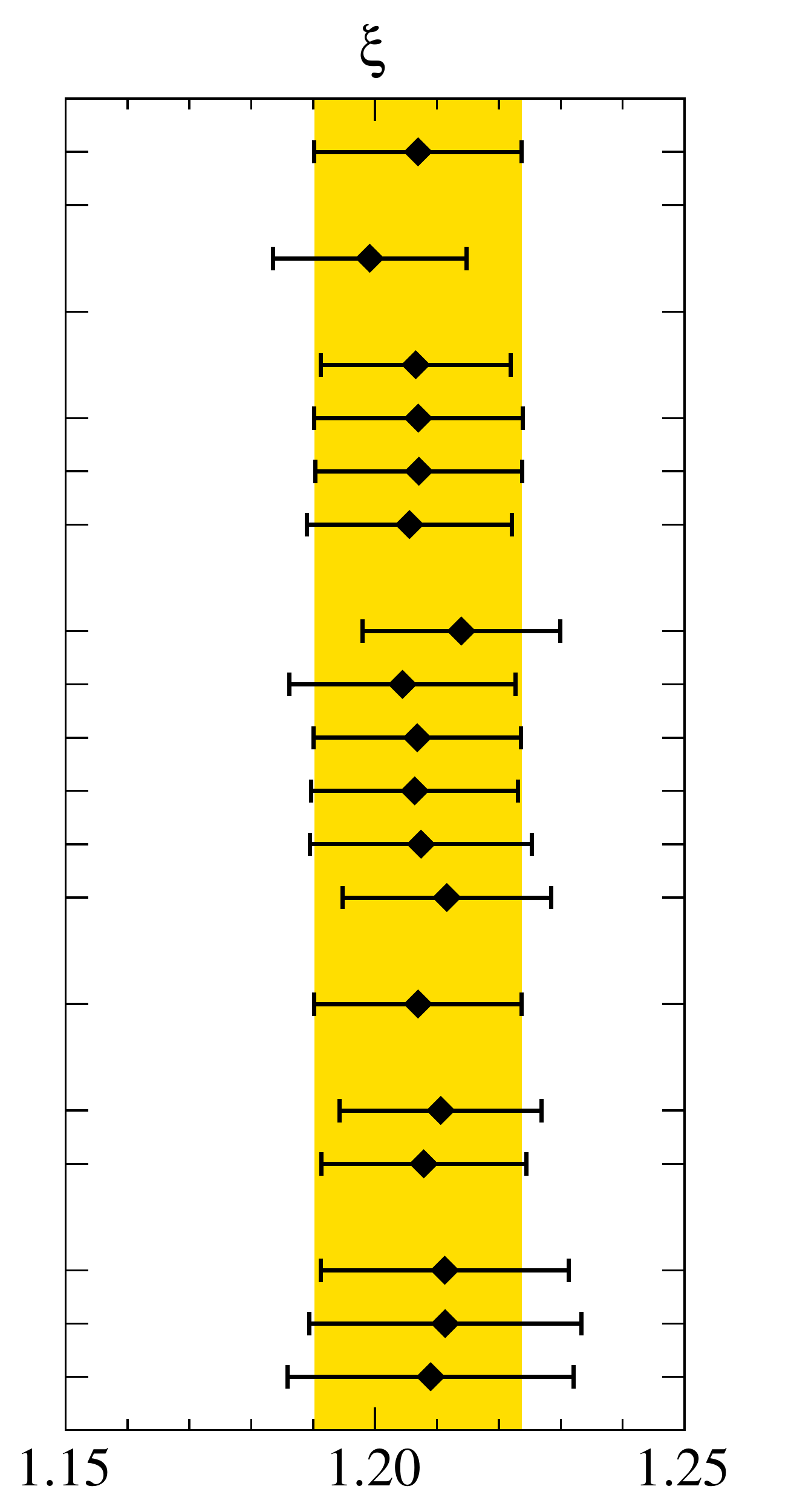}
    \caption{(\emph{left}) Results for the matrix elements $\langle \op_1^s \rangle$ (green squares) and 
    $\langle \op_1^d \rangle$ (blue circles) from the base chiral-continuum fit and the alternative fits 
    described in the text.
    (\emph{right})  Results for the SU(3)-breaking ratio $\xi$ from the same fits.
    In both plots, the base-fit result is shown  as the top entry and also indicated by a solid band to 
    enable comparison with the other fit results, which are shown in the same order in which 
    they are discussed.}
    \label{fig:stabO1ANDxi}
\end{figure}

\begin{figure}[tbp]
    \vspace{-0.2in}
    \includegraphics[width=0.38\textwidth,angle=0,origin=c]{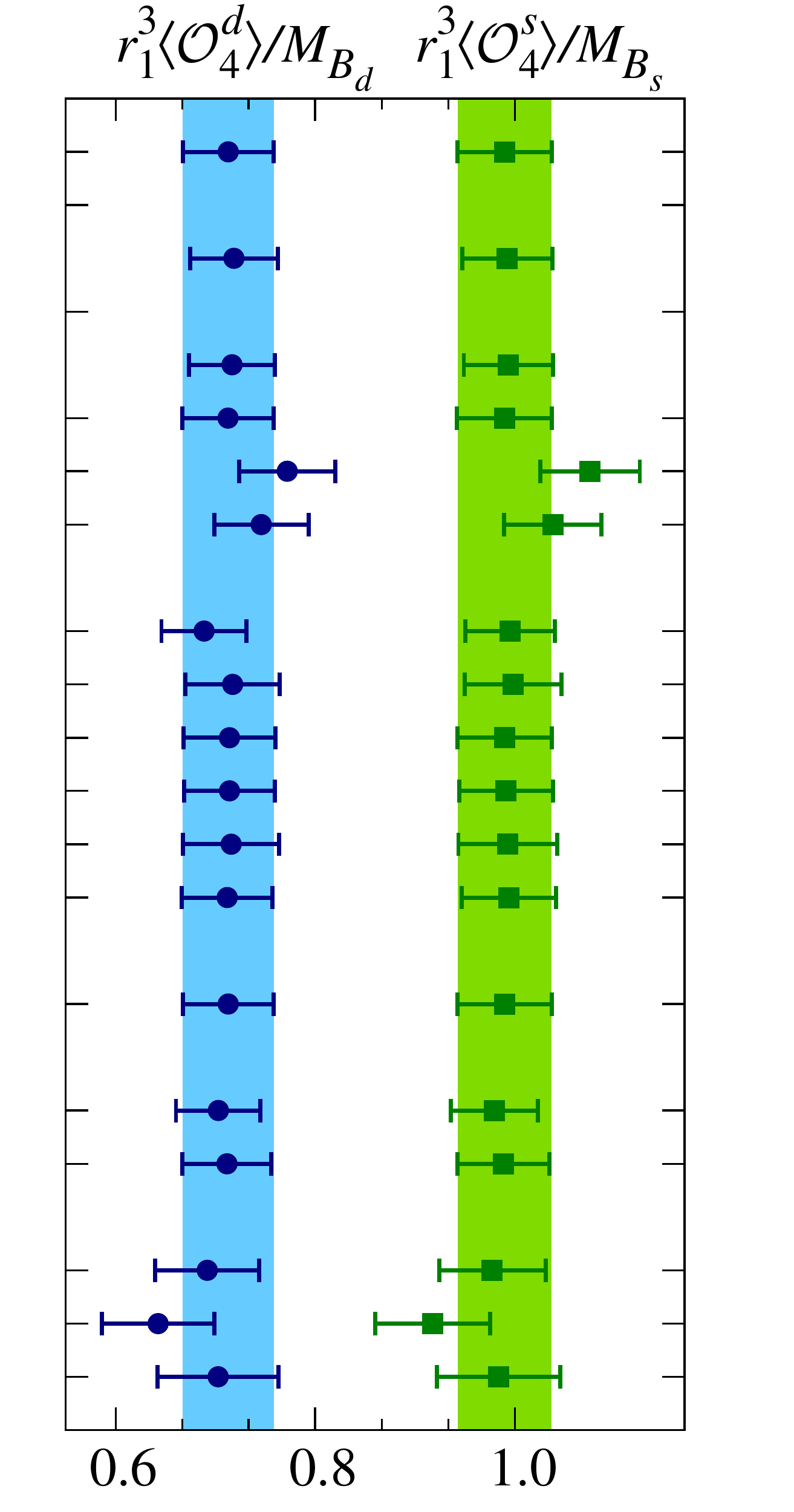}
    \hspace{-0.8in}
    \includegraphics[width=0.38\textwidth,angle=0,origin=c]{figs/stab_label.pdf}
    \hspace{-0.8in}
    \includegraphics[width=0.38\textwidth,angle=0,origin=c]{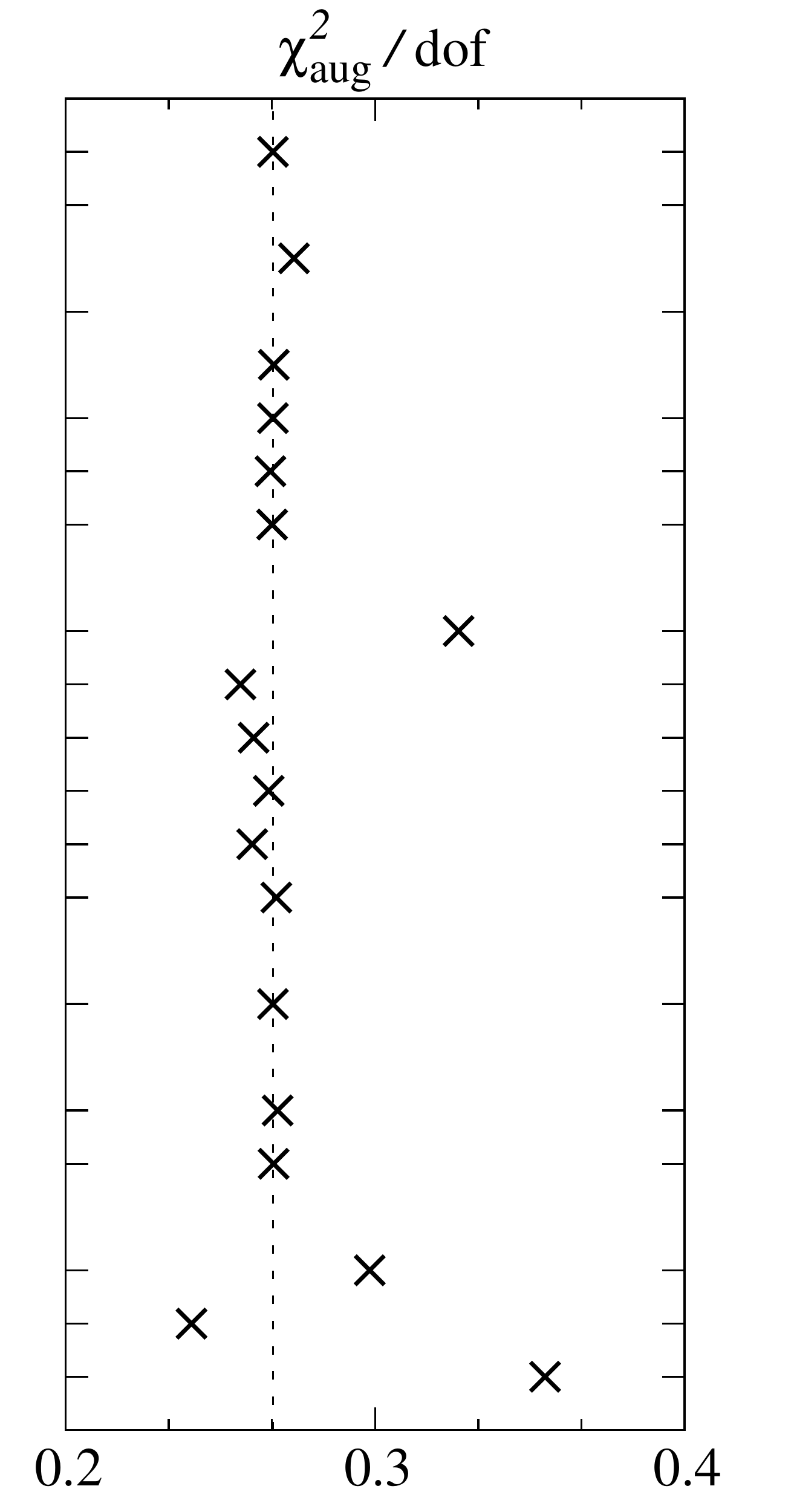}
    \caption{(\emph{left}) Results for the matrix elements $\langle\op_4^s\rangle$ (green squares) and
    $\langle \op_4^d \rangle$ (blue circles) from the base chiral-continuum fit and the alternative fits 
    described in the text.
    (\emph{right}) Relative fit quality for the same fits as indicated by the augmented 
    $\chi^2_\text{aug}$/dof.
    The base-fit $\chi^2_\text{aug}$/dof is also shown as a dashed vertical line to enable comparison with 
    the other fit results. }
    \label{fig:stabO4ANDchisq}
\end{figure}

We check the consistency between fit results with different data subsets by performing two additional fits
in which we drop either the data on our largest or smallest lattice spacing.
The results of these fits are labeled ``no $a \approx 0.12$~fm'' and ``no $a \approx 0.045$~fm'' in
Figs.~\ref{fig:stabO1ANDxi}--\ref{fig:stabO4ANDchisq}, respectively.
As expected, the errors increase when data is omitted.
In most cases, the new central values differ from those of the base fit by less than one $\sigma_{\rm fit}$,
with one observed difference of almost 2$\sigma_{\rm fit}$ as would be expected given so many fits.
The consistency between the fit results with different subsets of data is further evidence that our
chiral-continuum fit provides a good description of the observed discretization effects.

Finally, to study the effect of correlations between the operators that mix under renormalization and due to
the choice of local four-quark operators, we perform separate fits to the data for each operator.
The results of these fits are labeled ``individual" in Figs.~\ref{fig:stabO1ANDxi}--\ref{fig:stabO4ANDchisq}
and are very close to those of the base fit.
As expected, the errors on the individual fit results are larger because the contributions from the other
off-diagonal matrix elements cannot be resolved, so their uncertainties are governed by the prior widths on
the associated LECs ($\beta_j^{(\prime)}$s).

\subsection{Omission of the charm sea} \label{sec:charm}

The MILC asqtad ensembles do not include charm sea quarks.
One can expand the charm-quark determinant $\Det(\Dslash+m_c)$ in powers of $1/m_c$ to obtain an estimate of
the leading effects of this omission~\cite{Nobes:2005yh}.
The first nonvanishing contribution is proportional to $g_0^2\Tr[F^{\mu\nu}F_{\mu\nu}]$, which is absorbed
into the bare gauge coupling.
The next contribution---after using identities and eliminating the redundant operator $\Tr[(D^\mu
F_{\mu\nu})(D_\rho F^{\rho\nu})]$---is
$g_0^3m_c^{-2}\Tr[F_\mu^{\phantom{\mu}\nu}F_\nu^{\phantom{\nu}\rho}F_\rho^{\phantom{\rho}\mu}]$.
Thus, the ensembles lack a contribution of order $\alpha_s(\LamQCD/2m_c)^2$.
Taking $\overline m_c(\overline m_c)= 1.275$~GeV from the PDG~\cite{Agashe:2014kda} and reasonable values
for $\alpha_s$ and $\LamQCD$ leads to error estimates of 1\%--2\% for the matrix elements, which should be
suppressed by $(m_s-m_d)/\LamQCD$ in the SU(3)-breaking ratio $\xi$.

To check this power-counting estimate, we have examined results for related 
matrix elements from calculations with different numbers of sea quarks.
A comparison of four- and three-flavor results for kaon, $D$-, and $B$-meson decay
constants~\cite{Rosner:2015wva,Aoki:2013ldr} yields differences that are consistent with zero within errors,
but that could allow for effects as large as 1\%--2\%.
There are also no significant differences between three- and two-flavor results~\cite{Aoki:2013ldr}, which
one might expect to be larger, since contributions from the strange sea are entirely nonperturbative, unlike
the charm sea case.
In the kaon system, where lattice results with total errors at the sub-percent level exist for both the
three- and four-flavor cases, the differences are much smaller than the sub-percent level errors.

The scale dependence of the $B_q$-mixing matrix elements gives rise to another charm-loop effect that must
be considered.
Because we obtain the matrix elements at the scale $\mu=m_b$, charm-loop contributions to the scale
dependence of the matrix elements between $\mu=m_c$ and $m_b$ are omitted in a three-flavor calculation.
We estimate this effect by taking our result for $\langle \op_1 \rangle(m_b)$ and evolving it to
$\mu=\overline{m}_c$ using the two-loop $\beta$-function and anomalous dimension given in Eq.~(\ref{eq:RGI})
evaluated with $N_f = 3$, and then evolving the matrix element back to $\mu=\overline{m}_b$ with $N_f=4$.
This procedure yields a shift of $0.3\%$.

In summary, it is possible that the effects on $B_q$-mixing matrix elements from omitting charm sea quarks
are as small as in the kaon system, \emph{i.e.}, at-or-below the sub-percent level.
In the absence of similarly precise results in the $B$- and $D$-meson systems, however, we conservatively
use the power counting estimate discussed above and assign an error of $2\%$ on the $B_q$-mixing matrix
elements from the omission of charm sea quarks.
This estimate accounts for both effects of omitting charm-sea quarks considered here.
For $\xi$, we consider the additional SU(3) breaking suppression factor $(m_s-m_d)/\LamQCD$ in our estimate,
and assign a similarly conservative error of $0.5\%$ due to omitted charm-sea effects.
The bag parameters are ratios of mixing matrix elements and decay constants, and we assume that the remaining 
contributions from charm loops are negligible compared to other uncertainties.
Hence, we do not assign an error due to the omission of the charm sea to our bag parameter results.
Finally, because the estimated charm sea error is much less quantitative than all
our other uncertainties, we do not add it in quadrature to the total error.
Instead we list it as a separate ``charm sea" error in Table~\ref{tbl:ME_tot_error} and also show it
separately in our results presented in Sec.~\ref{sec:results}.

\subsection{Error summary}
\label{sec:errsum}

In this subsection we present a summary of all systematic errors in our lattice-QCD calculation, and then
combine them to obtain the total errors in the matrix elements and~$\xi$.
Our base chiral-continuum fit function includes higher-order terms constrained with Bayesian priors to
account for the dominant sources of systematic error.
As illustrated in Figs.~\ref{fig:stabO1ANDxi}--\ref{fig:stabO4ANDchisq}, we consider over a dozen fit
variations to study residual truncation effects for the dominant sources of uncertainty in our lattice
calculation, including chiral extrapolation, light- and heavy-quark discretization, and renormalization
effects.
We conclude that the fit error from our base chiral-continuum fit properly accounts for these effects.
Table~\ref{tbl:ME_fit_error} gives an approximate breakdown of the chiral-continuum fit error into separate
contributions for each matrix element and the ratio $\xi$ as described in Sec.~\ref{sec:SystErrorA}.
We then add in quadrature to the ``fit total" error all the significant contributions that are not already
included in the chiral-continuum fit uncertainty (i.e., those from finite-volume effects, $r_1/a$
uncertainties, the physical scale $r_1$, and electromagnetic effects) to obtain the ``Total" error in
Table~\ref{tbl:ME_tot_error}.
Errors associated with isospin breaking are estimated to be negligible.
In summary, the ``Total'' error of Table~\ref{tbl:ME_tot_error} includes all significant contributions to
the matrix elements and $\xi$ after all possible sources of uncertainty have been considered with the
exception of our estimate of dynamical charm effects (discussed in Sec.~\ref{sec:charm}), which is listed
separately in the last column of the table.
This separation will enable the errors on our results to be easily adjusted in the future if more reliable
estimates of the size of charm sea-quark contributions become available.

\begin{table}[tp]
\caption{Total error budget for matrix elements converted to physical units of GeV${}^3$ and for the
dimensionless ratio $\xi$.
The error from isospin breaking, which is estimated to be negligible at our current level of precision is
not shown.
Entries are in percent.}
\label{tbl:ME_tot_error}
\begin{tabular}{c@{\quad}c@{\quad}r@{\quad}c@{\quad}c@{\quad}r@{\quad}r@{\quad}c}\hline\hline
\hspace{0.5in}              & Fit total & FV     & $r_1/a$ & $r_1$ &  EM  & Total & Charm sea \\
\hline
$\langle\op_1^d\rangle/M_{B_d}$  & 7.7 	& 0.2       	& 2.5		& 2.1   & 0.2  & 8.3  	& 2.0 \\
$\langle\op_2^d\rangle/M_{B_d}$  & 8.0 	& 0.3        	& 2.8     	& 2.1   & 0.2  & 8.8  	& 2.0 \\
$\langle\op_3^d\rangle/M_{B_d}$  & 19.0 & $<0.1$ 	& 2.5		& 2.1   & 0.2  & 19.3	& 2.0 \\
$\langle\op_4^d\rangle/M_{B_d}$  & 6.4 	& $<0.1$ 	& 2.1		& 2.1   & 0.2  & 7.1  	& 2.0 \\
$\langle\op_5^d\rangle/M_{B_d}$  & 9.1 	& $<0.1$ 	& 2.2		& 2.1   & 0.2  & 9.6 	& 2.0 \\
$\langle\op_1^s\rangle/M_{B_s}$  & 5.4 	& 0.1		& 1.9		& 2.1   & 0.2  & 6.1 	& 2.0 \\
$\langle\op_2^s\rangle/M_{B_s}$  & 5.5 	& 0.1		& 2.1     	& 2.1   & 0.2  & 6.2  	& 2.0 \\
$\langle\op_3^s\rangle/M_{B_s}$  & 13.0 	& $<0.1$ 	& 1.9		& 2.1   & 0.2  & 13.3	& 2.0 \\
$\langle\op_4^s\rangle/M_{B_s}$  & 4.8 	& $<0.1$ 	& 1.7		& 2.1   & 0.2  & 5.5  	& 2.0 \\
$\langle\op_5^s\rangle/M_{B_s}$  & 6.7 	& $<0.1$ 	& 1.8     	& 2.1   & 0.2  & 7.2  	& 2.0 \\
$\xi$                            		    & 1.4 	& $<0.1$ 	& 0.6     	&  0    & 0.04 & 1.5  	& 0.5 \\
\hline\hline
\end{tabular}
\end{table}

\section{Results}
    \label{sec:results}
    Here we present our final results with total uncertainties that include all contributions to the errors
considered in the preceding section.
As discussed there, we report the charm-sea error separately from the total of the statistical and all other
systematic uncertainties.
First, in Sec.~\ref{sec:MEResults}, we give the $B_q$-mixing matrix elements.
Next, in Sec.~\ref{sec:DerivedQuantites}, we provide quantities that are derived from the $B_q$-mixing
matrix elements, such as the SU(3)-breaking ratio $\xi$ and the bag parameters ${B}_q^{(i)}$.
In both Secs.~\ref{sec:MEResults} and~\ref{sec:DerivedQuantites}, we highlight a few main results---such as
for the renormalization-group-invariant quantities for the Standard-Model operator $\op_q^{(1)}$---and
compare our results with those from other calculations.
Appendix~\ref{app:Results} provides our complete results for all matrix elements and bag parameters,
including their correlations.
Finally, in Sec.~\ref{sec:SMPheno} we use the $B_q$-mixing matrix elements and combinations thereof to
compute phenomenologically-interesting observables and CKM matrix elements within the Standard Model.

The tables in Appendix~\ref{app:Results} present the matrix elements and bag parameters for operators
$\op^q_{2,3}$ ($q=d,s$) corresponding to both the BMU and BBGLN evanescent-operator schemes 
discussed in Sec.~\ref{sec:PT}.
These results are obtained from separate chiral-continuum fits which employ different matching coefficients
for $\me{2,3}$.
Although the matching coefficients for $\op^q_{1,4,5}$ are identical in the two schemes, the fitted matrix
elements $\me{1,4,5}$ can differ slightly due to correlations with $\me{2,3}$.
In practice, the results for $\me{1,4,5}$ differ by at most 1 in the least significant digit reported.
We therefore present results for $\me{1,4,5}$ and the corresponding bag parameters from only the BBGLN fit.
 
\subsection{Matrix elements}
\label{sec:MEResults}

We convert our final results for the $B_q$-mixing matrix elements $\langle \op_i^q \rangle$ ($q=d,s$;
$i$=1--5), obtained in the $\MSbar$-NDR scheme, to the combination $f^2_{B_q} B_{B_q}^{(i)}(\overline{m}_b)$
via Eqs.~(\ref{eq:Bq_1})--(\ref{eq:Bq_45}), taking the $\MSbar$ quark masses in Eqs.~(\ref{eq:Bq_23}) and
(\ref{eq:Bq_45}) from Table~\ref{tab:inputs}.
Here, and in the rest of the paper, we choose $\mu = \overline m_b(\overline m_b)=4.18(3)$~GeV.
Our final results for $f^2_{B_q} B_{B_q}^{(i)}(\overline{m}_b)$ are listed in Table~\ref{tbl:ME_results}
of Appendix~\ref{app:Results}.
They are the complete set needed to describe neutral $B$-meson mixing in the Standard Model and all
extensions thereof.
To facilitate their use in other phenomenological analyses, we provide the matrix of correlations between
their values in Table~\ref{tbl:ME_cor}.

\begin{figure}
\includegraphics[width=0.48\textwidth, angle=0]{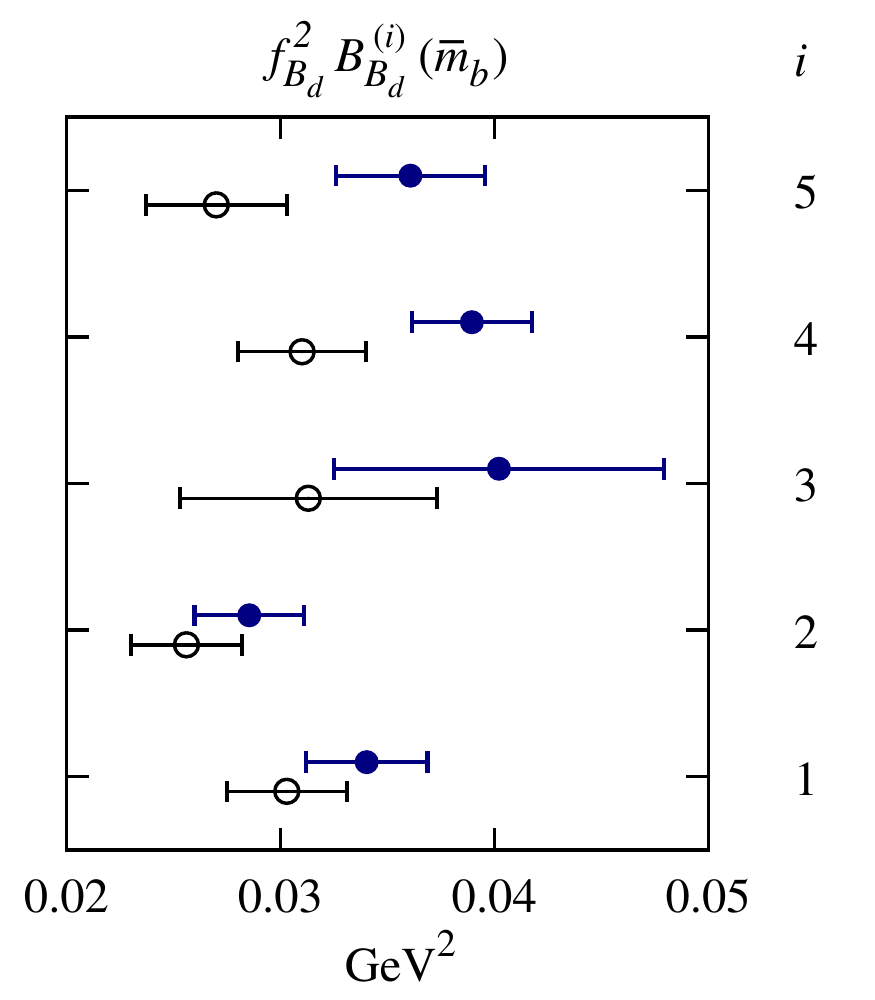} \hspace{-0.2in}
\includegraphics[width=0.48\textwidth, angle=0]{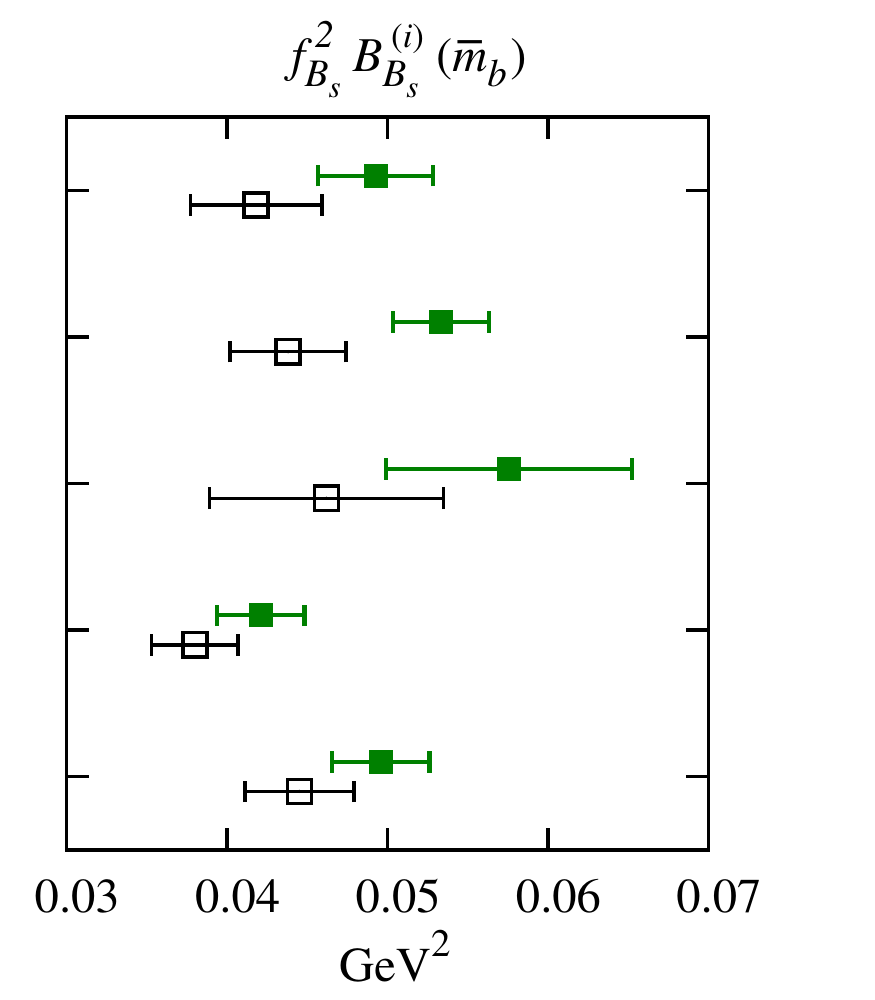} \hspace{-0.4in}
\caption{Comparison of our full set of $B_q$-mixing matrix elements (filled symbols) with the
two-flavor results from ETM~\cite{Carrasco:2013zta} (open symbols) converted to our definition of the bag parameters in Eqs.~(\ref{eq:Bq_1})--(\ref{eq:Bq_45}); the quoted ETM uncertainties do not
include an error from omitting strange sea quarks.
For $f^2_{B_q} B_{B_q}^{(2,3)} (\overline m_b)$, we use the BMU scheme to enable direct comparison with the
values in Table~4 of that paper.
The error bars on our results do not include the estimated charm-sea uncertainties, which are too small to
be visible.}
\label{fig:ME_results}
\end{figure}

Figure~\ref{fig:ME_results} compares our results for the full set of $B_q$-mixing matrix elements with those
of the ETM collaboration~\cite{Carrasco:2013zta}, which were obtained using two sea-quark flavors.
Our matrix-element errors range from about 5--15\%, and are larger for $B_d$ operators due to the need to
extrapolate to the physical $d$-quark mass.
The uncertainties quoted by ETM are similar, but do not include an uncertainty due to omitting the strange
sea quark in their error budget.
Our results for $\langle \op_{1,2,3}^q \rangle$ agree with those of ETM, while those for
$\langle\op_{4,5}^q\rangle$ differ by about 2$\sigma$.

\begin{figure}
\hspace{0.2in}
\includegraphics[width=0.485\textwidth, angle=0]{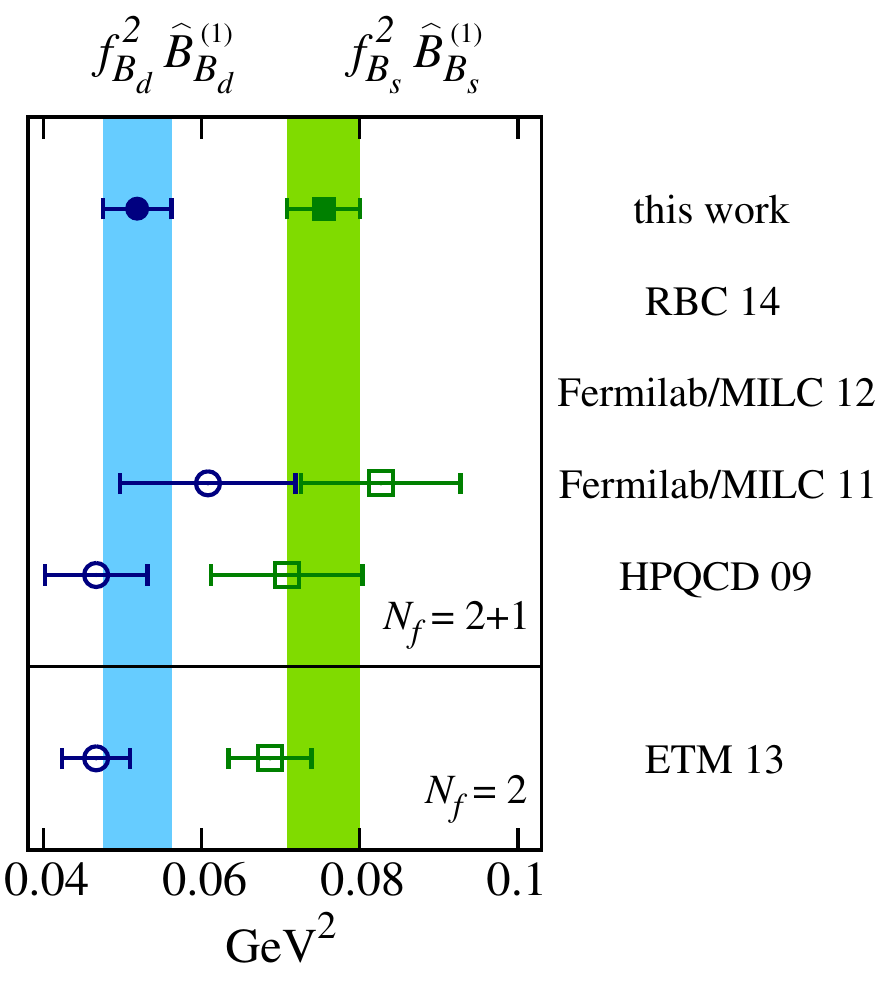} 
\hspace{-0.14in}
\includegraphics[width=0.485\textwidth, angle=0]{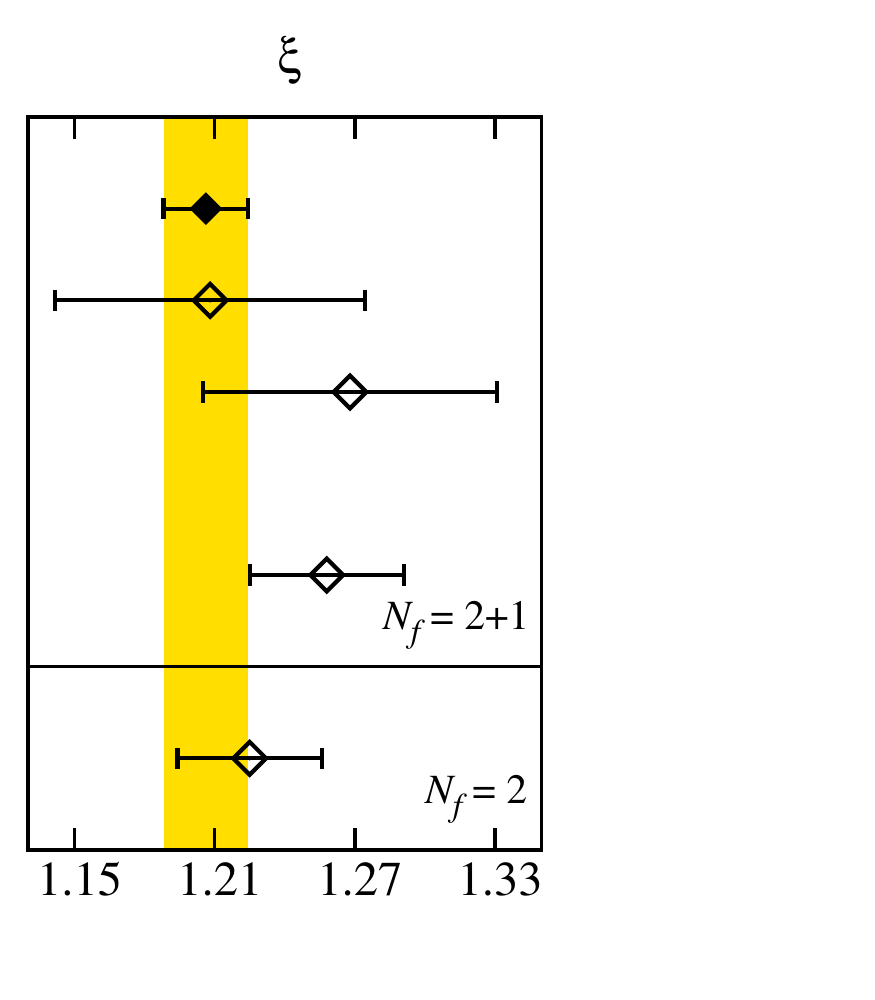} 
\hspace{-0.9in}
\caption{Comparison of the $B_q$-mixing matrix elements for $\op_1^q$ obtained in this work (filled symbols 
and vertical bands) with other unquenched lattice-QCD calculations~\cite{Gamiz:2009ku,Bouchard:2011xj,%
Bazavov:2012zs,Carrasco:2013zta,Aoki:2014nga}. 
(\emph{left}) the RGI combination $f^2_{B_q} \hat B_{B_q}^{(1)}$ for $q=d,s$.
We do not show the matrix-element results from RBC because their estimated uncertainties due to the use of
static $b$ quarks are larger than the displayed range.
(\emph{right}) the SU(3)-breaking ratio $\xi$.
The quoted ETM uncertainties do not include an error due to quenching the strange sea quark, while the
quoted HPQCD and \FerMILC~11 uncertainties do not include a contribution from the omission of wrong-spin
operators in the chiral-continuum extrapolation.
The error bars on our results do not include the estimated charm-sea uncertainties, which are too small to
be visible.}
\label{fig:O1_results}
\end{figure}

It is common to express the Standard-Model oscillation frequency $\Delta M_q$ in terms of the
renormalization-group-invariant (RGI) combination $f^2_{B_q} \hat B_{B_q}^{(1)}$, which is related to
$f^2_{B_q} B_{B_q}^{(1)}(\mu)$ at two loops by Eq.~(\ref{eq:RGI}).
Taking $\overline \alpha_s({\overline m_b}) = 0.2271(22)$ from Table~\ref{tab:inputs}, we obtain
\bea
	f_{B_d}^2 \hat B_{B_d}^{(1)} & = & 0.0518(43)(10)~\text{GeV}^2, \label{eq:fSqBhat_d} \\		
	f_{B_s}^2 \hat B_{B_s}^{(1)} & = & 0.0754(46)(15)~\text{GeV}^2, \label{eq:fSqBhat_s} 
\eea
where the first error is from the column labeled ``Total" in Table~\ref{tbl:ME_tot_error} and the second is
our estimated error from the omission of charm sea quarks.%
\footnote{The first error also includes a 0.2\% uncertainty estimate for converting from $\MSbar$-NDR to
RGI.} %
As shown in Fig.~\ref{fig:O1_results}, left, our results in Eqs.~(\ref{eq:fSqBhat_d})
and~(\ref{eq:fSqBhat_s}) agree with those from previous unquenched lattice-QCD calculations, but have
smaller uncertainties.
Compared with the other three-flavor calculations by the HPQCD~\cite{Gamiz:2009ku} and RBC/UKQCD
collaborations~\cite{Aoki:2014nga}, our data set includes lighter pions and finer lattice spacings.
The RBC/UKQCD errors are dominated by their estimate of the neglected contributions of order~$1/m_b$ due to
their use of the static approximation.
We note that the HPQCD results~\cite{Gamiz:2009ku} and the preliminary matrix elements from
\FerMILC~\cite{Bouchard:2011xj} are compatible with our new values in
Eqs.~(\ref{eq:fSqBhat_d})--(\ref{eq:fSqBhat_s}), despite the fact that these earlier works do not include
contributions from wrong-spin operators in the chiral-continuum extrapolation or include an uncertainty due
to this omission in their error budgets, suggesting that the earlier uncertainty estimates were sufficiently
conservative.

\subsection{Derived quantities}
\label{sec:DerivedQuantites}

We now present quantities that are derived from the $B_q$-mixing matrix element results in
Table~\ref{tbl:ME_results} that are especially useful for Standard-Model phenomenology.

\subsubsection{Ratios and other combinations of matrix elements}

The SU(3)-breaking ratio $\xi$, defined in Eq.~(\ref{eq:xidef}), is needed for obtaining the ratio of CKM
matrix elements $|V_{td} / V_{ts}|$ from experimental measurements of the oscillation frequencies.
The uncertainties on the individual matrix elements $\langle \op_1^q \rangle$ ($q=d,s$) in
Table~\ref{tbl:ME_results} are about 6--9~\%, and are still substantially larger than the experimental
errors on $\Delta M_q$.
Because the results for $\langle \op_1^d \rangle$ and $\langle \op_1^s \rangle$ are correlated, however,
both the statistical and systematic uncertainties largely cancel in their ratio.
We obtain
\be
	\xi = 1.206(18)(6), \label{eq:xiResult}
\ee 
where again the first error is from the quadrature sum of the statistical and systematic uncertainties (the
column labeled ``Total" in Table~\ref{tbl:ME_tot_error}) except for the charm-sea error, which is listed
separately as a second error.
Our result in Eq.~(\ref{eq:xiResult}) is consistent with the previous \FerMILC{}
result~\cite{Bazavov:2012zs}, $\xi = 1.268(63)$, but about three times more precise due to the substantially
increased data set and the inclusion of wrong-spin operators in the chiral-continuum extrapolation.
Figure~\ref{fig:O1_results}, right, compares our result with other unquenched lattice-QCD calculations.

We also present matrix-element combinations that enter the Standard-Model expression for the width
difference $\Delta \Gamma_q$.
Following Ref.~\cite{Lenz:2006hd}, we compute the $1/m_b$-suppressed quantity $\langle R_0\rangle$, which is
a linear combination of the matrix elements $\langle \op_{1,2,3}^q \rangle$.
The expression for $\langle R_0\rangle$ is given in Eq.~(\ref{eq:R0def}), where the NLO perturbative
coefficients evaluated at the renormalization scale $\mu=\overline m_b$ in the $\MSbar$-BBGLN scheme
are~\cite{Lenz:2006hd}
\bea
	\alpha_1(\overline{m}_b) & = & 1 + 2\frac{\overline \alpha_s(\overline m_b)}{\pi} ,
    \label{eq:alpha1} \\
	\alpha_2(\overline{m}_b) & = & 1 + \frac{13}{6}\frac{\overline{\alpha}_s(\overline m_b)}{\pi}.
    \label{eq:alpha2} 
\eea
We obtain
\bea
	\langle R_0^d \rangle (\overline m_b) & = & -0.09(21)~\text{GeV}^4, \label{eq:R0Result_d} \\		
	\langle R_0^s \rangle (\overline m_b) & = & -0.21(21)~\text{GeV}^4, \label{eq:R0Result_s} 
\eea
where the errors are from the hadronic mixing matrix elements.
The uncertainties from the parametric inputs $\overline\alpha_s(\overline m_b)$ and $M_{B_q}$ and the
omission of charm sea quarks are negligible.
Because the combination of operators in Eq.~(\ref{eq:R0def}) is by construction $1/m_b$-suppressed, our
results for $\langle R_0^q\rangle$ have larger relative errors than $\me{1,2,3}$ (which are of order 1 in
the heavy-quark expansion), and may also be more sensitive to the systematic-uncertainty contributions
considered previously.
The stability of our results for $\langle R_0^q\rangle$ under the fit modifications outlined in
Sec.~\ref{sec:altdatasets}, however, indicates that the quoted uncertainties adequately
accommodate all sources of error.
In Refs.~\cite{Lenz:2006hd,Artuso:2015swg}, the width difference $\Delta \Gamma_q$ is expressed in terms of
bag parameters $B^{(R_0)}_{B_q}$, rather than the matrix elements above.
Our results for $B^{(R_0)}_{B_q}$ are given in Table~\ref{tbl:BR0_results} of Appendix~\ref{app:Results},
which provides our complete set of bag-parameter results.

\begin{table} 
\centering
\caption{Numerical inputs used in the calculations of observables in this paper.
Top panel: the masses, lifetimes, and coupling constants are taken from the PDG~\cite{Agashe:2014kda} unless
otherwise specified.
The $\MSbar$ top-quark mass is obtained from the pole mass $m_{t,\text{pole}}=173.21(87)~$GeV
using the four-loop relation from Ref.~\cite{Marquard:2015qpa}.
The $\MSbar$ down- and strange-quark masses are obtained by four-loop
running~\cite{Chetyrkin:1997dh,Vermaseren:1997fq} of the values $\overline m_d(2~{\rm GeV})=4.68(16)$
MeV and $\overline m_s(2\ {\rm GeV})=93.8(2.4)$~MeV~\cite{Aoki:2013ldr} to the scale
$\overline m_b({\overline m_b})$.
The strong coupling $\overline \alpha_s({\overline m_b})$ is obtained by four-loop
running~\cite{vanRitbergen:1997va,Czakon:2004bu} of $\overline\alpha_s(M_Z) = 0.1185(6)$ to the scale
$\overline m_b{(\overline m_b)}$.
Middle panel: the coefficient $c_{\rm RGI}$ is the combination of factors in Eq.~(\ref{eq:RGI}) needed to
convert the bag parameters in the $\MSbar${-NDR} scheme $B_{B_q}(\overline m_b)$ to the RGI values
$\hat{B}_{B_q}$.
The expressions for the electroweak loop function $S_{0}(x_t)$ and the QCD factor $\eta_{2B}$ in the
$\MSbar${-NDR} scheme are given in Eqs.~(XII.4) and~(XIII.3) of Ref.~\cite{Buchalla:1995vs}, respectively.
The Wilson coefficient $C_A(\mu_b)$ includes both NLO electroweak and NNLO QCD corrections, and is given at
the scale $\mu_b = 5$~GeV in Eq.~(4) of Ref.~\cite{Bobeth:2013uxa}.
Bottom panel: the CKM combinations are obtained using the most recent determinations of the Wolfenstein
parameters from CKMfitter group's unitarity-triangle analysis including results through EPS
2015~\cite{Charles:2004jd}, where the errors have been symmetrized.
For the calculations of $B_q$-mixing observables, we take the parameters from the fit including only
tree-level quantities which excludes $\Delta M_q$~\cite{SebastienPC}:
$\{\lambda, A, \bar\rho, \bar\eta\}_{\rm tree} =
\{0.22541(^{+30}_{-21}), 0.8212(^{+66}_{-338}), 0.132(^{+21}_{-21}), 0.383(^{+22}_{-22})\}$.
For the calculations of the $B_q\to\mu^+\mu^-$ branching fractions, we take the parameters from the full fit:
$\{\lambda, A, \bar\rho, \bar\eta \}_{\rm full} =
\{0.22543(^{+42}_{-31}), 0.8227(^{+66}_{-136}), 0.1504(^{+121}_{-62}), 0.3540(^{+69}_{-76}) \}$.
%
}
\label{tab:inputs}
\begin{tabular}{l@{\hskip 10mm}l}
\hline 
\hline
\spp
$m_W = {80.385{(15)}}\;\gev$ & $m_Z = {91.1876{(21)}}\;\gev$ \\ 
\spp
$M_{B_d} = {5.27961(16)}\ {\rm GeV}$ & $M_{B_s} = {5.36679(23)}\ {\rm GeV}$ \\
\spp
$\overline m_d(\overline m_b) = 3.93(13) \times 10^{-3}\ {\rm GeV}$ 
& $\overline m_s(\overline m_b) = 79.1(2.0) \times 10^{-3}\ {\rm GeV} $  \\  
\spp
$\overline m_b({\overline m_b}) = {4.18(3)}\ {\rm GeV}$  &  $\overline m_t({\overline m_t}) = 163.53(83) \ {\rm GeV}$ \\
\spp
$\tau_{B_d} = {1.520(4)}\; \text{ps}$ & $\tau_{H_s} = {1.604(10)}  \; \text{ps}$~\cite{Amhis:2014hma} \\
\spp
$ \overline \alpha_s(\overline m_b) =  0.2271(22)$  & $G_F = {1.166 378 7(6)\times 10^{-5}}\ {\rm GeV}^{-2}$ \\
\spp
$\hbar = {6.582 119 28(15) \times 10^{-25}}\ {\rm GeV\cdot s}$  & $m_\mu = 105.6583715(35) \times 10^{-3}\ {\rm GeV}$ \\
\hline
\spp
$c_{\rm RGI} = 1.5158(36)$ &  $C_A(\mu_b) = 0.4694(36)$  \\
\spp
$S_{0}(x_t) = 2.322(18)$  & $\eta_{2B} =  0.55210(62)$ \\
\hline
\spp
$|V_{ts}^* V_{tb}^{}|_{\rm tree} = 40.9(1.0) \times 10^{-3}$ &  $|V_{td}^* V_{tb}^{}|_{\rm tree} = 8.92(30) \times 10^{-3} $ \\
\spp
$|V_{td}/V_{ts}|_{\rm tree} = 0.2180(51)$ &  \\
$|V_{ts}^* V_{tb}^{}|_{\rm full} = 41.03(52) \times 10^{-3}$ &  $|V_{td}^* V_{tb}^{}|_{\rm full} = 8.67(14) \times 10^{-3} $ \\
\spp
$|V_{td}/V_{ts}|_{\rm full} = 0.2113(22)$ &  \\
\hline
\hline
\end{tabular}
\end{table}


%
%
%
%
%

%
%
%
%
%
%


\subsubsection{Bag parameters}
\label{sec:bags}

The $B_q$-mixing bag parameters are defined in Eqs.~(\ref{eq:Bq_1})--(\ref{eq:Bq_45}).
We obtain their values from the matrix-element results in Table~\ref{tbl:ME_results} by dividing by the
appropriate factors of the leptonic decay constants $f_{B_d}$ and $f_{B_s}$.
We take the $N_f = 2+1$ decay-constant averages from the recent PDG review~\cite{Rosner:2015wva}:
\begin{equation}
	f_{B_d} = 193.6(4.2)~{\rm MeV} \,, \;\;  f_{B_s} = 228.6(3.8)~{\rm MeV} \,, \;\;  \frac{f_{B_s}}{f_{B_d}} = 1.187(15) \,, \qquad \label{eq:fB_2p1} 
\end{equation}
which include lattice-QCD results from Refs.~\cite{Bazavov:2011aa,McNeile:2011ng,Carrasco:2013naa,Christ:2014uea,Aoki:2014nga}.  
We use the three-flavor averages because we expect partial cancellations between the charm-sea effects in the decay constants and matrix elements to yield smaller overall errors on the bag parameters.

For the Standard-Model operator, we obtain the RGI bag parameters 
\bea
	 \hat B_{B_d}^{(1)} & = &1.38(12)(6),  \label{eq:Bdhat} \\		
	 \hat B_{B_s}^{(1)} & = &1.443(88)(48),  \label{eq:Bshat}
\eea
where the errors are from the matrix elements and the decay constants, respectively.
Despite the reduction in errors on the matrix elements attained in this paper, the errors on our bag
parameters are still similar in size to those of previous calculations because we use the PDG averages for
the decay constants.
Thus, we do not take advantage of the correlations and potential error cancellations that are possible in a
combined analysis of $B_q$-mixing matrix elements with the decay constants from the same lattice calculation.
The SU(3)-breaking ratio of bag parameters for the Standard-Model operator $\op_1^q$ is often used as an
input for global CKM-unitarity-triangle fits~\cite{Charles:2004jd,Bona:2006ah}.
We obtain
\bea
	\frac{\hat B^{(1)}_{B_s}}{\hat B^{(1)}_{B_d}} =  1.033(31)(26), 
 	\label{eq:Bshat-to-Bdhat}
\eea
where the errors are from $\xi$ and $f_{B_s}/f_{B_d}$, respectively.
Here the bag-parameter ratio takes advantage of correlations between the $B_s$- and $B_d$-mixing matrix
elements, making the error cancellation between the matrix elements and decay constants less important.
Thus the decreased uncertainty on our result for $\xi$ in Eq.~(\ref{eq:xiResult}) translates into a
commensurate error reduction on the bag-parameter ratio in Eq.~(\ref{eq:Bshat-to-Bdhat}).
Our error for $\hat B^{(1)}_{B_s}/\hat B^{(1)}_{B_d} $ is 
2--3 times smaller than the uncertainties quoted
for previous three-flavor results \cite{Gamiz:2009ku,Bazavov:2012zs,Aoki:2014nga}.

Table~\ref{tbl:Bhat_results} in Appendix~\ref{app:Results} gives results for the complete set of bag
parameters $B_{B_q}^{(i)}$ ($i$=1--5, $q=d,s$) in the $\MSbar$-NDR scheme evaluated at the scale
$\mu=\overline m_b$.
It also provides the ratios $B_{B_q}^{(i)}/B_{B_q}^{(1)}$ for $i$=2--5.
The errors on the $B$-parameters stem primarily from the matrix elements, and range from about 5--15\%.
To reduce the uncertainties, we are now performing a correlated, combined analysis of the
mixing-matrix-elements in Table~\ref{tbl:ME_results} with our own collaboration's calculation of the decay
constants using the same lattice ensembles and parameters~\cite{Kronfeld:2015xka}.
The results will be reported in a future paper.

The values of the $B$-parameters are all within about 20\% of the vacuum-saturation-approximation
expectation $B_{B_q}^{(i)} =1$.
A different definition of the $B$-parameters for the mixed-chirality operators $\op_4$ and $\op_5$ is
employed in Refs.~\cite{Lenz:2006hd,Lenz:2011ti,Artuso:2015swg}, which simplifies the expression for
$\Delta\Gamma_q$ in the Standard Model, but results in bag parameters that are not equal to one in the VSA.
Specifically, the term $M^2_{B_q}/(m_b + m_q)^2 \approx 1.6$ inside the brackets of Eq.~(\ref{eq:Bq_45}) is
set to one.
To facilitate the use of our results in determinations of $\Delta \Gamma_q$ based on the expressions in
Ref.~\cite{Lenz:2006hd}, we provide the bag parameters $B_{B_q}^{(R_1)}$ and $B_{B_q}^{(\tilde{R}_1)}$ in
the other convention, which are proportional to $B_{B_q}^{(4)}$ and $B_{B_q}^{(5)}$, respectively, in Table~\ref{tbl:BR0_results} of Appendix~\ref{app:Results}.
The observed deviations from unity are greater than for $B_{B_q}^{(4)}$ and $B_{B_q}^{(5)}$, being as much
as 30\% for $B_{B_d}^{(R_1)}$.

The contributions to the errors on $B_{B_q}^{(i)}$ ($i$=1--5) and $B_{B_q}^{(R_i)}$ ($i$=0,1) from the decay
constants are 100\% correlated.
Using this information, plus the correlations between the matrix elements given in Table~\ref{tbl:ME_cor},
we obtain the correlations between the $B$-parameters given in Table~\ref{tbl:B_cor}.
This enables the calculation of observables that depend upon all possible combinations of bag parameters in
Tables~\ref{tbl:Bhat_results} and~\ref{tbl:BR0_results}.

\subsection{Implications for Standard-Model phenomenology}
\label{sec:SMPheno}

We now illustrate the utility of the $B_q$-mixing matrix elements given in Table~\ref{tbl:ME_results} for
Standard-Model phenomenology.
First, we compute experimental observables associated with $B_q$--$\bar B_q$ mixing within the
Standard Model.
Next, assuming that the Standard Model is a complete description of Nature, we use the
experimentally-measured oscillation frequencies to determine the associated CKM matrix elements $|V_{td}|$,
$|V_{ts}|$, and their ratio.
Finally, we compute the total Standard-Model branching fractions for the rare decays $B_q \to \mu^+ \mu^-$.
Table~\ref{tab:inputs} provides the numerical inputs used for the calculations in this section.
 
\subsubsection{Oscillation frequencies \texorpdfstring{$\Delta M_q$}{} }
\label{sec:DeltaMq}

The physical observables associated with the neutral $B_q$-meson system are the mass difference $\Delta M_q$
and the decay-width difference $\Delta \Gamma_q$ between the two mass eigenstates.
Of these, the former have been measured at the sub-percent level~\cite{Amhis:2014hma}:
\be \label{eq:DMav}
    \Delta M_d = ( 0.5055 \pm 0.0020 )~\text{ps}^{-1} ,   
    \qquad
    \Delta M_s = (17.757 \pm 0.021 )~\text{ps}^{-1} , 
\ee
while the width differences have been measured with larger uncertainties~\cite{Amhis:2014hma},
\be \label{eq:DGav}
    \frac{\Delta \Gamma_d}{\Gamma_d}= ( 0.001 \pm 0.010 ) ,
    \qquad \qquad
    \frac{\Delta \Gamma_s}{\Gamma_s} = (0.124 \pm 0.009 )  . 
\ee
The average for $\Delta M_d$ from the Heavy Flavor Averaging Group (HFAG) in Eq.~(\ref{eq:DMav}) is based on
over 30 measurements and is dominated by results from the Belle, BaBar, and LHCb experiments
\cite{Abe:2004mz,Aubert:2005kf,Aaij:2012nt}, while the average for $\Delta M_s$ is obtained from
measurements from the CDF and LHCb experiments
\cite{Abulencia:2006ze,Aaij:2011qx,Aaij:2013mpa,Aaij:2013gja,Aaij:2014zsa}.
The $\Delta\Gamma_d$ average is obtained from measurements by the Delphi, BaBar, Belle, D0, and LHCb
experiments \cite{Abdallah:2002mr,Aubert:2003hd,Aubert:2004xga,Higuchi:2012kx,Abazov:2013uma,Aaij:2014owa}
and the $\Delta\Gamma_s$ average is based on measurements by the CDF, ATLAS, CMS, and LHCb experiments
\cite{Aaltonen:2012ie,Abazov:2011ry,Aad:2014cqa,Khachatryan:2015nza,Aaij:2014zsa}.

Equation~(\ref{eq:DM}) gives the expression for the mass difference in the Standard Model.
Using the hadronic matrix elements from Table~\ref{tbl:ME_results} and $\xi$ from Eq.~(\ref{eq:xiResult})
and other numerical inputs from Table~\ref{tab:inputs}, we obtain
\bea
	\Delta M_d^{\rm SM} & = & 0.630(53)(42)(5)(13)~\text{ps}^{-1} ,
    \label{eq:DeltaMd_SM} \\
	\Delta M_s^{\rm SM} & = & 19.6(1.2)(1.0)(0.2)(0.4)~\text{ps}^{-1}  ,
    \label{eq:DeltaMs_SM} \\
	\left( \frac{\Delta M_d}{\Delta M_s} \right)^{\rm SM} & = & 0.0321(10)(15)(0)(3)
    \label{eq:DeltaM_Ratio_SM} ,
\eea
where the errors shown are from the theoretical errors on the mixing matrix elements, the CKM matrix
elements, the remaining parametric inputs to Eq.~(\ref{eq:DM}), and the omission of charm sea quarks,
respectively.
Note that, for the CKM parameters, we use the determination from the CKMfitter Group's analysis including
only tree-level observables because it does not utilize the constraints from $B_q$-meson mixing.

The Standard-Model values for $\Delta M_q$ above have significantly larger uncertainties than the
experimental averages quoted in Eqs.~(\ref{eq:DMav}).
The dominant error in the theoretical calculation of $\Delta M_d$ stems from the hadronic mixing matrix
elements.
For $\Delta M_s$ and $\Delta M_d/\Delta M_s$, the hadronic matrix-element and CKM uncertainties are
commensurate.
Our results for $\Delta M_d$, $\Delta M_{s}$, and $\Delta M_d/\Delta M_s$ differ from experiment by 1.8$\sigma$, 1.1$\sigma$, and 2.0$\sigma$, respectively.

\subsubsection{CKM matrix elements \texorpdfstring{$|V_{td}|$ and $|V_{ts}|$}{Vtd and Vts}}
\label{sec:CKM}

In the Standard Model, the neutral $B_q$-meson oscillation frequencies $\Delta M_q$ [Eq.~(\ref{eq:DM})] are
proportional to the combination of CKM matrix elements $|V_{tq}^* V_{tb}|$.
Therefore experimental measurements of $\Delta M_d$ and $\Delta M_s$ enable determinations of $|V_{td}|$,
$|V_{ts}|$, and their ratio, within the Standard-Model CKM framework.
At present, the errors on these determinations~\cite{Agashe:2014kda} are limited by the theoretical
uncertainties on the hadronic $B_q$-mixing matrix elements~\cite{Gamiz:2009ku,Bazavov:2012zs}.
Here we improve upon them using our new, more precise matrix-element calculations.

We take the results for $f_{B_q}^2 \hat B_{B_q}^{(1)}$ ($q=d,s$) from Table~\ref{tbl:ME_results} and for
$\xi$ from Eq.~(\ref{eq:xiResult}), and use the experimental averages for $\Delta M_{d,s}$ in
Eq.~(\ref{eq:DMav})~\cite{Amhis:2014hma}.
The other numerical inputs that enter the expressions for $\Delta M_q$ are given in Table~\ref{tab:inputs}.
To infer values for the individual matrix elements $|V_{td}|$ and $|V_{ts}|$, we take $|V_{tb}| = 0.99912$
from CKM unitarity~\cite{Charles:2004jd}, where the error is of $\order(10^{-5})$ and hence negligible.
We obtain
\bea
	|V_{td}| & = & 8.00(33)(2)(3)(8) \times 10^{-3},  \label{eq:Vtd} \\
	|V_{ts}| & = & 39.0(1.2)(0.0)(0.2)(0.4) \times 10^{-3},  \label{eq:Vts} \\
	|V_{td}/V_{ts}| & = & 0.2052(31)(4)(0)(10), \label{eq:VtdOverVts} 
\eea
where the errors are from 
the lattice mixing matrix elements, 
the measured $\Delta M_q$,
the remaining parametric inputs to Eq.~(\ref{eq:DM}), 
and the omission of charm sea quarks, respectively.  
The uncertainty on $|V_{td} / V_{ts}|$ is 2--3 times smaller than those on $|V_{td}|$ and $|V_{ts}|$
individually because the hadronic uncertainties are suppressed in the ratio.
The theoretical uncertainties from the $B_q$-mixing matrix elements are still, however, the dominant sources
of error in all three results in Eqs.~(\ref{eq:Vtd})--(\ref{eq:VtdOverVts}).

\begin{figure}
	\includegraphics[width=0.49\textwidth]{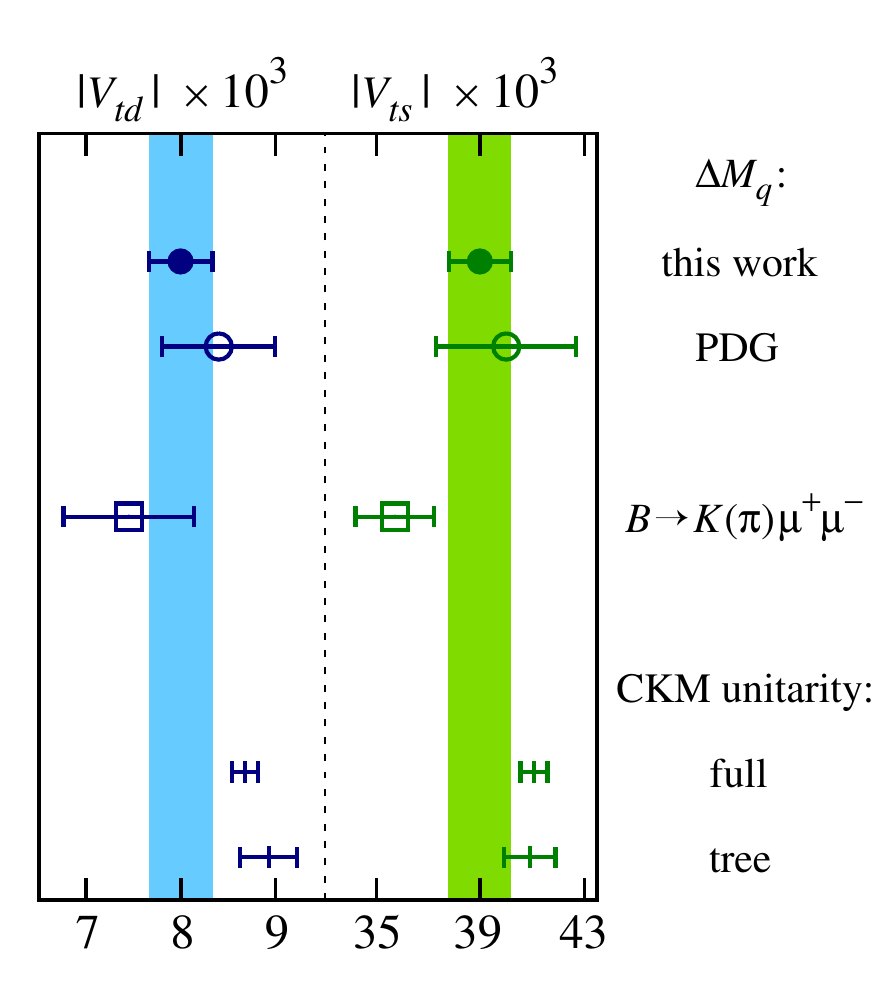} \hspace{-0.2in}
	\includegraphics[width=0.49\textwidth]{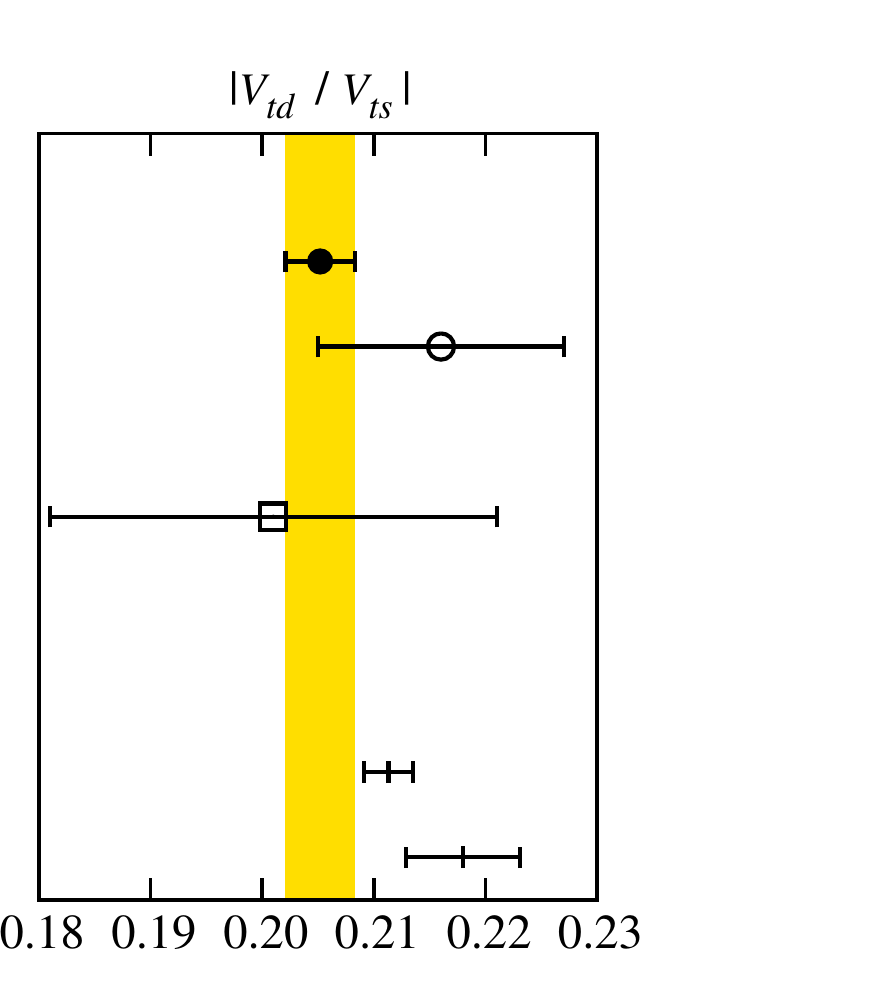} \hspace{-0.8in}
	\caption{(\emph{left}) Recent determinations $|V_{td}|$ and $|V_{ts}|$, and (\emph{right}) their ratio.
    The filled circles and vertical bands show our new results in Eqs.~(\ref{eq:Vtd})--(\ref{eq:VtdOverVts}), while the open circles show the previous values from $B_q$-mixing~\cite{Agashe:2014kda}.
    The squares show the determinations from semileptonic $B\to \pi \mu^+\mu^-$ and $B\to K \mu^+\mu^-$ 
    decays~\cite{Du:2015tda}, while the plus symbols show the values inferred from CKM
    unitarity~\cite{Charles:2004jd}.   
    The error bars on our results do not include the estimated charm-sea uncertainties, which are too small 
    to be visible.}
    \label{fig:VtxCompare}
\end{figure}

Figure~\ref{fig:VtxCompare} compares our results for $|V_{td}|$, $|V_{ts}|$, and their ratio in
Eqs.~(\ref{eq:Vtd})--(\ref{eq:VtdOverVts}) with other determinations.
Our results are consistent with the values from $B_q$-meson mixing in the PDG review~\cite{Agashe:2014kda},
which are obtained using approximately the same experimental inputs, and lattice-QCD calculations of the
$f_{B_q}^2 \hat B_{B_q}^{(1)}$ and $\xi$ from Refs.~\cite{Gamiz:2009ku} and~\cite{Bazavov:2012zs},
respectively.
Our errors on $|V_{td}|$, $|V_{ts}|$ are about two times smaller, however, and on $|V_{td}/V_{ts}|$ they are
more than three times smaller, due to the reduced theoretical errors on the hadronic matrix elements.

The CKM matrix elements $|V_{td}|$ and $|V_{ts}|$ can be obtained independently from rare semileptonic
$B$-meson decays because the Standard-Model rates for $\BR(B\to \pi (K) \mu^+\mu^-)$ are proportional to the
same combination $|V_{td(s)}^* V_{tb}|$.
Until recently, these determinations were not competitive with those from $B_q$-meson mixing due to both
large experimental and theoretical uncertainties.
In the past year, however, the LHCb collaboration published new measurements of
$\BR(B\to\pi\mu^+\mu^-)$ and $\BR(B\to K \mu^+\mu^-)$~\cite{Aaij:2014pli,Aaij:2015nea}, and we calculated
the full set of $B\to\pi$ and $B\to K$ form factors in three-flavor lattice
QCD~\cite{Lattice:2015tia,Bailey:2015nbd}.
Using these results, Ref.~\cite{Du:2015tda} obtains
\bea
	|V_{td}|^{B\to\pi\mu\mu} & = & 7.45(69) \times10^{-3} , \\
	|V_{ts}|^{B\to K \mu\mu} & = & 35.7(1.5) \times10^{-3} , \label{eq:Vts_BtoKmumu}\\
	|V_{td} / V_{ts}|^{\frac{B\to\pi \mu\mu}{B\to K\mu\mu}} & = & 0.201(20)  , 
\eea
where the errors include all sources of uncertainty.
These determinations of $|V_{td}|$ and $|V_{td}/V_{ts}|$ agree with our $B_q$-mixing results in
Eqs.~(\ref{eq:Vtd}) and~(\ref{eq:VtdOverVts}), while $|V_{ts}|$ above differs from Eq.~(\ref{eq:Vts}) by
about 1.6$\sigma$.

It is instructive to compare these results for $|V_{td}|$, $|V_{ts}|$, and $|V_{td}/V_{ts}|$ from
$B_q$ mixing and rare semileptonic $B$ decays with expectations from CKM unitarity.
These processes, being mediated by flavor-changing-neutral currents (FCNCs), may 
receive observable contributions from new physics. 
Figure~\ref{fig:VtxCompare} shows two sets of CKM elements inferred from unitarity, labeled ``full'' and 
``tree,'' 
which are obtained, respectively, from CKMfitter's full global unitarity-triangle fit using 
all inputs~\cite{Charles:2004jd} and from a fit including only observables that are mediated at the tree 
level of the weak interactions~\cite{SebastienPC}.
With the improvements in this paper for the $B_q$-mixing matrix elements, and in Refs.~\cite{Aaij:2014pli,%
Aaij:2015nea,Lattice:2015tia,Bailey:2015nbd,Du:2015tda} for rare semileptonic $B$ decay form factors, a
discrepancy between FCNC and tree-level processes may be emerging.
Quantitatively, the $B_q$-mixing results for $|V_{td}|$ and $|V_{ts}|$ differ from the
tree-fit values by $2.0\sigma$ and $1.2\sigma$, and their ratio by $2.1\sigma$.
\footnote{These CKM discrepancies are not distinct from those observed for
$\Delta M_q$ in Eqs.~(\ref{eq:DeltaMd_SM})--(\ref{eq:DeltaM_Ratio_SM}), but the CKM perspective is
convenient for contrasting the role of different observables.} %
Similarly, the FCNC semileptonic $B$ decays yield values of $|V_{td}|$ and $|V_{ts}|$ that lie,
respectively, $2.0\sigma$ and $2.9\sigma$ below the CKM tree-fit results. 
(The comparison of $|V_{td}/V_{ts}|$ from FCNC semileptonic $B$ decays is not yet useful because of the
large experimental uncertainties.) %
The overall impression does not change qualitatively when comparing with the results of the full CKM
unitarity-triangle fit: $|V_{td}|$, $|V_{ts}|$, and $|V_{td}/V_{ts}|$ from $B_q$ mixing differ by
1.8$\sigma$, 1.5$\sigma$, and 1.5$\sigma$, respectively; $|V_{td}|$ and $|V_{ts}|$ from FCNC semileptonic
$B$ decays differ by $1.7\sigma$ and $3.4\sigma$, respectively.
It would be interesting to see whether new flavor-changing neutral currents could explain this pattern, but
such a study lies beyond the scope of this work.

\subsubsection{Branching ratios for \texorpdfstring{$B_q\to\mu^+ \mu^-$}{B2mumu}}
\label{sec:Btomumu}

The rare decays $B_q \to \mu^+ \mu^-$ ($q=d,s$) also proceed via flavor-changing neutral currents, and are
therefore similarly sensitive to physics beyond the Standard Model. 
Revealing the presence of such effects, however, requires both precise experimental measurements and
reliable theoretical predictions with commensurate uncertainties.
Here we calculate the Standard-Model rates for $B_q \to \mu^+ \mu^-$ using our calculations of the
$B$-mixing matrix elements in Table~\ref{tbl:ME_results}.

The expression for the Standard-Model rate $\BR(B_{d(s)} \to \mu^+ \mu^-)$ is given in, \emph{e.g.}, Eq.~(3)
of Ref.~\cite{Bobeth:2013uxa}, and depends upon the leptonic decay constant $f_{B_q}$ and the combination of
CKM matrix elements $|V_{tq}^* V_{tb}|$.
The Standard-Model predictions receive substantial error contributions from both the CKM matrix elements and
the decay constants, as shown in Eqs.~(\ref{eq:Bdtomumu_fBd}) and~(\ref{eq:Bstomumu_fBs}) below.
Buras pointed out~\cite{Buras:2003td}, however, that the same CKM matrix elements also enter the
Standard-Model expression for the neutral $B_q$-mixing oscillation frequency $\Delta M_q$.
Thus, the ratio $\BR(B_{q} \to \mu^+ \mu^-)/\Delta M_q$ is independent of both $|V_{tq}^* V_{tb}|$ and
$f_{B_q}$, although it still depends upon the hadronic bag parameter $\hat B_{B_q}^{(1)}$.
Assuming that new physics does not alter the frequency of $B_q$-meson oscillations, this cancellation
potentially enables more precise Standard-Model predictions for $\BR(B_d \to \mu^+ \mu^-)$ and $\BR(B_s \to
\mu^+ \mu^-)$, since the relevant CKM combinations have large uncertainties, whereas the oscillation
frequencies have been measured to better than a percent.

In the Standard Model, the ratio of the $B_{d(s)} \to \mu^+ \mu^-$ decay rate over the $B_q$-meson mass
difference is given by~\cite{Buras:2003td,Bobeth:2013uxa}
\be
    \left( \frac{\Gamma(B_q \to \mu^+ \mu^-)}{ \Delta M_q} \right)^{\rm SM} = \frac{3}{\pi^3}
        \frac{(G_F M_W m_\mu)^2}{\eta_{2B} S_{0}(x_t)} \frac{C_A^2(\mu_b)}{\hat{B}^{(1)}_{B_q}} 
        \sqrt{1-\frac{4m_\mu^2}{M^2_{B_q}}}.
    \label{eq:Btomumu_DeltaMq}
\ee
where the Wilson coefficient $C_A(\mu_b)$ includes NLO electroweak and NNLO QCD
corrections~\cite{Hermann:2013kca,Bobeth:2013tba} and the remaining quantities are the same as in the expression for $\Delta M_q$ [Eq.~(\ref{eq:DM})].
For the $B_s$ decay mode, the nonzero $\Delta \Gamma_s$ gives rise to a difference between the
experimentally measured, time-integrated decay rate and the theoretical, CP-averaged rate
\cite{DeBruyn:2012wj,DeBruyn:2012wk,Buras:2013uqa}.
Following Ref.~\cite{Bobeth:2013uxa}, we use the lifetime of the heavy $B_s$-meson eigenstate $\tau_{H_s}$
to compute the time-averaged branching fraction as
$\oBR(B_s\to\mu^+\mu^-)^\text{SM}=\tau_{H_s}\Gamma(B_s\to\mu^+\mu^-)^\text{SM}$,
which holds when only the heavy eigenstate can decay to $\mu^+\mu^-$, as in the Standard Model.
(In the case of the $B_d$ system, $\oBR(B_d\to\mu^+\mu^-)=\BR(B_d \to \mu^+\mu^-)$, because 
$\Delta\Gamma_d\ll\Gamma_d$.)
For the double ratio of $B_d$-to-$B_s$ processes, the expression simplifies to
\be
\left(\frac{\oBR(B_d \to \mu^+ \mu^-)}{\oBR(B_s \to \mu^+ \mu^-)} \frac{\Delta M_s}{ \Delta M_d} \right)^{\rm SM} =
\frac{\tau_{B_d}}{\tau_{H_s}} \frac{\hat{B}^{{(1)}}_{B_s}}{\hat{B}^{{(1)}}_{B_d}} 
\frac{M_{B_s}}{M_{B_d}} \left(\frac{M^2_{B_d}-4m_\mu^2}{M^2_{B_s}-4m_\mu^2} \right)^{1/2}.
\ee
Using the bag-parameter results from Eqs.~(\ref{eq:Bdhat})--(\ref{eq:Bshat-to-Bdhat}), and the numerical
inputs from Table~\ref{tab:inputs}, we obtain
\bea
	\left(\frac{\oBR(B_d \to \mu^+ \mu^-)}{\Delta M_d}\right)^{\rm SM} & = &
        1.79(17)(3) \times 10^{-10}~\text{ps},
    \label{eq:Bdtomumu_Over_DeltaMd}  \\
	\left(\frac{\oBR(B_s \to \mu^+ \mu^-)}{\Delta M_s}\right)^{\rm SM} & = &
        1.81(13)(3) \times 10^{-10}~\text{ps},
    \label{eq:Bstomumu_Over_DeltaMs} 
\eea
and 
\be 
    \left(\frac{\oBR(B_d\to\mu^+\mu^-)}{\oBR(B_s \to \mu^+ \mu^-)}\frac{\Delta M_s}{\Delta M_d}\right)^{\rm SM} =
        0.978(38)(7),
        \label{eq:Bds-over-DeltaMds}
\ee
where the errors shown are from the theoretical errors on the bag parameters, and all other uncertainties
added in quadrature, respectively.
Multiplying the ratios in Eqs.~(\ref{eq:Bdtomumu_Over_DeltaMd})--(\ref{eq:Bds-over-DeltaMds}) by the
oscillation frequencies $\Delta M_q$ from Eq.~(\ref{eq:DMav})~\cite{Amhis:2014hma}, we obtain the following
Standard-Model total rates:
\bea
	\oBR(B_d \to \mu^+ \mu^-)^{{\rm SM},\, \Delta M_d} & = & 9.06(85)(4)(16) \times 10^{-11},
    \label{eq:Bdtomumu_DeltaMd} \\
	\overline{\BR}(B_s \to \mu^+ \mu^-)^{{\rm SM},\, \Delta M_s} & = & 3.22(22)(0)(6) \times 10^{-9},
    \label{eq:Bstomumu_DeltaMs} \\
	\bigg(\frac{\oBR(B_d \to \mu^+ \mu^-)}{\oBR(B_s \to \mu^+ \mu^-)}\bigg)^{{\rm SM},\,
        \Delta M_q} & = & 0.02786(109)(12)(19), 
	\label{eq:Bds_DeltaM}
\eea
where the errors shown are from the theoretical errors on the bag parameters, the measured $\Delta M_q$, and
the quadrature sum of all other sources of uncertainty, respectively.
The bag parameters are the dominant source of uncertainty.
Because our bag-parameter errors are comparable to those of previous calculations, our decay-rate results in
Eqs.~(\ref{eq:Bdtomumu_DeltaMd}) and (\ref{eq:Bstomumu_DeltaMs}) are similar in precision to those presented
in Ref.~\cite{Bobeth:2013uxa}.

We can also compare our predictions with the Standard-Model $B_{d(s)}\to \mu^+ \mu^-$ decay rates calculated
via Eqs.~(6) and~(7) of Ref.~\cite{Bobeth:2013uxa}:
\bea
	\oBR(B_d \to \mu^+ \mu^-)^{{\rm SM},\, f_{B_d}} & = & 10.36(44)(33)(27) \times 10^{-11},
    \label{eq:Bdtomumu_fBd} \\
	\oBR(B_s \to \mu^+ \mu^-)^{{\rm SM},\, f_{B_s}} & = & 3.53(11)(9)(9) \times 10^{-9} , 
    \label{eq:Bstomumu_fBs} \\
    \bigg(\frac{\oBR(B_d \to \mu^+ \mu^-)}{\oBR(B_s \to \mu^+ \mu^-)}\bigg)^{{\rm SM},\, f_{B_q}} & = &
        0.02929(29)(61)(76),
    \label{eq:Bds_fBq}
\eea
where the errors are from the decay constants, CKM matrix elements, and the quadrature sum of all other contributions, respectively.  For the calculations of Eqs.~(\ref{eq:Bdtomumu_fBd})--(\ref{eq:Bds_fBq}), we use the PDG's preferred decay-constant averages
\begin{equation}
	f_{B_d} = 190.9(4.1)~{\rm MeV} \,, \;\;  f_{B_s} = 227.2(3.4)~{\rm MeV} \,, \;\;  \frac{f_{B_s}}{f_{B_d}} = 1.192(6) \,, \qquad \label{eq:fB_2p1p1}
\end{equation}
which include both three- and four-flavor lattice-QCD results~\cite{Bazavov:2011aa,McNeile:2011ng,Dowdall:2013tga,Carrasco:2013naa,Christ:2014uea,Aoki:2014nga}, because they are the most precise available.
In this case, the contributions to the error on the total decay rates from the three sources of uncertainty
are similar in size.  The uncertainties in Eqs.~(\ref{eq:Bdtomumu_DeltaMd})--(\ref{eq:Bds_DeltaM}) are slightly larger
than those in Eqs.~(\ref{eq:Bdtomumu_fBd})--(\ref{eq:Bds_fBq}).
The central values differ only slightly by 1.3$\sigma$, 1.2$\sigma$, and 1.2$\sigma$, respectively.

Recently the LHCb and CMS experiments reported the first observation of $B_s \to \mu^+ \mu^-$ decay, as well
as 3$\sigma$ evidence for the process $B_d \to \mu^+ \mu^-$~\cite{CMS:2014xfa}, obtaining
\bea
	\oBR(B_d \to \mu^+ \mu^-)^{\rm exp} & = & 3.9(^{+1.6}_{-1.4})\times10^{-10} , \\
	\oBR(B_s \to \mu^+ \mu^-)^{\rm exp} & = & 2.8(^{+0.7}_{-0.6})\times10^{-9}  , \\
	\left(\frac{\oBR(B_d \to \mu^+ \mu^-)}{\oBR(B_s \to \mu^+ \mu^-)} \right)^{\rm exp} & = &
        0.14(^{+0.08}_{-0.06}) . 
\eea
The measured branching fraction for $B_s \to \mu^+ \mu^-$ is compatible with our Standard-Model value in
Eq.~(\ref{eq:Bstomumu_DeltaMs}), within present uncertainties, but there is still ample room for new-physics
contributions of a size that may be observable with improved experimental measurements after the LHC
luminosity upgrade.
The measured branching fraction for $B_d \to \mu^+ \mu^-$ is 2.0$\sigma$ above the Standard-Model
expectation in Eq.~(\ref{eq:Bdtomumu_DeltaMd}) after averaging the asymmetric experimental errors, while the
measured ratio $\oBR(B_d \to \mu^+ \mu^-)/\oBR(B_s \to \mu^+ \mu^-)$ lies 1.6$\sigma$ above the result in
Eq.~(\ref{eq:Bds_DeltaM}).
The measurement errors must be reduced, however, before one can draw meaningful conclusions.

\section{Conclusions and outlook}
    \label{sec:conclusions}
    We have presented the first three-flavor lattice-QCD results for the $B_{q}$-meson ($q=s,d$) mixing matrix
elements of the full set of dimension-six $\Delta B = 2$ four-fermion operators in the electroweak effective
Hamiltonian.
The first matrix element $\langle \op_1^q \rangle$ is needed to calculate the mass differences $\Delta M_q$
in the Standard Model, while the remaining four matrix elements $\langle \op_i^q \rangle$ ($i$=2--5) are
sufficient to parameterize the hadronic contributions to $\Delta M_q$ in any Standard-Model extension.
These matrix elements are also sufficient to obtain the leading-order contributions to the Standard-Model
width differences $\Delta \Gamma_q$, as well as some of the corrections of order $1/m_b$.
For the Standard-Model matrix element $\langle \op_1^q \rangle$, we obtain the
renormalization-group-invariant combinations
\begin{align}
	f_{B_d} \sqrt{\hat B_{B_d}^{(1)}} &= 227.7(9.5)(2.3)~\text{MeV}, \\		
	f_{B_s} \sqrt{\hat B_{B_s}^{(1)}} &= 274.6(8.4)(2.7)~\text{MeV} , 
\end{align}
where the first error includes statistical and all systematic uncertainties except for the charm-sea error,
which is shown separately.
Our results for the complete set of matrix elements $f_{B_q}^2 B_{B_q}^{(i)}(\overline m_b)$ ($i=$1--5) are
given in Table~\ref{tbl:ME_results}.
To enable the use of our lattice $B_q$-meson mixing matrix elements for additional phenomenological studies,
we provide the correlations between them in Table~\ref{tbl:ME_cor}.

Although there have been previous three-flavor lattice-QCD calculations of the matrix elements of the
Standard-Model operator $f_{B_q}^2 B_{B_q}^{(1)}$, ours are the first with all sources of systematic
uncertainty controlled.
In particular, compared with Refs.~\cite{Gamiz:2009ku,Bouchard:2011xj,Bazavov:2012zs}, we include the
contributions from wrong-spin operators in the chiral continuum-extrapolation.
Our analysis also includes ensembles with finer lattice spacings and lighter pions than in previous works.
We obtain the SU(3)-breaking ratio
\begin{equation}
    \xi = 1.206(18)(6)
\end{equation}
to 1.6\% precision, which is the most precise lattice-QCD determination of this quantity to-date.
The reduction in errors by about a factor of three compared with our earlier work~\cite{Bazavov:2012zs} is due
in large part to correctly handling the wrong-spin operators in the chiral-continuum extrapolation.
The total uncertainty is now definitively smaller than that quoted in many early
estimates---see, for example, Refs.~\cite{Ciuchini:2000de,Hocker:2001xe}---in which the uncertainty from the chiral
extrapolation was underestimated~\cite{Kronfeld:2002ab}.

To illustrate the phenomenological utility of our $B_q$-mixing matrix elements, we have used them to
calculate the mass differences $\Delta M_q$ and the total branching fractions for
the rare-decay processes $B_q \to \mu^+\mu^-$ in the Standard Model, where we highlight the ratios
$\BR(B_q\to\mu^+\mu^-)/\Delta M_q$ in which the CKM factors and decay constants cancel.
The Standard-Model expectations for $\Delta M_d$, $\Delta M_s$, and their ratio are all greater than the
experimental averages, where the differences are 1.8$\sigma$, 1.1$\sigma$, and 2.0$\sigma$,
respectively.
The Standard-Model rate for $\BR(B_d \to \mu^+\mu^-)$ lies 2.0$\sigma$ below experiment, although one should
bear in mind that the experimental observation of $B_d \to \mu^+\mu^-$ decay has yet to reach 5$\sigma$
significance.

We also obtain the CKM matrix elements $|V_{td}|$, $|V_{ts}|$, and their ratio assuming that there are no
significant new-physics contributions to $B_q$-meson oscillations.
The results
\begin{align}
	|V_{td}| &= 8.00(34)(8)  \times 10^{-3} , \label{eq:Vtd_conc} \\
	|V_{ts}| &= 39.0(1.2)(0.4) \times 10^{-3}  ,  \\
	|V_{td} / V_{ts}| &= 0.2052(31)(10) \label{eq:VtdVts_conc} \,, 
\end{align}
are the single-most precise determinations of these quantities, and differ with expectations from CKM
unitarity~\cite{Charles:2004jd,Bona:2006ah}.
(As above, the semiquantitative charm-sea error is listed separately from the other uncertainties.)
The CKM-element results from $B_q$-meson mixing lie below the determinations from CKMfitter's full
global unitarity-triangle fit using all inputs by 1.5--1.8$\sigma$, and below the determinations using only tree-level inputs by 1.2--2.1$\sigma$~\cite{Charles:2004jd}.
Because our new $B_q$-mixing matrix elements imply lower values for $|V_{td}|$, $|V_{ts}|$, and their ratio,
they enhance the observed tension between tree-level and loop-induced
processes~\cite{Charles:2004jd}.
\begin{figure}[t]
	\includegraphics[width=0.8\textwidth]{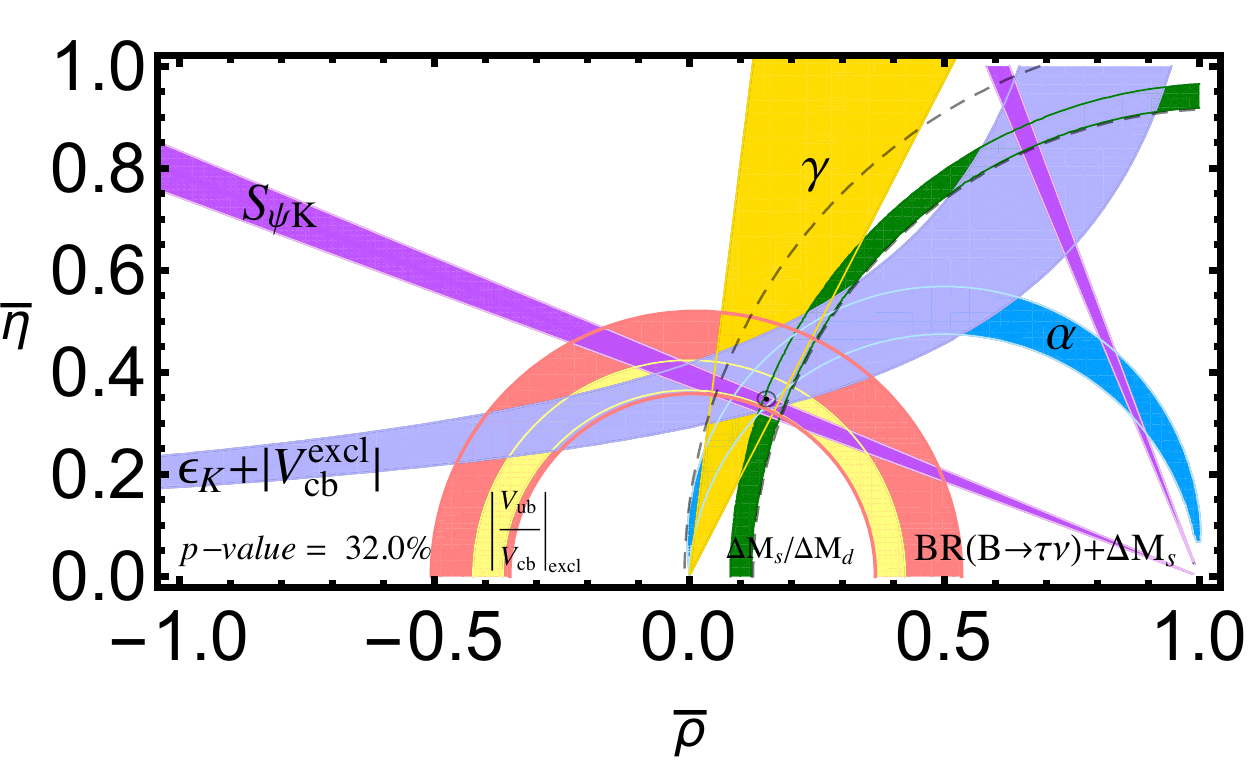}
    \caption{Global CKM unitarity-triangle fit using the new determination of $\xi$ from this work as well 
    as $|V_{ub}|$ and $|V_{cb}|$ based on our calculations of the $B\to \pi\ell\nu$ and $B\to D\ell\nu$ form 
    factors on the same gauge-field configurations~\cite{Lattice:2015rga,Lattice:2015tia,DeTar:2015orc}.
    The constraint from $B$-meson mixing (solid green band) is approximately three times smaller than that 
    obtained using our previous result for $\xi$~\cite{Bazavov:2012zs} (dashed gray lines).
    For the remaining hadronic matrix elements, we use the preliminary $(2+1)$-flavor FLAG~III average
    $\hat{B}_K=0.7627(97)$~\cite{Vladikas:2015bra}, and the 2015 PDG averages $f_{K^\pm}=155.6(0.4)$~MeV and
    $f_{B^\pm}=187.1(4.2)$~MeV~\cite{Rosner:2015wva}, which include $(2+1)$- and $(2+1+1)$-flavor 
    lattice-QCD results. 
    The QCD contributions to $\varepsilon_K$ from charm- and top-quark loops are taken from 
    Refs.~\cite{Buras:1990fn,Brod:2010mj,Brod:2011ty}, while all experimental inputs are from the 
    PDG~\cite{Agashe:2014kda}.
    Plot courtesy E.~Lunghi~\cite{Enrico} using the procedures of Ref.~\cite{Laiho:2009eu}.}
	\label{fig:utfit}
\end{figure}
Figure~\ref{fig:utfit} shows the current status of the CKM unitarity triangle using our new result for $\xi$
plus recent determinations of $|V_{ub}|=3.72(16)\times10^{-3}$ and $|V_{cb}|=40.8(1.0)\times10^{-3}$ from
our calculations of the $B\to \pi\ell\nu$ and $B\to D\ell\nu$ form
factors~\cite{Lattice:2015rga,Lattice:2015tia}.%
\footnote{We obtain this value of $|V_{cb}|$ from a fit similar to the one in 
Ref.~\cite{DeTar:2015orc} including the recent $B\to D\ell\nu$ measurements from
Belle~\cite{Glattauer:2015teq}, earlier measurements from BaBar~\cite{Aubert:2008yv}, and our lattice-QCD 
form factors from Ref.~\cite{Lattice:2015tia}.}
At present, the experimental measurements are compatible with the Standard Model at $p=0.32$.
The overall precision still leaves ample room for BSM flavor-changing neutral currents that
may be observable with anticipated theoretical improvements such as those discussed below, in conjunction
with more precise experimental measurements expected from the LHC upgrade~\cite{Bediaga:2012py,CMS:2015iha}
and Belle~II~\cite{Aushev:2010bq}.

Despite the improvement in the $B_q$-mixing elements obtained in this paper, the theoretical hadronic errors
are still the limiting source of uncertainty in all calculations of observables in Sec.~\ref{sec:results}.
In a forthcoming paper, we will report bag-parameter results from a combined analysis of the
mixing-matrix-elements presented here with our collaboration's companion decay-constant calculation using
the same lattice ensembles and parameters~\cite{Kronfeld:2015xka}.
We anticipate that the inclusion of statistical and systematic correlations between the matrix elements and
decay constants will reduce the bag-parameter errors, thereby enabling better predictions of the
$B_q\to\mu^+\mu^-$ decay rates and other observables.

Additional work is still needed, however, to reduce the QCD uncertainties to the level of experimental
measurements.
The dominant errors in our current matrix-element results stem from statistics and heavy-quark
discretization errors.
The contribution from the chiral extrapolation is also significant for the $B_d$ matrix elements and~$\xi$.
We plan to reduce these uncertainties by using the newly generated gauge-field ensembles by the MILC
Collaboration~\cite{Bazavov:2010ru,Bazavov:2012xda} with four flavors of highly-improved
staggered quarks (HISQ)~\cite{Follana:2006rc}.
Ensembles with physical-mass pions at four lattice spacings are already available, the use of which will
render the chiral extrapolation unnecessary and eliminate the associated systematic uncertainty.
In addition, the inclusion of charm quarks in the sea will eliminate the least-well quantified source of
error in our current calculation.
The inclusion of ensembles at an even finer lattice spacing will be particularly important for reducing
heavy-quark discretization effects.
Such a fine lattice spacing will also be useful in calculations that employ the HISQ action for the $b$
quark~\cite{McNeile:2011ng}.
Finally, because the matrix elements are dimensionful, they also receive error contributions from the
lattice-spacing uncertainty, which will become relatively more important as the other errors discussed above
are reduced.
Recently several new scale-setting quantities have been
introduced~\cite{Luscher:2010iy,Borsanyi:2012zs,Bazavov:2012xda,Bazavov:2015yea} that can be obtained more
directly and precisely than $r_1$, one of which has already been employed in our calculation of charmed and
light pseudoscalar-meson decay constants on the HISQ ensembles~\cite{Bazavov:2014wgs}.

Once the matrix-element errors reach the few-percent level, it will be important to provide the $B_d$-mixing matrix elements at the correct physical $d$-quark mass.  This can be done straightforwardly, as we have shown in this work, by including partially-quenched data in the analysis, and then evaluating the chiral fit function with the valence-quark mass equal to $m_d$ and the sea-quark masses equal to the average up-down quark mass.  For calculations where only full-QCD data is available, the corrections listed in Table~\ref{tab:val_isospin} can be used to adjust the isospin-averaged $B$-mixing matrix elements for valence isospin breaking.

Neutral $B$-meson mixing provides a powerful test of the Standard Model and stringent constraints on new,
high-scale physics~\cite{Isidori:2010kg,Lenz:2010gu,Carrasco:2013zta}.
Our work reveals several $\sim 2\sigma$ deviations between the Standard Model and experiment in 
$B$-meson oscillations and rare leptonic $B$ decays.
Similar-sized tensions have been observed in rare semileptonic $B$ decays~\cite{Aaij:2013qta,Aaij:2014ora,%
Altmannshofer:2015sma,ChristophLangenbruchonbehalfoftheLHCb:2015iha,Du:2015tda}, which also proceed via
$b\to d$ and $b\to s$ flavor-changing neutral currents.
The full basis of $\Delta B=2$ mixing matrix elements, including their correlations, provided here will
allow these interesting tensions to be better explored, leading to sharper tests of the Standard Model and
tighter constraints on allowed new physics.
    
\begin{acknowledgments}

We thank S\'ebastien Descotes-Genon for helpful discussions and supplementary information on CKMfitter's
results, Enrico Lunghi for providing the global unitarity-triangle fit plot and assisting in calculations of
$B\to\mu\mu$ observables, and Andrzej Buras, Martin Jung, and Alexander Lenz for helpful discussions and valuable feedback on the manuscript.
Computations for this work were carried out with resources provided by the USQCD Collaboration, the National
Energy Research Scientific Computing Center and the Argonne Leadership Computing Facility, which are funded
by the Office of Science of the U.S.\ Department of Energy; and with resources provided by the National
Institute for Computational Science and the Texas Advanced Computing Center, which are funded through the
National Science Foundation's Teragrid/XSEDE Program.
This work was supported in part by the U.S.\ Department of Energy under grants
No.~DE-FG02-91ER40628 (C.B.)
No.~DE-FC02-06ER41446 (C.D.)
No.~DE-SC0010120 (S.G.),
No.~DE-SC0010005 (E.T.N.),
No.~DE-FG02-91ER40661 (S.G., R.Z.), 
No. DE-FG02-13ER42001 (C.C.C., D.D., A.X.K.),
No.~DE-FG02-13ER41976 (D.T.); 
by the U.S.\ National Science Foundation under grants 
PHY10-67881 and PHY10-034278 (C.D.),
PHY14-17805~(D.D., J.L.),
and PHY13-16748 (R.S.);
by the Fermilab Fellowship in Theoretical Physics (C.M.B., C.C.C.);
by the URA Visiting Scholars' program (C.M.B., C.C.C., D.D., A.X.K.);
by the MICINN (Spain) under grant FPA2010-16696 and Ram\'on y Cajal program (E.G.);
by the Junta de Andaluc\'ia (Spain) under Grants No.~FQM-101 and No.~FQM-6552 (E.G.);
by the European Commission (EC) under Grant No.~PCIG10-GA-2011-303781 (E.G.);
by the German Excellence Initiative and the European Union Seventh Framework Program under grant agreement
No.~291763 as well as the European Union's Marie Curie COFUND program (A.S.K.).
Fermilab is operated by Fermi Research Alliance, LLC, under Contract No.\ DE-AC02-07CH11359 with the United
States Department of Energy.  Brookhaven National Laboratory is supported by the DoE under contract no. DE-SC0012704.

\end{acknowledgments}

\appendix

\section{Numerical results for $B_q$-mixing matrix elements and bag parameters}
    \label{app:Results}
    Tables~\ref{tbl:ME_results}--\ref{tbl:B_cor} in this Appendix present our results for the complete set of
$B_q$-mixing matrix elements and bag parameters with total statistical plus systematic uncertainties, as
well as the correlations between them.
The bag parameters are defined in Eqs.~(\ref{eq:Bq_1})--(\ref{eq:Bq_45}) to be one in the vacuum saturation approximation.
This information is sufficient to use our results in calculations of $B_q$-mixing observables both within
and beyond the Standard Model.

Because the estimated uncertainty due to the omission of charm sea quarks is less quantitative than the
other uncertainties, we do not provide correlations for this error.
Rather we suggest that an error of 2\% be taken on all sums or differences of matrix elements $\me{i}$
($q=d,s$; $i$=1--5), and that an error of 0.5\% be taken on all ratios of (sums or differences of)
$B_s$-to-$B_d$ matrix elements.
As discussed earlier, we consider the charm-sea error in the bag parameters and same-flavor matrix-element
ratios to be negligible.


\begin{table}[h]
\caption{$B_q$-mixing matrix elements $f^2_{B_q} B_{B_q}^{(i)}$ in the $\MSbar$-NDR scheme evaluated at the
scale $\mu = \overline m_b$, with total statistical plus systematic uncertainties.
The first error is the ``Total'' error listed in Table~\ref{tbl:ME_tot_error} and the second is the ``charm
sea'' error listed in the last column of that table.
For operators $\op_2^q$ and $\op_3^q$, results for both the BMU~\cite{Buras:2000if} and
BBGLN~\cite{Beneke:1998sy,Beneke:2002rj} evanescent-operator conventions are shown.
Entries are in GeV$^2$.
\label{tbl:ME_results}}
\vspace{0.1in}
\begin{tabular}{c@{\quad}cc@{\quad}cc}
\hline\hline
\hspace{0.5in}  & \multicolumn{2}{c}{$B_d$--$\bar B_d$}            & \multicolumn{2}{c}{$B_s$--$\bar B_s$} \\
                                & BMU           & BBGLN         & BMU           & BBGLN \\ \hline 
\vspace{-0.18in}\\
$f_{B_q}^2 B_{B_q}^{(1)}(\overline m_b)$ & \multicolumn{2}{c}{0.0342(29)(7)} & \multicolumn{2}{c}{0.0498(30)(10)}    \vspace{0.07in}\\
$f_{B_q}^2 B_{B_q}^{(2)}(\overline m_b)$ & 0.0285(26)(6)    & 0.0303(27)(6)  & 0.0421(27)(8)     &  0.0449(29)(9)        \vspace{0.07in}\\
$f_{B_q}^2 B_{B_q}^{(3)}(\overline m_b)$ & 0.0402(77)(8)    & 0.0399(77)(8)  & 0.0576(77)(12)    & 0.0571(77)(11)        \vspace{0.07in}\\
$f_{B_q}^2 B_{B_q}^{(4)}(\overline m_b)$ & \multicolumn{2}{c}{0.0390(28)(8)} & \multicolumn{2}{c}{0.0534(30)(11)}    \vspace{0.07in}\\
$f_{B_q}^2 B_{B_q}^{(5)}(\overline m_b)$ & \multicolumn{2}{c}{0.0361(35)(7)} & \multicolumn{2}{c}{0.0493(36)(10)}    \vspace{0.07in}\\ 
\hline\hline
\end{tabular}
\end{table}

\begin{table}[h]
\caption{Correlations between the matrix-element results in the $\overline{\text{MS}}$-NDR-BBGLN scheme presented in Table~\ref{tbl:ME_results}.  
Correlations between the BMU-scheme results differ by $<1\%$.  
The entries of the correlation matrix are symmetric across the diagonal. 
The contributions from the ``charm sea'' error are not included in this table, because they are not as well quantified as the other uncertainties.
 \label{tbl:ME_cor}}
 \vspace{0.1in}
\begin{tabular}{c|cccccccccc}
\hline\hline
& & & & & & & & & & \vspace{-0.15in} \\
\hspace{0.2in}  & $f^2_{B_d} B^{(1)}_{B_d}$ & $f^2_{B_d} B^{(2)}_{B_d}$ & $f^2_{B_d} B^{(3)}_{B_d}$ & $f^2_{B_d} B^{(4)}_{B_d}$ & $f^2_{B_d} B^{(5)}_{B_d}$ & $f^2_{B_s} B^{(1)}_{B_s}$ & $f^2_{B_s} B^{(2)}_{B_s}$ & $f^2_{B_s} B^{(3)}_{B_s}$ & $f^2_{B_s} B^{(4)}_{B_s}$ & $f^2_{B_s} B^{(5)}_{B_s}$ \\ 
& & & & & & & & & & \vspace{-0.15in} \\ \hline
& & & & & & & & & & \vspace{-0.15in} \\
$f^2_{B_d} B^{(1)}_{B_d}$& 1    & 0.378 & 0.070 & 0.336     & 0.287 & 0.968     & 0.395     & 0.089     & 0.346     & 0.305 \\
& & & & & & & & & & \vspace{-0.15in} \\
$f^2_{B_d} B^{(2)}_{B_d}$&      & 1     & 0.212     & 0.348 & 0.255     & 0.394     & 0.961     & 0.230     & 0.365 & 0.277 \\
& & & & & & & & & & \vspace{-0.15in} \\
$f^2_{B_d} B^{(3)}_{B_d}$&      &       & 1         & 0.134     & 0.065 & 0.079 & 0.207     & 0.980 & 0.137     & 0.071  \\
& & & & & & & & & & \vspace{-0.15in} \\
$f^2_{B_d} B^{(4)}_{B_d}$&      &       &       & 1         & 0.404     & 0.371     & 0.391     & 0.162     & 0.955     & 0.426 \\ 
& & & & & & & & & & \vspace{-0.15in} \\
$f^2_{B_d} B^{(5)}_{B_d}$&      &       &       &       & 1         & 0.309     & 0.281     & 0.084 & 0.404     & 0.962 \\ 
& & & & & & & & & & \vspace{-0.15in} \\
$f^2_{B_s} B^{(1)}_{B_s}$   &   &       &       &       &       & 1     & 0.455 & 0.117     & 0.419 & 0.359 \\ 
& & & & & & & & & & \vspace{-0.15in} \\
$f^2_{B_s} B^{(2)}_{B_s}$   &   &       &       &       &       &       & 1     & 0.253 & 0.453     & 0.339 \\
& & & & & & & & & & \vspace{-0.15in} \\
$f^2_{B_s} B^{(3)}_{B_s}$   &   &       &       &       &       &       &       & 1         & 0.186     & 0.107 \\ 
& & & & & & & & & & \vspace{-0.15in} \\
$f^2_{B_s} B^{(4)}_{B_s}$   &   &       &       &       &       &       &       &       & 1     & 0.471 \\ 
& & & & & & & & & & \vspace{-0.15in} \\
$f^2_{B_s} B^{(5)}_{B_s}$   &   &       &       &       &       &       &       &       &       & 1  \\[0.1in] 
\hline\hline
\end{tabular}
\end{table}

\begin{table}[h]
\caption{Upper panel: $B_{B_q}^{(i)}(\mu)$ in the $\MSbar$-NDR scheme evaluated at the scale $\mu = \overline m_b$ with evanescent operator scheme specified by BMU or BBGLN.  Errors shown are from the matrix elements in Table~\ref{tbl:ME_results} and from the decay constants, respectively.  
Lower panel: ratios of bag parameters $B_{B_q}^{(i)} (\overline m_b) /B_{B_q}^{(1)} (\overline m_b)$ ($i$=2--5).  Errors are from the matrix elements in Table~\ref{tbl:ME_results}, and include correlations between the chiral-continuum fit and $r_1/a$ errors on the bag parameters in the numerator and denominator.
The remaining subleading errors added after the chiral-continuum fit are treated as uncorrelated between the bag parameters.
In both panels, the error on the bag parameters due to the omission of the charm sea is considered to be negligible.
\label{tbl:Bhat_results}}
\vspace{0.1in}
\begin{tabular}{c@{\quad}cc@{\quad}cc}\hline\hline
\hspace{0.5in}                              & \multicolumn{2}{c}{$B_d$--$\bar B_d$}    & \multicolumn{2}{c}{$B_s$--$\bar B_s$} \\
                                        						& BMU               	& BBGLN         & BMU               & BBGLN \\ \hline
\vspace{-0.12in}\\
$ B_{B_q}^{(1)} (\overline m_b)$ \vspace{0.07in}    	& \multicolumn{2}{c}{0.913(76)(40)}	& \multicolumn{2}{c}{0.952(58)(32)}  \\ 
$ B_{B_q}^{(2)} (\overline m_b)$ \vspace{0.07in}    	& 0.761(68)(33)		& 0.808(72)(35)	& 0.806(52)(27)& 0.859(55)(29)     \\
$ B_{B_q}^{(3)} (\overline m_b)$ \vspace{0.07in}    	& 1.07(21)(5)           	& 1.07(21)(5)   	& 1.10(15)(4) 	& 1.09(15)(4)       \\
$ B_{B_q}^{(4)} (\overline m_b)$ \vspace{0.07in}    	& \multicolumn{2}{c}{1.040(75)(45)}	& \multicolumn{2}{c}{1.022(57)(34)}     \\
$ B_{B_q}^{(5)} (\overline m_b)$ \vspace{0.07in}    	& \multicolumn{2}{c}{0.964(93)(42)}	& \multicolumn{2}{c}{0.943(68)(31)}     \\ 
\hline \vspace{-0.12in}\\
$ B_{B_q}^{(2)} / B_{B_q}^{(1)} $               		& 0.838(81)        	& 0.885(73)    	& 0.849(56)      	& 0.902(59)  \vspace{0.07in}\\
$ B_{B_q}^{(3)} / B_{B_q}^{(1)} $               		& 1.18(24)              	& 1.17(24)     	& 1.16(16)      	& 1.15(16)   \vspace{0.07in}\\
$ B_{B_q}^{(4)} / B_{B_q}^{(1)} $               		& \multicolumn{2}{c}{1.14(10)}        	& \multicolumn{2}{c}{1.073(68)} \vspace{0.07in}\\
$ B_{B_q}^{(5)} / B_{B_q}^{(1)} $\vspace{0.05in}    	& \multicolumn{2}{c}{1.06(11)}       	& \multicolumn{2}{c}{0.990(75)} \\ 
\hline\hline
\end{tabular}
\end{table}

\begin{table}[h]
\caption{Bag parameters that enter the expression for the width difference $\Delta \Gamma_q$ at $\order{(1/m_b)}$ in Ref.~\cite{Lenz:2006hd} (upper panel), and their ratios with respect to $B_{B_q}^{(1)}$  (lower panel).  
The bag parameters are obtained in the $\MSbar$-NDR-BBGLN scheme evaluated at the scale $\mu = \overline m_b$ using Eq.~(28) of Ref.~\cite{Lenz:2006hd}. 
(Note that $B_{B_q}^{(R_1)}$ and $B_{B_q}^{(\tilde{R}_1)}$ are proportional to our $B_{B_q}^{(4)}$ and $B_{B_q}^{(5)}$, respectively.)
The errors in the upper panel are from the matrix elements in Table~\ref{tbl:ME_results} and the decay constants, respectively.  
The errors on the ratios in the lower panel include correlations between the chiral-continuum fit and $r_1/a$ errors in the numerator and denominator, while the subleading errors added after the chiral-continuum fit are treated as uncorrelated.
In both panels, the errors from parametric inputs and the omission of the charm sea are negligible.
\label{tbl:BR0_results}}
\vspace{0.2in}
\begin{tabular}{ccc@{\quad}cc}\hline\hline
\hspace{0.5in} 
& & \vspace{-0.17in}\\  
                                    	& \multicolumn{2}{c}{$B_d$--$\bar B_d$}    & \multicolumn{2}{c}{$B_s$--$\bar B_s$}\vspace{0.05in}\\ \hline \vspace{-0.12in}\\
$B_{B_q}^{(R_0^q)} $                    		& \multicolumn{2}{c}{0.32(73)(1)}       & \multicolumn{2}{c}{0.52(52)(2)}       \vspace{0.07in}\\ 
$B_{B_q}^{(R_1^q)}$                 		& \multicolumn{2}{c}{1.57(11)(7)}       & \multicolumn{2}{c}{1.536(84)(51)}     \vspace{0.07in}\\ 
$B_{B_q}^{(\tilde R_1^q)} $             		& \multicolumn{2}{c}{1.19(11)(5)}	& \multicolumn{2}{c}{1.165(84)(39)} \vspace{0.07in}\\ \hline
\vspace{-0.12in}\\
$B_{B_q}^{(R_0^q)} / B_{B_q}^{(1)} $      	& \multicolumn{2}{c}{0.35(80)} 		& \multicolumn{2}{c}{0.54(55)} \vspace{0.07in}\\ 
$B_{B_q}^{(R_1^q)} / B_{B_q}^{(1)} $      	& \multicolumn{2}{c}{1.72(15)} 		& \multicolumn{2}{c}{1.61(10)}   \vspace{0.07in}\\ 
$B_{B_q}^{(\tilde R_1^q)} /B_{B_q}^{(1)}$	& \multicolumn{2}{c}{1.31(14)}		& \multicolumn{2}{c}{1.223(93)}    \vspace{0.07in}\\ 
\hline\hline
\end{tabular}
\end{table}

\begin{sidewaystable}[h]
\caption{Correlations between the bag-parameters presented in Tables~\ref{tbl:Bhat_results}
and~~\ref{tbl:BR0_results}.
The entries of the correlation matrix are symmetric across the diagonal.
The correlations between the BMU-scheme results for $B_{B_q}^{(2,3)}$ and the other $B$-parameters differ by
$<1\%$.
We note that, although $R_0$ is a linear combination of matrix elements $\me{1,2,3}$, the coefficients in
Eq.~(\ref{eq:R0def}) are given only in the BBGLN scheme~\cite{Lenz:2006hd}.}
\label{tbl:B_cor}
\vspace{0.1in}
\begin{tabular}{c|cccccccccccccccc}
\hline\hline
& & & & & & & & & & \vspace{-0.15in} \\
\hspace{0.0in}  & $B_{B_d}^{(1)}$ & $B_{B_d}^{(2)}$ & $B_{B_d}^{(3)}$ & $B_{B_d}^{(4)}$ & $B_{B_d}^{(5)}$ & $B_{B_d}^{(R_0)}$ & $B_{B_d}^{(R_1)}$ & $B_{B_d}^{(\tilde R_1)}$ & $B_{B_s}^{(1)}$ & $B_{B_s}^{(2)}$ & $B_{B_s}^{(3)}$ & $B_{B_s}^{(4)}$ & $B_{B_s}^{(5)}$ & $B_{B_s}^{(R_0)}$ & $B_{B_s}^{(R_1)}$ & $B_{B_s}^{(\tilde R_1)}$\\
& & & & & & & & & & & & & & & & \vspace{-0.15in} \\ 
\hline
& & & & & & & & & & & & & & & & \vspace{-0.15in} \\ 
$ B_{B_d}^{(1)} $       & 1     & 0.504 & 0.162     & 0.494 & 0.422     & -0.231    & 0.500 & 0.423     & 0.754 & 0.311 & 0.077 & 0.264 & 0.246 & -0.235    & 0.269 & 0.247  \\
& & & & & & & & & & & & & & & & \vspace{-0.15in} \\ 
$ B_{B_d}^{(2)} $       &   & 1     &  0.282    & 0.494 & 0.389     & 0.528     & 0.477 & 0.380     & 0.311 & 0.766 & 0.201 & 0.283 & 0.226 & 0.509 & 0.259 & 0.214  \\
& & & & & & & & & & & & & & & & \vspace{-0.15in} \\ 
$ B_{B_d}^{(3)} $       &   &       & 1         & 0.225     & 0.148 & -0.456    & 0.216     & 0.143 & 0.068     & 0.179 & 0.929 & 0.115 & 0.063 & -0.454    & 0.103 & 0.057  \\
& & & & & & & & & & & & & & & & \vspace{-0.15in} \\ 
$ B_{B_d}^{(4)} $       &   &       &       & 1         & 0.528     & 0.036     & 0.988     & 0.518     & 0.279     & 0.297 & 0.134 & 0.705 & 0.332 & 0.030 & 0.688 & 0.319 \\
& & & & & & & & & & & & & & & & \vspace{-0.15in} \\ 
$ B_{B_d}^{(5)} $       &   &       &       &       & 1         & 0.032     & 0.523     & 0.998 & 0.247     & 0.228 & 0.074 & 0.318 & 0.797 & 0.026 & 0.310 & 0.794  \\
& & & & & & & & & & & & & & & & \vspace{-0.15in} \\ 
$ B_{B_d}^{(R_0)} $     &   &       &       &       &           & 1         & 0.036 & 0.031     & -0.212    & 0.469 & -0.443    & 0.031 & 0.027 & 0.982 & 0.031 & 0.027 \\
& & & & & & & & & & & & & & & & \vspace{-0.15in} \\ 
$ B_{B_d}^{(R_1)} $     &   &       &       &       &           &           & 1         & 0.524 & 0.283     & 0.270 & 0.120 & 0.683 & 0.322 & 0.030 & 0.696 & 0.323  \\
& & & & & & & & & & & & & & & & \vspace{-0.15in} \\ 
$ B_{B_d}^{(\tilde{R}_1)}$  &   &       &       &       &           &           &       & 1     & 0.248     & 0.214 & 0.067 & 0.305 & 0.793 & 0.026 & 0.311 & 0.796  \\
& & & & & & & & & & & & & & & & \vspace{-0.15in} \\ 
$ B_{B_s}^{(1)}  $      &   &       &       &       &       &       &       &       & 1     & 0.575 & 0.215 & 0.560 & 0.486 & -0.196    & 0.571 & 0.488  \\
& & & & & & & & & & & & & & & & \vspace{-0.15in} \\ 
$ B_{B_s}^{(2)}  $      &   &       &       &       &       &       &       &       &       & 1     & 0.329 & 0.581 & 0.466 & 0.498 & 0.553 & 0.450  \\
& & & & & & & & & & & & & & & & \vspace{-0.15in} \\ 
$ B_{B_s}^{(3)}  $      &   &       &       &       &       &       &       &       &       &       & 1     & 0.278 & 0.195 & -0.437    & 0.263 & 0.186  \\
& & & & & & & & & & & & & & & & \vspace{-0.15in} \\ 
$ B_{B_s}^{(4)}  $      &   &       &       &       &       &       &       &       &       &       &       & 1     & 0.581 & 0.057 & 0.981 & 0.565  \\
& & & & & & & & & & & & & & & & \vspace{-0.15in} \\ 
$ B_{B_s}^{(5)}  $      &   &       &       &       &       &       &       &       &       &       &       &       & 1     & 0.048 & 0.574 & 0.996   \\ 
& & & & & & & & & & & & & & & & \vspace{-0.15in} \\ 
$ B_{B_s}^{(R_0)} $     &   &       &       &       &           &           &       &       &       &       &       &       &       & 1     & 0.058 & 0.048  \\
& & & & & & & & & & & & & & & & \vspace{-0.15in} \\ 
$ B_{B_s}^{(R_1)} $     &   &       &       &       &           &           &       &       &       &       &       &       &       &       & 1     & 0.577  \\
& & & & & & & & & & & & & & & & \vspace{-0.15in} \\ 
$ B_{B_s}^{(\tilde{R}_1)}$  &   &       &       &       &           &           &       &       &       &       &       &       &       &       &       & 1  \\[0.1in] 
\hline\hline
\end{tabular}
\end{sidewaystable}

\clearpage	 
\section{Evaluating quality of fits}
    \label{app:pval}
    In this Appendix, we discuss the measures of goodness of fit used to guide and scrutinize the analysis.
All fits presented in this work take the correlations between the data into account.
As discussed below, it is (conceptually) necessary to constrain the fit parameters, and we do so with
Gaussian priors.

We evaluate the quality of our fit results with two test statistics denoted $Q$ and $p$, both of which are
defined via the (complementary) cumulative $\chi^2$ distribution with $\nu$ degrees of
freedom~\cite{Agashe:2014kda}:
\begin{equation}
    F_{\chi^2}(\chi^2_\text{min},\nu) = \int_{\chi^2_\text{min}}^\infty
        \frac{x^{2(\nu/2-1)} e^{-x^2/2}}{2^{\nu/2} \Gamma(\nu/2)} dx^2,
    \label{eq:pvalue}
\end{equation}
which gives the probability that a random $\nu$-component vector $\bm{x}$ satisfies
$x^2\ge\chi^2_\text{min}$.
The statistics $Q$ and $p$ differ in the choices made for $\chi^2_\text{min}$ and $\nu$ in
this formula.

With a \emph{finite} set of parameters in a theoretical model, it is customary to form the function
\begin{equation}
    \chi^2(P) = \sum_{\alpha,\beta=1}^{N_D}
        [f_\alpha(P)-D_\alpha] \left( \sigma^{2} \right)^{-1}_{\alpha\beta} [f_\beta(P)-D_\beta],
    \label{eq:chi2}
\end{equation}
where $P$ is the parameter vector, $f$ denotes the fit model, $D$ the data, and $\sigma^2$ is the covariance
matrix between the data.
One then finds the parameter vector $P^*$ that minimizes $\chi^2(P)$.
A good model, fit to many similar data sets, should yield for
$F_{\chi^2}(\chi^2(P^*),N_D-N_P)$ a uniform distribution over $[0,1]$, where $N_D$ and $N_P$
are the numbers of data points and fitted parameters, respectively.

In the cases at hand---correlator fits and the chiral-continuum extrapolation---there are, in principle,
\emph{infinitely many} parameters.
Because only the first few terms in the tower of states in Eqs.~(\ref{eq:2ptfitfn}) and~(\ref{eq:3ptfitfn})
or in the $\chi$PT expansion in Sec.~\ref{sec:ChPT} can be determined by the data, one must in practice
constrain all but a few parameters.
Note that truncating the tower of states or the order in $\chi$PT is the same as constraining all omitted
terms to vanish.
Fits are more stable over a wider range of data when one includes several higher-order parameters that are
constrained moderately with priors instead of sharply truncating the series.

We thus introduce an ``augmented'' $\chi^2$~function~\cite{Lepage:2001ym}:
\begin{equation}
    \chi^2_\text{aug}(P) = \chi^2(P) +
        \sum_{m,n=1}^{N_P} (P_m-\tilde{P}_m) \left( \tilde{\sigma}^{2} \right)^{-1}_{mn} (P_n-\tilde{P}_n) ,
    \label{eq:augchi2}
\end{equation}
imposing prior distributions for $N_P$ parameters with prior central values $\tilde{P}$ and covariance
matrix $\tilde{\sigma}^2$.
In practice, we take the priors to be uncorrelated, \emph{i.e.}, $\tilde{\sigma}^2$ is diagonal.
Further, we impose a prior for every parameter actively fit, including those well-constrained by the data.
Our best-fit parameter vector $\hat{P}$ is the one that minimizes $\chi^2_\text{aug}(P)$.

Each prior is an additional piece of information,  so we examine
\begin{equation}
    Q \equiv F_{\chi^2}\left(\chi^2_\text{aug}(\hat{P}),N_D\right)     
    \label{eq:Qval}
\end{equation}
to rank the quality of the fits.
When carrying out many similar fits, the resulting $Q$~values need not follow a uniform distribution,
because $\chi^2_\text{aug}(\hat{P})$ need not follow the $\chi^2$ distribution for $N_D$ degrees of freedom.
A uniform distribution for $Q$ arises only when the priors have been chosen such that the extra terms in
Eq.~(\ref{eq:augchi2}) make a contribution $\chi^2_\text{aug}(\hat{P})-\chi^2(\hat{P})\approx N_P$,
\emph{i.e.}, when the prior widths are narrow.

Our priors are not chosen to yield such an outcome.
Instead, the priors for the essential fit parameters are effectively unconstraining, and the others are
chosen to promote fit stability.
Nevertheless, we find $Q$ to be useful when ranking fits with different $N_P$.
To test the influence of the priors on the best fit, we introduce a $p$~value via
\begin{equation}
    p \equiv F_{\chi^2}\left(\chi^2(\hat{P}), N_D-N_P\right), 
    \label{eq:unaugpval}
\end{equation}
omitting from $\chi^2_\text{aug}$ the terms corresponding to the priors, but still using the best-fit parameter
vector $\hat{P}$ that minimizes $\chi^2_\text{aug}$.
If the priors do not influence the fits in an undesirable way, then $\chi^2(\hat{P})$ should be close enough
to $\chi^2(P^*)$ that the $p$~value in Eq.~(\ref{eq:unaugpval}) is uniformly distributed between 0~and~1.
Figure~\ref{fig:pvalues} shows this to be the case for the correlator fits.

\section{Chiral logarithms}
    \label{app:log}
    For reasons discussed in Sec.~\ref{sec:chptLogs}, we use a notation for the loop-diagram functions that is 
slightly different than in the original paper~\cite{Bernard:2013dfa}.
Here we provide a dictionary to translate between this work and Ref.~\cite{Bernard:2013dfa}.

In this paper, the light valence quark ($d$ or $s$) is labeled with $q$ but in Ref.~\cite{Bernard:2013dfa}
with $x$.
Here, the index $i$ denotes the five operators, but in Ref.~\cite{Bernard:2013dfa} $n$ is used.
Thus, when a label on the functions corresponds to these properties, the notation will differ in a trivial 
way.

We follow the notation of Ref.~\cite{Bernard:2013dfa} for correct-spin contributions.
The self-energy function $\mathcal{W}_{q\bar{b}}=\mathcal{W}_{b\bar{q}}$ is given explicitly in Eq.~(62) of 
Ref.~\cite{Bernard:2013dfa}. 
The correct-spin tadpole functions depend on whether $i\in\{1,2,3\}$ or $\{4,5\}$;
$\mathcal{T}_q^{(1,2,3)}$ and $\mathcal{T}_q^{(4,5)}$ are given in 
Eqs.~(82) and~(83), respectively, of Ref.~\cite{Bernard:2013dfa}.
The sunset function is the same for all $i$ (or $n$); $\mathcal{Q}_q^{(i)}$ is given in Eq.~(89) of 
Ref.~\cite{Bernard:2013dfa}.

For the wrong-spin contributions, we find it more transparent to separate the LECs $\beta_i^{(\prime)}$ from
the loop-diagram functions.
The relation between our notation (left-hand side) and that of Ref.~\cite{Bernard:2013dfa} (right-hand side)
is as follows:
\begin{subequations}\label{eq:Ttilde}
\begin{align}
      \beta_1\tilde{T}_q^\text{(1a)} + (\beta_2+\beta_3)\tilde{T}_q^\text{(1b)} &=
        \beta_1 \tilde{\mathcal{T}}_q^{(1)}, \\ 
    \beta_2\tilde{T}_q^\text{(23a)} + \beta_1\tilde{T}_q^\text{(23b)} + \beta_3\tilde{T}_q^\text{(23c)} &=
        \beta_2 \tilde{\mathcal{T}}_q^{(2)}, \\ 
     \beta_3\tilde{T}_q^\text{(23a)} + \beta_1\tilde{T}_q^\text{(23b)} + \beta_2\tilde{T}_q^\text{(23c)} &=
        \beta_3 \tilde{\mathcal{T}}_q^{(3)}, \\ 
    \beta_4\tilde{T}_q^\text{(45a)} + \beta_5\tilde{T}_q^\text{(45b)}  &=
        \beta_4 \tilde{\mathcal{T}}_q^{(4)}, \\ 
    \beta_5\tilde{T}_q^\text{(45a)} + \beta_4\tilde{T}_q^\text{(45b)}  &=
        \beta_5 \tilde{\mathcal{T}}_q^{(5)},    
\end{align}
\end{subequations}
and
\begin{subequations}\label{eq:Qtilde}
\begin{align}
     \beta_1 \tilde{Q}_q^\text{(1a)} + (\beta'_2+\beta'_3)\tilde{Q}_q^\text{(1b)}  &=
        \beta_1  \tilde{\mathcal{Q}}_q^{(1)}, \\ 
     \beta'_2\tilde{Q}_q^\text{(23a)} + \beta_1 \tilde{Q}_q^\text{(23b)} + \beta'_3\tilde{Q}_q^\text{(23c)}&=
        \beta'_2 \tilde{\mathcal{Q}}_q^{(2)}, \\  
     \beta'_3\tilde{Q}_q^\text{(23a)} + \beta_1 \tilde{Q}_q^\text{(23b)} + \beta'_2\tilde{Q}_q^\text{(23c)}&=
        \beta'_3 \tilde{\mathcal{Q}}_q^{(3)}, \\ 
     \beta'_4\tilde{Q}_q^\text{(45a)} + \beta'_5\tilde{Q}_q^\text{(45b)}  &=
        \beta'_4 \tilde{\mathcal{Q}}_q^{(4)}, \\ 
    \beta'_5\tilde{Q}_q^\text{(45a)} + \beta'_4\tilde{Q}_q^\text{(45b)}  &=
        \beta'_5 \tilde{\mathcal{Q}}_q^{(5)} . 
\end{align}
\end{subequations}
Equations~(\ref{eq:Ttilde}) can be obtained from Eqs.~(84)--(88) of Ref.~\cite{Bernard:2013dfa}, and
Eqs.~(\ref{eq:Qtilde}) from Eqs.~(90)--(94) of that paper.

\bibliography{BBbar,asqtad_ensemb,chipt}

\end{document}